\documentclass{pasj00}
\usepackage{bm}

\def\commenta{$^*$}
\def\commentb{$^\dagger$}
\def\commentc{$^\ddagger$}
\def\commentd{$^\S$}

\def\submitted{submitted}
\def\inpress{in press}

\def\arxiv#1{ (arXiv astro-ph/#1)}

\DeclareAbbreviation\AAHam{Astron. Abh. Hamburg. Sternw.}
\DeclareAbbreviation\AARv{Astron. Astrophys. Rev.}
\DeclareAbbreviation\AAS{American Astron. Soc. Meeting Abstracts}
\DeclareAbbreviation\AcA{Acta Astron.}
\DeclareAbbreviation\actaa{Acta Astron.}
\DeclareAbbreviation\Afz{Astrofizika}
\DeclareAbbreviation\AGAb{Astronomische Gesellschaft Abstract Ser.}
\DeclareAbbreviation\an{Astron. Nachr.}
\DeclareAbbreviation\AnAp{Annales d'Astrophysique}
\DeclareAbbreviation\AnTok{Tokyo Astron. Obs. Annals, Sec. Ser.}
\DeclareAbbreviation\Ap{Astrophysics}
\DeclareAbbreviation\ARep{Astron. Rep.}
\DeclareAbbreviation\ATel{Astron. Telegram}
\DeclareAbbreviation\ATsir{Astron. Tsirk.}
\DeclareAbbreviation\AcApS{Acta Astrophys. Sinica}
\DeclareAbbreviation\AstL{Astron. Lett.}
\DeclareAbbreviation\BaltA{Baltic Astron.}
\DeclareAbbreviation\BANS{Bull. of the Astron. Institutes of the Netherlands Suppl. Ser.}
\DeclareAbbreviation\BASI{Bull. Astron. Soc. India}
\DeclareAbbreviation\BeSN{Be Newslett.}
\DeclareAbbreviation\BHarO{Harvard Coll. Obs. Bull.}
\DeclareAbbreviation\CBET{Cent. Bur. Electron. Telegrams}
\DeclareAbbreviation\ChJAA{Chinese J. of Astron. and Astrophys.}
\DeclareAbbreviation\caa{Chinese J. of Astron. and Astrophys.}
\DeclareAbbreviation\CoAsi{Asiago Contr.}
\DeclareAbbreviation\CoSka{Contributions of the Astronomical Observatory Skalnat\'e Pleso}
\DeclareAbbreviation\GCN{GRB Coord. Netw. Circ.}
\DeclareAbbreviation\ibvs{IBVS}
\DeclareAbbreviation\IEEEP{IEEE Proc.}
\DeclareAbbreviation\JAD{J. Astron. Data}
\DeclareAbbreviation\JAVSO{J. American Assoc. Variable Star Obs.}
\DeclareAbbreviation\JBAA{J. Br. Astron. Assoc.}
\DeclareAbbreviation\JPhCS{J. of Physics Conference Series}
\DeclareAbbreviation\JPSJ{J. Phys. Soc. Japan}
\DeclareAbbreviation\JSARA{J. of the Southeastern Assoc. for Research in Astron.}
\DeclareAbbreviation\LowOB{Lowell Obs. Bull.}
\DeclareAbbreviation\MitAG{Mitteil. der Astronom. Gesell. Hamburg}
\DeclareAbbreviation\MitVS{Mitteil. Ver\"{a}nderl. Sterne}
\DeclareAbbreviation\MmSAI{Mem. Soc. Astron. Ital.}
\DeclareAbbreviation\Msngr{Messenger}
\DeclareAbbreviation\NewA{New Astron.}
\DeclareAbbreviation\na{New Astron.}
\DeclareAbbreviation\NewAR{New Astron. Rev.}
\DeclareAbbreviation\NInfo{Nauchnye Informatsii}
\DeclareAbbreviation\OAP{Odessa Astron. Publ.}
\DeclareAbbreviation\Obs{Observatory}
\DeclareAbbreviation\OEJV{Open Eur. J. on Variable Stars}
\DeclareAbbreviation\PASA{Publ. Astron. Soc. Australia}
\DeclareAbbreviation\PASAu{Publ. Astron. Soc. Australia}
\DeclareAbbreviation\PCCP{Phys. Chem. Chem. Phys.}
\DeclareAbbreviation\PAZh{Pis'ma AZh}
\DeclareAbbreviation\PhR{Phys. Rep.}
\DeclareAbbreviation\PVSS{Publ. Variable Stars Sect. R. Astron. Soc. New Zealand}
\DeclareAbbreviation\PZ{Perem. Zvezdy}
\DeclareAbbreviation\PZP{Perem. Zvezdy, Prilozh.}
\DeclareAbbreviation\QJRAS{QJRAS}
\DeclareAbbreviation\RA{Ricerche Astronomiche}
\DeclareAbbreviation\RMxAA{Rev. Mexicana Astron. Astrof.}
\DeclareAbbreviation\RvMA{Reviews of Modern Astron.}
\DeclareAbbreviation\SASS{Society for Astronom. Sciences Ann. Symp.}
\DeclareAbbreviation\Sci{Science}
\DeclareAbbreviation\SPIE{SPIE Proc.}
\DeclareAbbreviation\SvA{Soviet Astronomy}
\DeclareAbbreviation\SvAL{Soviet Astronomy Letters}
\DeclareAbbreviation\VeSon{Ver\"{o}ff. Sternw. Sonneberg}
\DeclareAbbreviation\VSOLJBul{VSOLJ Variable Star Bull.}
\DeclareAbbreviation\yCat{VizieR Online Data Catalog}
\DeclareAbbreviation\ZA{Z. Astrophys.}

\def\ASPConf#1#2{ASP Conf. Ser. #1, #2}

\def\PublisherCambridge{Cambridge: Cambridge University Press}

\def\PublisherASP{San Francisco: ASP}
\def\PublisherAIP{Maryland: AIP}
\def\PublisherReidel{Dordrecht: D. Reidel Publishing Company}

\newcounter{author}
\def\authorcount#1#2{\refstepcounter{author}\label{#1}
                     \altaffiltext{\ref{#1}}{#2}}

\begin{document}
\SetRunningHead{T. Kato et al.}{Period Variations in SU UMa-Type Dwarf Novae III}

\Received{201X/XX/XX}
\Accepted{201X/XX/XX}

\title{Survey of Period Variations of Superhumps in SU UMa-Type Dwarf Novae.
    III: The Third Year (2010--2011)}

\author{Taichi~\textsc{Kato},\altaffilmark{\ref{affil:Kyoto}*}
        Hiroyuki~\textsc{Maehara},\altaffilmark{\ref{affil:HidaKwasan}}
        Ian~\textsc{Miller},\altaffilmark{\ref{affil:Miller}}
        Tomohito~\textsc{Ohshima},\altaffilmark{\ref{affil:Kyoto}}
        Enrique~de~\textsc{Miguel},\altaffilmark{\ref{affil:Miguel}}$^,$\altaffilmark{\ref{affil:Miguel2}}
        Kenji~\textsc{Tanabe},\altaffilmark{\ref{affil:OUS}}
        Kazuyoshi~\textsc{Imamura},\altaffilmark{\ref{affil:OUS}}
        Hidehiko~\textsc{Akazawa},\altaffilmark{\ref{affil:OUS}}
        Nanae~\textsc{Kunitomi},\altaffilmark{\ref{affil:OUS}}
        Ryosuke~\textsc{Takagi},\altaffilmark{\ref{affil:OUS}}
        Mikiha~\textsc{Nose},\altaffilmark{\ref{affil:OUS}}
        Franz-Josef~\textsc{Hambsch},\altaffilmark{\ref{affil:GEOS}}$^,$\altaffilmark{\ref{affil:BAV}}$^,$\altaffilmark{\ref{affil:Hambsch}}
        Seiichiro~\textsc{Kiyota},\altaffilmark{\ref{affil:Kis}}
        Elena~P.~\textsc{Pavlenko},\altaffilmark{\ref{affil:CrAO}}
        Aleksei~V.~\textsc{Baklanov},\altaffilmark{\ref{affil:CrAO}}
        Oksana~I.~\textsc{Antonyuk},\altaffilmark{\ref{affil:CrAO}}
        Denis~\textsc{Samsonov},\altaffilmark{\ref{affil:CrAO}}
        Aleksei~\textsc{Sosnovskij},\altaffilmark{\ref{affil:CrAO}}
        Kirill~\textsc{Antonyuk},\altaffilmark{\ref{affil:CrAO}}
        Maksim~V.~\textsc{Andreev},\altaffilmark{\ref{affil:Terskol}}$^,$\altaffilmark{\ref{affil:ICUkraine}}
        Etienne~\textsc{Morelle},\altaffilmark{\ref{affil:Morelle}}
        Pavol~A.~\textsc{Dubovsky},\altaffilmark{\ref{affil:Dubovsky}}
        Igor~\textsc{Kudzej},\altaffilmark{\ref{affil:Dubovsky}}
        Arto~\textsc{Oksanen},\altaffilmark{\ref{affil:Nyrola}}
        Gianluca~\textsc{Masi},\altaffilmark{\ref{affil:Masi}} 
        Thomas~\textsc{Krajci},\altaffilmark{\ref{affil:Krajci}}
        Roger~D.~\textsc{Pickard},\altaffilmark{\ref{affil:BAAVSS}}$^,$\altaffilmark{\ref{affil:Pickard}}
        Richard~\textsc{Sabo},\altaffilmark{\ref{affil:Sabo}}
        Hiroshi~\textsc{Itoh},\altaffilmark{\ref{affil:Ioh}}
        William~\textsc{Stein},\altaffilmark{\ref{affil:Stein}}
        Shawn~\textsc{Dvorak},\altaffilmark{\ref{affil:Dvorak}}
        Arne~\textsc{Henden},\altaffilmark{\ref{affil:AAVSO}}
        Shinichi~\textsc{Nakagawa},\altaffilmark{\ref{affil:OKU}}
        Ryo~\textsc{Noguchi},\altaffilmark{\ref{affil:OKU}}
        Eriko~\textsc{Iino},\altaffilmark{\ref{affil:OKU}}
        Katsura~\textsc{Matsumoto},\altaffilmark{\ref{affil:OKU}}
        Hiroki~\textsc{Nishitani},\altaffilmark{\ref{affil:OKU}}
        Tomoya~\textsc{Aoki},\altaffilmark{\ref{affil:OKU}}
        Hiroshi~\textsc{Kobayashi},\altaffilmark{\ref{affil:OKU}}
        Chihiro~\textsc{Akasaka},\altaffilmark{\ref{affil:OKU}}
        Greg~\textsc{Bolt},\altaffilmark{\ref{affil:Bolt}}
        Jeremy~\textsc{Shears},\altaffilmark{\ref{affil:Shears}}
        Javier~\textsc{Ruiz},\altaffilmark{\ref{affil:Ruiz1}}$^,$\altaffilmark{\ref{affil:Ruiz2}}
        Sergey~Yu.~\textsc{Shugarov},\altaffilmark{\ref{affil:Sternberg}}$^,$\altaffilmark{\ref{affil:Slovak}}
        Drahomir~\textsc{Chochol},\altaffilmark{\ref{affil:Slovak}}
        Nikolai~A.~\textsc{Parakhin},\altaffilmark{\ref{affil:ICUkraine}}
        Berto~\textsc{Monard},\altaffilmark{\ref{affil:Monard}}
        Kazuhiko~\textsc{Shiokawa},\altaffilmark{\ref{affil:Siz}}
        Kiyoshi~\textsc{Kasai},\altaffilmark{\ref{affil:Kai}}
        Bart~\textsc{Staels},\altaffilmark{\ref{affil:AAVSO}}$^,$\altaffilmark{\ref{affil:Staels}}
        Atsushi~\textsc{Miyashita},\altaffilmark{\ref{affil:Seikei}}
        Donn~R.~\textsc{Starkey},\altaffilmark{\ref{affil:Starkey}}
        Yenal~\textsc{\"Ogmen},\altaffilmark{\ref{affil:Ogmen}}
        Colin~\textsc{Littlefield},\altaffilmark{\ref{affil:LCO}}
        Natalia~\textsc{Katysheva},\altaffilmark{\ref{affil:Sternberg}}
        Ivan~M.~\textsc{Sergey},\altaffilmark{\ref{affil:Sergey}}
        Denis~\textsc{Denisenko},\altaffilmark{\ref{affil:Denisenko}}
        Tamas~\textsc{Tordai},\altaffilmark{\ref{affil:Polaris}}
        Robert~\textsc{Fidrich},\altaffilmark{\ref{affil:Polaris}}
        Vitaly~P.~\textsc{Goranskij},\altaffilmark{\ref{affil:Sternberg}}
        Jani~\textsc{Virtanen},\altaffilmark{\ref{affil:Virtanen}}
        Tim~\textsc{Crawford},\altaffilmark{\ref{affil:Crawford}}
        Jochen~\textsc{Pietz},\altaffilmark{\ref{affil:Pietz}}
        Robert~A.~\textsc{Koff},\altaffilmark{\ref{affil:Koff}} 
        David~\textsc{Boyd},\altaffilmark{\ref{affil:DavidBoyd}} 
        Steve~\textsc{Brady},\altaffilmark{\ref{affil:Brady}}
        Nick~\textsc{James},\altaffilmark{\ref{affil:James}}
        William~N.~\textsc{Goff},\altaffilmark{\ref{affil:Goff}}
        Koh-ichi~\textsc{Itagaki},\altaffilmark{\ref{affil:Itagaki}}
        Hideo~\textsc{Nishimura},\altaffilmark{\ref{affil:Nmh}}
        Youichirou~\textsc{Nakashima},\altaffilmark{\ref{affil:Nakashima}}
        Seiichi~\textsc{Yoshida},\altaffilmark{\ref{affil:MISAO}}
        Rod~\textsc{Stubbings},\altaffilmark{\ref{affil:Stubbings}}
        Gary~\textsc{Poyner},\altaffilmark{\ref{affil:Poyner}}
        Yutaka~\textsc{Maeda},\altaffilmark{\ref{affil:Mdy}}
        Stanislav~A.~\textsc{Korotkiy},\altaffilmark{\ref{affil:KaDar}}
        Kirill~V.~\textsc{Sokolovsky},\altaffilmark{\ref{affil:Sokolovsky}}$^,$\altaffilmark{\ref{affil:Sokolovsky2}}
        Seiji~\textsc{Ueda},\altaffilmark{\ref{affil:SeijiUeda}}
}

\authorcount{affil:Kyoto}{
     Department of Astronomy, Kyoto University, Kyoto 606-8502}
\email{$^*$tkato@kusastro.kyoto-u.ac.jp}

\authorcount{affil:HidaKwasan}{
     Kwasan and Hida Observatories, Kyoto University, Yamashina,
     Kyoto 607-8471}

\authorcount{affil:Miller}{
     Furzehill House, Ilston, Swansea, SA2 7LE, UK}

\authorcount{affil:Miguel}{
     Departamento de F\'isica Aplicada, Facultad de Ciencias
     Experimentales, Universidad de Huelva,
     21071 Huelva, Spain}

\authorcount{affil:Miguel2}{
     Center for Backyard Astrophysics, Observatorio del CIECEM,
     Parque Dunar, Matalasca\~nas, 21760 Almonte, Huelva, Spain}

\authorcount{affil:OUS}{
     Department of Biosphere-Geosphere System Science, Faculty of Informatics,
     Okayama University of Science, 1-1 Ridai-cho, Okayama, Okayama 700-0005}

\authorcount{affil:GEOS}{
     Groupe Europ\'een d'Observations Stellaires (GEOS),
     23 Parc de Levesville, 28300 Bailleau l'Ev\^eque, France}

\authorcount{affil:BAV}{
     Bundesdeutsche Arbeitsgemeinschaft f\"ur Ver\"anderliche Sterne
     (BAV), Munsterdamm 90, 12169 Berlin, Germany}

\authorcount{affil:Hambsch}{
     Vereniging Voor Sterrenkunde (VVS), Oude Bleken 12, 2400 Mol, Belgium}

\authorcount{affil:Kis}{
     Variable Star Observers League in Japan (VSOLJ), 405-1003 Matsushiro,
     Tsukuba, Ibaraki 305-0035}

\authorcount{affil:CrAO}{
     Crimean Astrophysical Observatory, 98409, Nauchny, Crimea, Ukraine}

\authorcount{affil:Terskol}{
     Institute of Astronomy, Russian Academy of Sciences, 361605 Peak Terskol,
     Kabardino-Balkaria, Russia}

\authorcount{affil:ICUkraine}{
     International Center for Astronomical, Medical and Ecological Research
     of NASU, Ukraine 27 Akademika Zabolotnoho Str. 03680 Kyiv,
     Ukraine}

\authorcount{affil:Morelle}{
     9 rue Vasco de GAMA, 59553 Lauwin Planque, France}

\authorcount{affil:Dubovsky}{
     Vihorlat Observatory, Mierova 4, Humenne, Slovakia}

\authorcount{affil:Nyrola}{
     Nyrola observatory, Jyvaskylan Sirius ry, Vertaalantie
     419, FI-40270 Palokka, Finland}

\authorcount{affil:Masi}{
     The Virtual Telescope Project, Via Madonna del Loco 47, 03023
     Ceccano (FR), Italy}

\authorcount{affil:Krajci}{
     Astrokolkhoz Observatory,
     Center for Backyard Astrophysics New Mexico, PO Box 1351 Cloudcroft,
     New Mexico 83117, USA}

\authorcount{affil:BAAVSS}{
     The British Astronomical Association, Variable Star Section (BAA VSS),
     Burlington House, Piccadilly, London, W1J 0DU, UK}

\authorcount{affil:Pickard}{
     3 The Birches, Shobdon, Leominster, Herefordshire, HR6 9NG, UK}

\authorcount{affil:Sabo}{
     2336 Trailcrest Dr., Bozeman, Montana 59718, USA}

\authorcount{affil:Ioh}{
     VSOLJ, 1001-105 Nishiterakata, Hachioji, Tokyo 192-0153}

\authorcount{affil:Stein}{
     6025 Calle Paraiso, Las Cruces, New Mexico 88012, USA}

\authorcount{affil:Dvorak}{
     Rolling Hills Observatory, 1643 Nightfall Drive,
     Clermont, Florida 34711, USA}

\authorcount{affil:AAVSO}{
     American Association of Variable Star Observers, 49 Bay State Rd.,
     Cambridge, MA 02138, USA}

\authorcount{affil:OKU}{
     Osaka Kyoiku University, 4-698-1 Asahigaoka, Osaka 582-8582}

\authorcount{affil:Bolt}{
     Camberwarra Drive, Craigie, Western Australia 6025, Australia}

\authorcount{affil:Shears}{
     ``Pemberton'', School Lane, Bunbury, Tarporley, Cheshire, CW6 9NR, UK}

\authorcount{affil:Ruiz1}{
     Observatorio de C\'antabria, Ctra. de Rocamundo s/n, Valderredible, 
     Cantabria, Spain}

\authorcount{affil:Ruiz2}{
     Agrupaci\'on Astron\'omica C\'antabra, Apartado 573,
     39080-Santander, Spain}

\authorcount{affil:Sternberg}{
     Sternberg Astronomical Institute, Moscow University, Universitetsky
     Ave., 13, Moscow 119992, Russia}

\authorcount{affil:Slovak}{
     Astronomical Institute of the Slovak Academy of Sciences, 05960,
     Tatranska Lomnica, the Slovak Republic}

\authorcount{affil:Monard}{
     Bronberg Observatory, Center for Backyard Astronomy Pretoria,
     PO Box 11426, Tiegerpoort 0056, South Africa}

\authorcount{affil:Siz}{
     Moriyama 810, Komoro, Nagano 384-0085}

\authorcount{affil:Kai}{
     Baselstrasse 133D, CH-4132 Muttenz, Switzerland}

\authorcount{affil:Staels}{
     Center for Backyard Astrophysics (Flanders),
     American Association of Variable Star Observers (AAVSO),
     Alan Guth Observatory, Koningshofbaan 51, Hofstade, Aalst, Belgium}

\authorcount{affil:Seikei}{
     Seikei Meteorological Observatory, Seikei High School,
     3-3-1, Kichijoji-Kitamachi, Musashino-shi, Tokyo 180-8633}

\authorcount{affil:Starkey}{
     DeKalb Observatory, H63, 2507 County Road 60, Auburn, Indiana 46706, USA}

\authorcount{affil:Ogmen}{
     Green Island Observatory, Ge\c{c}itkale, Magosa, via Mersin, North Cyprus}

\authorcount{affil:LCO}{
     Department of Physics, University of Notre Dame, Notre Dame,
     Indiana 46556, USA}

\authorcount{affil:Sergey}{
     Group Betelgeuse, Republic Center of Technical Creativity of Pupils,
     Minsk, Belarus}

\authorcount{affil:Denisenko}{
     Space Research Institute (IKI), Russian Academy of Sciences, Moscow,
     Russia}

\authorcount{affil:Polaris}{
     Polaris Observatory, Hungarian Astronomical Association,
     Laborc utca 2/c, 1037 Budapest, Hungary}

\authorcount{affil:Virtanen}{
     Ollilantie 98, 84880 Ylivieska, Finland}

\authorcount{affil:Crawford}{
     Arch Cape Observatory, 79916 W. Beach Road, Arch Cape, Oregon 97102, USA}

\authorcount{affil:Pietz}{
     Nollenweg 6, 65510 Idstein, Germany}

\authorcount{affil:Koff}{
     Antelope Hills Observatory, 980 Antelope Drive West
     Bennett, CO 80102, USA}

\authorcount{affil:DavidBoyd}{
     Silver Lane, West Challow, Wantage, OX12 9TX, UK}

\authorcount{affil:Brady}{
     5 Melba Drive, Hudson, New Hampshire 03051, USA}

\authorcount{affil:James}{
     11 Tavistock Road, Chelmsford, Essex CM1 6JL, UK}

\authorcount{affil:Goff}{
     13508 Monitor Ln., Sutter Creek, California 95685, USA}

\authorcount{affil:Itagaki}{
     Itagaki Astronomical Observatory, Teppo-cho, Yamagata 990-2492}

\authorcount{affil:Nmh}{
     Miyawaki 302-6, Kakegawa, Shizuoka 436-0086}

\authorcount{affil:Nakashima}{
     968-4 Yamadanoshou, Oku-cho, Setouchi-City, Okayama 701-4246}

\authorcount{affil:MISAO}{
     2-4-10-708 Tsunashima-nishi, Kohoku-ku, Yokohama-City, Kanagawa 223-0053}

\authorcount{affil:Stubbings}{
     Tetoora Observatory, Tetoora Road, Victoria, Australia}

\authorcount{affil:Poyner}{
     BAA Variable Star Section, 67 Ellerton Road, Kingstanding,
     Birmingham B44 0QE, UK}

\authorcount{affil:Mdy}{
     Kaminishiyamamachi 12-14, Nagasaki, Nagasaki 850-0006}

\authorcount{affil:KaDar}{
     Ka-Dar Scientific Center and Public Observatory,
     Neopalimovskij 1-j per. d. 16/13, 119121 Moscow, Russia}

\authorcount{affil:Sokolovsky}{
     Max-Planck-Institute f\"ur Radioastronomie, Auf dem H\"ugel 69,
     53121 Bonn, Germany}

\authorcount{affil:Sokolovsky2}{
     Astro-Space Center, Lebedev Physical Institute of Russian Academy of
     Sciences, Profsoyuznaya 84/32, 117997 Moscow, Russia}

\authorcount{affil:SeijiUeda}{
     6-23-9 Syowa-Minami, Kushiro City, Hokkaido 084-0909
     }


\KeyWords{accretion, accretion disks
          --- stars: novae, cataclysmic variables
          --- stars: dwarf novae
         }

\maketitle

\begin{abstract}
   Continuing the project described by \citet{Pdot}, we collected
times of superhump maxima for 51 SU UMa-type dwarf novae mainly observed
during the 2010--2011 season.  Although most of the new data for
systems with short superhump periods basically confirmed the findings
by \citet{Pdot} and \citet{Pdot2}, the long-period system GX Cas
showed an exceptionally large positive period derivative.
An analysis of public Kepler data of V344 Lyr and V1504 Cyg yielded
less striking stage transitions.
In V344 Lyr, there was prominent secondary component growing during
the late stage of superoutbursts, and the component persisted at least
for two more cycles of successive normal outbursts.
We also investigated the superoutbursts of two conspicuous eclipsing objects:
HT Cas and the WZ Sge-type object SDSS J080434.20$+$510349.2.
Strong beat phenomena were detected in both objects, and late-stage
superhumps in the latter object had an almost constant luminosity
during the repeated rebrightenings.  The WZ Sge-type object
SDSS J133941.11$+$484727.5 showed a phase reversal around the rapid
fading from the superoutburst.  The object showed a prominent beat
phenomenon even after the end of the superoutburst.
A pilot study of superhump amplitudes
indicated that the amplitudes of superhumps are strongly correlated
with orbital periods, and the dependence on the inclination is
weak in systems with inclinations smaller than 80$^{\circ}$.
\end{abstract}

\newpage

\section{Introduction}

   In papers \citet{Pdot} and \citet{Pdot2}, we systematically surveyed
period variations of superhumps in SU UMa-type dwarf novae (for general
information of SU UMa-type dwarf novae and superhumps, see \cite{war95book}).
These works indicated that evolution of superhump period ($P_{\rm SH}$)
is generally composed of three distinct stages: early evolutionary
stage with a longer superhump period (stage A), middle stage with
systematically varying periods (stage B), final stage with a shorter,
stable superhump period (stage C).  In short-$P_{\rm SH}$ systems,
these stages are often observed as distinct segments (e.g. see
a representative example in figure \ref{fig:j1025comp}:
$E \le 25$ in this figure corresponds to stage A,
$25 < E < 140$ with a parabolic $O-C$ curve corresponds to
stage B with an increasing period, and $E \ge 140$ corresponds to
stage C).  Although systems with longer $P_{\rm SH}$ tend to show less
distinct stages, there is often a systematic ``break'' in the $O-C$
diagram, which we attribute to a stage B-C transition based on
the knowledge in the systematic variation of the $O-C$ diagrams
against $P_{\rm SH}$ (cf. subsection 3.2 and figure 4 in \cite{Pdot}).
It was also shown that the period
derivatives ($P_{\rm dot} = \dot{P}/P$)
during stage B is correlated with $P_{\rm SH}$,
or binary mass-ratios ($q = M_2/M_1$).

   We used similar, but either phenomenologically or conceptually
different, terminologies referring the superhumps observed during
the late stages of superoutbursts or post-superoutburst stages.
For reader's convenience, we summarized them in table
\ref{tab:latesuperhumps}.  Note that the use of the terminology is
not the same between different authors.  Although some authors use
the term ``late superhumps'' for our stage C superhumps, we restrict
the usage of (traditional) late superhumps to superhumps with
an $\sim$0.5 phase shift.

\begin{table}
\caption{Types of late-stage superhumps.}\label{tab:latesuperhumps}
\begin{center}
\begin{tabular}{c|l}
\hline
Superhump type & Usage in this paper \\
\hline
traditional late   & Superhumps with a $\sim$0.5 phase \\
superhumps         & shift seen during the very late or \\
                   & post-superoutburst stages. \\
                   & At least some of these traditional \\
                   & late superhumps reported in past \\
                   & literature may have been mis- \\
                   & identification of stage C super- \\
                   & humps.   We used this term only \\
                   & if $\sim$0.5 phase shift is \\
                   & confirmed. \\
\hline
post-superoutburst & Superhumps during the post- \\ 
superhumps         & superoutburst stages (persisting \\
                   & superhumps).  This term is mostly \\
                   & used in WZ Sge-type dwarf novae \\
                   & and related systems, in which \\
                   & clear stage B-C transitions are \\
                   & often missing or continuity of \\
                   & the phase becomes unclear during \\
                   & the rapid fading stage, making it \\
                   & difficult to assign our stage- \\
                   & based classification. \\
\hline
late-stage         & General term referring to super- \\
superhumps         & humps seen during the late stage \\
                   & of superoutbursts. \\
\hline
stage C superhumps & Superhumps after stage B--C \\
                   & transition based on our staging \\
                   & scheme.  No $\sim$0.5 phase \\
                   & shift is recognized. \\
\hline
\end{tabular}
\end{center}
\end{table}

   In the present study, we extended the survey to newly recorded objects
and superoutbursts since the publication of \citet{Pdot2}.  Some of new
observations have led to revisions of analysis in the previous studies.
We also include a few past superoutbursts not analyzed in the previous studies.

   The present study is particularly featured with extensively observed
rare phenomena, including a superoutburst of the eclipsing dwarf nova HT Cas,
first ever since 1985, the eclipsing WZ Sge-type dwarf nova
SDSS J080434.20$+$510349.2, and the bright WZ Sge-type dwarf nova
with a long-lasting post-superoutburst state SDSS J133941.11$+$484727.5,
all of which provided a significant contribution to our knowledge
in the evolution of superhumps.

   Two new topics will be discussed: comparison of Kepler observations
with ground-based observations, and a pilot study of superhump amplitudes,
the latter having been motivated by a recent work by \citet{sma10SHamp}.

   The structure of the paper follows the scheme in \citet{Pdot},
in which we mostly restricted ourselves to superhump timing analysis.
We also include some more details (evidence for an SU UMa-type dwarf nova
by presenting period analysis and averaged superhump profile)
if the paper provides the first solid presentation of individual objects.

\section{Observation and Analysis}\label{sec:obs}

   The data were obtained under campaigns led by the VSNET Collaboration
\citep{VSNET}.  In some objects, we used archival data for published
papers, and the public data from the AAVSO International Database\footnote{
$<$http://www.aavso.org/data-download$>$.
}.
The majority of the data were acquired
by time-resolved CCD photometry by using 30 cm-class telescopes, whose
observational details on individual objects will be presented in
future papers dealing with analysis and discussion on individual objects.
The list of outbursts and observers is summarized in table \ref{tab:outobs}.
The data analysis was performed just in the same way described
in \citet{Pdot}.  The times of all observations are expressed in
Barycentric Julian Dates (BJD).
We also use the same abbreviations $P_{\rm orb}$ for
the orbital period and $\epsilon = P_{\rm SH}/P_{\rm orb}-1$ for 
the fractional superhump excess.

The derived $P_{\rm SH}$, $P_{\rm dot}$ and other parameters
are listed in table \ref{tab:perlist} in same format as in
\citet{Pdot}.  The definitions of parameters $P_1, P_2, E_1, E_2$
and $P_{\rm dot}$ are the same as in \citet{Pdot}.
We also present comparisons of $O-C$ diagrams between different
superoutbursts since this has been one of the motivations of
these surveys (cf. \cite{uem05tvcrv}).

   While most of the analyses were performed using the same technique
as in \citet{Pdot}, we introduced a variety of bootstrapping in
estimating the robustness of the result of Phase Dispersion Minimization
(PDM; \cite{PDM}).
We typically analyzed 100 samples which randomly contain 50 \% of
observations, and performed PDM analysis for these samples.
The bootstrap result is shown as a form of 90 \% confidence intervals
in the resultant $\theta$ statistics.

   We also employed locally-weighted polynomial regression (LOWESS,
\cite{LOWESS}) in removing trends resulting from outbursts for highly
structured and uninterrupted light curves like Kepler observations.

\begin{table*}
\caption{List of Superoutbursts.}\label{tab:outobs}
\begin{center}
\begin{tabular}{ccccl}
\hline
Subsection & Object & Year & Observers or references\commenta & ID\commentb \\
\hline
\ref{obj:foand}    & FO And     & 2010 & deM, IMi, OUS, HMB & \\
                   & FO And     & 2011 & PXR, deM, Ogm & \\
\ref{obj:v402and}  & V402 And   & 2011 & IMi, MEV, PXR, SRI & \\
\ref{obj:bgari}    & BG Ari     & 2010 & SRI, IMi, MEV, Boy, JSh & \\
\ref{obj:v496aur}  & V496 Aur   & 2010 & IMi, KU & \\
\ref{obj:ttboo}    & TT Boo     & 2007 & Mhh & \\
                   & TT Boo     & 2010 & Ioh, Mhh & \\
\ref{obj:gxcas}    & GX Cas     & 2010 & OUS, DPV & \\
\ref{obj:htcas}    & HT Cas     & 2010 & OKU, Mhh, KU, OUS, AAVSO, Ioh, deM, & \\
                   &            &      & CRI, Mas, IMi, MEV, Rui, Tze, SRI, & \\
                   &            &      & DPV, Ter, HMB, Pol, Kai, NDJ, Nyr, & \\
                   &            &      & PXR, LCO, Ogm, DRS, Kis, SAc & \\
\ref{obj:v1504cyg} & V1504 Cyg  & 2007 & CRI & \\
                   & V1504 Cyg  & 2009b & Kepler & \\
\ref{obj:awgem}    & AW Gem     & 2011 & OUS & \\
\ref{obj:v844her}  & V844 Her   & 2010b & DPV, deM & \\
\ref{obj:mmhya}    & MM Hya     & 2011 & GBo, OUS, Mhh, IMi & \\
\ref{obj:v406hya}  & V406 Hya   & 2010 & Ter, Kis, Mhh & \\
\ref{obj:v344lyr}  & V344 Lyr   & 2009 & Kepler & \\
                   & V344 Lyr   & 2009b & Kepler & \\
\ref{obj:v1195oph} & V1195 Oph  & 2011 & SWI, Mhh & \\
\ref{obj:v1212tau} & V1212 Tau  & 2011 & deM, Mhh, Kra, MEV, IMi, KU, SRI, SWI & \\
\ref{obj:swuma}    & SW UMa     & 2010 & OUS, Rui, DPV & \\
\ref{obj:ciuma}    & CI UMa     & 2011 & IMi, DPV, HMB & \\
\ref{obj:dvuma}    & DV UMa     & 2011 & Mhh, LCO, IMi & \\
\ref{obj:j0038}    & 1RXS J0038 & 2010 & PIE, KU, CRI & \\
\ref{obj:j2224}    & 2QZ J2224  & 2010 & MLF, Mhh & \\
\ref{obj:asas0918} & ASAS J0918 & 2010 & KU, Kis, DKS, OUS & \\
\ref{obj:asas1025} & ASAS J1025 & 2011 & DKS, Kis, Mas & \\
\ref{obj:misv1443} & MisV 1443  & 2011 & OUS, Ioh, deM, Mhh, Kis, PXR & \\
\ref{obj:j1715}    & RX J1715   & 2010 & IMi, KU & \\
\ref{obj:j0732}    & SDSS J0732 & 2011 & Kra, KU, IMi & \\
\ref{obj:j0803}    & SDSS J0803 & 2011 & Mhh, IMi, Mas & \citet{wil10newCVs} \\
\ref{obj:j0804}    & SDSS J0804 & 2010 & Ter, KU, deM, IMi, OKU, OUS, Mhh, & \\
                   &            &      & AKz, KRV, NKa, HMB, CRI, Nyr, DPV, & \\
                   &            &      & MEV, Shu, Den, Ser, SRI, Vir, AAVSO & \\
\ref{obj:j0812}    & SDSS J0812 & 2008 & \citet{Pdot} & \\
                   & SDSS J0812 & 2011 & OUS, Mhh, IMi & \\
\ref{obj:j0932}    & SDSS J0932 & 2011 & OKU, deM, CRI, MEV, Nyr, SRI, DKS & \\
\ref{obj:j1120}    & SDSS J1120 & 2011 & SWI, JSh & \citet{wil10newCVs} \\
\ref{obj:j1146}    & SDSS J1146 & 2011 & KU, Mhh, BSt, IMi, HMB, JSh, Nyr, & \\
                   &            &      & deM, BXS, DRS, AAVSO & \\
\ref{obj:j1227}    & SDSS J1227 & 2007 & \citet{Pdot}, Gor & \\
                   & SDSS J1227 & 2011 & Nyr, Ter, CRI, deM & \\
\ref{obj:j1250}    & SDSS J1250 & 2011 & Rui, Mhh, IMi & \\
\ref{obj:j1339}    & SDSS J1339 & 2011 & Ter, Mhh, OUS, Nyr, HMB, KU, SRI, & \\
                   &            &      & AAVSO, IMi, Mas, DPV, GFB, DKS, PXR, & \\
                   &            &      & CTX, BSt, JSh, MEV & \\
\ref{obj:j1605}    & SDSS J1605 & 2010 & KU, Mhh, Kis & Itagaki (vsnet-alert 12414) \\
\hline
  \multicolumn{5}{l}{\commenta Key to observers:
AKz (Astrokolkhoz Obs.),
Boy\commentc (D. Boyd),
BSt (B. Staels),
BXS\commentc (S. Brady),
}\\ \multicolumn{5}{l}{
CRI (Crimean Astrophys. Obs.),
CTX\commentc (T. Crawford), 
Den (D. Denisenko),
deM (E. de Miguel),
DKS\commentc (S. Dvorak),
}\\ \multicolumn{5}{l}{
DPV (P. Dubovsky),
DRS\commentc (D. Starkey),
GBo (G. Bolt),
GFB\commentc (W. Goff),
Gor (V. Goranskij),
HMB (F.-J. Hambsch),
}\\ \multicolumn{5}{l}{
IMi\commentc (I. Miller),
Ioh (H. Itoh),
JSh\commentc (J. Shears),
Kai (K. Kasai),
Kis (S. Kiyota),
Kra (T. Krajci),
KRV\commentc (R. Koff),
}\\ \multicolumn{5}{l}{
KU (Kyoto U., campus obs.),
LCO\commentc (C. Littlefield),
Mas (G. Masi),
MEV\commentc (E. Morelle),
Mhh (H. Maehara),
}\\ \multicolumn{5}{l}{
MLF (B. Monard),
NDJ (N. James),
NKa (N. Katysheva),
Nyr (Nyrola and Hankasalmi Obs.),
Ogm\commentc (Y. Ogmen),
}\\ \multicolumn{5}{l}{
OKU (Osaya Kyoiku U.),
OUS (Okayama U. of Science),
PIE (J. Pietz),
Pol (Polaris Obs.),
PXR\commentc (R. Pickard),
}\\ \multicolumn{5}{l}{
Rui (J. Ruiz),
SAc (Seikei High School),
Ser (I. Sergey),
Shu (S. Shugarov),
SRI\commentc (R. Sabo),
SWI\commentc (W. Stein),
}\\ \multicolumn{5}{l}{
Ter (Terskol Obs.),
Tze (Tzec Maun Obs., remotely operated by HMB),
Vir\commentc (J. Virtanen),
AAVSO (AAVSO database)
} \\
  \multicolumn{5}{l}{\commentb Original identifications or discoverers.} \\
  \multicolumn{5}{l}{\commentc Inclusive of observations from the AAVSO database.} \\
\end{tabular}
\end{center}
\end{table*}

\addtocounter{table}{-1}
\begin{table*}
\caption{List of Superoutbursts (continued).}
\begin{center}
\begin{tabular}{ccccl}
\hline
Subsection & Object & Year & Observers or references\commenta & ID\commentb \\
\hline
\ref{obj:j2100}    & SDSS J2100 & 2010 & IMi & \\
\ref{obj:j0009}    & OT J0009   & 2010 & GBo, KU, Mhh & CSS101007:000938$-$121017\\
\ref{obj:j0120}    & OT J0120   & 2010 & Ter, Mhh, OUS, KU, Ioh, Siz, CRI, & Itagaki (vsnet-alert 12431) \\
                   &            &      & AAVSO, deM, MEV, IMi, DKS, Shu, & \\
                   &            &      & Nyr, SAc, PXR, SWI, Kai & \\
\ref{obj:j0141}    & OT J0141   & 2010 & Mhh, Kis & CSS101127:014150$+$090822 \\
\ref{obj:j0413}    & OT J0413   & 2011 & Kra, Mhh & CSS110125:041350$+$094515 \\
\ref{obj:j0431}    & OT J0431   & 2011 & SWI, Mas, KU, Kis, Kra, Mhh & CSS110113:043112$-$031452 \\
\ref{obj:j0442}    & OT J0442   & 2011 & Mas & CSS071115:044216$-$002334 \\
\ref{obj:j0648}    & OT J0648   & 2011 & Mhh, Kra, deM, Ter & CSS091026:064805$+$414702 \\
\ref{obj:j0754}    & OT J0754   & 2011 & deM, HMB & CSS110414:075414$+$313216 \\
\ref{obj:j102616}  & OT J102616 & 2010 & Ter, KU, Mhh & CSS101130:102616$+$192045 \\
\ref{obj:j1027}    & OT J1027   & 2011 & GBo & SSS110314:102706$-$434341 \\
\ref{obj:j1200}    & OT J1200   & 2011 & Mas & CSS110205:120053$-$152620 \\
\ref{obj:j1329}    & OT J1329   & 2011 & MLF & SSS110403:132901$-$365859 \\
\ref{obj:j1545}    & OT J1545   & 2011 & OKU, HMB & CSS110428:154545+442830 \\
\ref{obj:j2234}    & OT J2234   & 2009 & Mhh & CSS090910:223418$-$035530 \\
                   & OT J2234   & 2010 & GBo & \\
\ref{obj:j2304}    & OT J2304   & 2011 & Siz, Ioh, Mhh, CRI & Nishimura \citep{nak11j2304cbet2616} \\
\hline
  \multicolumn{5}{l}{\commenta Key to observers.} \\
  \multicolumn{5}{l}{\commentb Original identifications or discoverers.} \\
  \multicolumn{5}{l}{\commentc Inclusive of observations from the AAVSO database.} \\
\end{tabular}
\end{center}
\end{table*}

\begin{table*}
\caption{Superhump Periods and Period Derivatives}\label{tab:perlist}
\begin{center}
\begin{tabular}{cccccccccccccc}
\hline
Object & Year & $P_1$ (d) & err & \multicolumn{2}{c}{$E_1$\commenta} & $P_{\rm dot}$\commentb & err\commentb & $P_2$ (d) & err & \multicolumn{2}{c}{$E_2$\commenta} & $P_{\rm orb}$ (d) & Q\commentc \\
\hline
FO And & 2010 & 0.074512 & 0.000017 & 0 & 50 & 4.6 & 3.5 & 0.074117 & 0.000030 & 59 & 131 & 0.07161 & B \\
FO And & 2011 & -- & -- & -- & -- & -- & -- & 0.074266 & 0.000114 & 39 & 54 & 0.07161 & C \\
V402 And & 2011 & 0.063489 & 0.000030 & 0 & 125 & 7.7 & 0.8 & 0.063128 & 0.000071 & 125 & 158 & -- & B \\
BG Ari & 2010 & 0.084897 & 0.000025 & 32 & 91 & 4.9 & 3.0 & 0.084683 & 0.000081 & 102 & 221 & -- & B \\
V496 Aur & 2010 & 0.061162 & 0.000076 & 0 & 20 & -- & -- & -- & -- & -- & -- & -- & C \\
TT Boo & 2007 & 0.077876 & 0.000116 & 0 & 26 & -- & -- & -- & -- & -- & -- & -- & C2 \\
TT Boo & 2010 & 0.078115 & 0.000025 & 0 & 104 & 5.4 & 1.4 & -- & -- & -- & -- & -- & B \\
GX Cas & 2010 & 0.092998 & 0.000054 & 0 & 60 & 25.0 & 4.6 & -- & -- & -- & -- & -- & B \\
HT Cas & 2010 & 0.076349 & 0.000016 & 12 & 58 & 3.0 & 3.6 & 0.075900 & 0.000011 & 57 & 126 & 0.073647 & A \\
V1504 Cyg & 2007 & 0.072323 & 0.000146 & 0 & 55 & -- & -- & 0.071865 & 0.000040 & 55 & 139 & 0.069549 & C \\
V1504 Cyg & 2009b & 0.072221 & 0.000009 & 21 & 89 & $-$1.0 & 1.4 & 0.071883 & 0.000009 & 89 & 163 & 0.069549 & A \\
AW Gem & 2011 & 0.079376 & 0.000037 & 0 & 27 & -- & -- & 0.078627 & 0.000013 & 38 & 102 & 0.07621 & B \\
V844 Her & 2010b & 0.056115 & 0.000041 & 0 & 72 & 8.9 & 5.8 & 0.055906 & 0.000044 & 70 & 107 & 0.054643 & C \\
MM Hya & 2011 & 0.058855 & 0.000017 & 16 & 120 & 7.2 & 0.9 & -- & -- & -- & -- & 0.057590 & B \\
V344 Lyr & 2009 & 0.091583 & 0.000009 & 26 & 79 & $-$0.3 & 1.3 & 0.091277 & 0.000016 & 80 & 156 & -- & A \\
V344 Lyr & 2009b & 0.091614 & 0.000012 & 30 & 72 & $-$1.4 & 2.3 & 0.091259 & 0.000010 & 73 & 170 & -- & A \\
V1195 Oph & 2011 & 0.067038 & 0.000070 & 0 & 25 & -- & -- & -- & -- & -- & -- & -- & C2 \\
V1212 Tau & 2011 & 0.070115 & 0.000018 & 20 & 90 & 7.0 & 2.7 & 0.069778 & 0.000032 & 89 & 167 & -- & B \\
SW UMa & 2010 & 0.058208 & 0.000034 & 52 & 174 & 12.9 & 0.9 & 0.057978 & 0.000035 & 171 & 277 & 0.056815 & B \\
CI UMa & 2011 & 0.062690 & 0.000029 & 0 & 66 & 16.8 & 3.2 & -- & -- & -- & -- & -- & B \\
DV UMa & 2011 & 0.088933 & 0.000045 & 0 & 52 & 14.6 & 6.6 & -- & -- & -- & -- & 0.085853 & C \\
1RXS J0038 & 2010 & 0.097080 & 0.000080 & 0 & 42 & -- & -- & -- & -- & -- & -- & -- & CG \\
2QZ J2224 & 2010 & -- & -- & -- & -- & -- & -- & 0.058243 & 0.000011 & 0 & 106 & -- & C \\
ASAS J0918 & 2010 & -- & -- & -- & -- & -- & -- & 0.062618 & 0.000036 & 20 & 86 & -- & C \\
ASAS J1025 & 2011 & 0.063538 & 0.000023 & 0 & 72 & 9.9 & 2.4 & 0.063271 & 0.000083 & 72 & 103 & 0.06136 & CE \\
MisV 1443 & 2011 & 0.056725 & 0.000014 & 19 & 110 & 5.5 & 1.5 & -- & -- & -- & -- & -- & B \\
RX J1715 & 2010 & -- & -- & -- & -- & -- & -- & 0.070752 & 0.000102 & 0 & 30 & 0.0683 & C \\
SDSS J0732 & 2011 & -- & -- & -- & -- & -- & -- & 0.079325 & 0.000042 & 0 & 72 & -- & C \\
SDSS J0803 & 2011 & 0.075100 & 0.000033 & 0 & 50 & 13.6 & 4.2 & -- & -- & -- & -- & -- & C \\
SDSS J0804 & 2010 & 0.059630 & 0.000016 & 31 & 159 & 9.6 & 1.1 & -- & -- & -- & -- & 0.059005 & A \\
SDSS J0812 & 2008 & 0.077559 & 0.000148 & 0 & 103 & $-$24.8 & 6.2 & -- & -- & -- & -- & -- & BG \\
SDSS J0812 & 2011 & 0.077892 & 0.000029 & 13 & 53 & $-$7.6 & 6.3 & 0.077337 & 0.000072 & 77 & 131 & -- & B \\
SDSS J0932 & 2011 & -- & -- & -- & -- & -- & -- & 0.068110 & 0.000016 & 0 & 161 & 0.066304 & B \\
SDSS J1120 & 2011 & 0.070550 & 0.000054 & 0 & 22 & -- & -- & -- & -- & -- & -- & -- & C \\
SDSS J1146 & 2011 & 0.063326 & 0.000017 & 9 & 83 & 10.9 & 3.0 & 0.063055 & 0.000086 & 83 & 105 & -- & BG \\
SDSS J1227 & 2007 & 0.064604 & 0.000029 & 33 & 129 & 10.5 & 2.4 & 0.064399 & 0.000046 & 126 & 216 & 0.062950 & B \\
SDSS J1227 & 2011 & 0.064883 & 0.000032 & 0 & 29 & -- & -- & 0.064494 & 0.000021 & 43 & 184 & 0.062950 & C \\
SDSS J1250 & 2011 & 0.060261 & 0.000057 & 0 & 59 & -- & -- & -- & -- & -- & -- & 0.058736 & CGM \\
SDSS J1339 & 2011 & 0.058094 & 0.000007 & 34 & 205 & 5.4 & 0.2 & -- & -- & -- & -- & 0.057289 & A \\
SDSS J2100 & 2010 & 0.087295 & 0.000186 & 0 & 69 & $-$34.1 & 2.2 & -- & -- & -- & -- & -- & CG \\
OT J0009 & 2010 & 0.090005 & 0.000055 & 0 & 14 & -- & -- & 0.088851 & 0.000049 & 12 & 48 & -- & C \\
OT J0120 & 2010 & 0.057833 & 0.000009 & 0 & 174 & 4.3 & 0.5 & -- & -- & -- & -- & 0.057157 & AE \\
OT J0141 & 2010 & 0.062488 & 0.000040 & 0 & 115 & 10.1 & 1.6 & 0.062060 & 0.000086 & 114 & 131 & -- & B \\
OT J0413 & 2011 & 0.054830 & 0.000007 & 0 & 76 & 4.2 & 0.8 & -- & -- & -- & -- & -- & B \\
OT J0431 & 2011 & 0.067583 & 0.000026 & 0 & 112 & 8.4 & 1.2 & 0.067309 & 0.000043 & 125 & 200 & 0.066050 & B \\
OT J0442 & 2011 & 0.076761 & 0.000037 & 0 & 27 & -- & -- & -- & -- & -- & -- & -- & CG \\
OT J0648 & 2011 & 0.066324 & 0.000033 & 0 & 70 & 14.5 & 2.9 & -- & -- & -- & -- & -- & B \\
OT J0754 & 2011 & 0.063080 & 0.000081 & 0 & 33 & -- & -- & 0.062682 & 0.000044 & 31 & 96 & -- & C \\
OT J102616 & 2010 & 0.082830 & 0.000013 & 13 & 36 & -- & -- & 0.082526 & 0.000017 & 36 & 107 & -- & B \\
OT J1027 & 2011 & 0.080887 & 0.000034 & 0 & 15 & -- & -- & -- & -- & -- & -- & -- & C \\
\hline
  \multicolumn{13}{l}{\commenta Interval used for calculating the period (corresponding to $E$ in section \ref{sec:individual}).} \\
  \multicolumn{13}{l}{\commentb Unit $10^{-5}$.} \\
  \multicolumn{13}{l}{\commentc Data quality and comments. A: excellent, B: partial coverage or slightly low quality, C: insufficient coverage or}\\
  \multicolumn{13}{l}{\phantom{\commentc} observations with large scatter, G: $P_{\rm dot}$ denotes global $P_{\rm dot}$, M: observational gap in middle stage,}\\
  \multicolumn{13}{l}{\phantom{\commentc} 2: late-stage coverage, the listed period may refer to $P_2$, E: $P_{\rm orb}$ refers to the period of early superhumps.} \\
\end{tabular}
\end{center}
\end{table*}

\addtocounter{table}{-1}
\begin{table*}
\caption{Superhump Periods and Period Derivatives (continued)}
\begin{center}
\begin{tabular}{cccccccccccccc}
\hline
Object & Year & $P_1$ & err & \multicolumn{2}{c}{$E_1$} & $P_{\rm dot}$ & err & $P_2$ & err & \multicolumn{2}{c}{$E_2$} & $P_{\rm orb}$ & Q \\
\hline
OT J1200 & 2011 & 0.097710 & 0.000058 & 0 & 11 & -- & -- & -- & -- & -- & -- & -- & CG \\
OT J1329 & 2011 & 0.070999 & 0.000066 & 0 & 16 & -- & -- & -- & -- & -- & -- & -- & C \\
OT J1545 & 2011 & -- & -- & -- & -- & -- & -- & 0.076930 & 0.000022 & 0 & 45 & -- & C \\
OT J2234 & 2009 & 0.092192 & 0.000020 & 0 & 109 & -- & -- & -- & -- & -- & -- & -- & CG \\
OT J2234 & 2010 & 0.092000 & 0.000400 & 0 & 2 & -- & -- & -- & -- & -- & -- & -- & C \\
OT J2304 & 2010 & 0.067194 & 0.000030 & 0 & 123 & $-$3.9 & 2.4 & 0.066281 & 0.000063 & 118 & 254 & -- & C \\
\hline
\end{tabular}
\end{center}
\end{table*}

\section{Individual Objects}\label{sec:individual}

\subsection{FO Andromedae}\label{obj:foand}

\begin{figure}
  \begin{center}
    \FigureFile(88mm,70mm){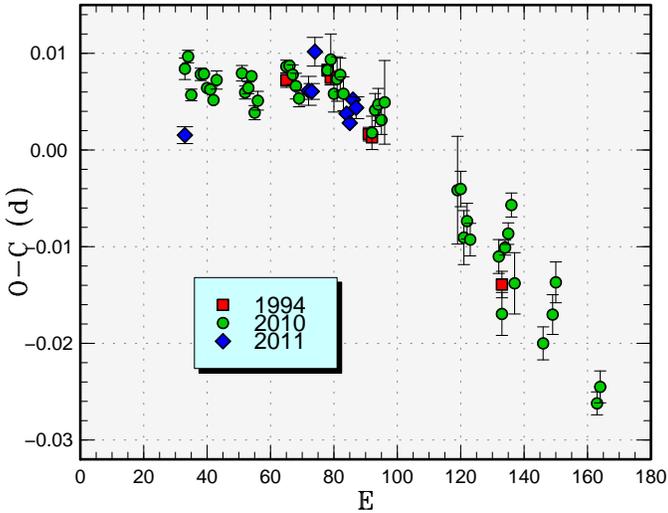}
  \end{center}
  \caption{Comparison of $O-C$ diagrams of FO And between different
  superoutbursts.  A period of 0.07451 d was used to draw this figure.
  Approximate cycle counts ($E$) after the start of the superoutburst
  were used.  Since the starts of the 1994 and 2011 superoutbursts
  were not well constrained, we shifted the $O-C$ diagrams
  to best fit the best-recorded 2010 one.
  }
  \label{fig:foandcomp}
\end{figure}

   We observed two superoutbursts of this object in 2010 and 2011.
The 2010 superoutburst was well-observed except for the initial part
(table \ref{tab:foandoc2010}).  The 2011 superoutburst was only
partly observed (table \ref{tab:foandoc2011}).
The 2010 observation confirmed the presence of a stage B-C transition.
The later part of the 2011 observation probably caught a part of
stage C, and we listed values based on this interpretation
(figure \ref{fig:foandcomp}).

\begin{figure}
  \begin{center}
    \FigureFile(88mm,110mm){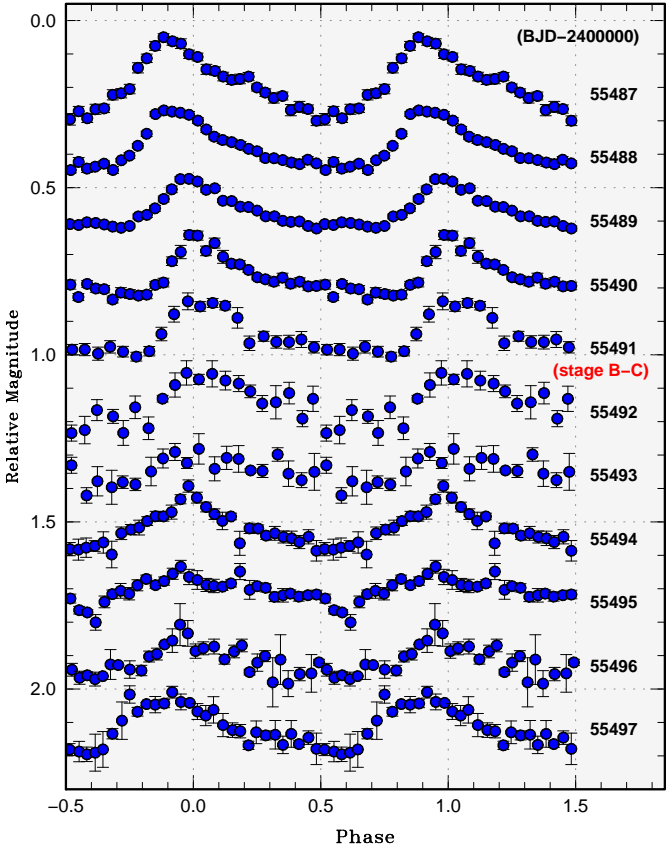}
  \end{center}
  \caption{Superhump profiles of FO And (2010).
     There was no apparent growth of secondary maxima as in ER UMa
     \citep{kat03erumaSH}.  The apparent regrowth of the amplitudes
     of superhump at BJD 2455494 corresponds to 3 d after the
     stage B--C transition.
     The figure was drawn against a mean period of 0.074283 d.}
  \label{fig:foandprof}
\end{figure}

   FO And has a short supercycle of 100--140 d and has similar outburst
properties to V503 Cyg, V344 Lyr \citep{kat02v344lyr}
and MN Dra \citep{pav10mndraproc}, all of which are known to show distinct
negative superhumps in quiescence
(\cite{har95v503cyg}; \cite{sti10v344lyr}; \cite{pav10mndra}).
The recent discovery of long-persisting negative superhumps
in ER UMa \citep{ohs11eruma} and its unusual development of
superhumps \citep{kat03erumaSH}
has led to a question whether these system show unusual behavior similar
to ER UMa.  In the case of FO And, the $O-C$ behavior and variation
of the superhump profile (figure \ref{fig:foandprof}) of the 2010
superoutburst looked relatively normal for an ordinary SU UMa-type
dwarf nova.

\begin{table}
\caption{Superhump Maxima of FO And (2010).}\label{tab:foandoc2010}
\begin{center}
\begin{tabular}{ccccc}
\hline
$E$ & max\commenta & error & $O-C$\commentb & $N$\commentc \\
\hline
0 & 55486.9737 & 0.0011 & $-$0.0041 & 108 \\
1 & 55487.0494 & 0.0006 & $-$0.0026 & 108 \\
2 & 55487.1200 & 0.0006 & $-$0.0064 & 107 \\
5 & 55487.3456 & 0.0006 & $-$0.0036 & 77 \\
6 & 55487.4202 & 0.0005 & $-$0.0033 & 74 \\
7 & 55487.4933 & 0.0006 & $-$0.0045 & 71 \\
8 & 55487.5676 & 0.0005 & $-$0.0044 & 73 \\
9 & 55487.6410 & 0.0006 & $-$0.0053 & 74 \\
10 & 55487.7176 & 0.0009 & $-$0.0030 & 45 \\
18 & 55488.3144 & 0.0008 & $-$0.0005 & 64 \\
19 & 55488.3869 & 0.0007 & $-$0.0022 & 76 \\
20 & 55488.4619 & 0.0006 & $-$0.0015 & 77 \\
21 & 55488.5376 & 0.0005 & $-$0.0001 & 77 \\
22 & 55488.6084 & 0.0007 & $-$0.0036 & 77 \\
23 & 55488.6841 & 0.0010 & $-$0.0022 & 73 \\
32 & 55489.3582 & 0.0007 & 0.0034 & 158 \\
33 & 55489.4328 & 0.0005 & 0.0037 & 153 \\
34 & 55489.5064 & 0.0010 & 0.0030 & 102 \\
35 & 55489.5797 & 0.0013 & 0.0021 & 77 \\
36 & 55489.6530 & 0.0009 & 0.0010 & 77 \\
45 & 55490.3265 & 0.0005 & 0.0060 & 113 \\
46 & 55490.4021 & 0.0026 & 0.0073 & 37 \\
47 & 55490.4731 & 0.0019 & 0.0040 & 34 \\
48 & 55490.5491 & 0.0023 & 0.0057 & 38 \\
49 & 55490.6240 & 0.0017 & 0.0064 & 38 \\
50 & 55490.6966 & 0.0018 & 0.0047 & 31 \\
59 & 55491.3631 & 0.0017 & 0.0027 & 35 \\
60 & 55491.4400 & 0.0017 & 0.0053 & 34 \\
61 & 55491.5151 & 0.0017 & 0.0061 & 37 \\
62 & 55491.5880 & 0.0015 & 0.0047 & 39 \\
63 & 55491.6643 & 0.0043 & 0.0067 & 28 \\
86 & 55493.3690 & 0.0056 & 0.0029 & 23 \\
87 & 55493.4436 & 0.0018 & 0.0032 & 25 \\
88 & 55493.5131 & 0.0028 & $-$0.0016 & 25 \\
89 & 55493.5893 & 0.0018 & 0.0004 & 25 \\
90 & 55493.6619 & 0.0017 & $-$0.0013 & 25 \\
99 & 55494.3307 & 0.0018 & $-$0.0010 & 27 \\
100 & 55494.3993 & 0.0022 & $-$0.0067 & 34 \\
101 & 55494.4807 & 0.0008 & 0.0003 & 70 \\
102 & 55494.5566 & 0.0011 & 0.0020 & 91 \\
103 & 55494.6341 & 0.0012 & 0.0052 & 59 \\
104 & 55494.7005 & 0.0032 & $-$0.0027 & 19 \\
113 & 55495.3649 & 0.0017 & $-$0.0068 & 33 \\
116 & 55495.5914 & 0.0021 & $-$0.0032 & 25 \\
117 & 55495.6692 & 0.0021 & 0.0004 & 26 \\
130 & 55496.6254 & 0.0012 & $-$0.0092 & 38 \\
131 & 55496.7016 & 0.0017 & $-$0.0072 & 21 \\
\hline
  \multicolumn{5}{l}{\commenta BJD$-$2400000.} \\
  \multicolumn{5}{l}{\commentb Against max $= 2455486.9778 + 0.074283 E$.} \\
  \multicolumn{5}{l}{\commentc Number of points used to determine the maximum.} \\
\end{tabular}
\end{center}
\end{table}

\begin{table}
\caption{Superhump Maxima of FO And (2011).}\label{tab:foandoc2011}
\begin{center}
\begin{tabular}{ccccc}
\hline
$E$ & max\commenta & error & $O-C$\commentb & $N$\commentc \\
\hline
0 & 55578.4067 & 0.0009 & $-$0.0017 & 77 \\
39 & 55581.3172 & 0.0015 & 0.0012 & 81 \\
40 & 55581.3916 & 0.0008 & 0.0011 & 158 \\
41 & 55581.4702 & 0.0015 & 0.0052 & 61 \\
51 & 55582.2089 & 0.0006 & $-$0.0017 & 55 \\
52 & 55582.2825 & 0.0004 & $-$0.0027 & 89 \\
53 & 55582.3594 & 0.0006 & $-$0.0003 & 40 \\
54 & 55582.4331 & 0.0011 & $-$0.0012 & 32 \\
\hline
  \multicolumn{5}{l}{\commenta BJD$-$2400000.} \\
  \multicolumn{5}{l}{\commentb Against max $= 2455578.4084 + 0.074553 E$.} \\
  \multicolumn{5}{l}{\commentc Number of points used to determine the maximum.} \\
\end{tabular}
\end{center}
\end{table}

\subsection{V402 Andromedae}\label{obj:v402and}

\begin{figure}
  \begin{center}
    \FigureFile(88mm,70mm){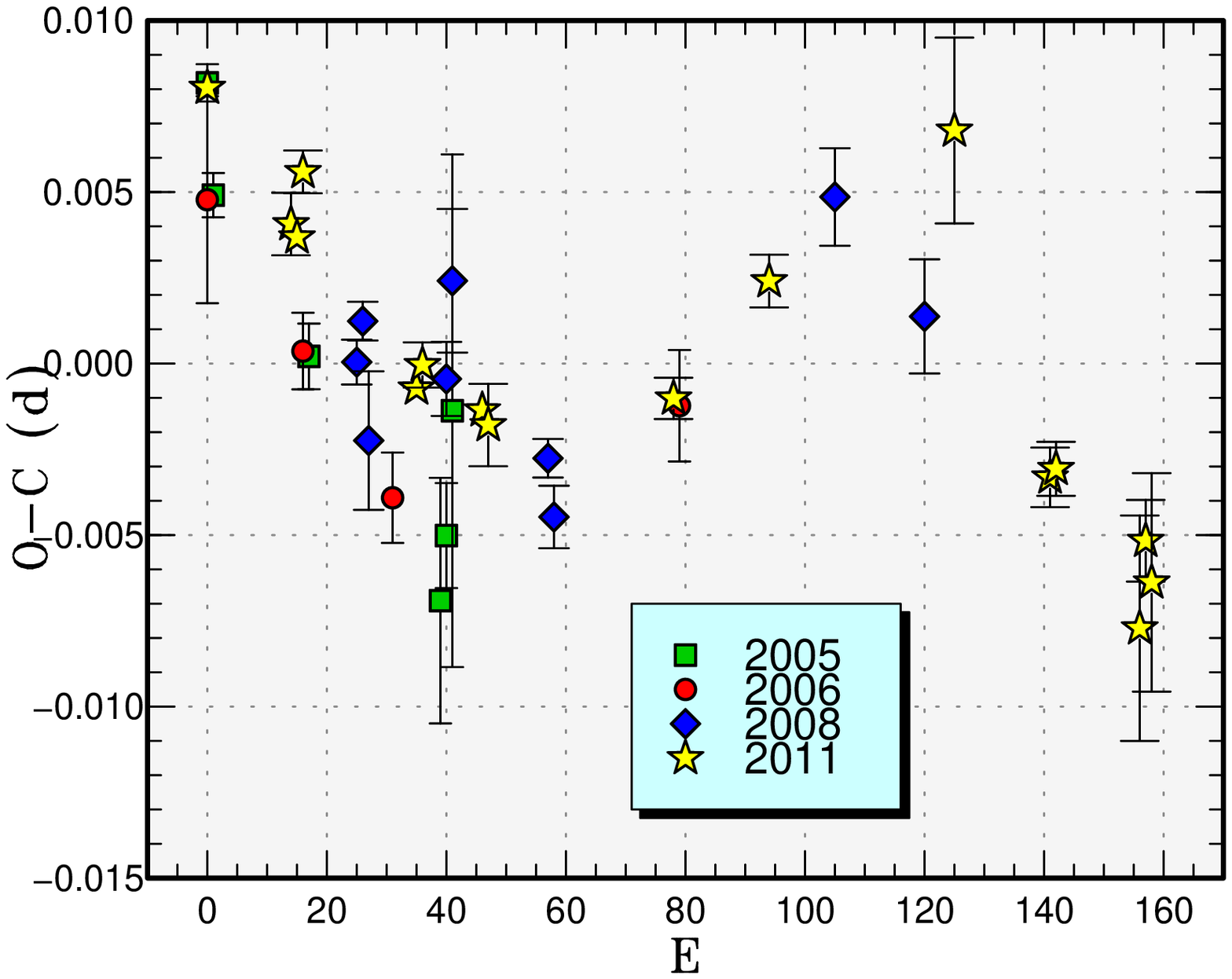}
  \end{center}
  \caption{Comparison of $O-C$ diagrams of V402 And between different
  superoutbursts.  A period of 0.06350 d was used to draw this figure.
  Approximate cycle counts ($E$) after the start of the
  superoutburst were used.
  }
  \label{fig:v402andcomp2}
\end{figure}

   We observed the 2011 superoutburst (table \ref{tab:v402andoc2011}).
The data demonstrated stage B with a positive $P_{\rm dot}$ of
$+7.7(0.8) \times 10^{-5}$ and stage C first time for this object.
A comparison of $O-C$ diagrams for different superoutbursts is shown
in figure \ref{fig:v402andcomp2}.

\begin{table}
\caption{Superhump maxima of V402 And (2011).}\label{tab:v402andoc2011}
\begin{center}
\begin{tabular}{ccccc}
\hline
$E$ & max\commenta & error & $O-C$\commentb & $N$\commentc \\
\hline
0 & 55582.3645 & 0.0004 & 0.0037 & 49 \\
14 & 55583.2496 & 0.0009 & 0.0005 & 38 \\
15 & 55583.3127 & 0.0003 & 0.0002 & 116 \\
16 & 55583.3781 & 0.0006 & 0.0021 & 63 \\
35 & 55584.5783 & 0.0004 & $-$0.0031 & 46 \\
36 & 55584.6425 & 0.0007 & $-$0.0024 & 37 \\
46 & 55585.2761 & 0.0005 & $-$0.0032 & 59 \\
47 & 55585.3392 & 0.0012 & $-$0.0035 & 52 \\
78 & 55587.3085 & 0.0006 & $-$0.0011 & 58 \\
94 & 55588.3279 & 0.0008 & 0.0032 & 78 \\
125 & 55590.3008 & 0.0027 & 0.0093 & 17 \\
141 & 55591.3067 & 0.0009 & 0.0001 & 65 \\
142 & 55591.3704 & 0.0008 & 0.0004 & 39 \\
156 & 55592.2548 & 0.0033 & $-$0.0035 & 39 \\
157 & 55592.3208 & 0.0012 & $-$0.0009 & 126 \\
158 & 55592.3831 & 0.0032 & $-$0.0020 & 37 \\
\hline
  \multicolumn{5}{l}{\commenta BJD$-$2400000.} \\
  \multicolumn{5}{l}{\commentb Against max $= 2455582.3608 + 0.063445 E$.} \\
  \multicolumn{5}{l}{\commentc Number of points used to determine the maximum.} \\
\end{tabular}
\end{center}
\end{table}

\subsection{BG Arietis}\label{obj:bgari}

   We observed the 2010 superoutburst of this object (=PG 0149$+$138,
SDSS J015151.87$+$140047.2).  The outburst was well-observed since
its earliest stage.  The times of superhump maxima are listed in
table \ref{tab:bgarioc2010}.  All A--C stages are clearly present.
There was a marginal hint of positive $P_{\rm dot}$ during stage B,
as in the 2009 superoutburst \citep{Pdot2}.

\begin{figure}
  \begin{center}
    \FigureFile(88mm,70mm){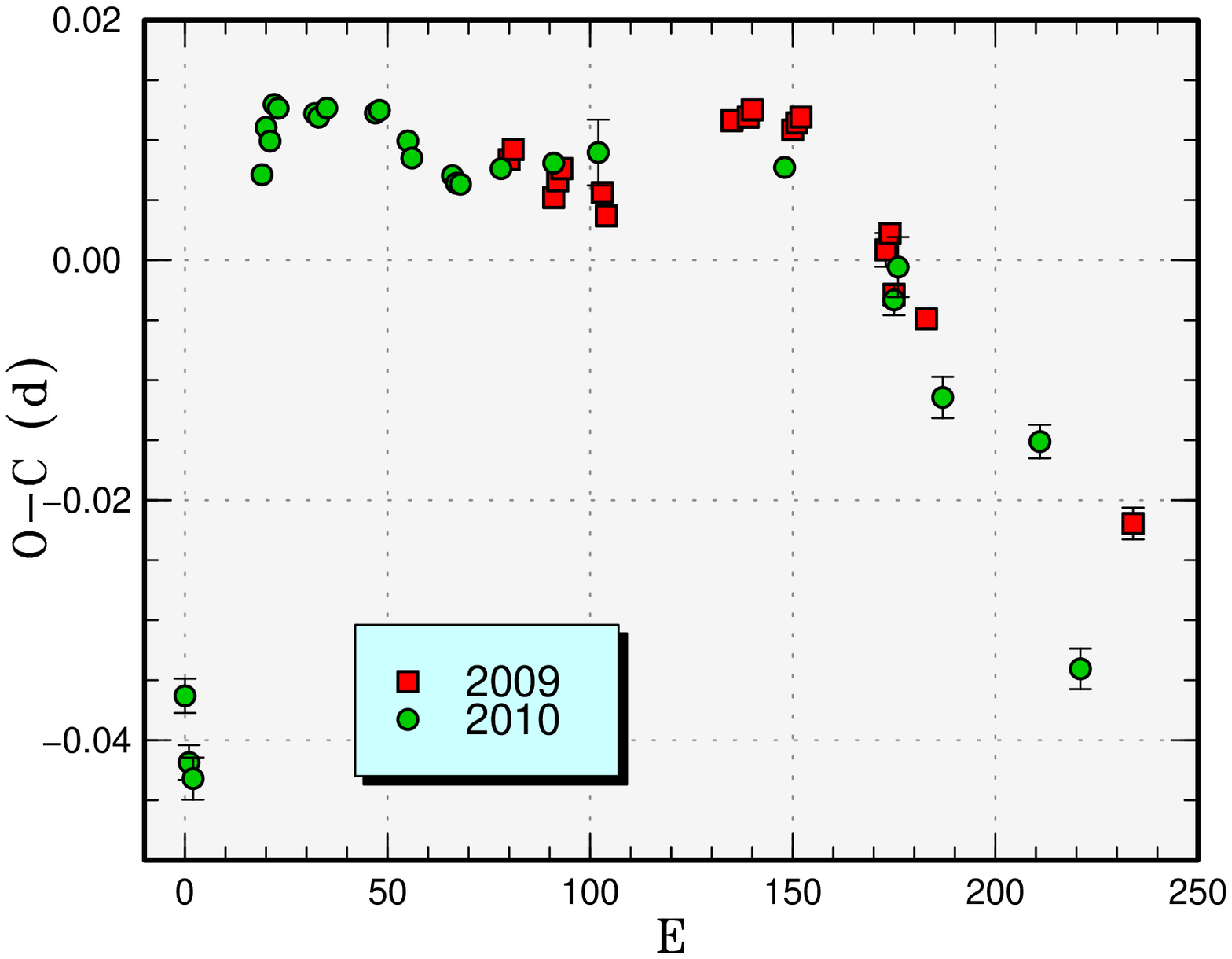}
  \end{center}
  \caption{Comparison of $O-C$ diagrams of BG Ari between different
  superoutbursts.  A period of 0.08501 d was used to draw this figure.
  Approximate cycle counts ($E$) after the start of the superoutburst
  were used, assuming that the 2010 outburst was detected in its
  earliest stage.
  Since the start of the 2009 superoutburst was not well constrained,
  we shifted the $O-C$ diagram to best fit [in this and many other cases
  involving shifting $O-C$ diagrams, we adjusted the location of
  the ``break'' (stage B--C transition) in the $O-C$ diagram]
  the best-recorded 2010 one.
  }
  \label{fig:bgaricomp}
\end{figure}

   A comparison of $O-C$ diagrams is shown in figure \ref{fig:bgaricomp}.

\begin{table}
\caption{Superhump maxima of BG Ari (2010).}\label{tab:bgarioc2010}
\begin{center}
\begin{tabular}{ccccc}
\hline
$E$ & max\commenta & error & $O-C$\commentb & $N$\commentc \\
\hline
0 & 55536.6520 & 0.0014 & $-$0.0391 & 79 \\
1 & 55536.7314 & 0.0014 & $-$0.0446 & 70 \\
2 & 55536.8151 & 0.0018 & $-$0.0459 & 64 \\
19 & 55538.3106 & 0.0005 & 0.0051 & 87 \\
20 & 55538.3996 & 0.0007 & 0.0090 & 89 \\
21 & 55538.4834 & 0.0005 & 0.0079 & 87 \\
22 & 55538.5715 & 0.0004 & 0.0110 & 134 \\
23 & 55538.6562 & 0.0004 & 0.0107 & 69 \\
32 & 55539.4208 & 0.0003 & 0.0106 & 92 \\
33 & 55539.5055 & 0.0003 & 0.0104 & 80 \\
35 & 55539.6763 & 0.0003 & 0.0112 & 69 \\
47 & 55540.6960 & 0.0006 & 0.0112 & 76 \\
48 & 55540.7813 & 0.0007 & 0.0115 & 58 \\
55 & 55541.3738 & 0.0002 & 0.0092 & 91 \\
56 & 55541.4574 & 0.0002 & 0.0078 & 56 \\
66 & 55542.3060 & 0.0004 & 0.0067 & 54 \\
67 & 55542.3904 & 0.0005 & 0.0061 & 53 \\
68 & 55542.4753 & 0.0004 & 0.0061 & 56 \\
78 & 55543.3267 & 0.0006 & 0.0078 & 83 \\
91 & 55544.4323 & 0.0004 & 0.0087 & 56 \\
102 & 55545.3683 & 0.0027 & 0.0100 & 29 \\
148 & 55549.2775 & 0.0007 & 0.0105 & 48 \\
175 & 55551.5617 & 0.0013 & 0.0004 & 52 \\
176 & 55551.6495 & 0.0025 & 0.0032 & 41 \\
187 & 55552.5737 & 0.0017 & $-$0.0072 & 54 \\
211 & 55554.6103 & 0.0014 & $-$0.0100 & 51 \\
221 & 55555.4415 & 0.0017 & $-$0.0286 & 82 \\
\hline
  \multicolumn{5}{l}{\commenta BJD$-$2400000.} \\
  \multicolumn{5}{l}{\commentb Against max $= 2455536.6911 + 0.084973 E$.} \\
  \multicolumn{5}{l}{\commentc Number of points used to determine the maximum.} \\
\end{tabular}
\end{center}
\end{table}

\subsection{V496 Aurigae}\label{obj:v496aur}

   V496 Aur is a dwarf nova discovered by \citet{qiu97v496auriauc6758},
reporting the maximum unfiltered CCD magnitude of 15.7, and the absence of
the quiescent counterpart on the Palomer Observatory Sky Survey plates.
Based on spectroscopy during outburst, they confirmed the dwarf nova-type
classification.
\citet{wei98v496auriauc7068} reported another outburst in 1998.
Unpublished photometry by one of the authors (T. Kato) during this 1998
outburst did not yield significant superhumps.

\begin{figure}
  \begin{center}
    \FigureFile(88mm,110mm){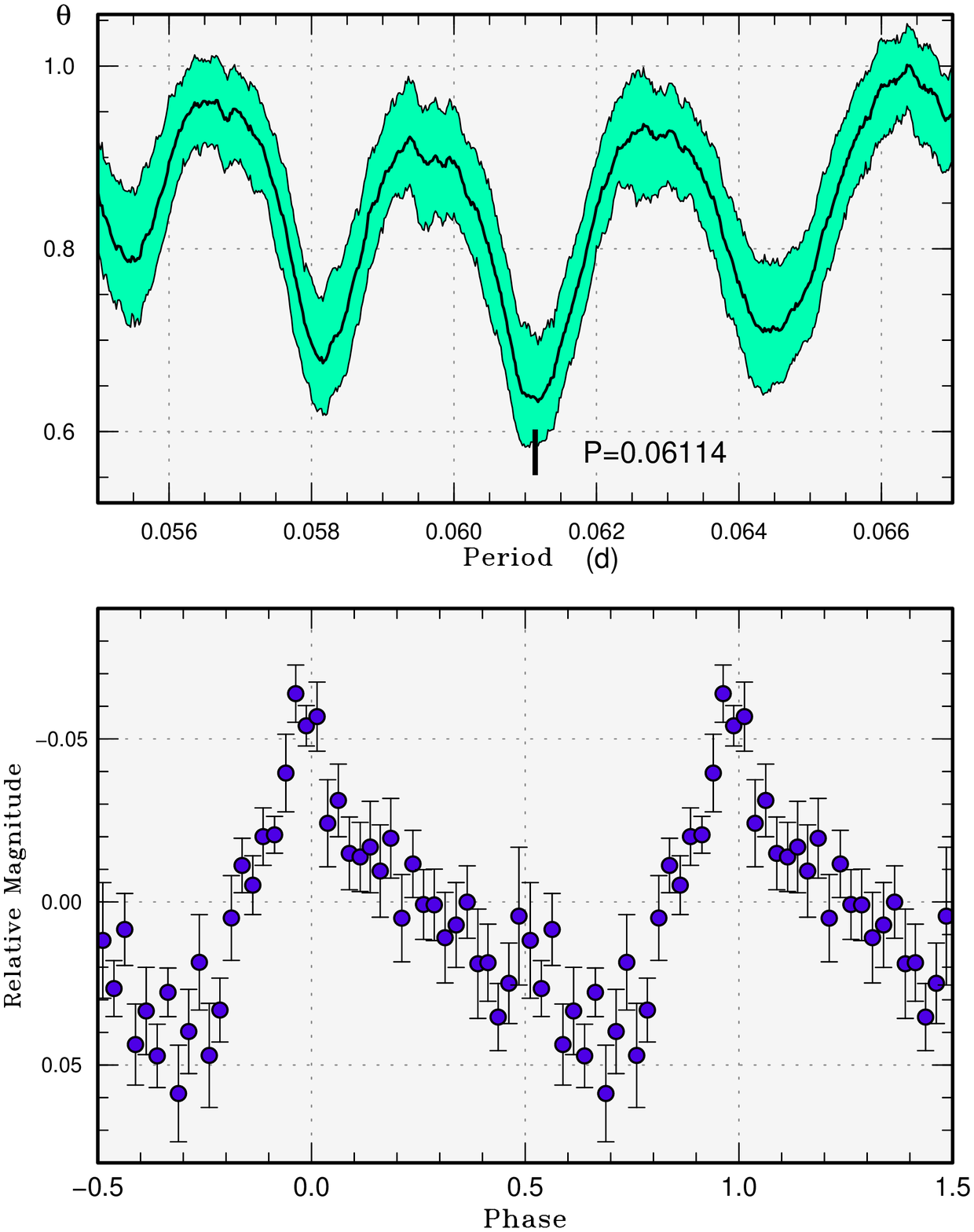}
  \end{center}
  \caption{Superhumps in V496 Aur (2010). (Upper): PDM analysis.
     (Lower): Phase-averaged profile.}
  \label{fig:v496aurshpdm}
\end{figure}

   The 2010 outburst of this object was by chance detected by the Catalina
Real-time Transient Survey
(CRTS, \cite{CRTS}).\footnote{
   $<$http://nesssi.cacr.caltech.edu/catalina/$>$.
   For the information of the individual Catalina CVs, see
   $<$http://nesssi.cacr.caltech.edu/catalina/AllCV.html$>$.
}
Subsequent observations confirmed the presence of superhumps
(vsnet-alert 12430, 12439; figure \ref{fig:v496aurshpdm}).
The times of superhump maxima are listed in table \ref{tab:v496auroc2010}.
The relatively small amplitudes might suggest that the object had
already entered the late stage (possibly stage C) of the superoutburst.
Observations taken four days later (December 5) did not detect
significant superhump signals, and were excluded from this analysis.
The improved astrometry by the CRTS is
\timeform{07h 27m 52.23s}, \timeform{+40D 46' 52.5''} (J2000.0).

\begin{table}
\caption{Superhump maxima of V496 Aur (2010).}\label{tab:v496auroc2010}
\begin{center}
\begin{tabular}{ccccc}
\hline
$E$ & max\commenta & error & $O-C$\commentb & $N$\commentc \\
\hline
0 & 55530.4937 & 0.0011 & 0.0007 & 60 \\
1 & 55530.5521 & 0.0012 & $-$0.0021 & 67 \\
2 & 55530.6167 & 0.0012 & 0.0014 & 52 \\
19 & 55531.6556 & 0.0012 & 0.0005 & 48 \\
20 & 55531.7158 & 0.0015 & $-$0.0005 & 61 \\
\hline
  \multicolumn{5}{l}{\commenta BJD$-$2400000.} \\
  \multicolumn{5}{l}{\commentb Against max $= 2455530.4930 + 0.061162 E$.} \\
  \multicolumn{5}{l}{\commentc Number of points used to determine the maximum.} \\
\end{tabular}
\end{center}
\end{table}

\subsection{TT Bootis}\label{obj:ttboo}

\begin{figure}
  \begin{center}
    \FigureFile(88mm,70mm){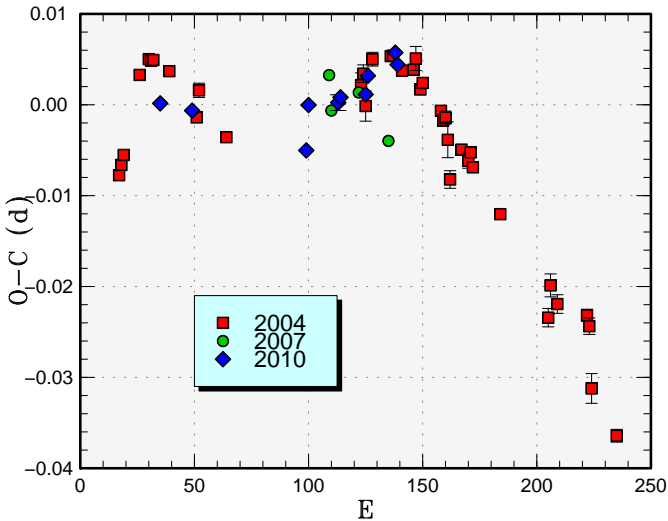}
  \end{center}
  \caption{Comparison of $O-C$ diagrams of TT Boo between different
  superoutbursts.  A period of 0.07807 d was used to draw this figure.
  Approximate cycle counts ($E$) after the start of the superoutburst
  were used.  Since the starts of the 2010 superoutburst
  was not well constrained, we shifted the $O-C$ diagrams
  to best fit the best-recorded 2004 one.
  }
  \label{fig:ttboocomp}
\end{figure}

   We report on two more superoutbursts in 2007 and 2010
(tables \ref{tab:ttboooc2007} and \ref{tab:ttboooc2010}).
The 2007 observations were performed during the middle stage of the
superoutburst.
The 2010 observations fairly well covered the superoutburst except
the final stage.  The resultant $P_{\rm dot}$ of $+5.4(1.4) \times 10^{-5}$
confirms the rather large positive $P_{\rm dot}$ for this $P_{\rm SH}$
or $P_{\rm orb}$ \citep{Pdot}.
A comparison of $O-C$ diagrams is presented in figure \ref{fig:ttboocomp}.
The figure might suggest that a stage B--C transition already occurred
before the 2007 observation.

\begin{table}
\caption{Superhump maxima of TT Boo (2007).}\label{tab:ttboooc2007}
\begin{center}
\begin{tabular}{ccccc}
\hline
$E$ & max\commenta & error & $O-C$\commentb & $N$\commentc \\
\hline
0 & 54200.1584 & 0.0006 & 0.0013 & 168 \\
1 & 54200.2326 & 0.0006 & $-$0.0024 & 98 \\
13 & 54201.1714 & 0.0005 & 0.0019 & 167 \\
26 & 54202.1810 & 0.0005 & $-$0.0009 & 168 \\
\hline
  \multicolumn{5}{l}{\commenta BJD$-$2400000.} \\
  \multicolumn{5}{l}{\commentb Against max $= 2454200.1571 + 0.077876 E$.} \\
  \multicolumn{5}{l}{\commentc Number of points used to determine the maximum.} \\
\end{tabular}
\end{center}
\end{table}

\begin{table}
\caption{Superhump maxima of TT Boo (2010).}\label{tab:ttboooc2010}
\begin{center}
\begin{tabular}{ccccc}
\hline
$E$ & max\commenta & error & $O-C$\commentb & $N$\commentc \\
\hline
0 & 55311.0879 & 0.0001 & 0.0022 & 232 \\
14 & 55312.1801 & 0.0003 & 0.0008 & 161 \\
64 & 55316.0792 & 0.0008 & $-$0.0058 & 140 \\
65 & 55316.1623 & 0.0007 & $-$0.0009 & 163 \\
78 & 55317.1774 & 0.0009 & $-$0.0012 & 146 \\
79 & 55317.2561 & 0.0006 & $-$0.0006 & 152 \\
90 & 55318.1152 & 0.0006 & $-$0.0008 & 152 \\
91 & 55318.1953 & 0.0009 & 0.0012 & 154 \\
103 & 55319.1347 & 0.0007 & 0.0032 & 148 \\
104 & 55319.2114 & 0.0006 & 0.0018 & 156 \\
\hline
  \multicolumn{5}{l}{\commenta BJD$-$2400000.} \\
  \multicolumn{5}{l}{\commentb Against max $= 2455311.0857 + 0.078115 E$.} \\
  \multicolumn{5}{l}{\commentc Number of points used to determine the maximum.} \\
\end{tabular}
\end{center}
\end{table}

\subsection{GX Cassiopeiae}\label{obj:gxcas}

   Although GX Cas was observed in multiple instances \citep{Pdot},
no significant variation of period was recorded, mostly due to the short
baseline of observations.

\begin{figure}
  \begin{center}
    \FigureFile(88mm,110mm){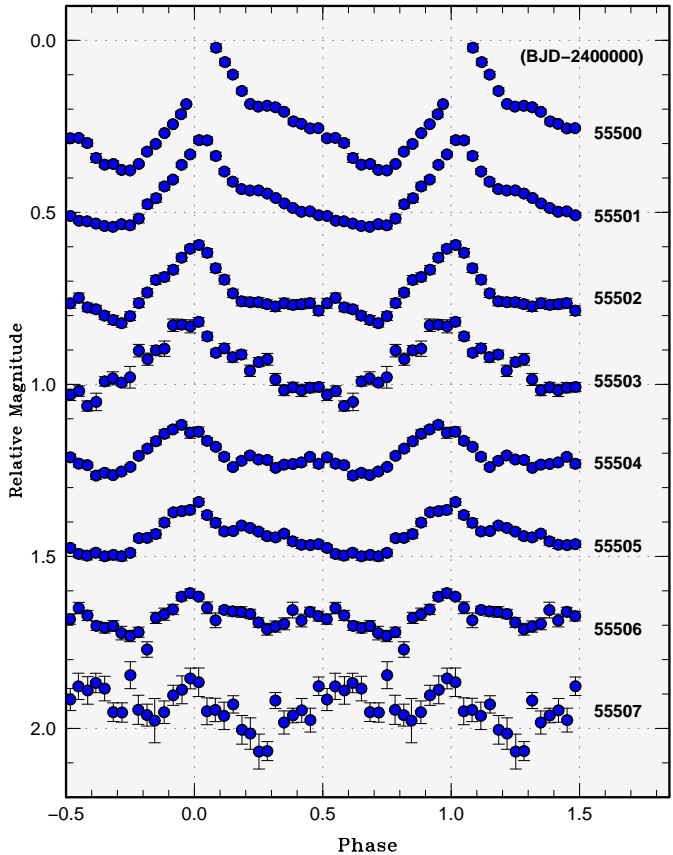}
  \end{center}
  \caption{Superhump profiles of GX Cas (2010).
     The positive $P_{\rm dot}$ is also evident from these phase-averaged
     nightly light curves except for the final night during the rapid
     fading stage.
     The figure was drawn against a mean period of 0.092959 d.}
  \label{fig:gxcasprof}
\end{figure}

\begin{figure}
  \begin{center}
    \FigureFile(88mm,70mm){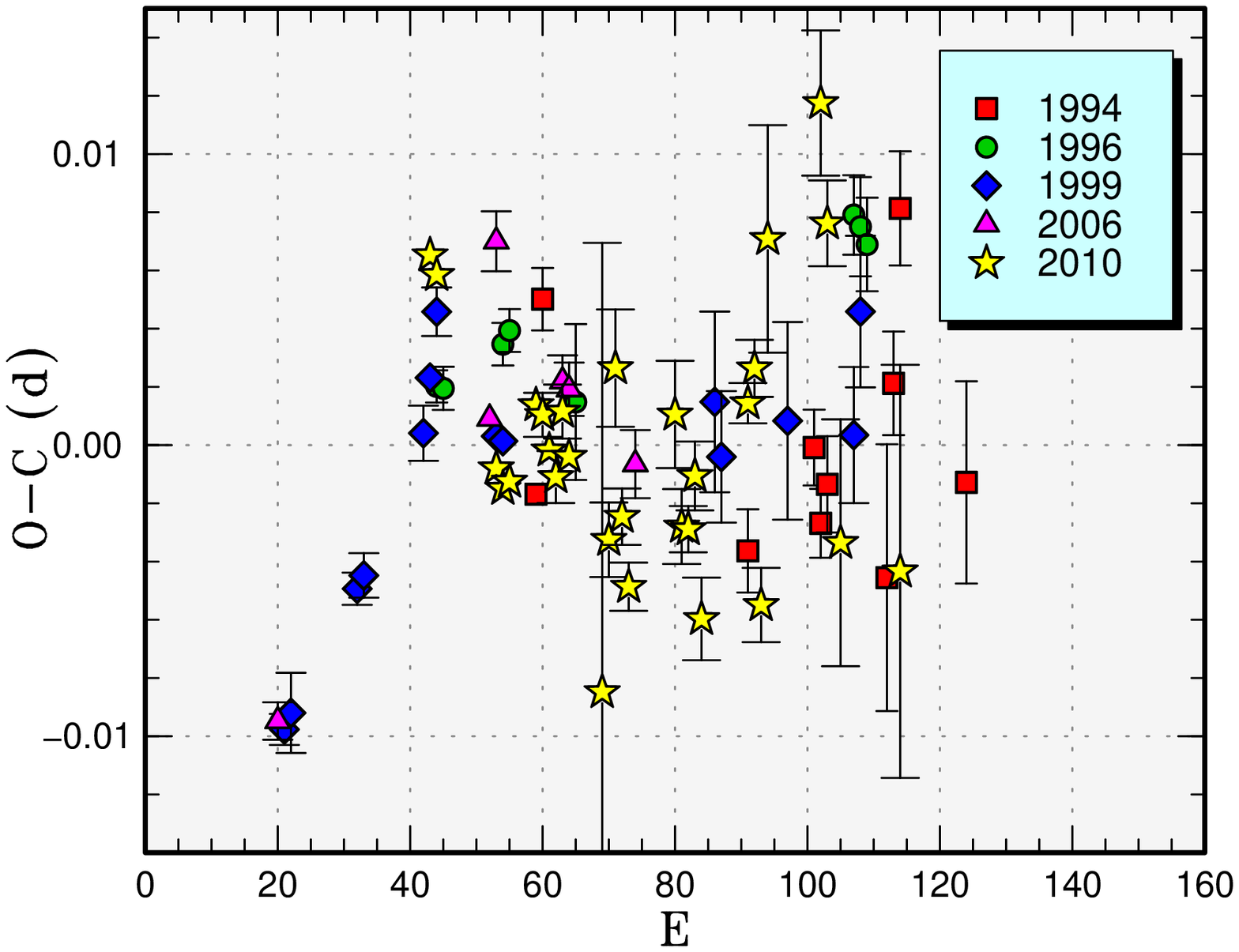}
  \end{center}
  \caption{Comparison of $O-C$ diagrams of GX Cas between different
  superoutbursts.  A period of 0.092959 d was used to draw this figure.
  Approximate cycle counts ($E$) after the start of the superoutburst
  were used.  Since the starts of the 1994 and 2011 superoutbursts
  was not well constrained, we shifted the $O-C$ diagrams
  to best fit the best-recorded 2010 one.
  }
  \label{fig:gxcascomp2}
\end{figure}

   During the 2010 superoutburst, we finally obtained a relatively dense
set of observations during the later half of the superoutburst.
The times of superhump maxima are listed in table \ref{tab:gxcasoc2010}.
Surprisingly, they suggest a positive $P_{\rm dot}$ of
$+25(5) \times 10^{-5}$ for $E \le 60$ (see also figure
\ref{fig:gxcasprof}; the conclusion is unchanged
even using a rather discrepant point at $E=62$; $E=71$ corresponds to
a point during the rapid decline phase).
A combined $O-C$ diagram (figure \ref{fig:gxcascomp2}) also seems
to support a general trend of a positive $P_{\rm dot}$.

   As discussed in \citet{Pdot} (subsection 4.10), most
long-$P_{\rm orb}$ objects generally show relatively large negative
$P_{\rm dot}$, while some objects (mostly rarely outbursting objects) have
nearly zero $P_{\rm dot}$.  GX Cas is clearly an exception.
Although this conclusion may have been affected by the lack of observations
during the early stage of the superoutburst, the object will be worth
studying in more details.
Being a bright object, the object would be a good target for
a radial-velocity study in calibrating $P_{\rm orb}$--$\epsilon$ relation
in long-period systems.

\begin{table}
\caption{Superhump maxima of GX Cas (2010).}\label{tab:gxcasoc2010}
\begin{center}
\begin{tabular}{ccccc}
\hline
$E$ & max\commenta & error & $O-C$\commentb & $N$\commentc \\
\hline
0 & 55500.5225 & 0.0004 & 0.0065 & 154 \\
1 & 55500.6148 & 0.0003 & 0.0058 & 196 \\
10 & 55501.4448 & 0.0003 & $-$0.0008 & 203 \\
11 & 55501.5370 & 0.0003 & $-$0.0015 & 202 \\
12 & 55501.6303 & 0.0005 & $-$0.0013 & 206 \\
16 & 55502.0047 & 0.0005 & 0.0014 & 285 \\
17 & 55502.0973 & 0.0008 & 0.0010 & 250 \\
18 & 55502.1891 & 0.0006 & $-$0.0002 & 217 \\
19 & 55502.2811 & 0.0009 & $-$0.0011 & 230 \\
20 & 55502.3763 & 0.0009 & 0.0011 & 208 \\
21 & 55502.4677 & 0.0004 & $-$0.0004 & 207 \\
26 & 55502.9245 & 0.0154 & $-$0.0085 & 59 \\
27 & 55503.0226 & 0.0013 & $-$0.0033 & 158 \\
28 & 55503.1215 & 0.0020 & 0.0026 & 234 \\
29 & 55503.2094 & 0.0010 & $-$0.0025 & 112 \\
30 & 55503.2999 & 0.0008 & $-$0.0049 & 135 \\
37 & 55503.9565 & 0.0018 & 0.0011 & 181 \\
38 & 55504.0456 & 0.0013 & $-$0.0028 & 273 \\
39 & 55504.1385 & 0.0008 & $-$0.0029 & 271 \\
40 & 55504.2333 & 0.0012 & $-$0.0011 & 130 \\
41 & 55504.3213 & 0.0014 & $-$0.0060 & 108 \\
48 & 55504.9795 & 0.0007 & 0.0014 & 205 \\
49 & 55505.0736 & 0.0010 & 0.0026 & 219 \\
50 & 55505.1585 & 0.0013 & $-$0.0055 & 141 \\
51 & 55505.2640 & 0.0039 & 0.0071 & 99 \\
59 & 55506.0123 & 0.0025 & 0.0118 & 209 \\
60 & 55506.1012 & 0.0015 & 0.0076 & 206 \\
62 & 55506.2761 & 0.0042 & $-$0.0033 & 100 \\
71 & 55507.1118 & 0.0071 & $-$0.0043 & 119 \\
\hline
  \multicolumn{5}{l}{\commenta BJD$-$2400000.} \\
  \multicolumn{5}{l}{\commentb Against max $= 2455500.5160 + 0.092959 E$.} \\
  \multicolumn{5}{l}{\commentc Number of points used to determine the maximum.} \\
\end{tabular}
\end{center}
\end{table}

\subsection{HT Cassiopeiae}\label{obj:htcas}

   HT Cas is renowned for its deep eclipses and for its historical
superoutburst in 1985.  There were only fragmentary observations of superhumps
during the 1985 superoutburst (\cite{zha86htcas}; reanalyzed by \cite{Pdot}).
Although there were known normal outbursts (1987, 1989, 1995, 1997, 1998,
1999, 2002, 2008) since then, there have been no confirmed superoutburst.
The 1995 normal outburst was particularly well observed \citep{ioa99htcas}.

\begin{figure*}
  \begin{center}
    \FigureFile(160mm,190mm){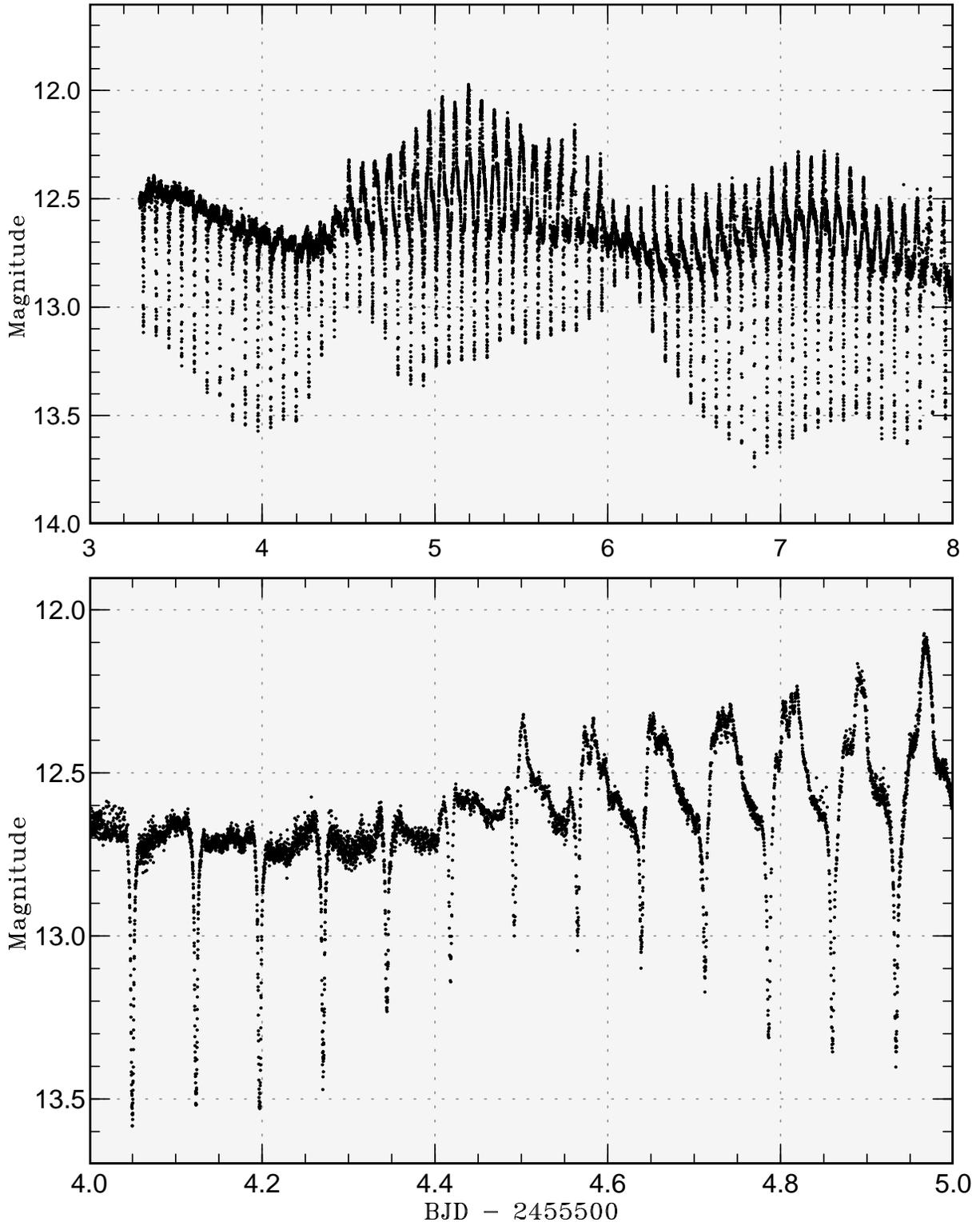}
  \end{center}
  \caption{Superhumps in HT Cas (2010) at the early stage.
     (Upper): Evolution of superhumps and beat phenomenon.
     (Lower): Enlargement of the evolutionary phase of superhumps.}
  \label{fig:htcasearly}
\end{figure*}

\begin{figure}
  \begin{center}
    \FigureFile(88mm,110mm){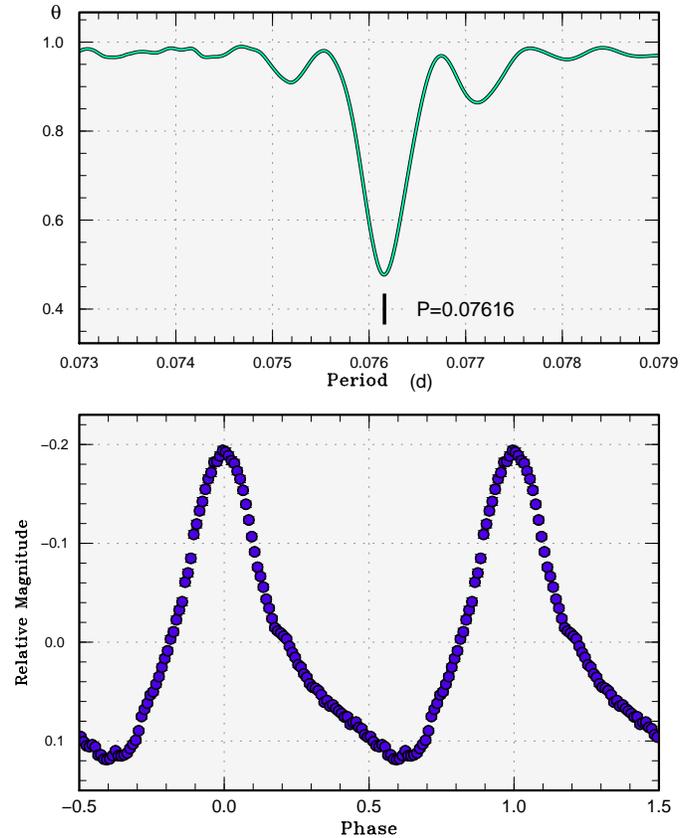}
  \end{center}
  \caption{Superhumps in HT Cas (2010) during the superoutburst plateau.
     The averaged profile outside eclipses is shown.
     (Upper): PDM analysis.
     (Lower): Phase-averaged profile.}
  \label{fig:htcasshpdm}
\end{figure}

   The 2010 superoutburst was detected by T. Parsons on November 2.43 UT
at a visual magnitude of 12.9 (cvnet-discussion 1404).  The object
further brightened, and growing superhumps were detected $\sim$1.5 d
later (vsnet-alert 12353, 12359; figure \ref{fig:htcasearly}).
The superhumps (figure \ref{fig:htcasshpdm}) quickly developed to
the full amplitude and slowly decayed.

\begin{figure}
  \begin{center}
    \FigureFile(88mm,110mm){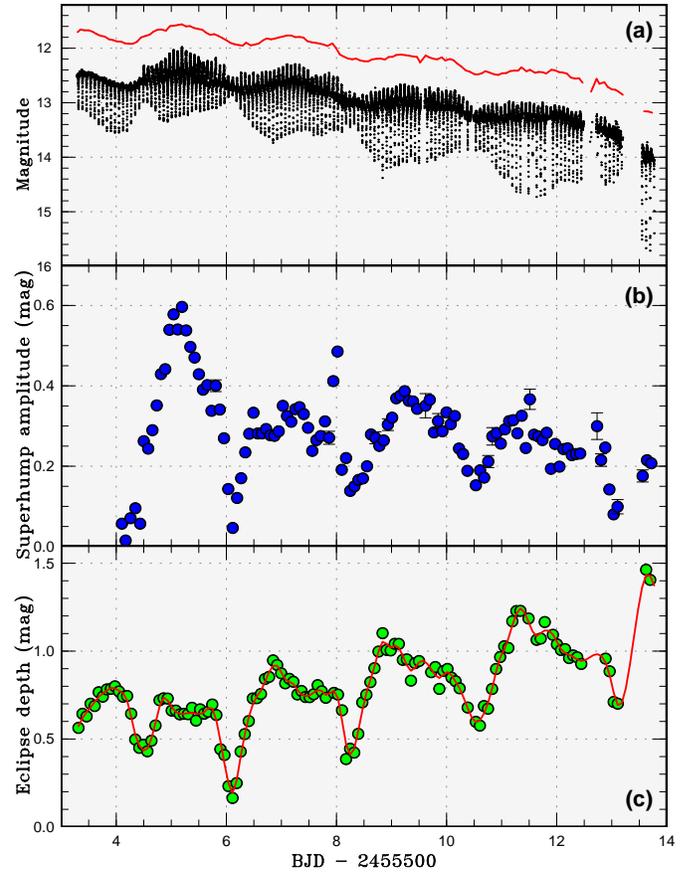}
  \end{center}
  \caption{Beat phenomenon in HT Cas (2010).
    (a) Light curve.  The solid curve represent averaged magnitudes
    outside eclipses (offset by 0.8 mag).
    (b) Amplitude of superhumps.
    (c) Depth of eclipses.  The curve represent the globally smoothed
    depths of eclipses used in subtracting the orbital variation.
  }
  \label{fig:htcasbeat}
\end{figure}

   Due to the overlapping eclipses and very strong effect of beat phenomenon
(figure \ref{fig:htcasbeat}),
we first subtracted the template orbital variation
(cf. figure \ref{fig:htcasorbprof}) scaled by fitting
individual eclipses (the results were globally smoothed;
cf. figure \ref{fig:htcasbeat}c) and removed the
part (within 0.11 $P_{\rm orb}$ of eclipses) most strongly affected by
eclipses, and applied the usual method of fitting maxima.
The ephemeris of eclipse used in this procedure is given in equation
\ref{equ:htcasecl}, which were determined by one of the authors (H. Maehara)
using published and newly obtained eclipse timings since
2009 January 1.\footnote{
   Since the orbital period of HT Cas is known to vary significantly
   (e.g. \cite{bor08htcas}), we only used recent timing data for calculating
   the orbital phase.  This ephemeris is not designed for long-term
   prediction of eclipses.
}

\begin{equation}
{\rm Min(BJD)} = 2443727.93823(20) + 0.0736471875(13) E
\label{equ:htcasecl}.
\end{equation}

\begin{figure}
  \begin{center}
    \FigureFile(88mm,110mm){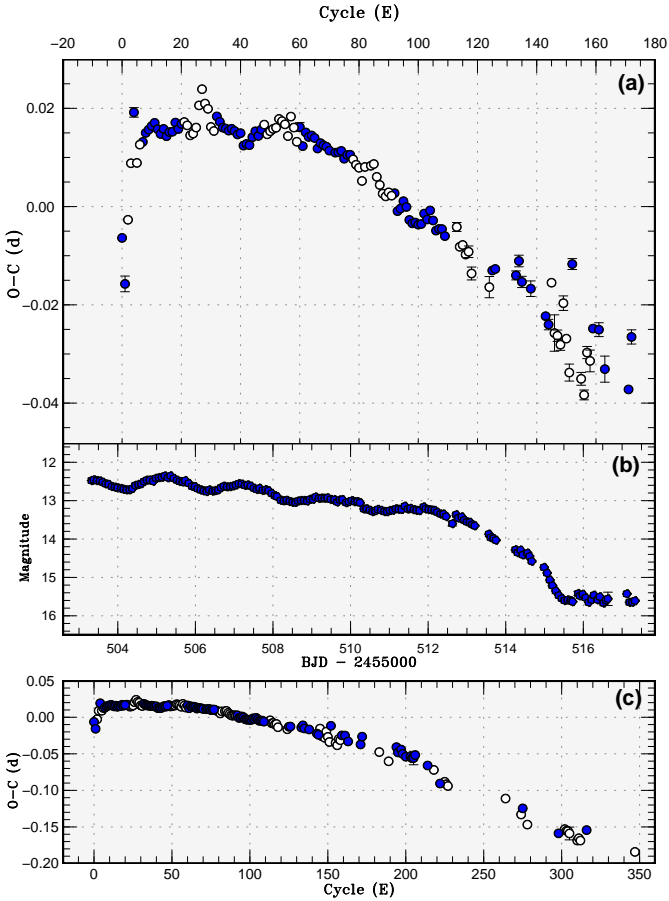}
  \end{center}
  \caption{$O-C$ diagram of superhumps in HT Cas (2010).
     (a) $O-C$ diagram indicating stage A--C evolution.
     Open circle indicate humps within 0.2 orbital phase of eclipses.
     Filled circles are humps outside the phase of eclipses.
     We used a period of 0.07635 d for calculating the $O-C$'s.
     (b) Light curve.  Each point represents an orbital average of
     magnitudes outside the eclipses.  Strong beat modulations are present.
     (c) $O-C$ diagram of the entire observation.
     Superhumps with a slightly shorter period persisted even after
     panel (a).  There was no phase jump as expected for ``traditional''
     late superhumps.
  }
  \label{fig:htcashumpall}
\end{figure}

\begin{figure}
  \begin{center}
    \FigureFile(88mm,172mm){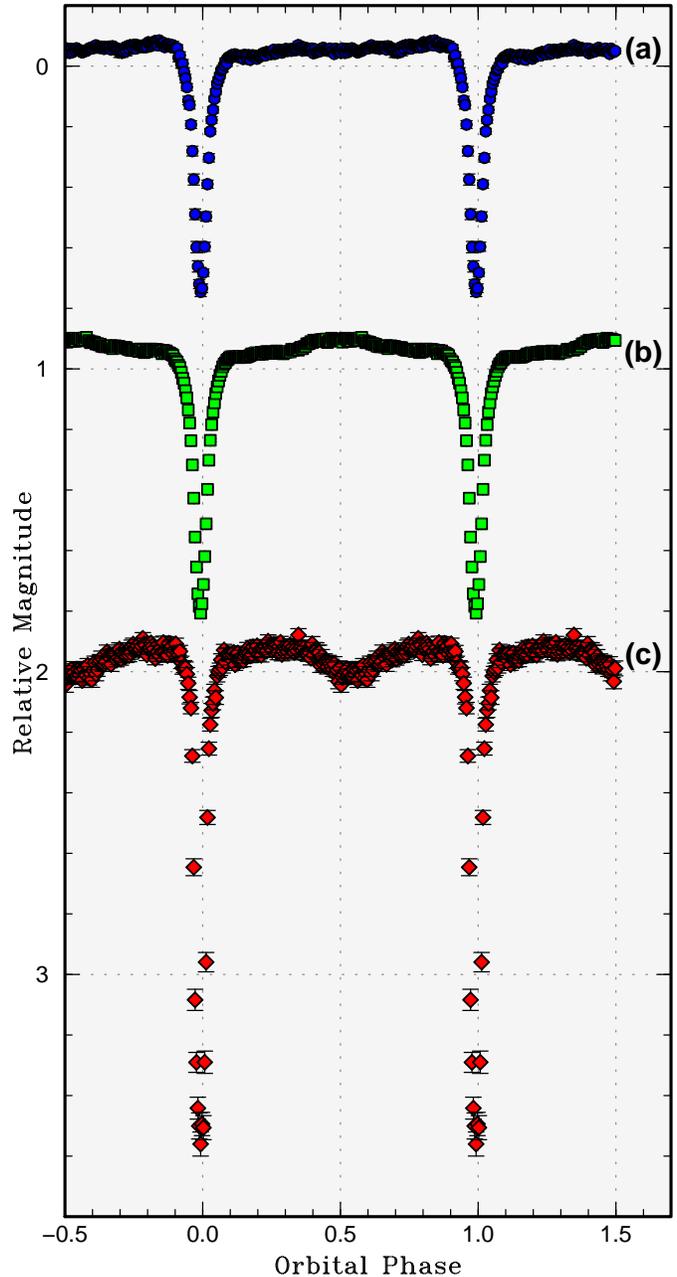}
  \end{center}
  \caption{Variation of orbital profile HT Cas (2010).
     The phases were defined against equation \ref{equ:htcasecl}.
     (a) Before superhumps appeared.  Although likely orbital humps were
     present, no indication of early superhump was present.
     (b) During the superoutburst plateau.  There was a slight bump
     around phase 0.5 (see text for details).
     (c) Post-superoutburst.  Eclipses became deeper and double-wave
     modulations were present.
  }
  \label{fig:htcasorbprof}
\end{figure}

Although this procedure lacks the detailed treatment of variations in
eclipse profiles, it will be sufficient for the present purpose of
measuring the global behavior of superhumps.
The times of superhump maxima are listed in table \ref{tab:htcasoc2010}.
Three stages of A--C are apparently present (figure \ref{fig:htcashumpall}).
Stage A and the middle of
stage B were very strongly affected by overlapping eclipses.
By rejecting the most severely affected maxima ($26 \le E \le 28$),
we obtained an almost zero $+3.0(3.6) \times 10^{-5}$ period derivative
for stage B ($12 \le E \le 58$).  The other values are listed in
table \ref{tab:perlist}.
As judged from the recorded period, \citet{zha86htcas} seems to have
observed stage C superhumps.

   The resultant period variation was generally normal for an SU UMa-type
dwarf nova with $P_{\rm orb}$ of 0.07365 d, despite the long lack of
superoutbursts.  There were no detectable features of early superhumps,
which likely excludes the WZ Sge-like phenomenon suggested for the 1985
superoutburst (cf. \cite{kat01hvvir}).  The very large amplitude of
the outburst reported during the 1985 outburst may have resulted
from an old AAVSO scale of comparison stars, combined with strong
orbital/superhump modulations.

   Even after the superoutburst, superhumps with slightly shorter periods
persisted at least for 16 d (table \ref{tab:htcasoc2010post}; maxima
were measured in the same way as in the superoutburst).
There was no phase jump following the stage C superhumps
(figure \ref{fig:htcashumpall}a,c), and these
superhumps are not considered to be ``traditional" late superhumps
(see a discussion in \cite{Pdot}, subsection 4.2).

   Before superhumps appeared, there was likely an indication of
orbital humps (figure \ref{fig:htcasorbprof}a).
The orbital humps were not evident during the plateau phase of the
superoutburst, and there was a slight bump around orbital phase 0.5,
which might look like a reflection effect on the secondary
(figure \ref{fig:htcasorbprof}b).
The post-superoutburst phase was dominated by double-wave modulations,
which probably arise from ellipsoidal variation of the secondary
(figure \ref{fig:htcasorbprof}c).  No evident orbital humps were
detected during this phase.

\begin{figure}
  \begin{center}
    \FigureFile(88mm,110mm){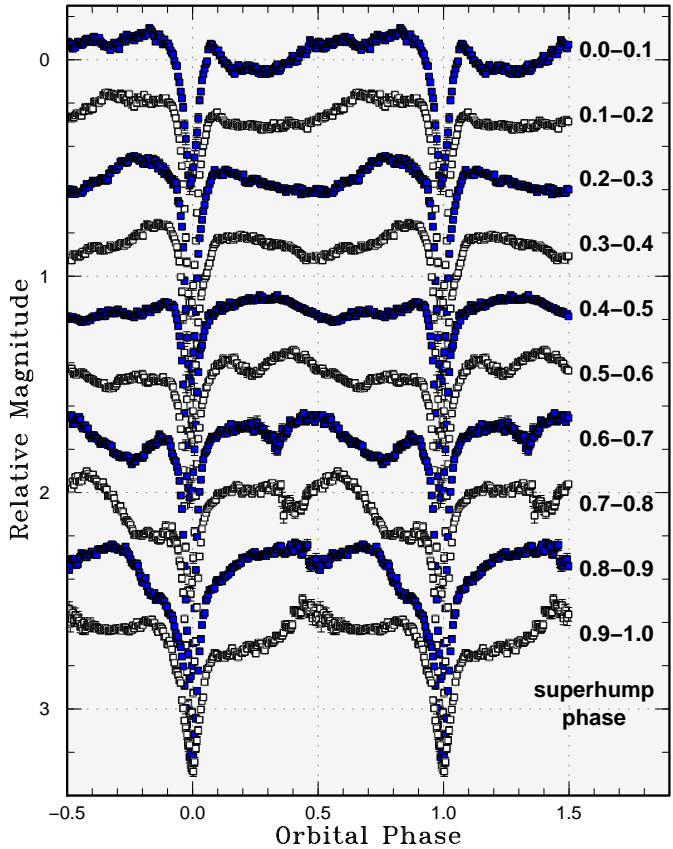}
  \end{center}
  \caption{Dependence of orbital profile HT Cas on superhump phase
     during the plateau phase of the superoutburst (2010).
     The data approximately covered three beat periods.
     There is no strong indication of persistent orbital humps
     nor reflection effect.  The profiles outside the eclipses are highly
     dependent on superhump phase and highly structured.
  }
  \label{fig:htcasorbdep}
\end{figure}

\begin{figure}
  \begin{center}
    \FigureFile(88mm,110mm){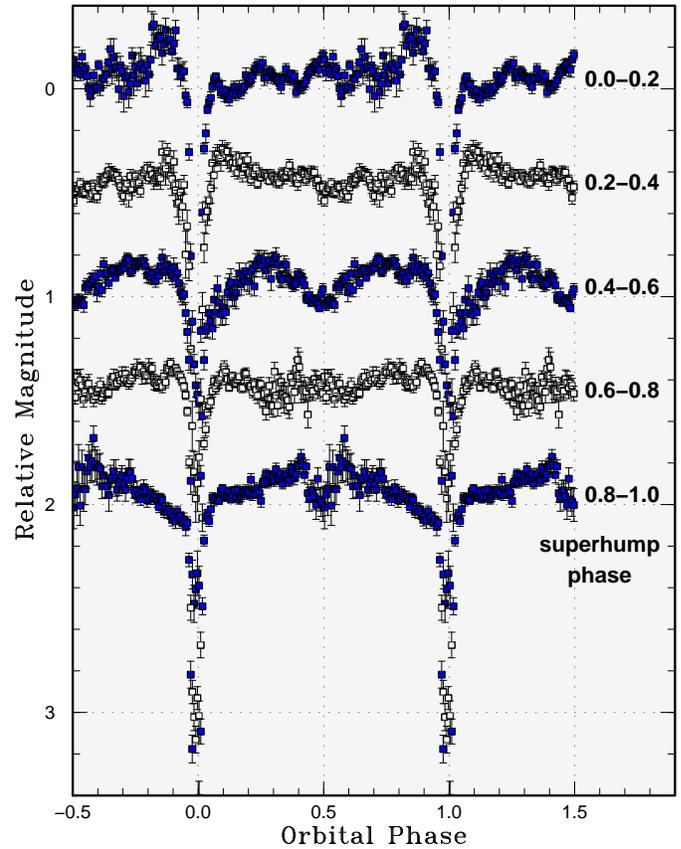}
  \end{center}
  \caption{Dependence of orbital profile HT Cas on superhump phase
     after the rapid fading of the superoutburst (2010).
     The features outside the eclipses have a similar tendency to
     figure \ref{fig:htcasorbdep}.
  }
  \label{fig:htcasorbdeplate}
\end{figure}

   Figure \ref{fig:htcasorbdep} shows a more detailed dependence of
orbital profiles during the plateau phase on the superhump phase 
($\phi_{\rm SH}$).
The superhump phases were defined so that $\phi_{\rm SH}=0$ corresponds
to superhump maxima.  Although there was a feature resembling
orbital humps for $0 \le \phi_{\rm SH} \le 0.4$, this feature was not
persistently present at other $\phi_{\rm SH}$.
The phase 0.5 bump was not clearly present independent of $\phi_{\rm SH}$,
and this bump (in averaged profile) appears to have originated from
complex features outside the eclipses for $0.7 \le \phi_{\rm SH} \le 1.0$.
Since these features have different profiles and times of maxima for
different $\phi_{\rm SH}$, we regard it less likely that they arise from
the reflection effect on the secondary (cf. \cite{sma11suumareflection}).
These features more likely represent the geometrical effect arising from
the non-axisymmetrically extended vertical structure resulting from
the heating by the superhump source.  Figure \ref{fig:htcasorbdeplate}
illustrate the same dependence after the rapid decline from
the superoutburst (BJD 2455514--2455522).  Although the general tendency
of variations outside the eclipses were similar to those during the
plateau phase, eclipses became sharper for $0.8 \le \phi_{\rm SH} \le 1.0$
and a reflection-like effect became prominent.

\begin{table}
\caption{Superhump maxima of HT Cas (2010).}\label{tab:htcasoc2010}
\begin{center}
\begin{tabular}{cccccc}
\hline
$E$ & max\commenta & error & $O-C$\commentb & phase\commentc & $N$\commentd \\
\hline
0 & 55504.1036 & 0.0004 & $-$0.0263 & 0.72 & 1374 \\
1 & 55504.1706 & 0.0016 & $-$0.0355 & 0.63 & 1133 \\
2 & 55504.2600 & 0.0006 & $-$0.0223 & 0.85 & 668 \\
3 & 55504.3479 & 0.0004 & $-$0.0106 & 0.04 & 587 \\
4 & 55504.4346 & 0.0010 & $-$0.0000 & 0.22 & 281 \\
5 & 55504.5006 & 0.0003 & $-$0.0101 & 0.11 & 188 \\
6 & 55504.5807 & 0.0004 & $-$0.0062 & 0.20 & 335 \\
\end{tabular}
\end{center}
\end{table}

\addtocounter{table}{-1}
\begin{table}
\caption{Superhump maxima of HT Cas (2010) (continued).}
\begin{center}
\begin{tabular}{cccccc}
\hline
$E$ & max\commenta & error & $O-C$\commentb & phase\commentc & $N$\commentd \\
\hline
53 & 55508.1744 & 0.0004 & 0.0079 & 1.00 & 238 \\
54 & 55508.2503 & 0.0006 & 0.0077 & 0.03 & 359 \\
55 & 55508.3260 & 0.0003 & 0.0072 & 0.05 & 541 \\
56 & 55508.4000 & 0.0002 & 0.0050 & 0.06 & 535 \\
57 & 55508.4803 & 0.0004 & 0.0092 & 0.15 & 528 \\
58 & 55508.5544 & 0.0003 & 0.0071 & 0.16 & 634 \\
59 & 55508.6278 & 0.0003 & 0.0044 & 0.15 & 504 \\
\end{tabular}
\end{center}
\end{table}

\addtocounter{table}{-1}
\begin{table}
\caption{Superhump maxima of HT Cas (2010) (continued).}
\begin{center}
\begin{tabular}{cccccc}
\hline
$E$ & max\commenta & error & $O-C$\commentb & phase\commentc & $N$\commentd \\
\hline
108 & 55512.3512 & 0.0003 & $-$0.0040 & 0.71 & 427 \\
109 & 55512.4261 & 0.0002 & $-$0.0053 & 0.73 & 275 \\
113 & 55512.7334 & 0.0009 & $-$0.0026 & 0.90 & 37 \\
114 & 55512.8057 & 0.0008 & $-$0.0065 & 0.88 & 54 \\
115 & 55512.8824 & 0.0003 & $-$0.0059 & 0.92 & 470 \\
116 & 55512.9568 & 0.0005 & $-$0.0077 & 0.93 & 802 \\
117 & 55513.0338 & 0.0012 & $-$0.0069 & 0.98 & 582 \\
118 & 55513.1057 & 0.0013 & $-$0.0112 & 0.95 & 425 \\
124 & 55513.5610 & 0.0022 & $-$0.0128 & 0.14 & 130 \\
125 & 55513.6407 & 0.0007 & $-$0.0092 & 0.22 & 163 \\
126 & 55513.7174 & 0.0007 & $-$0.0087 & 0.26 & 94 \\
\hline
  \multicolumn{6}{l}{\commenta BJD$-$2400000.} \\
  \multicolumn{6}{l}{\commentb Against max $= 2455504.1300 + 0.076160 E$.} \\
  \multicolumn{6}{l}{\commentc Orbital phase.} \\
  \multicolumn{6}{l}{\commentd Number of points used to determine the maximum.} \\
\end{tabular}
\end{center}
\end{table}

\begin{table}
\caption{Superhump maxima of HT Cas (2010) (post-superoutburst).}\label{tab:htcasoc2010post}
\begin{center}
\begin{tabular}{ccccc}
\hline
$E$ & max\commenta & error & $O-C$\commentb & $N$\commentc \\
\hline
0 & 55514.2505 & 0.0009 & $-$0.0074 & 51 \\
1 & 55514.3298 & 0.0012 & $-$0.0036 & 92 \\
2 & 55514.4019 & 0.0011 & $-$0.0070 & 138 \\
5 & 55514.6296 & 0.0016 & $-$0.0058 & 73 \\
10 & 55515.0057 & 0.0008 & $-$0.0071 & 116 \\
11 & 55515.0804 & 0.0011 & $-$0.0080 & 235 \\
12 & 55515.1653 & 0.0006 & 0.0014 & 511 \\
\end{tabular}
\end{center}
\end{table}

\subsection{V1504 Cygni}\label{obj:v1504cyg}

   We report additional superoutburst in 2007 (table \ref{tab:v1504cygoc2007}).
Although the data were not abundant, stages B and C can be recognized.
The determined periods are in good agreement with the following
Kepler observations.
The times of maxima derived from Kepler observations discussed in
subsection \ref{obj:kepv1504cyg} are presented in table
\ref{tab:v1504cygoc2009b}.

\begin{table}
\caption{Superhump maxima of V1504 Cyg (2007).}\label{tab:v1504cygoc2007}
\begin{center}
\begin{tabular}{ccccc}
\hline
$E$ & max\commenta & error & $O-C$\commentb & $N$\commentc \\
\hline
0 & 54324.4787 & 0.0018 & $-$0.0061 & 13 \\
27 & 54326.4283 & 0.0013 & $-$0.0019 & 29 \\
41 & 54327.4339 & 0.0026 & $-$0.0049 & 32 \\
42 & 54327.5122 & 0.0016 & 0.0013 & 33 \\
55 & 54328.4608 & 0.0039 & 0.0133 & 19 \\
84 & 54330.5368 & 0.0017 & $-$0.0002 & 15 \\
96 & 54331.4028 & 0.0016 & 0.0013 & 18 \\
97 & 54331.4754 & 0.0039 & 0.0018 & 19 \\
98 & 54331.5491 & 0.0023 & 0.0034 & 18 \\
110 & 54332.4129 & 0.0020 & 0.0026 & 13 \\
111 & 54332.4855 & 0.0032 & 0.0032 & 13 \\
112 & 54332.5530 & 0.0018 & $-$0.0014 & 11 \\
124 & 54333.4125 & 0.0053 & $-$0.0064 & 13 \\
125 & 54333.4911 & 0.0020 & 0.0000 & 12 \\
126 & 54333.5634 & 0.0025 & 0.0004 & 11 \\
139 & 54334.4933 & 0.0022 & $-$0.0064 & 12 \\
\hline
  \multicolumn{5}{l}{\commenta BJD$-$2400000.} \\
  \multicolumn{5}{l}{\commentb Against max $= 2454324.4848 + 0.072050 E$.} \\
  \multicolumn{5}{l}{\commentc Number of points used to determine the maximum.} \\
\end{tabular}
\end{center}
\end{table}

\begin{table}
\caption{Superhump maxima of V1504 Cyg (2009b).}\label{tab:v1504cygoc2009b}
\begin{center}
\begin{tabular}{ccccc}
\hline
$E$ & max\commenta & error & $O-C$\commentb & $N$\commentc \\
\hline
0 & 55076.2646 & 0.0009 & $-$0.0344 & 84 \\
1 & 55076.3457 & 0.0014 & $-$0.0254 & 84 \\
2 & 55076.4227 & 0.0008 & $-$0.0205 & 84 \\
3 & 55076.4962 & 0.0004 & $-$0.0192 & 84 \\
4 & 55076.5663 & 0.0006 & $-$0.0212 & 85 \\
5 & 55076.6365 & 0.0003 & $-$0.0232 & 85 \\
6 & 55076.7075 & 0.0004 & $-$0.0243 & 85 \\
\end{tabular}
\end{center}
\end{table}

\addtocounter{table}{-1}
\begin{table}
\caption{Superhump maxima of V1504 Cyg (2009b) (continued).}
\begin{center}
\begin{tabular}{ccccc}
\hline
$E$ & max\commenta & error & $O-C$\commentb & $N$\commentc \\
\hline
52 & 55080.0552 & 0.0004 & 0.0049 & 85 \\
53 & 55080.1261 & 0.0004 & 0.0036 & 85 \\
54 & 55080.1989 & 0.0003 & 0.0043 & 85 \\
55 & 55080.2704 & 0.0005 & 0.0037 & 84 \\
56 & 55080.3446 & 0.0005 & 0.0058 & 84 \\
57 & 55080.4193 & 0.0005 & 0.0083 & 84 \\
58 & 55080.4909 & 0.0003 & 0.0078 & 85 \\
\end{tabular}
\end{center}
\end{table}

\addtocounter{table}{-1}
\begin{table}
\caption{Superhump maxima of V1504 Cyg (2009b) (continued).}
\begin{center}
\begin{tabular}{ccccc}
\hline
$E$ & max\commenta & error & $O-C$\commentb & $N$\commentc \\
\hline
101 & 55083.5903 & 0.0003 & 0.0051 & 85 \\
102 & 55083.6613 & 0.0003 & 0.0039 & 74 \\
103 & 55083.7337 & 0.0005 & 0.0042 & 84 \\
104 & 55083.8059 & 0.0004 & 0.0043 & 84 \\
105 & 55083.8770 & 0.0003 & 0.0032 & 84 \\
106 & 55083.9501 & 0.0005 & 0.0042 & 85 \\
107 & 55084.0222 & 0.0004 & 0.0041 & 85 \\
\end{tabular}
\end{center}
\end{table}

\addtocounter{table}{-1}
\begin{table}
\caption{Superhump maxima of V1504 Cyg (2009b) (continued).}
\begin{center}
\begin{tabular}{ccccc}
\hline
$E$ & max\commenta & error & $O-C$\commentb & $N$\commentc \\
\hline
150 & 55087.1146 & 0.0003 & $-$0.0055 & 85 \\
151 & 55087.1876 & 0.0004 & $-$0.0046 & 84 \\
152 & 55087.2567 & 0.0003 & $-$0.0077 & 85 \\
153 & 55087.3278 & 0.0004 & $-$0.0087 & 85 \\
154 & 55087.3972 & 0.0005 & $-$0.0115 & 85 \\
155 & 55087.4706 & 0.0006 & $-$0.0102 & 84 \\
156 & 55087.5428 & 0.0008 & $-$0.0102 & 85 \\
157 & 55087.6146 & 0.0004 & $-$0.0105 & 85 \\
158 & 55087.6897 & 0.0005 & $-$0.0075 & 85 \\
159 & 55087.7620 & 0.0010 & $-$0.0073 & 84 \\
160 & 55087.8269 & 0.0005 & $-$0.0146 & 85 \\
161 & 55087.9015 & 0.0005 & $-$0.0122 & 85 \\
162 & 55087.9788 & 0.0009 & $-$0.0070 & 85 \\
163 & 55088.0412 & 0.0012 & $-$0.0168 & 85 \\
164 & 55088.1033 & 0.0012 & $-$0.0268 & 84 \\
\hline
  \multicolumn{5}{l}{\commenta BJD$-$2400000.} \\
  \multicolumn{5}{l}{\commentb Against max $= 2455076.2990 + 0.072141 E$.} \\
  \multicolumn{5}{l}{\commentc Number of points used to determine the maximum.} \\
\end{tabular}
\end{center}
\end{table}

\subsection{AW Geminorum}\label{obj:awgem}

\begin{figure}
  \begin{center}
    \FigureFile(88mm,70mm){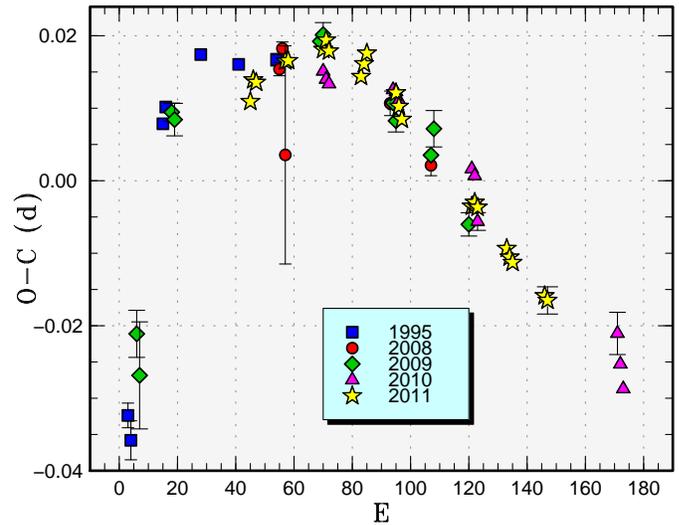}
  \end{center}
  \caption{Comparison of $O-C$ diagrams of AW Gem between different
  superoutbursts.  A period of 0.07915 d was used to draw this figure.
  Approximate cycle counts ($E$) after the start of the
  superoutburst were used.
  Since the start of 2008, 2010 and 2011 superoutburst were poorly constrained,
  we shifted the $O-C$ diagrams to best match other other superoutbursts.
  }
  \label{fig:awgemcomp2}
\end{figure}

   We observed the middle-to-late stage of the 2011 superoutburst.
The stage B--C transition was well-recorded (table \ref{tab:awgemoc2011}).
A comparison of $O-C$ diagrams between different superoutbursts
is shown in figure \ref{fig:awgemcomp2}.

\begin{table}
\caption{Superhump maxima of AW Gem (2011).}\label{tab:awgemoc2011}
\begin{center}
\begin{tabular}{ccccc}
\hline
$E$ & max\commenta & error & $O-C$\commentb & $N$\commentc \\
\hline
0 & 55571.1836 & 0.0002 & $-$0.0112 & 112 \\
1 & 55571.2657 & 0.0002 & $-$0.0079 & 111 \\
2 & 55571.3445 & 0.0005 & $-$0.0079 & 83 \\
12 & 55572.1390 & 0.0003 & $-$0.0017 & 108 \\
13 & 55572.2182 & 0.0003 & $-$0.0013 & 112 \\
25 & 55573.1695 & 0.0003 & 0.0042 & 108 \\
26 & 55573.2500 & 0.0003 & 0.0058 & 111 \\
27 & 55573.3276 & 0.0005 & 0.0046 & 84 \\
38 & 55574.1947 & 0.0003 & 0.0048 & 165 \\
39 & 55574.2757 & 0.0003 & 0.0068 & 168 \\
40 & 55574.3563 & 0.0006 & 0.0086 & 84 \\
50 & 55575.1423 & 0.0003 & 0.0065 & 112 \\
51 & 55575.2196 & 0.0003 & 0.0049 & 112 \\
52 & 55575.2969 & 0.0004 & 0.0035 & 111 \\
76 & 55577.1845 & 0.0006 & $-$0.0006 & 115 \\
77 & 55577.2642 & 0.0006 & 0.0003 & 114 \\
78 & 55577.3427 & 0.0013 & $-$0.0001 & 77 \\
88 & 55578.1285 & 0.0005 & $-$0.0025 & 84 \\
89 & 55578.2065 & 0.0007 & $-$0.0033 & 87 \\
90 & 55578.2849 & 0.0006 & $-$0.0037 & 87 \\
101 & 55579.1509 & 0.0013 & $-$0.0047 & 88 \\
102 & 55579.2295 & 0.0019 & $-$0.0050 & 87 \\
\hline
  \multicolumn{5}{l}{\commenta BJD$-$2400000.} \\
  \multicolumn{5}{l}{\commentb Against max $= 2455571.1948 + 0.078820 E$.} \\
  \multicolumn{5}{l}{\commentc Number of points used to determine the maximum.} \\
\end{tabular}
\end{center}
\end{table}

\subsection{V844 Herculis}\label{obj:v844her}

   We reported on the 2010 April--May superoutburst of this object
in \citet{Pdot2}.  The object underwent another superoutburst
in 2010 October--November.  The interval since the April--May
superoutburst was only 179 d.  The current state with frequent
superoutbursts started in 2009, in contrast to earlier state with
relatively infrequent superoutbursts (\cite{kat00v844her}; \cite{oiz07v844her}).

   Since the present observations were obtained at high airmasses,
we corrected observations by using a second-order atmospheric extinction
whose coefficients were experimentally determined.
The times of superhump maxima are listed in table \ref{tab:v844heroc2010b}.
Despite rather unfavorable seasonal conditions, we have succeeded in
obtaining good timing data of superhumps.  The $O-C$ diagram indicates
the presence of a stage B--C transition at late epochs.

\begin{figure}
  \begin{center}
    \FigureFile(88mm,70mm){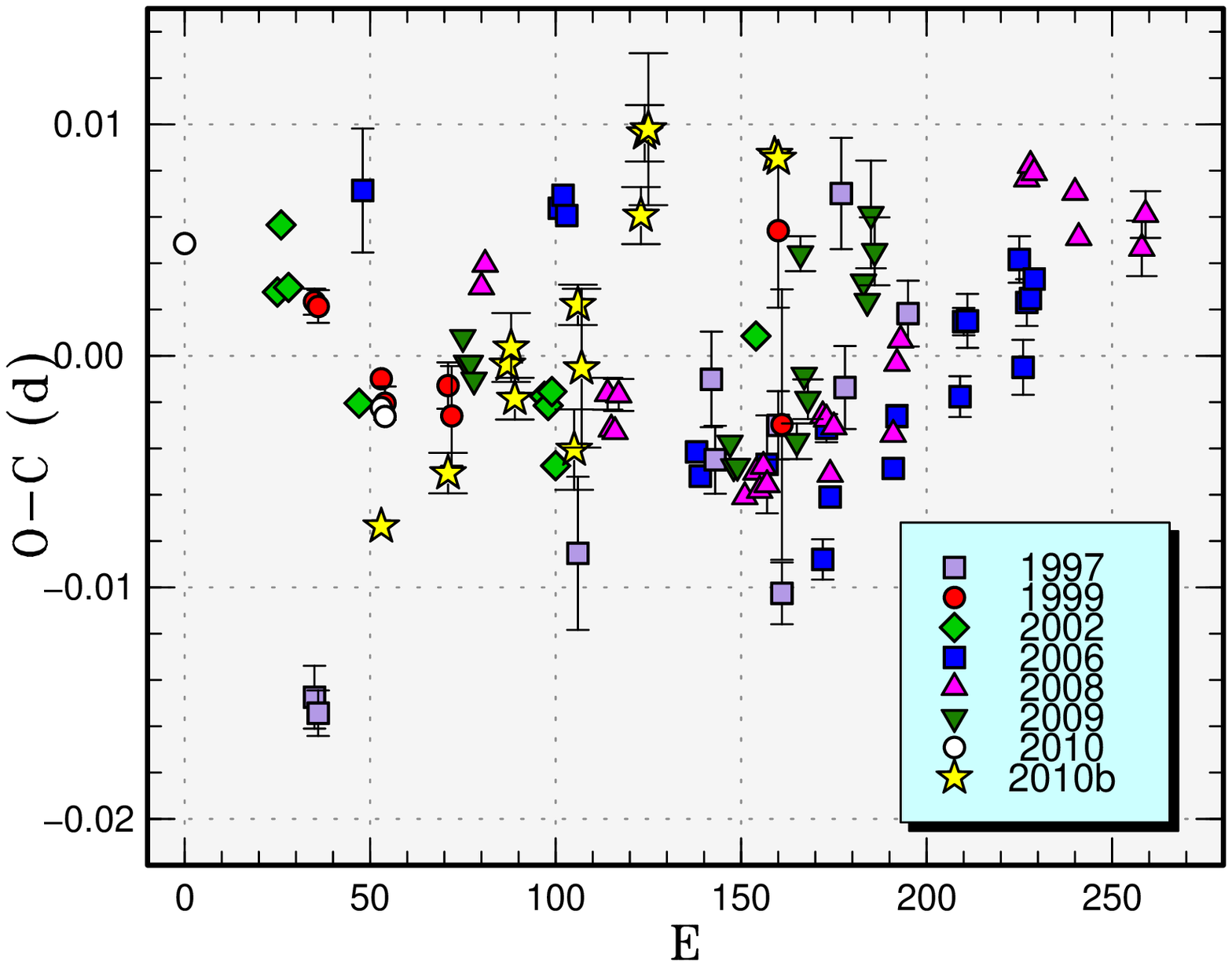}
  \end{center}
  \caption{Comparison of $O-C$ diagrams of V844 Her between different
  superoutbursts.  A period of 0.05590 d was used to draw this figure.
  Approximate cycle counts ($E$) after the start of the
  superoutburst were used.
  Since the start of the 2009 superoutburst was not well constrained,
  we shifted the $O-C$ diagram to best match the others.
  The initial superhump maximum of the 2010 superoutburst may have
  a one-cycle ambiguity in the cycle count, and the large $O-C$ may
  have been a result of the rapidly evolving stage A superhumps. 
  There was only 2 d gap of observation before the initial detection
  of the 2010b superoutburst, and the cycle counts drawn in this
  figure has a ambiguity less than $\sim$35.  The stage B--C transition
  in this small-scale superoutburst occurred much earlier than in past
  large-scale superoutbursts.
  }
  \label{fig:v844hercomp3}
\end{figure}

   Figure \ref{fig:v844hercomp3} illustrates a comparison of $O-C$
diagrams in different superoutbursts.  It is noteworthy that
the stage B--C transition in this small-scale (faint and short)
superoutburst occurred much earlier than in past large-scale superoutbursts.
The period of stage B superhumps appears to be noticeably longer
than in other superoutbursts.  Although this may have been a result of
the lack of the early stage observations, this result suggests that
the evolution of small-scale superoutbursts can be much faster
than in large-scale ones.

\begin{table}
\caption{Superhump maxima of V844 Her (2010 Oct.).}\label{tab:v844heroc2010b}
\begin{center}
\begin{tabular}{ccccc}
\hline
$E$ & max\commenta & error & $O-C$\commentb & $N$\commentc \\
\hline
0 & 55495.2921 & 0.0005 & $-$0.0003 & 38 \\
18 & 55496.3006 & 0.0009 & $-$0.0010 & 60 \\
34 & 55497.1996 & 0.0008 & 0.0010 & 112 \\
35 & 55497.2563 & 0.0015 & 0.0016 & 123 \\
36 & 55497.3100 & 0.0009 & $-$0.0008 & 124 \\
52 & 55498.2022 & 0.0017 & $-$0.0056 & 120 \\
53 & 55498.2643 & 0.0009 & 0.0004 & 123 \\
54 & 55498.3175 & 0.0034 & $-$0.0025 & 92 \\
70 & 55499.2185 & 0.0012 & 0.0015 & 119 \\
71 & 55499.2779 & 0.0012 & 0.0049 & 121 \\
72 & 55499.3340 & 0.0033 & 0.0049 & 57 \\
106 & 55501.2335 & 0.0007 & $-$0.0019 & 126 \\
107 & 55501.2892 & 0.0008 & $-$0.0022 & 122 \\
\hline
  \multicolumn{5}{l}{\commenta BJD$-$2400000.} \\
  \multicolumn{5}{l}{\commentb Against max $= 2455495.2923 + 0.056067 E$.} \\
  \multicolumn{5}{l}{\commentc Number of points used to determine the maximum.} \\
\end{tabular}
\end{center}
\end{table}

\subsection{MM Hydrae}\label{obj:mmhya}

   In \citet{Pdot}, we reported on rather fragmentary observations of MM Hya
during the 1998 and 2001 superoutbursts.  Since MM Hya was initially
suggested to be related to WZ Sge-type dwarf novae \citep{mis95PGCV},
more detailed observations were needed to examine this possibility.
There has been no report on measurement of $P_{\rm dot}$ for stage B
superhumps, which is expected to be useful in characterizing the object
(cf. \cite{Pdot}, subsection 5.3).

   The 2011 superoutburst was fortunately detected during its rising stage
(R. Stubbings, vsnet-alert 13113).  This was the first recorded
superoutburst since the last one in 2006 March.
Early observations detected long-period superhumps (vsnet-alert 13130, 13131),
rather than early superhumps in WZ Sge-type dwarf novae,
which further developed into full superhumps within 1 d (vsnet-alert 12132).

   The times of superhump maxima are listed in table \ref{tab:mmhyaoc2011}.
Due to a short gap in observation between the first and second nights,
the cycle count is somewhat ambiguous in the earliest phase.
We identified the cycle numbers given in the table since observations
on the first night apparently recorded the growing stage (stage A)
of superhumps and there must have been a large period change before
the system started showing fully developed stage B superhumps.
The adopted cycle counts gives a mean period of 0.0629(2) d for $E \le 16$,
which is close to the period [0.0617(3) d] obtained from the observations
on the first night.  The fractional excess (7 \%) of the period of
stage A superhumps to that of stage B superhumps is exceptionally
large (cf. \cite{Pdot}, subsection 3.7).  If we assume an additional
cycle count between the first and second nights, we get a period of
0.0590(4) d on these nights.  Although this value is close to
that of stage B superhumps, the $O-C$ variations within the first night
become inconsistent with the later evolution.  We therefore favor the
former assumption of the cycle counts.

   The resultant $P_{\rm dot}$ for stage B superhumps was
$+7.2(0.9) \times 10^{-5}$, which also supports the classification
as an usual SU UMa-type dwarf nova with a short $P_{\rm orb}$.

\begin{table}
\caption{Superhump maxima of MM Hya (2011).}\label{tab:mmhyaoc2011}
\begin{center}
\begin{tabular}{ccccc}
\hline
$E$ & max\commenta & error & $O-C$\commentb & $N$\commentc \\
\hline
0 & 55660.0258 & 0.0018 & $-$0.0393 & 113 \\
1 & 55660.0952 & 0.0011 & $-$0.0291 & 113 \\
2 & 55660.1546 & 0.0006 & $-$0.0289 & 112 \\
3 & 55660.2149 & 0.0012 & $-$0.0277 & 59 \\
16 & 55661.0355 & 0.0003 & 0.0237 & 76 \\
17 & 55661.0944 & 0.0002 & 0.0235 & 113 \\
18 & 55661.1528 & 0.0002 & 0.0228 & 106 \\
22 & 55661.3880 & 0.0004 & 0.0213 & 58 \\
32 & 55661.9756 & 0.0013 & 0.0173 & 47 \\
33 & 55662.0327 & 0.0014 & 0.0151 & 64 \\
39 & 55662.3856 & 0.0005 & 0.0131 & 60 \\
49 & 55662.9702 & 0.0012 & 0.0060 & 47 \\
50 & 55663.0342 & 0.0010 & 0.0108 & 118 \\
51 & 55663.0898 & 0.0009 & 0.0073 & 125 \\
66 & 55663.9713 & 0.0008 & 0.0013 & 159 \\
67 & 55664.0301 & 0.0008 & 0.0009 & 181 \\
68 & 55664.0915 & 0.0022 & 0.0032 & 67 \\
83 & 55664.9723 & 0.0015 & $-$0.0035 & 101 \\
84 & 55665.0337 & 0.0007 & $-$0.0012 & 108 \\
85 & 55665.0925 & 0.0008 & $-$0.0016 & 113 \\
86 & 55665.1519 & 0.0010 & $-$0.0014 & 110 \\
117 & 55666.9788 & 0.0010 & $-$0.0086 & 113 \\
118 & 55667.0397 & 0.0010 & $-$0.0069 & 106 \\
119 & 55667.0972 & 0.0014 & $-$0.0085 & 113 \\
120 & 55667.1553 & 0.0010 & $-$0.0096 & 87 \\
\hline
  \multicolumn{5}{l}{\commenta BJD$-$2400000.} \\
  \multicolumn{5}{l}{\commentb Against max $= 2455660.0651 + 0.059165 E$.} \\
  \multicolumn{5}{l}{\commentc Number of points used to determine the maximum.} \\
\end{tabular}
\end{center}
\end{table}

\subsection{V406 Hydrae}\label{obj:v406hya}

   V406 Hya is an AM CVn-type CV (for a recent review, see
\cite{sol10amcvnreview}) which was initially discovered as
a possible supernova (\cite{wou03sn2003aw} and references therein).
\citet{nog04v406hya} reported on the unusual behavior of its
2004 superoutburst.
\citet{roe06v406hya} further reported phase-resolved spectroscopy
and obtained an orbital period of 2027.8(5) s = 0.023470(6) d.
A bright ($\sim$15.0 mag) outburst of this objected in 2010 December
was detected by CRTS.  Superhumps with amplitudes of 0.065 mag
were detected (vsnet-alert 12483).
The times of superhump maxima are listed in table \ref{tab:v406hyaoc2010}.
The mean period of superhumps was 0.02365(5) d (PDM analysis),
which is in agreement with previous measurements
[0.023628(4) d, \citet{wou03sn2003aw}; 0.02357(4) d, \citet{nog04v406hya}].

The known outbursts of this object are listed in table \ref{tab:v406hyaout}.
Based on magnitudes and durations, all known outbursts appear to be
superoutbursts, and the intervals between superoutbursts
are apparently longer than the one (60 d) in KL Dra \citep{ram10kldra}
or $\sim$100 d in CP Eri based on CRTS data (vsnet-alert 11504)\footnote{
$<$http://nesssi.cacr.caltech.edu/catalina/20090925/909250090174119865.html$>$.
}.

\begin{table}
\caption{Superhump maxima of V406 Hya (2010).}\label{tab:v406hyaoc2010}
\begin{center}
\begin{tabular}{ccccc}
\hline
$E$ & max\commenta & error & $O-C$\commentb & $N$\commentc \\
\hline
0 & 55539.5314 & 0.0007 & $-$0.0004 & 34 \\
1 & 55539.5564 & 0.0004 & 0.0008 & 49 \\
2 & 55539.5800 & 0.0005 & 0.0005 & 47 \\
3 & 55539.6026 & 0.0006 & $-$0.0007 & 47 \\
4 & 55539.6267 & 0.0004 & $-$0.0004 & 48 \\
26 & 55540.1552 & 0.0038 & 0.0042 & 62 \\
28 & 55540.1955 & 0.0011 & $-$0.0032 & 43 \\
29 & 55540.2201 & 0.0011 & $-$0.0024 & 72 \\
30 & 55540.2405 & 0.0013 & $-$0.0059 & 87 \\
31 & 55540.2776 & 0.0070 & 0.0074 & 50 \\
\hline
  \multicolumn{5}{l}{\commenta BJD$-$2400000.} \\
  \multicolumn{5}{l}{\commentb Against max $= 2455539.5318 + 0.023817 E$.} \\
  \multicolumn{5}{l}{\commentc Number of points used to determine the maximum.} \\
\end{tabular}
\end{center}
\end{table}

\begin{table}
\caption{List of known outbursts of V406 Hya.}\label{tab:v406hyaout}
\begin{center}
\begin{tabular}{cccc}
\hline
Month & max\commenta & magnitude & source \\
\hline
2003 2 & 52677 & 17.8 & \citet{wou03sn2003awiauc8077} \\
2004 5 & 53143 & 14.7--15.2 & \citet{nog04v406hya} \\
2005 4 & 53465 & 15.2 & CRTS \\
2006 11 & 54057 & 17.5 & CRTS \\
2008 4 & 54565 & 15.3 & CRTS \\
2009 12 & 55181 & 15.8 & CRTS \\
2010 12 & 55536 & 15.0 & CRTS \\
\hline
  \multicolumn{4}{l}{\commenta JD$-$2400000.} \\
\end{tabular}
\end{center}
\end{table}

\subsection{V344 Lyrae}\label{obj:v344lyr}

   The data discussed in subsection \ref{obj:kepv344lyr} are presented.
The object showed prominent double peaks during the late course
of the superoutbursts.  We therefore used phases $-$0.2 to 0.2
(instead of $-$0.4 to 0.4, cf. \cite{Pdot}) for fitting the template
superhump profile in order to separate individual peaks.
The times of maxima of main peaks are listed in table \ref{tab:v344lyroc2009}.
The times of secondary maxima, which smoothly evolved into persisting
superhumps during the next normal outburst and subsequent interval
of quiescence, are listed in table \ref{tab:v344lyroc2009sec}.
The object underwent another superoutburst in 2009 November--December.
The data for this superoutburst are listed in tables \ref{tab:v344lyroc2009b}
and \ref{tab:v344lyroc2009bsec}.

\begin{table}
\caption{Superhump maxima of V344 Lyr (2009).}\label{tab:v344lyroc2009}
\begin{center}
\begin{tabular}{ccccc}
\hline
$E$ & max\commenta & error & $O-C$\commentb & $N$\commentc \\
\hline
0 & 55057.6071 & 0.0008 & $-$0.0327 & 54 \\
1 & 55057.7010 & 0.0024 & $-$0.0303 & 53 \\
2 & 55057.7969 & 0.0011 & $-$0.0258 & 54 \\
3 & 55057.8860 & 0.0006 & $-$0.0283 & 54 \\
4 & 55057.9779 & 0.0008 & $-$0.0279 & 54 \\
5 & 55058.0721 & 0.0003 & $-$0.0253 & 53 \\
6 & 55058.1630 & 0.0002 & $-$0.0259 & 54 \\
\end{tabular}
\end{center}
\end{table}

\addtocounter{table}{-1}
\begin{table}
\caption{Superhump maxima of V344 Lyr (2009) (continued).}
\begin{center}
\begin{tabular}{ccccc}
\hline
$E$ & max\commenta & error & $O-C$\commentb & $N$\commentc \\
\hline
54 & 55062.5902 & 0.0003 & 0.0083 & 54 \\
55 & 55062.6838 & 0.0002 & 0.0104 & 54 \\
56 & 55062.7741 & 0.0004 & 0.0092 & 54 \\
57 & 55062.8649 & 0.0003 & 0.0084 & 54 \\
58 & 55062.9561 & 0.0002 & 0.0081 & 54 \\
59 & 55063.0462 & 0.0003 & 0.0067 & 54 \\
60 & 55063.1397 & 0.0002 & 0.0087 & 54 \\
\end{tabular}
\end{center}
\end{table}

\addtocounter{table}{-1}
\begin{table}
\caption{Superhump maxima of V344 Lyr (2009) (continued).}
\begin{center}
\begin{tabular}{ccccc}
\hline
$E$ & max\commenta & error & $O-C$\commentb & $N$\commentc \\
\hline
120 & 55068.6203 & 0.0006 & $-$0.0020 & 54 \\
121 & 55068.7128 & 0.0005 & $-$0.0011 & 54 \\
122 & 55068.8044 & 0.0006 & $-$0.0009 & 54 \\
123 & 55068.8988 & 0.0010 & 0.0019 & 54 \\
124 & 55068.9884 & 0.0020 & 0.0000 & 54 \\
125 & 55069.0782 & 0.0005 & $-$0.0017 & 54 \\
126 & 55069.1682 & 0.0006 & $-$0.0033 & 53 \\
\end{tabular}
\end{center}
\end{table}

\begin{table}
\caption{Secondary superhump maxima of V344 Lyr (2009).}\label{tab:v344lyroc2009sec}
\begin{center}
\begin{tabular}{ccccc}
\hline
$E$ & max\commenta & error & $O-C$\commentb & $N$\commentc \\
\hline
0 & 55065.9135 & 0.0009 & 0.0029 & 54 \\
1 & 55066.0041 & 0.0021 & 0.0018 & 54 \\
3 & 55066.1883 & 0.0012 & 0.0028 & 53 \\
4 & 55066.2770 & 0.0009 & $-$0.0001 & 54 \\
5 & 55066.3677 & 0.0011 & $-$0.0011 & 53 \\
6 & 55066.4627 & 0.0008 & 0.0022 & 54 \\
7 & 55066.5528 & 0.0008 & 0.0008 & 54 \\
\end{tabular}
\end{center}
\end{table}

\addtocounter{table}{-1}
\begin{table}
\caption{Secondary superhump maxima of V344 Lyr (2009) (continued).}
\begin{center}
\begin{tabular}{ccccc}
\hline
$E$ & max\commenta & error & $O-C$\commentb & $N$\commentc \\
\hline
57 & 55071.1300 & 0.0006 & $-$0.0035 & 54 \\
58 & 55071.2262 & 0.0010 & 0.0011 & 54 \\
59 & 55071.3175 & 0.0011 & 0.0008 & 54 \\
60 & 55071.4121 & 0.0005 & 0.0037 & 54 \\
61 & 55071.5063 & 0.0010 & 0.0063 & 54 \\
62 & 55071.5987 & 0.0007 & 0.0071 & 54 \\
63 & 55071.6912 & 0.0006 & 0.0079 & 54 \\
\end{tabular}
\end{center}
\end{table}

\addtocounter{table}{-1}
\begin{table}
\caption{Secondary superhump maxima of V344 Lyr (2009) (continued).}
\begin{center}
\begin{tabular}{ccccc}
\hline
$E$ & max\commenta & error & $O-C$\commentb & $N$\commentc \\
\hline
112 & 55076.1611 & 0.0008 & $-$0.0120 & 54 \\
113 & 55076.2642 & 0.0010 & $-$0.0005 & 54 \\
117 & 55076.6187 & 0.0019 & $-$0.0126 & 54 \\
118 & 55076.7048 & 0.0015 & $-$0.0181 & 54 \\
119 & 55076.7971 & 0.0007 & $-$0.0174 & 53 \\
120 & 55076.8900 & 0.0007 & $-$0.0162 & 54 \\
121 & 55076.9811 & 0.0008 & $-$0.0168 & 54 \\
\end{tabular}
\end{center}
\end{table}

\addtocounter{table}{-1}
\begin{table}
\caption{Secondary superhump maxima of V344 Lyr (2009) (continued).}
\begin{center}
\begin{tabular}{ccccc}
\hline
$E$ & max\commenta & error & $O-C$\commentb & $N$\commentc \\
\hline
170 & 55081.5011 & 0.0059 & 0.0134 & 54 \\
171 & 55081.5750 & 0.0017 & $-$0.0043 & 53 \\
172 & 55081.6704 & 0.0008 & $-$0.0005 & 54 \\
173 & 55081.7619 & 0.0013 & $-$0.0006 & 54 \\
174 & 55081.8498 & 0.0011 & $-$0.0044 & 54 \\
175 & 55081.9463 & 0.0022 & 0.0005 & 54 \\
176 & 55082.0429 & 0.0020 & 0.0055 & 54 \\
177 & 55082.1342 & 0.0054 & 0.0052 & 54 \\
178 & 55082.2234 & 0.0091 & 0.0027 & 54 \\
179 & 55082.3130 & 0.0008 & 0.0007 & 54 \\
180 & 55082.4069 & 0.0015 & 0.0029 & 54 \\
181 & 55082.4974 & 0.0015 & 0.0018 & 54 \\
182 & 55082.5900 & 0.0007 & 0.0028 & 54 \\
183 & 55082.6700 & 0.0013 & $-$0.0088 & 54 \\
184 & 55082.7594 & 0.0012 & $-$0.0110 & 54 \\
185 & 55082.8545 & 0.0009 & $-$0.0076 & 54 \\
186 & 55082.9446 & 0.0018 & $-$0.0091 & 54 \\
187 & 55083.0394 & 0.0018 & $-$0.0059 & 53 \\
188 & 55083.1421 & 0.0022 & 0.0051 & 54 \\
\hline
  \multicolumn{5}{l}{\commenta BJD$-$2400000.} \\
  \multicolumn{5}{l}{\commentb Against max $= 2455065.9106 + 0.091629 E$.} \\
  \multicolumn{5}{l}{\commentc Number of points used to determine the maximum.} \\
\end{tabular}
\end{center}
\end{table}

\begin{table}
\caption{Superhump maxima of V344 Lyr (2009b).}\label{tab:v344lyroc2009b}
\begin{center}
\begin{tabular}{ccccc}
\hline
$E$ & max\commenta & error & $O-C$\commentb & $N$\commentc \\
\hline
0 & 55161.9728 & 0.0038 & $-$0.0600 & 54 \\
1 & 55162.0796 & 0.0015 & $-$0.0448 & 54 \\
2 & 55162.1711 & 0.0022 & $-$0.0447 & 54 \\
3 & 55162.2755 & 0.0011 & $-$0.0318 & 53 \\
4 & 55162.3654 & 0.0006 & $-$0.0335 & 54 \\
5 & 55162.4583 & 0.0020 & $-$0.0321 & 54 \\
6 & 55162.5509 & 0.0004 & $-$0.0310 & 54 \\
\end{tabular}
\end{center}
\end{table}

\addtocounter{table}{-1}
\begin{table}
\caption{Superhump maxima of V344 Lyr (2009b) (continued).}
\begin{center}
\begin{tabular}{ccccc}
\hline
$E$ & max\commenta & error & $O-C$\commentb & $N$\commentc \\
\hline
52 & 55166.8022 & 0.0003 & 0.0105 & 54 \\
53 & 55166.8921 & 0.0003 & 0.0089 & 54 \\
54 & 55166.9838 & 0.0002 & 0.0091 & 54 \\
55 & 55167.0757 & 0.0002 & 0.0094 & 53 \\
56 & 55167.1675 & 0.0004 & 0.0098 & 53 \\
57 & 55167.2611 & 0.0002 & 0.0118 & 54 \\
58 & 55167.3504 & 0.0002 & 0.0096 & 54 \\
\end{tabular}
\end{center}
\end{table}

\addtocounter{table}{-1}
\begin{table}
\caption{Superhump maxima of V344 Lyr (2009b) (continued).}
\begin{center}
\begin{tabular}{ccccc}
\hline
$E$ & max\commenta & error & $O-C$\commentb & $N$\commentc \\
\hline
103 & 55171.4609 & 0.0005 & 0.0018 & 54 \\
104 & 55171.5544 & 0.0005 & 0.0039 & 54 \\
105 & 55171.6454 & 0.0006 & 0.0033 & 54 \\
106 & 55171.7352 & 0.0005 & 0.0017 & 54 \\
107 & 55171.8287 & 0.0005 & 0.0036 & 54 \\
108 & 55171.9198 & 0.0011 & 0.0032 & 54 \\
109 & 55172.0102 & 0.0006 & 0.0021 & 54 \\
\end{tabular}
\end{center}
\end{table}

\addtocounter{table}{-1}
\begin{table}
\caption{Superhump maxima of V344 Lyr (2009b) (continued).}
\begin{center}
\begin{tabular}{ccccc}
\hline
$E$ & max\commenta & error & $O-C$\commentb & $N$\commentc \\
\hline
158 & 55176.4829 & 0.0006 & $-$0.0096 & 54 \\
159 & 55176.5708 & 0.0007 & $-$0.0132 & 53 \\
160 & 55176.6610 & 0.0009 & $-$0.0144 & 54 \\
161 & 55176.7532 & 0.0008 & $-$0.0138 & 54 \\
162 & 55176.8418 & 0.0008 & $-$0.0167 & 54 \\
163 & 55176.9339 & 0.0009 & $-$0.0161 & 53 \\
164 & 55177.0264 & 0.0005 & $-$0.0151 & 54 \\
165 & 55177.1167 & 0.0008 & $-$0.0163 & 54 \\
166 & 55177.2072 & 0.0008 & $-$0.0174 & 54 \\
167 & 55177.3015 & 0.0006 & $-$0.0146 & 54 \\
168 & 55177.3911 & 0.0006 & $-$0.0165 & 54 \\
169 & 55177.4827 & 0.0007 & $-$0.0164 & 54 \\
170 & 55177.5723 & 0.0015 & $-$0.0184 & 54 \\
\hline
  \multicolumn{5}{l}{\commenta BJD$-$2400000.} \\
  \multicolumn{5}{l}{\commentb Against max $= 2455162.0328 + 0.091517 E$.} \\
  \multicolumn{5}{l}{\commentc Number of points used to determine the maximum.} \\
\end{tabular}
\end{center}
\end{table}

\begin{table}
\caption{Secondary superhump maxima of V344 Lyr (2009b).}\label{tab:v344lyroc2009bsec}
\begin{center}
\begin{tabular}{ccccc}
\hline
$E$ & max\commenta & error & $O-C$\commentb & $N$\commentc \\
\hline
0 & 55169.2136 & 0.0006 & 0.0047 & 54 \\
1 & 55169.2986 & 0.0013 & $-$0.0018 & 54 \\
2 & 55169.4014 & 0.0008 & 0.0094 & 54 \\
3 & 55169.4837 & 0.0014 & 0.0001 & 54 \\
4 & 55169.5834 & 0.0009 & 0.0083 & 53 \\
5 & 55169.6736 & 0.0009 & 0.0070 & 53 \\
6 & 55169.7640 & 0.0008 & 0.0058 & 54 \\
\end{tabular}
\end{center}
\end{table}

\addtocounter{table}{-1}
\begin{table}
\caption{Secondary superhump maxima of V344 Lyr (2009b) (continued).}
\begin{center}
\begin{tabular}{ccccc}
\hline
$E$ & max\commenta & error & $O-C$\commentb & $N$\commentc \\
\hline
52 & 55173.9671 & 0.0005 & $-$0.0027 & 54 \\
53 & 55174.0576 & 0.0007 & $-$0.0038 & 54 \\
54 & 55174.1508 & 0.0004 & $-$0.0022 & 54 \\
55 & 55174.2407 & 0.0006 & $-$0.0038 & 54 \\
56 & 55174.3332 & 0.0005 & $-$0.0029 & 54 \\
57 & 55174.4235 & 0.0006 & $-$0.0042 & 54 \\
58 & 55174.5187 & 0.0004 & $-$0.0005 & 54 \\
\end{tabular}
\end{center}
\end{table}

\addtocounter{table}{-1}
\begin{table}
\caption{Secondary superhump maxima of V344 Lyr (2009b) (continued).}
\begin{center}
\begin{tabular}{ccccc}
\hline
$E$ & max\commenta & error & $O-C$\commentb & $N$\commentc \\
\hline
104 & 55178.7409 & 0.0007 & 0.0100 & 54 \\
105 & 55178.8317 & 0.0007 & 0.0092 & 54 \\
106 & 55178.9270 & 0.0005 & 0.0130 & 54 \\
107 & 55179.0168 & 0.0008 & 0.0113 & 54 \\
108 & 55179.1049 & 0.0005 & 0.0078 & 54 \\
109 & 55179.1960 & 0.0005 & 0.0074 & 53 \\
110 & 55179.2890 & 0.0007 & 0.0088 & 54 \\
\end{tabular}
\end{center}
\end{table}

\subsection{V1195 Ophiuchi}\label{obj:v1195oph}

   Very little has been known for this dwarf nova.  The object was originally
selected as a recurrent nova or a Mira-type star \citep{pla68VSsurvey}
based on two detections at a photographic magnitude of 16.2 on 1956 June 12
and 15.8 on 1959 April 29.
\citet{due87novaatlas} favored the latter classification based on its
red color.  P. Schmeer, however, noted that this object is a dwarf nova
and reported an outburst on 1999 June 16 at an unfiltered CCD magnitude
of 16.4 (vsnet-alert 3086).  J. Kemp conducted multicolor and time-resolved
photometry upon this information and reported the detection of superhumps
with an amplitude $\sim$0.20 mag and a period of $\sim$0.068 d
(vsnet-alert 3087).  There have been three other known outburst
detections (all observations gave unfiltered CCD magnitudes):
2004 February 8 (16.1 mag, B. Monard), 2004 September 2
(17.3, B. Monard) and 2005 April 17 (16.3, P. Schmeer).

   The object was again detected in outburst by M. Simonsen at a magnitude
of $V$=17.54 on 2011 April 3.390 and $V$=16.00 on April 9.374.
Subsequent observations indeed detected superhumps
(vsnet-alert 13144, 13151).
The times of superhump maxima are listed in table \ref{tab:v1195ophoc2011}.
Since the first observations were obtained 9 d after the recorded maximum,
the observation likely recorded only stage C superhumps.

\begin{table}
\caption{Superhump maxima of V1195 Oph (2011).}\label{tab:v1195ophoc2011}
\begin{center}
\begin{tabular}{ccccc}
\hline
$E$ & max\commenta & error & $O-C$\commentb & $N$\commentc \\
\hline
0 & 55664.2522 & 0.0026 & $-$0.0003 & 132 \\
9 & 55664.8595 & 0.0006 & 0.0018 & 70 \\
10 & 55664.9236 & 0.0009 & $-$0.0014 & 69 \\
24 & 55665.8666 & 0.0007 & 0.0002 & 69 \\
25 & 55665.9334 & 0.0005 & $-$0.0003 & 70 \\
\hline
  \multicolumn{5}{l}{\commenta BJD$-$2400000.} \\
  \multicolumn{5}{l}{\commentb Against max $= 2455664.2525 + 0.067245 E$.} \\
  \multicolumn{5}{l}{\commentc Number of points used to determine the maximum.} \\
\end{tabular}
\end{center}
\end{table}

\subsection{V1212 Tauri}\label{obj:v1212tau}

   V1212 Tau was discovered as an eruptive object near M45
\citep{par83v1212tau}, whose location might suggest one of Pleiades
flare stars.  Although the object remained unnamed for long,
members of the Variable Star Observers' League in Japan (VSOLJ)
suspected that the object is more likely a dwarf nova based on published
observations, and continued their own monitoring since 1987.
There was a visual record of a bright outburst in 1987 October,
which remained unconfirmed.
The object was detected in outburst in 2007 January by G. Gualdoni
(vsnet-alert 9190).  Although observations of this outburst was rather
fragmentary, J. Patterson reported the detection of superhumps
with a period of 0.075 d.\footnote{
  $<$http://cbastro.org/communications/news/messages/0518.html$>$.
}

\begin{figure}
  \begin{center}
    \FigureFile(88mm,110mm){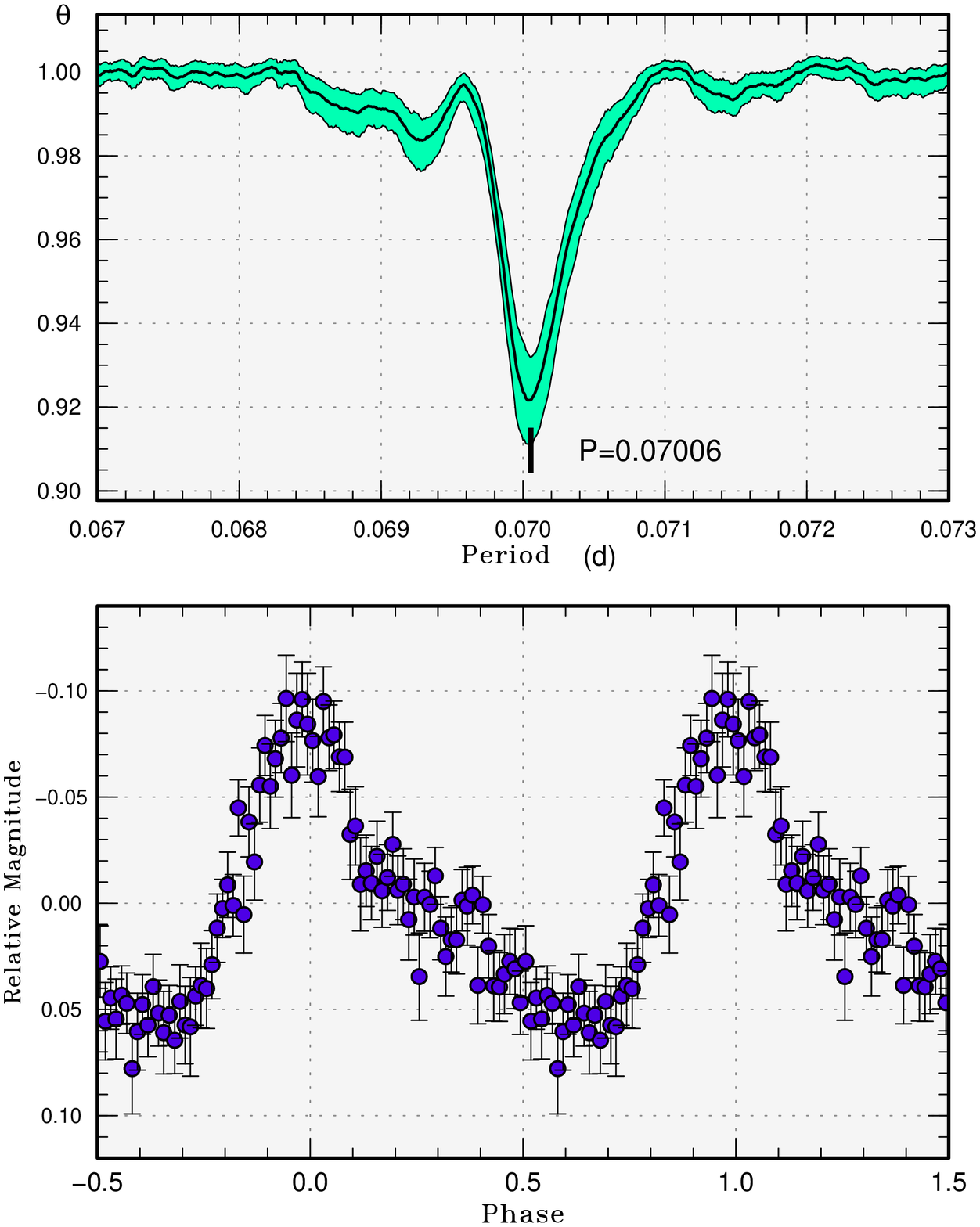}
  \end{center}
  \caption{Superhumps in V1212 Tau (2011). (Upper): PDM analysis
     excluding stage A (BJD 2455591--2455602).
     (Lower): Phase-averaged profile.}
  \label{fig:v1212taushpdm}
\end{figure}

   The object was again detected in outburst in 2011 January
(baavss-alert 2473, vsnet-alert 12718).  This outburst was detected
sufficiently early, and full evolution of superhumps were recorded
(vsnet-alert 12727, 12730, 12736, 12743; figure \ref{fig:v1212taushpdm}).
The time of superhump maxima are listed in table \ref{tab:v1212tauoc2011}.
Stages A--C can be well recognized.  The double-wave modulations reported
during the early stage were likely stage A superhumps having longer
period than stage B superhumps.\footnote{
  \citet{ole11j1625} discussed the possibility of early superhumps
  in SDSS J162520.29$+$120308.7.  The present case clearly excludes
  this possibility because of its long period (e.g. vsnet-alert 12727, 12730).
  The result and discussion by \citet{ole11j1625} probably need to be
  re-examined in the light of the knowledge from this object.
}
The evolution of superhumps is quite typical for a system with
this $P_{\rm SH}$.

\begin{table}
\caption{Superhump maxima of V1212 Tau (2011).}\label{tab:v1212tauoc2011}
\begin{center}
\begin{tabular}{ccccc}
\hline
$E$ & max\commenta & error & $O-C$\commentb & $N$\commentc \\
\hline
0 & 55589.6637 & 0.0028 & $-$0.0434 & 99 \\
1 & 55589.7459 & 0.0022 & $-$0.0313 & 98 \\
11 & 55590.4712 & 0.0011 & $-$0.0066 & 55 \\
12 & 55590.5479 & 0.0010 & 0.0000 & 68 \\
13 & 55590.6165 & 0.0003 & $-$0.0014 & 97 \\
14 & 55590.6863 & 0.0003 & $-$0.0017 & 99 \\
15 & 55590.7588 & 0.0003 & 0.0007 & 99 \\
20 & 55591.1074 & 0.0008 & $-$0.0010 & 131 \\
24 & 55591.3915 & 0.0003 & 0.0028 & 100 \\
25 & 55591.4617 & 0.0003 & 0.0030 & 126 \\
26 & 55591.5353 & 0.0007 & 0.0066 & 70 \\
27 & 55591.6027 & 0.0002 & 0.0039 & 94 \\
28 & 55591.6722 & 0.0002 & 0.0033 & 99 \\
29 & 55591.7429 & 0.0002 & 0.0040 & 97 \\
33 & 55592.0232 & 0.0015 & 0.0039 & 134 \\
37 & 55592.3031 & 0.0005 & 0.0036 & 50 \\
38 & 55592.3727 & 0.0004 & 0.0032 & 80 \\
39 & 55592.4430 & 0.0004 & 0.0034 & 135 \\
42 & 55592.6531 & 0.0004 & 0.0033 & 74 \\
43 & 55592.7228 & 0.0006 & 0.0029 & 49 \\
46 & 55592.9302 & 0.0017 & 0.0002 & 104 \\
47 & 55593.0036 & 0.0014 & 0.0035 & 128 \\
52 & 55593.3513 & 0.0005 & 0.0009 & 101 \\
53 & 55593.4217 & 0.0008 & 0.0012 & 106 \\
54 & 55593.4926 & 0.0009 & 0.0020 & 69 \\
56 & 55593.6330 & 0.0009 & 0.0023 & 32 \\
57 & 55593.7030 & 0.0010 & 0.0023 & 36 \\
66 & 55594.3340 & 0.0008 & 0.0027 & 100 \\
67 & 55594.4041 & 0.0007 & 0.0027 & 103 \\
68 & 55594.4734 & 0.0009 & 0.0020 & 82 \\
71 & 55594.6872 & 0.0023 & 0.0055 & 37 \\
72 & 55594.7543 & 0.0015 & 0.0026 & 36 \\
76 & 55595.0391 & 0.0042 & 0.0071 & 144 \\
80 & 55595.3159 & 0.0011 & 0.0036 & 90 \\
81 & 55595.3886 & 0.0007 & 0.0063 & 177 \\
82 & 55595.4588 & 0.0006 & 0.0064 & 165 \\
84 & 55595.5992 & 0.0009 & 0.0067 & 24 \\
85 & 55595.6695 & 0.0014 & 0.0070 & 35 \\
89 & 55595.9553 & 0.0022 & 0.0124 & 135 \\
90 & 55596.0145 & 0.0035 & 0.0016 & 147 \\
95 & 55596.3732 & 0.0009 & 0.0100 & 67 \\
96 & 55596.4406 & 0.0008 & 0.0074 & 68 \\
109 & 55597.3511 & 0.0010 & 0.0070 & 107 \\
110 & 55597.4186 & 0.0012 & 0.0044 & 107 \\
111 & 55597.4879 & 0.0026 & 0.0036 & 67 \\
118 & 55597.9668 & 0.0083 & $-$0.0079 & 119 \\
123 & 55598.3245 & 0.0019 & $-$0.0005 & 66 \\
124 & 55598.3951 & 0.0009 & 0.0001 & 104 \\
125 & 55598.4645 & 0.0011 & $-$0.0006 & 96 \\
137 & 55599.3017 & 0.0015 & $-$0.0042 & 40 \\
138 & 55599.3693 & 0.0012 & $-$0.0067 & 104 \\
139 & 55599.4410 & 0.0012 & $-$0.0051 & 97 \\
152 & 55600.3453 & 0.0013 & $-$0.0116 & 28 \\
153 & 55600.4174 & 0.0011 & $-$0.0096 & 36 \\
166 & 55601.3246 & 0.0022 & $-$0.0131 & 33 \\
167 & 55601.3966 & 0.0019 & $-$0.0112 & 27 \\
\hline
  \multicolumn{5}{l}{\commenta BJD$-$2400000.} \\
  \multicolumn{5}{l}{\commentb Against max $= 2455589.7071 + 0.070064 E$.} \\
  \multicolumn{5}{l}{\commentc Number of points used to determine the maximum.} \\
\end{tabular}
\end{center}
\end{table}

\subsection{SW Ursae Majoris}\label{obj:swuma}

   We observed the 2010 superoutburst.  This outburst occurred after
a relatively long ($\sim$890 d) interval following the 2008 superoutburst.
The outburst was relatively well observed despite the rather adverse
seasonal conditions.  Since some of the present observations were obtained
at high airmasses, we used a second-order correction for atmospheric
extinction as in V844 Her.
The times of superhump maxima are listed in table \ref{tab:swumaoc2010}.
Although not all parts were observed, all stages A--C are readily
identified.

\begin{figure}
  \begin{center}
    \FigureFile(88mm,70mm){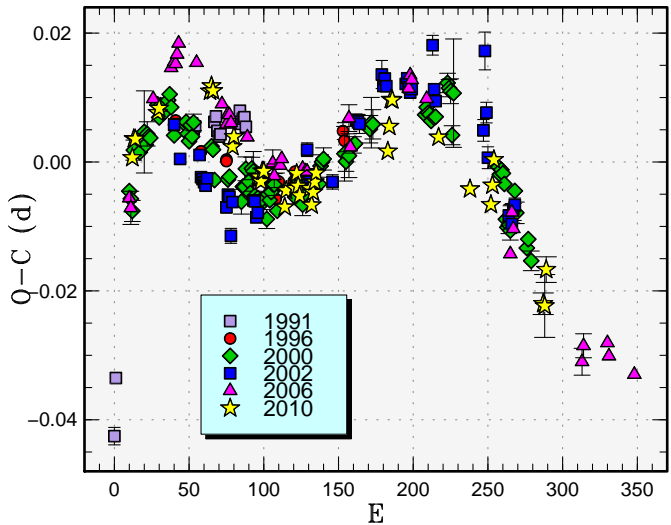}
  \end{center}
  \caption{Comparison of $O-C$ diagrams of SW UMa between different
  superoutbursts.  A period of 0.05822 d was used to draw this figure.
  Since the delay in the appearance of superhumps is known to vary
  in SW UMa, we shifted individual $O-C$ diagrams to get a best
  match (approximately corresponds to a definition of the appearance of
  superhumps to be $E=0$).
  Two relatively bright superoutburst in 2006 and 2010 showed a similar
  tendency (larger $O-C$) around the start of the stage B.
  }
  \label{fig:swumacomp2}
\end{figure}

   A comparison of $O-C$ diagrams in different superoutbursts is shown
in figure \ref{fig:swumacomp2}.  The 2010 superoutburst apparently followed
the similar trend to that of the 2006 superoutburst \citep{Pdot},
another bright superoutburst preceded by relatively long quiescence.

\begin{table}
\caption{Superhump maxima of SW UMa (2010).}\label{tab:swumaoc2010}
\begin{center}
\begin{tabular}{ccccc}
\hline
$E$ & max\commenta & error & $O-C$\commentb & $N$\commentc \\
\hline
0 & 55535.2103 & 0.0004 & $-$0.0056 & 234 \\
1 & 55535.2715 & 0.0005 & $-$0.0026 & 233 \\
2 & 55535.3296 & 0.0006 & $-$0.0025 & 149 \\
17 & 55536.2070 & 0.0002 & 0.0024 & 65 \\
18 & 55536.2659 & 0.0009 & 0.0032 & 73 \\
52 & 55538.2481 & 0.0003 & 0.0078 & 123 \\
53 & 55538.3071 & 0.0002 & 0.0087 & 114 \\
54 & 55538.3646 & 0.0002 & 0.0080 & 124 \\
67 & 55539.1128 & 0.0003 & 0.0001 & 233 \\
68 & 55539.1725 & 0.0003 & 0.0017 & 88 \\
84 & 55540.0979 & 0.0002 & $-$0.0035 & 82 \\
85 & 55540.1560 & 0.0002 & $-$0.0036 & 358 \\
86 & 55540.2135 & 0.0002 & $-$0.0042 & 357 \\
87 & 55540.2734 & 0.0001 & $-$0.0025 & 343 \\
88 & 55540.3316 & 0.0001 & $-$0.0025 & 349 \\
102 & 55541.1411 & 0.0006 & $-$0.0072 & 127 \\
103 & 55541.2019 & 0.0002 & $-$0.0046 & 222 \\
109 & 55541.5515 & 0.0002 & $-$0.0039 & 72 \\
110 & 55541.6121 & 0.0006 & $-$0.0016 & 47 \\
111 & 55541.6683 & 0.0003 & $-$0.0035 & 82 \\
112 & 55541.7251 & 0.0003 & $-$0.0049 & 67 \\
119 & 55542.1337 & 0.0009 & $-$0.0034 & 94 \\
120 & 55542.1894 & 0.0003 & $-$0.0059 & 206 \\
121 & 55542.2495 & 0.0002 & $-$0.0039 & 303 \\
122 & 55542.3094 & 0.0002 & $-$0.0022 & 289 \\
123 & 55542.3690 & 0.0004 & $-$0.0007 & 264 \\
171 & 55545.1669 & 0.0006 & 0.0055 & 179 \\
172 & 55545.2290 & 0.0004 & 0.0094 & 302 \\
173 & 55545.2914 & 0.0005 & 0.0136 & 303 \\
174 & 55545.3495 & 0.0007 & 0.0136 & 169 \\
205 & 55547.1486 & 0.0002 & 0.0096 & 140 \\
226 & 55548.3632 & 0.0008 & 0.0028 & 126 \\
240 & 55549.1758 & 0.0004 & 0.0012 & 221 \\
241 & 55549.2370 & 0.0004 & 0.0043 & 245 \\
242 & 55549.2992 & 0.0005 & 0.0083 & 248 \\
275 & 55551.1981 & 0.0017 & $-$0.0121 & 62 \\
276 & 55551.2560 & 0.0049 & $-$0.0124 & 82 \\
277 & 55551.3199 & 0.0020 & $-$0.0067 & 83 \\
\hline
  \multicolumn{5}{l}{\commenta BJD$-$2400000.} \\
  \multicolumn{5}{l}{\commentb Against max $= 2455535.2159 + 0.058161 E$.} \\
  \multicolumn{5}{l}{\commentc Number of points used to determine the maximum.} \\
\end{tabular}
\end{center}
\end{table}

\subsection{CI Ursae Majoris}\label{obj:ciuma}

\begin{figure}
  \begin{center}
    \FigureFile(88mm,70mm){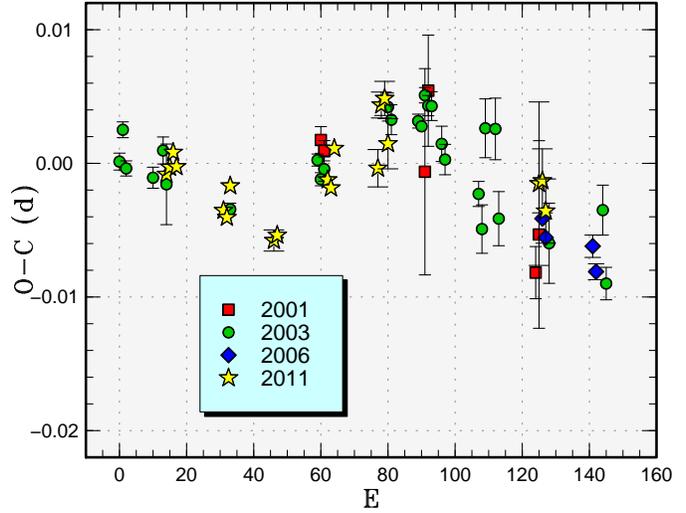}
  \end{center}
  \caption{Comparison of $O-C$ diagrams of CI UMa between different
  superoutbursts.  The $O-C$'s were calculated against a period of
  0.06264 d.  Approximate cycle counts ($E$) after the start of the
  superoutburst were used.
  }
  \label{fig:ciumacomp2}
\end{figure}

   We observed the 2011 superoutburst of this object
(table \ref{tab:ciumaoc2011}).  Stage B with positive $P_{\rm dot}$ of
$+16.8(3.2) \times 10^{-5}$ and a part of stage C were observed.
The behavior of $O-C$ variation is very similar between different
superoutbursts (figure \ref{fig:ciumacomp2}).

\begin{table}
\caption{Superhump maxima of CI UMa (2011).}\label{tab:ciumaoc2011}
\begin{center}
\begin{tabular}{ccccc}
\hline
$E$ & max\commenta & error & $O-C$\commentb & $N$\commentc \\
\hline
0 & 55660.4340 & 0.0002 & 0.0005 & 58 \\
1 & 55660.4971 & 0.0003 & 0.0011 & 64 \\
2 & 55660.5609 & 0.0003 & 0.0022 & 67 \\
3 & 55660.6224 & 0.0003 & 0.0011 & 68 \\
17 & 55661.4961 & 0.0004 & $-$0.0023 & 57 \\
18 & 55661.5582 & 0.0005 & $-$0.0028 & 68 \\
19 & 55661.6232 & 0.0006 & $-$0.0005 & 63 \\
32 & 55662.4335 & 0.0008 & $-$0.0047 & 127 \\
33 & 55662.4965 & 0.0005 & $-$0.0043 & 90 \\
48 & 55663.4403 & 0.0007 & $-$0.0003 & 66 \\
49 & 55663.5023 & 0.0006 & $-$0.0009 & 62 \\
50 & 55663.5679 & 0.0005 & 0.0021 & 55 \\
63 & 55664.3807 & 0.0014 & 0.0005 & 32 \\
64 & 55664.4481 & 0.0010 & 0.0052 & 34 \\
65 & 55664.5112 & 0.0013 & 0.0057 & 34 \\
66 & 55664.5705 & 0.0019 & 0.0023 & 24 \\
111 & 55667.3863 & 0.0061 & $-$0.0010 & 33 \\
112 & 55667.4491 & 0.0024 & $-$0.0008 & 34 \\
113 & 55667.5095 & 0.0024 & $-$0.0031 & 23 \\
\hline
  \multicolumn{5}{l}{\commenta BJD$-$2400000.} \\
  \multicolumn{5}{l}{\commentb Against max $= 2455660.4334 + 0.062648 E$.} \\
  \multicolumn{5}{l}{\commentc Number of points used to determine the maximum.} \\
\end{tabular}
\end{center}
\end{table}

\subsection{DV Ursae Majoris}\label{obj:dvuma}

\begin{figure}
  \begin{center}
    \FigureFile(88mm,70mm){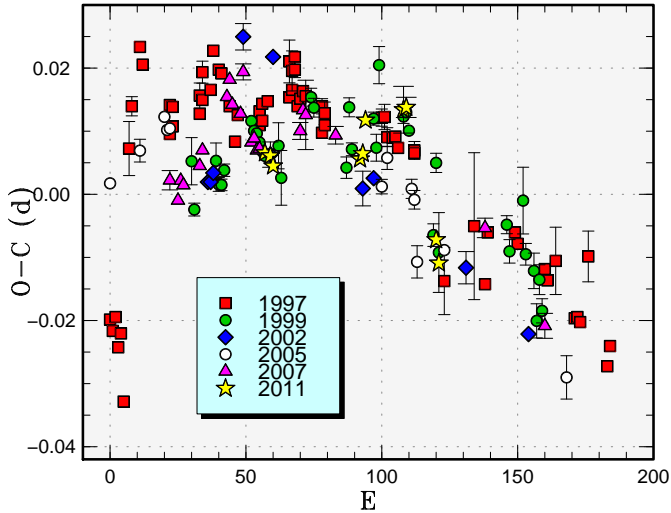}
  \end{center}
  \caption{Comparison of $O-C$ diagrams of DV UMa between different
  superoutbursts.  The $O-C$'s were calculated against a period of
  0.0888 d.  Approximate cycle counts ($E$) after the start of the
  superoutburst were used.
  }
  \label{fig:dvumacomp2}
\end{figure}

   We observed the 2011 superoutburst.  The maxima were determined from
observations outside the eclipses (table \ref{tab:dvumaoc2011}).
The 2011 observation only recorded the final portion of stage B
and partly stage C (cf. figure \ref{fig:dvumacomp2}), and the
resultant $P_{\rm dot}$ was rather unreliable.

\begin{table}
\caption{Superhump maxima of DV UMa (2011).}\label{tab:dvumaoc2011}
\begin{center}
\begin{tabular}{cccccc}
\hline
$E$ & max\commenta & error & $O-C$\commentb & phase\commentc & $N$\commentd \\
\hline
0 & 55608.4897 & 0.0004 & $-$0.0020 & 0.40 & 147 \\
2 & 55608.6668 & 0.0004 & $-$0.0023 & 0.46 & 138 \\
3 & 55608.7540 & 0.0007 & $-$0.0038 & 0.48 & 117 \\
35 & 55611.5967 & 0.0004 & 0.0007 & 0.59 & 138 \\
36 & 55611.6864 & 0.0006 & 0.0016 & 0.63 & 133 \\
37 & 55611.7805 & 0.0008 & 0.0070 & 0.73 & 97 \\
51 & 55613.0253 & 0.0037 & 0.0102 & 0.23 & 203 \\
52 & 55613.1145 & 0.0016 & 0.0106 & 0.27 & 178 \\
63 & 55614.0703 & 0.0012 & $-$0.0092 & 0.40 & 174 \\
64 & 55614.1554 & 0.0005 & $-$0.0128 & 0.39 & 174 \\
\hline
  \multicolumn{6}{l}{\commenta BJD$-$2400000.} \\
  \multicolumn{6}{l}{\commentb Against max $= 2455608.4917 + 0.088695 E$.} \\
  \multicolumn{6}{l}{\commentc Orbital phase.} \\
  \multicolumn{6}{l}{\commentd Number of points used to determine the maximum.} \\
\end{tabular}
\end{center}
\end{table}

\subsection{1RXS J003828.7$+$250920}\label{obj:j0038}

   This object was discovered as an eruptive object by K. Itagaki
in 2007 November (vsnet-outburst 8245).  The object was observed to fade
by 0.3 mag in seven days.  Because of the presence of a ROSAT X-ray
counterpart, we call this object in its ROSAT name (hereafter 1RXS J0038).

   The object was again reported to be in outburst on 2010 October 21
at an unfiltered CCD magnitude of 14.6 (E. Muyllaert, vsnet-alert 12295).
Early observations suggested the presence of superhumps
(vsnet-alert 12313).  The presence of long-period superhumps was finally
established (vsnet-alert 12318).  The long outburst in 2007 also must
have been a superoutburst.

\begin{figure}
  \begin{center}
    \FigureFile(88mm,110mm){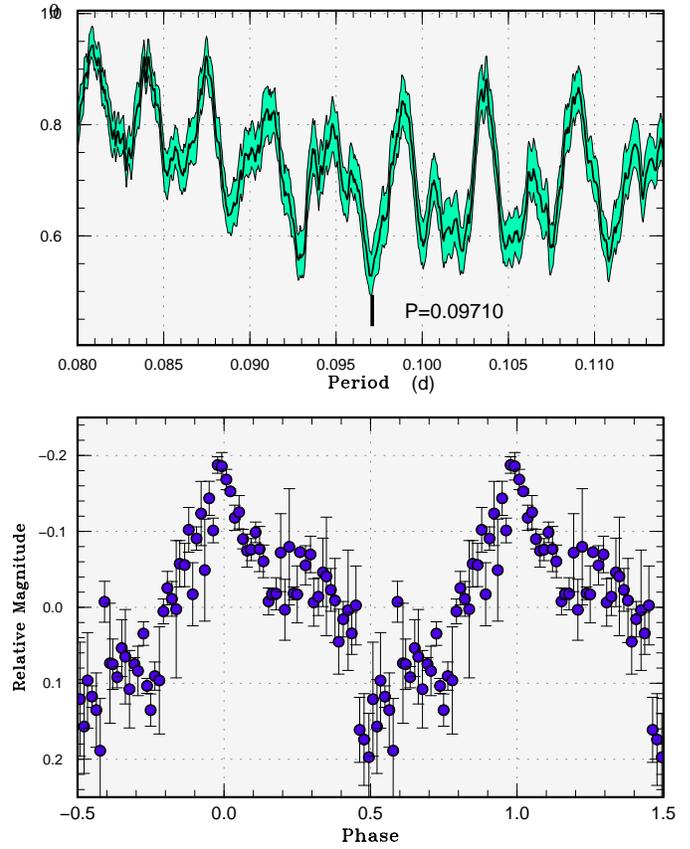}
  \end{center}
  \caption{Superhumps in 1RXS J0038 (2010). (Upper): PDM analysis.
     The rejection rate for bootstrapping was reduced to 0.2 for
     better visualization.
     (Lower): Phase-averaged profile.}
  \label{fig:j0038shpdm}
\end{figure}

   The times of superhump maxima are listed in table \ref{tab:j0038oc2010}.
The maximum at $E=42$ was recorded after the rapid fading, and is not
very reliable.  An analysis of the single-night continuous run yielded
a period of 0.0985(6) d.  The PDM analysis in figure \ref{fig:j0038shpdm}
used all observations, including those after the rapid fading, and yielded
a mean period of 0.09708(8) d.  We adopted the latter period in table
\ref{tab:perlist}.

\begin{table}
\caption{Superhump maxima of 1RXS J0038 (2010).}\label{tab:j0038oc2010}
\begin{center}
\begin{tabular}{ccccc}
\hline
$E$ & max\commenta & error & $O-C$\commentb & $N$\commentc \\
\hline
0 & 55495.3027 & 0.0008 & $-$0.0020 & 94 \\
1 & 55495.4040 & 0.0009 & 0.0020 & 97 \\
42 & 55499.3928 & 0.0032 & $-$0.0000 & 6 \\
\hline
  \multicolumn{5}{l}{\commenta BJD$-$2400000.} \\
  \multicolumn{5}{l}{\commentb Against max $= 2455495.3046 + 0.097339 E$.} \\
  \multicolumn{5}{l}{\commentc Number of points used to determine the maximum.} \\
\end{tabular}
\end{center}
\end{table}

\subsection{2QZ J222416.2$-$292421}\label{obj:j2224}

   This object (hereafter 2QZ J2224) was selected as a CV during the
2dF survey for quasars \citep{cro04qz7}.  The spectrum was typical for
a dwarf nova in quiescence with strong Balmer and He\textsc{I} emission
lines.  There is not a strong indication of an underlying white dwarf as in
many WZ Sge-type dwarf novae (see also vsnet-alert 12167).
The object was for the first time detected in outburst by
the CRTS Siding Spring Survey (SSS) (P. Wils, cvnet-outburst 3836).
Subsequent observations confirmed the presence of superhumps
(vsnet-alert 12171, 12180).

\begin{figure}
  \begin{center}
    \FigureFile(88mm,110mm){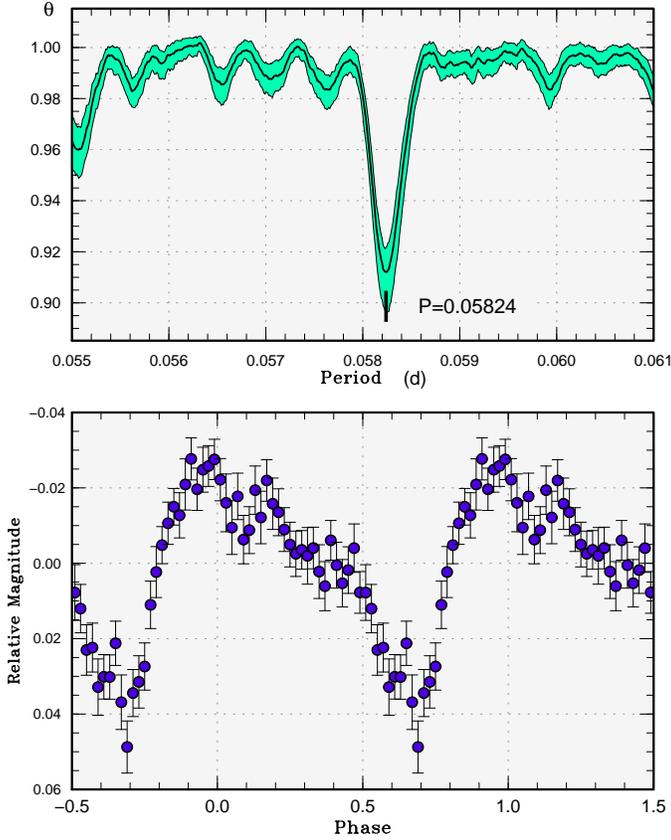}
  \end{center}
  \caption{Superhumps in 2QZ J2224 (2010). (Upper): PDM analysis.
     (Lower): Phase-averaged profile.}
  \label{fig:j2224shpdm}
\end{figure}

   The times of superhump maxima are listed in table \ref{tab:j2224oc2010}.
The scatter was relatively large due to the low amplitudes of superhumps.
Combined with the apparent lack of a large variation in the period,
it is likely we observed the stage C superhumps during their decay phase.
A PDM analysis (figure \ref{fig:j2224shpdm}) has yielded a mean period
of 0.05824(1) d, which is adopted in table \ref{tab:perlist}.

\begin{table}
\caption{Superhump maxima of 2QZ J2224 (2010).}\label{tab:j2224oc2010}
\begin{center}
\begin{tabular}{ccccc}
\hline
$E$ & max\commenta & error & $O-C$\commentb & $N$\commentc \\
\hline
0 & 55452.3543 & 0.0007 & 0.0006 & 127 \\
1 & 55452.4112 & 0.0007 & $-$0.0008 & 130 \\
2 & 55452.4665 & 0.0008 & $-$0.0037 & 129 \\
3 & 55452.5278 & 0.0009 & $-$0.0007 & 129 \\
4 & 55452.5829 & 0.0010 & $-$0.0039 & 115 \\
16 & 55453.2838 & 0.0013 & $-$0.0022 & 129 \\
17 & 55453.3415 & 0.0009 & $-$0.0027 & 129 \\
18 & 55453.3999 & 0.0011 & $-$0.0026 & 129 \\
19 & 55453.4587 & 0.0013 & $-$0.0021 & 129 \\
20 & 55453.5170 & 0.0012 & $-$0.0020 & 129 \\
35 & 55454.4010 & 0.0031 & 0.0081 & 120 \\
36 & 55454.4573 & 0.0033 & 0.0061 & 129 \\
37 & 55454.5074 & 0.0017 & $-$0.0021 & 114 \\
52 & 55455.3948 & 0.0038 & 0.0115 & 131 \\
53 & 55455.4424 & 0.0029 & 0.0007 & 132 \\
54 & 55455.5176 & 0.0051 & 0.0177 & 131 \\
55 & 55455.5533 & 0.0022 & $-$0.0049 & 132 \\
64 & 55456.0832 & 0.0014 & 0.0006 & 120 \\
65 & 55456.1285 & 0.0025 & $-$0.0123 & 121 \\
70 & 55456.4307 & 0.0017 & $-$0.0015 & 131 \\
71 & 55456.4994 & 0.0026 & 0.0090 & 116 \\
72 & 55456.5493 & 0.0026 & 0.0006 & 124 \\
85 & 55457.3044 & 0.0053 & $-$0.0017 & 77 \\
86 & 55457.3627 & 0.0031 & $-$0.0016 & 96 \\
88 & 55457.4791 & 0.0021 & $-$0.0017 & 56 \\
89 & 55457.5396 & 0.0028 & 0.0005 & 89 \\
104 & 55458.4094 & 0.0017 & $-$0.0036 & 129 \\
105 & 55458.4649 & 0.0047 & $-$0.0063 & 129 \\
106 & 55458.5300 & 0.0021 & 0.0004 & 129 \\
\hline
  \multicolumn{5}{l}{\commenta BJD$-$2400000.} \\
  \multicolumn{5}{l}{\commentb Against max $= 2455452.3538 + 0.058262 E$.} \\
  \multicolumn{5}{l}{\commentc Number of points used to determine the maximum.} \\
\end{tabular}
\end{center}
\end{table}

\subsection{ASAS J091858$-$2942.6}\label{obj:asas0918}

   We observed the 2010 superoutburst of this object discovered by
\citet{poj05dnpyx}.  The outburst, second known in its history since,
was detected on 2010 December 12 at
a visual magnitude of 12.8 (R. Stubbings).
The first CCD observation started on December 15.
Since some of the observations were recorded at high airmasses
(larger than 3), secondary corrections for atmospheric extinction
were applied to these observations.
The times of superhump maxima are listed in table \ref{tab:asas0918oc2010}.
The period was relatively constant except $E=0$.
Considering that the reported magnitude was fainter than that in 2005 and
the observation started relatively late, it is likely that we mostly observed
stage C superhumps.  The period was indeed similar to that of stage C
superhumps recorded in 2005.
The object underwent a rebrightening on December 30 (vsnet-alert 12544).

\begin{table}
\caption{Superhump maxima of ASAS J0918 (2010).}\label{tab:asas0918oc2010}
\begin{center}
\begin{tabular}{ccccc}
\hline
$E$ & max\commenta & error & $O-C$\commentb & $N$\commentc \\
\hline
0 & 55545.9210 & 0.0009 & $-$0.0044 & 63 \\
20 & 55547.1785 & 0.0013 & $-$0.0001 & 120 \\
21 & 55547.2384 & 0.0007 & $-$0.0028 & 285 \\
22 & 55547.3054 & 0.0007 & 0.0015 & 194 \\
32 & 55547.9344 & 0.0010 & 0.0040 & 51 \\
35 & 55548.1197 & 0.0015 & 0.0013 & 77 \\
52 & 55549.1855 & 0.0007 & 0.0019 & 112 \\
53 & 55549.2465 & 0.0008 & 0.0003 & 219 \\
54 & 55549.3125 & 0.0005 & 0.0036 & 196 \\
85 & 55551.2502 & 0.0014 & $-$0.0009 & 171 \\
86 & 55551.3094 & 0.0015 & $-$0.0044 & 165 \\
\hline
  \multicolumn{5}{l}{\commenta BJD$-$2400000.} \\
  \multicolumn{5}{l}{\commentb Against max $= 2455545.9255 + 0.062655 E$.} \\
  \multicolumn{5}{l}{\commentc Number of points used to determine the maximum.} \\
\end{tabular}
\end{center}
\end{table}

\subsection{ASAS J102522$-$1542.4}\label{obj:asas1025}

   This object (hereafter ASAS J1025) is a WZ Sge-type dwarf nova
discovered in 2006.  The history of observations can be found in
\citet{Pdot} and references therein.

\begin{figure}
  \begin{center}
    \FigureFile(88mm,70mm){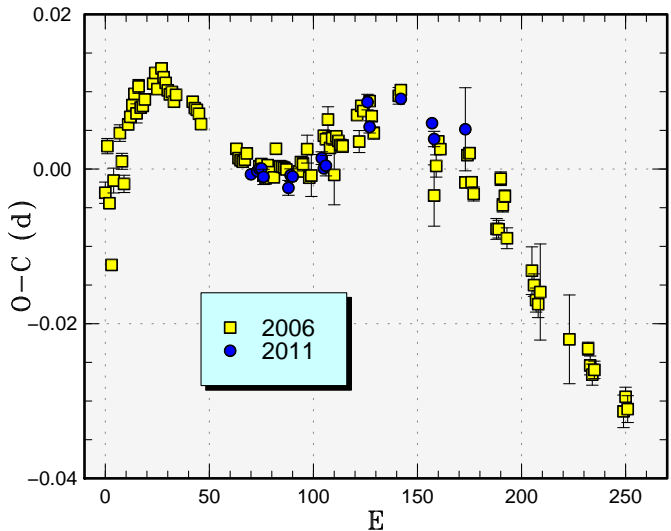}
  \end{center}
  \caption{Comparison of $O-C$ diagrams of ASAS J1025 between different
  superoutbursts.  A period of 0.06340 d was used to draw this figure.
  Approximate cycle counts ($E$) after the appearance of the superhumps
  were used.  Since the start of the 2011 superoutburst
  was not well constrained, we shifted the $O-C$ diagram
  to best fit the best-recorded 2006 one.
  }
  \label{fig:j1025comp}
\end{figure}

   The object was again detected in outburst in 2011 February by
the CRTS (=CSS110226:102522$-$154222).  Due to the gap in observation
before the CRTS detection, the outburst detection was apparently made
in a later stage than in the 2006 superoutburst.  The stage of
early superhumps was not recorded.
The times of superhump maxima are listed in table \ref{tab:asas1025oc2011},
which clearly demonstrates stage B superhumps with 
$P_{\rm dot} = +9.9(2.4) \times 10^{-5}$ and short segment of stage C.
This value is in good agreement with the 2006 measurement
(see also figure \ref{fig:j1025comp}).

   In \citet{Pdot}, we suggested that the object would be a ``borderline''
long-$P_{\rm SH}$ WZ Sge-like dwarf nova based in the large $P_{\rm dot}$
and the large fractional superhump excess.  The short interval of
superoutbursts (5 yr or shorter) for a WZ Sge-type dwarf nova also
strengthens this interpretation.

\begin{table}
\caption{Superhump maxima of ASAS J1025 (2011).}\label{tab:asas1025oc2011}
\begin{center}
\begin{tabular}{ccccc}
\hline
$E$ & max\commenta & error & $O-C$\commentb & $N$\commentc \\
\hline
0 & 55619.5018 & 0.0003 & 0.0005 & 95 \\
3 & 55619.6924 & 0.0006 & 0.0007 & 52 \\
4 & 55619.7561 & 0.0005 & 0.0009 & 52 \\
5 & 55619.8196 & 0.0006 & 0.0009 & 52 \\
6 & 55619.8819 & 0.0010 & $-$0.0003 & 50 \\
18 & 55620.6413 & 0.0010 & $-$0.0027 & 43 \\
19 & 55620.7063 & 0.0008 & $-$0.0012 & 43 \\
20 & 55620.7695 & 0.0005 & $-$0.0015 & 44 \\
34 & 55621.6595 & 0.0008 & $-$0.0003 & 51 \\
35 & 55621.7216 & 0.0007 & $-$0.0017 & 51 \\
36 & 55621.7853 & 0.0013 & $-$0.0014 & 49 \\
56 & 55623.0616 & 0.0010 & 0.0051 & 95 \\
57 & 55623.1217 & 0.0008 & 0.0018 & 90 \\
72 & 55624.0764 & 0.0005 & 0.0042 & 101 \\
87 & 55625.0242 & 0.0004 & $-$0.0003 & 106 \\
88 & 55625.0856 & 0.0010 & $-$0.0024 & 91 \\
103 & 55626.0378 & 0.0054 & $-$0.0024 & 113 \\
\hline
  \multicolumn{5}{l}{\commenta BJD$-$2400000.} \\
  \multicolumn{5}{l}{\commentb Against max $= 2455619.5013 + 0.063485 E$.} \\
  \multicolumn{5}{l}{\commentc Number of points used to determine the maximum.} \\
\end{tabular}
\end{center}
\end{table}

\subsection{MisV 1443}\label{obj:misv1443}

\begin{figure}
  \begin{center}
    \FigureFile(88mm,110mm){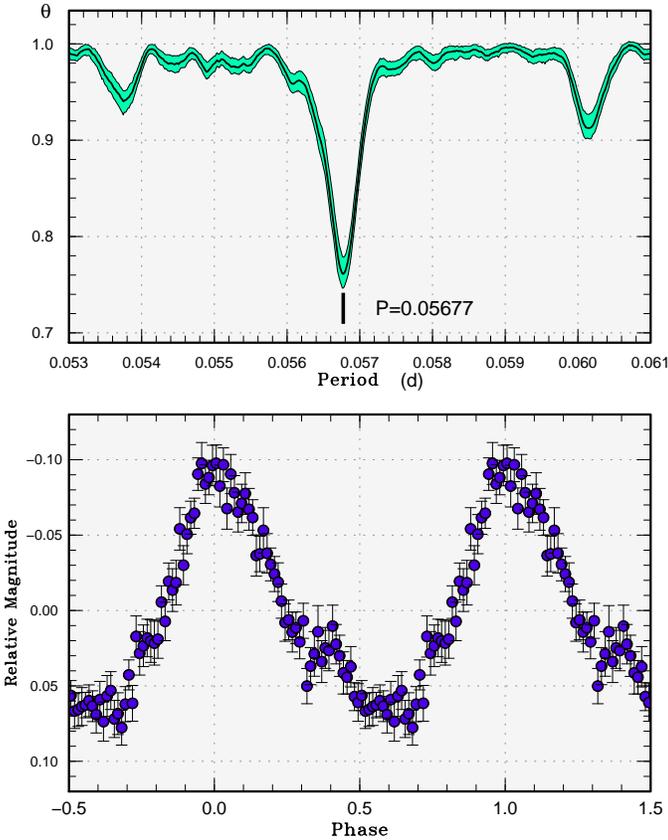}
  \end{center}
  \caption{Ordinary superhumps in MisV 1443 (2011). (Upper): PDM analysis
     for BJD 2455571.8--2455582 (excluding stage A superhumps)
     (Lower): Phase-averaged profile.}
  \label{fig:misv1443shpdm}
\end{figure}

   MisV 1443 was discovered as an eruptive object by Y. Nakashima
at an unfiltered CCD magnitude of 14.4 mag on 2011 January 8.59707 UT.
The position of the object is \timeform{06h 19m 59.96s},
\timeform{+19D 26' 59.0''} (J2000.0)
(vsnet-alert 12580; \cite{yos11misv1443cbet2633}).
There was a 20.4-mag quiescent counterpart
(vsnet-alert 12581).  Soon after this discovery announcement,
time-resolved photometry indicated the presence of double-wave modulations.
Although this signal was initially interpreted as early superhumps,
this identification requires confirmation because ordinary superhumps
quickly developed (vsnet-alert 12597; figure \ref{fig:misv1443shpdm}).
The dwarf nova nature of the object was spectroscopically confirmed
by \citet{yos11misv1443cbet2633} and A. Arai (vsnet-alert 12598).
A retrospective analysis of survey materials
revealed that the object was already in outburst at an unfiltered
CCD magnitude of 12.8 on 2011 January 2.019 UT
(S. Korotkiy and K. Sokolovsky, vsnet-alert 12592).

   The times of superhump maxima are listed in table \ref{tab:misv1443oc2011}.
Disregarding stage A superhumps ($E \le 5$), evidently positive
$P_{\rm dot}$ of $+5.5(1.5) \times 10^{-5}$ was recorded.
Based on the large outburst amplitude ($\ge7.6$ mag) and a long ($\ge$8 d)
delay before the appearance of ordinary superhumps, the object is likely
a WZ Sge-type dwarf nova.  The stage of early superhumps, however,
was not sufficiently observed.  The initial detection of double-wave
modulations may be interpreted as growing stage of ordinary
superhumps rather than genuine early superhumps.
The maximum amplitude of superhumps (0.30 mag) was, however, rather
exceptionally large for a WZ Sge-type dwarf nova
(cf. subsection \ref{sec:humpamp}).
MisV 1443 may bear intermediate characteristics
between WZ Sge-type dwarf novae and more usual SU UMa-type dwarf novae.

   The final stage of the superoutburst was observed by snapshot
observations.  The object was recorded at 15.1 on January 27 and
rapidly faded to 17.9 on January 28.
The entire duration of the outburst was thus longer than 25 d.
The object underwent a rebrightening at 15.3 on February 3
(vsnet-alert 12787, 12798; all observations are unfiltered CCD
magnitudes).  The object stayed at mag $\sim$18, 2.5 mag above
quiescence, at least until March 7.

\begin{table}
\caption{Superhump maxima of MisV 1443 (2011).}\label{tab:misv1443oc2011}
\begin{center}
\begin{tabular}{ccccc}
\hline
$E$ & max\commenta & error & $O-C$\commentb & $N$\commentc \\
\hline
0 & 55571.9590 & 0.0022 & $-$0.0046 & 100 \\
3 & 55572.1250 & 0.0005 & $-$0.0090 & 146 \\
4 & 55572.1842 & 0.0008 & $-$0.0065 & 113 \\
5 & 55572.2379 & 0.0010 & $-$0.0096 & 100 \\
19 & 55573.0464 & 0.0002 & 0.0038 & 100 \\
20 & 55573.1038 & 0.0003 & 0.0045 & 124 \\
21 & 55573.1609 & 0.0003 & 0.0048 & 117 \\
22 & 55573.2166 & 0.0004 & 0.0037 & 87 \\
23 & 55573.2735 & 0.0004 & 0.0038 & 130 \\
26 & 55573.4443 & 0.0007 & 0.0043 & 26 \\
27 & 55573.5006 & 0.0004 & 0.0038 & 47 \\
28 & 55573.5568 & 0.0003 & 0.0032 & 54 \\
29 & 55573.6135 & 0.0003 & 0.0031 & 51 \\
35 & 55573.9529 & 0.0009 & 0.0017 & 101 \\
37 & 55574.0662 & 0.0007 & 0.0015 & 52 \\
38 & 55574.1221 & 0.0004 & 0.0006 & 89 \\
39 & 55574.1793 & 0.0004 & 0.0010 & 88 \\
40 & 55574.2366 & 0.0005 & 0.0015 & 63 \\
41 & 55574.2916 & 0.0008 & $-$0.0003 & 62 \\
44 & 55574.4629 & 0.0004 & 0.0007 & 54 \\
45 & 55574.5202 & 0.0004 & 0.0012 & 41 \\
46 & 55574.5755 & 0.0007 & $-$0.0003 & 52 \\
53 & 55574.9743 & 0.0020 & 0.0011 & 61 \\
55 & 55575.0868 & 0.0004 & $-$0.0001 & 61 \\
56 & 55575.1431 & 0.0004 & $-$0.0006 & 109 \\
57 & 55575.1991 & 0.0004 & $-$0.0013 & 102 \\
58 & 55575.2551 & 0.0006 & $-$0.0021 & 59 \\
62 & 55575.4849 & 0.0005 & 0.0005 & 49 \\
63 & 55575.5408 & 0.0005 & $-$0.0004 & 48 \\
72 & 55576.0493 & 0.0034 & $-$0.0030 & 78 \\
73 & 55576.1125 & 0.0024 & 0.0035 & 105 \\
74 & 55576.1659 & 0.0039 & 0.0001 & 91 \\
75 & 55576.2154 & 0.0023 & $-$0.0072 & 99 \\
92 & 55577.1878 & 0.0008 & $-$0.0002 & 63 \\
93 & 55577.2434 & 0.0007 & $-$0.0013 & 62 \\
108 & 55578.0918 & 0.0025 & $-$0.0048 & 95 \\
109 & 55578.1533 & 0.0019 & $-$0.0000 & 88 \\
110 & 55578.2128 & 0.0017 & 0.0026 & 105 \\
\hline
  \multicolumn{5}{l}{\commenta BJD$-$2400000.} \\
  \multicolumn{5}{l}{\commentb Against max $= 2455571.9636 + 0.056787 E$.} \\
  \multicolumn{5}{l}{\commentc Number of points used to determine the maximum.} \\
\end{tabular}
\end{center}
\end{table}

\subsection{RX J1715.6$+$6856}\label{obj:j1715}

   The final stage of the 2009 superoutburst of this object
(hereafter RX J1715) was reported in \citet{she10j1715} and \citet{Pdot}.
I. Miller detected a new outburst on 2010 September 15 (baavss-alert 2386).
Subsequent observations confirmed superhumps (vsnet-alert 12182, 12183).
The times of superhump maxima are listed in table \ref{tab:j1715oc2010}.
The outburst was apparently observed in its late stage, and these
superhumps can be identified as stage C superhumps.

\begin{table}
\caption{Superhump maxima of RX J1715 (2010).}\label{tab:j1715oc2010}
\begin{center}
\begin{tabular}{ccccc}
\hline
$E$ & max\commenta & error & $O-C$\commentb & $N$\commentc \\
\hline
0 & 55455.4107 & 0.0010 & 0.0005 & 37 \\
14 & 55456.3996 & 0.0012 & $-$0.0011 & 77 \\
15 & 55456.4722 & 0.0007 & 0.0008 & 72 \\
16 & 55456.5439 & 0.0011 & 0.0018 & 18 \\
23 & 55457.0343 & 0.0051 & $-$0.0031 & 127 \\
24 & 55457.1051 & 0.0020 & $-$0.0030 & 112 \\
29 & 55457.4662 & 0.0016 & 0.0042 & 72 \\
30 & 55457.5325 & 0.0009 & $-$0.0001 & 76 \\
\hline
  \multicolumn{5}{l}{\commenta BJD$-$2400000.} \\
  \multicolumn{5}{l}{\commentb Against max $= 2455455.4101 + 0.070752 E$.} \\
  \multicolumn{5}{l}{\commentc Number of points used to determine the maximum.} \\
\end{tabular}
\end{center}
\end{table}

\subsection{SDSS J073208.11$+$413008.7}\label{obj:j0732}

   We observed the 2011 superoutburst of this object (hereafter SDSS J0732;
the object was selected by \cite{wil10newCVs}, see a comment in \cite{Pdot2}).
The times of superhump maxima are listed in table \ref{tab:j0732oc2011}.
The observation most likely recorded stage C superhumps, whose period
is in good agreement with the value in \citet{Pdot2}.

\begin{table}
\caption{Superhump maxima of SDSS J0732 (2011).}\label{tab:j0732oc2011}
\begin{center}
\begin{tabular}{ccccc}
\hline
$E$ & max\commenta & error & $O-C$\commentb & $N$\commentc \\
\hline
0 & 55582.2543 & 0.0021 & 0.0037 & 79 \\
12 & 55583.1975 & 0.0034 & $-$0.0050 & 159 \\
24 & 55584.1582 & 0.0022 & 0.0037 & 132 \\
30 & 55584.6289 & 0.0011 & $-$0.0015 & 153 \\
31 & 55584.7073 & 0.0010 & $-$0.0024 & 153 \\
68 & 55587.6474 & 0.0016 & 0.0026 & 152 \\
69 & 55587.7276 & 0.0014 & 0.0035 & 151 \\
70 & 55587.8042 & 0.0023 & 0.0008 & 152 \\
71 & 55587.8815 & 0.0032 & $-$0.0012 & 150 \\
72 & 55587.9580 & 0.0034 & $-$0.0041 & 122 \\
\hline
  \multicolumn{5}{l}{\commenta BJD$-$2400000.} \\
  \multicolumn{5}{l}{\commentb Against max $= 2455582.2506 + 0.079325 E$.} \\
  \multicolumn{5}{l}{\commentc Number of points used to determine the maximum.} \\
\end{tabular}
\end{center}
\end{table}

\subsection{SDSS J080306.99$+$284855.8}\label{obj:j0803}

\begin{figure}
  \begin{center}
    \FigureFile(88mm,110mm){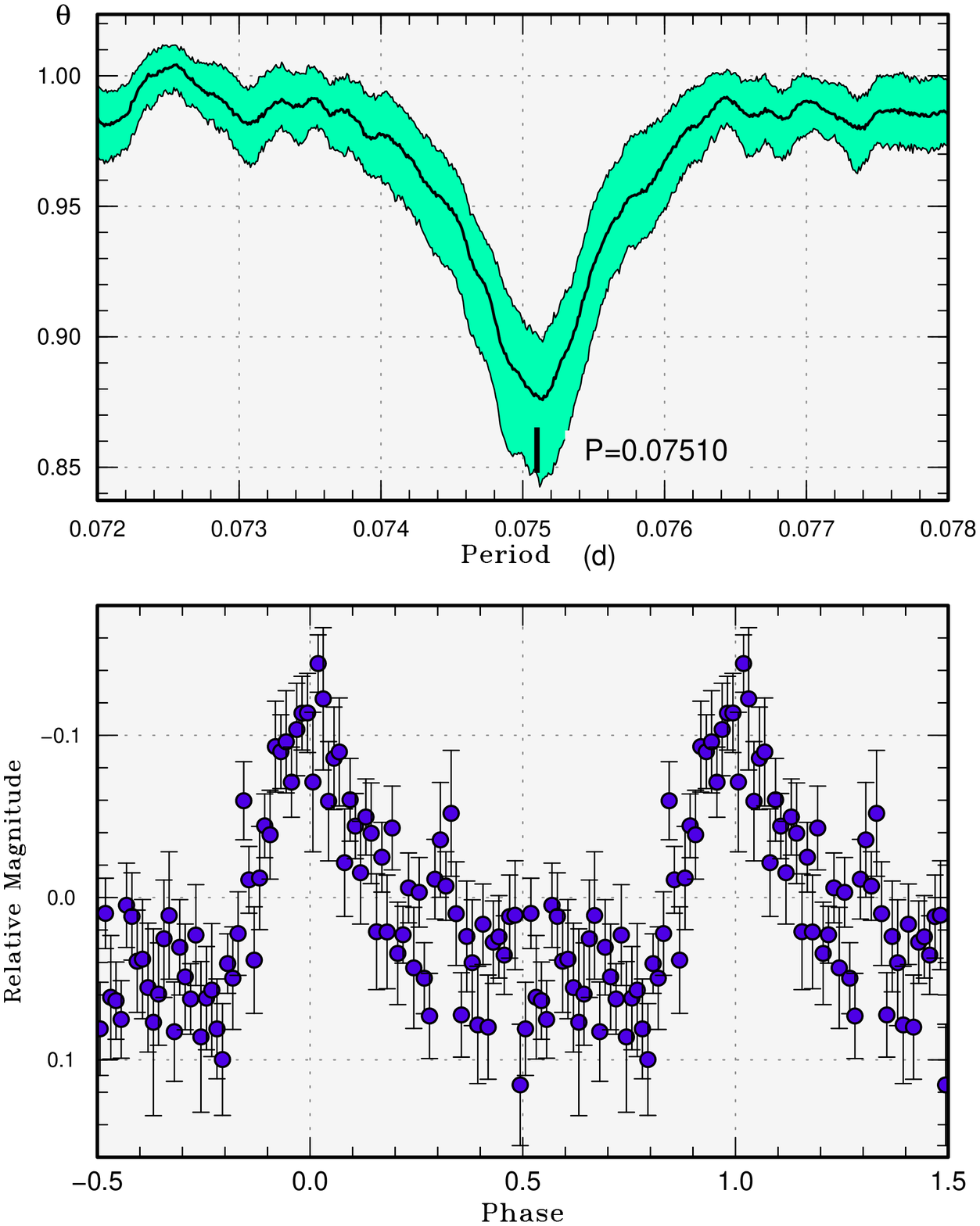}
  \end{center}
  \caption{Superhumps in SDSS J0803 (2011). (Upper): PDM analysis.
     (Lower): Phase-averaged profile.}
  \label{fig:j0803shpdm}
\end{figure}

   This object (hereafter SDSS J0803) was selected as a CV by
\citet{wil10newCVs}.  The 2011 outburst was detected by the CRTS, which was
the second brightest outbursts recorded by the CRTS.
Subsequent observations confirmed the presence of superhumps
(vsnet-alert 13070, 13073; figure \ref{fig:j0803shpdm}).
The times of superhump maxima are listed in table \ref{tab:j0803oc2011}.
As inferred from the initial large amplitudes of superhumps, the outburst
was apparently detected during its early stage (cf. vsnet-alert 13075).
We therefore identified these superhumps as stage B superhumps.
The large $P_{\rm dot}$ of $+13.6(4.2) \times 10^{-5}$ is consistent
with this interpretation.

\begin{table}
\caption{Superhump maxima of SDSS J0803 (2011).}\label{tab:j0803oc2011}
\begin{center}
\begin{tabular}{ccccc}
\hline
$E$ & max\commenta & error & $O-C$\commentb & $N$\commentc \\
\hline
0 & 55654.2962 & 0.0004 & 0.0017 & 58 \\
1 & 55654.3719 & 0.0005 & 0.0023 & 75 \\
11 & 55655.1193 & 0.0036 & $-$0.0013 & 133 \\
14 & 55655.3437 & 0.0005 & $-$0.0023 & 100 \\
15 & 55655.4210 & 0.0004 & $-$0.0000 & 143 \\
16 & 55655.4946 & 0.0008 & $-$0.0015 & 71 \\
23 & 55656.0223 & 0.0012 & 0.0005 & 154 \\
37 & 55657.0713 & 0.0025 & $-$0.0019 & 150 \\
49 & 55657.9750 & 0.0079 & 0.0005 & 125 \\
50 & 55658.0516 & 0.0040 & 0.0021 & 153 \\
\hline
  \multicolumn{5}{l}{\commenta BJD$-$2400000.} \\
  \multicolumn{5}{l}{\commentb Against max $= 2455654.2945 + 0.075100 E$.} \\
  \multicolumn{5}{l}{\commentc Number of points used to determine the maximum.} \\
\end{tabular}
\end{center}
\end{table}

\subsection{SDSS J080434.20$+$510349.2}\label{obj:j0804}

   SDSS J080434.20$+$510349.2 (hereafter SDSS J0804) is a CV selected
during the course of the Sloan Digital Sky Survey (SDSS)
\citep{szk06SDSSCV5}.  The first-ever superoutburst was recorded in 2006
\citep{pav07j0804}.  The detection of the 2006 superoutburst was only made
during the final stage of the plateau phase and the basic period of
superhumps was not very well established (\cite{she07j0804};
\cite{pav07j0804}; see \cite{kat09j0804} for a discussion of the
corrected superhump period).  This WZ Sge-type dwarf nova is
particularly notable in its grazing eclipses
(\cite{zha08j0804}; \cite{kat09j0804}), in the existence of eleven
rebrightenings during the 2006 superoutburst (\cite{pav07j0804};
\cite{kat09j0804}) and in the ZZ Cet-type pulsation of the white dwarf
(\cite{pav08j0804WD}; \cite{pav10j0804v1108her}).

   The object underwent an unexpected outburst in 2010 September,
only 4.5 yr after the preceding superoutburst (H. Maehara,
vsnet-alert 12184).  This outburst was fortunately detected in its
rising stage, and the presence of early superhumps and evolution
of ordinary superhumps were well recorded (vsnet-alert 12186, 12192,
12196; figure \ref{fig:j0804eshprof}).  The analysis
generally follows the method described in \citet{kat09j0804}.

\begin{figure}
  \begin{center}
    \FigureFile(88mm,110mm){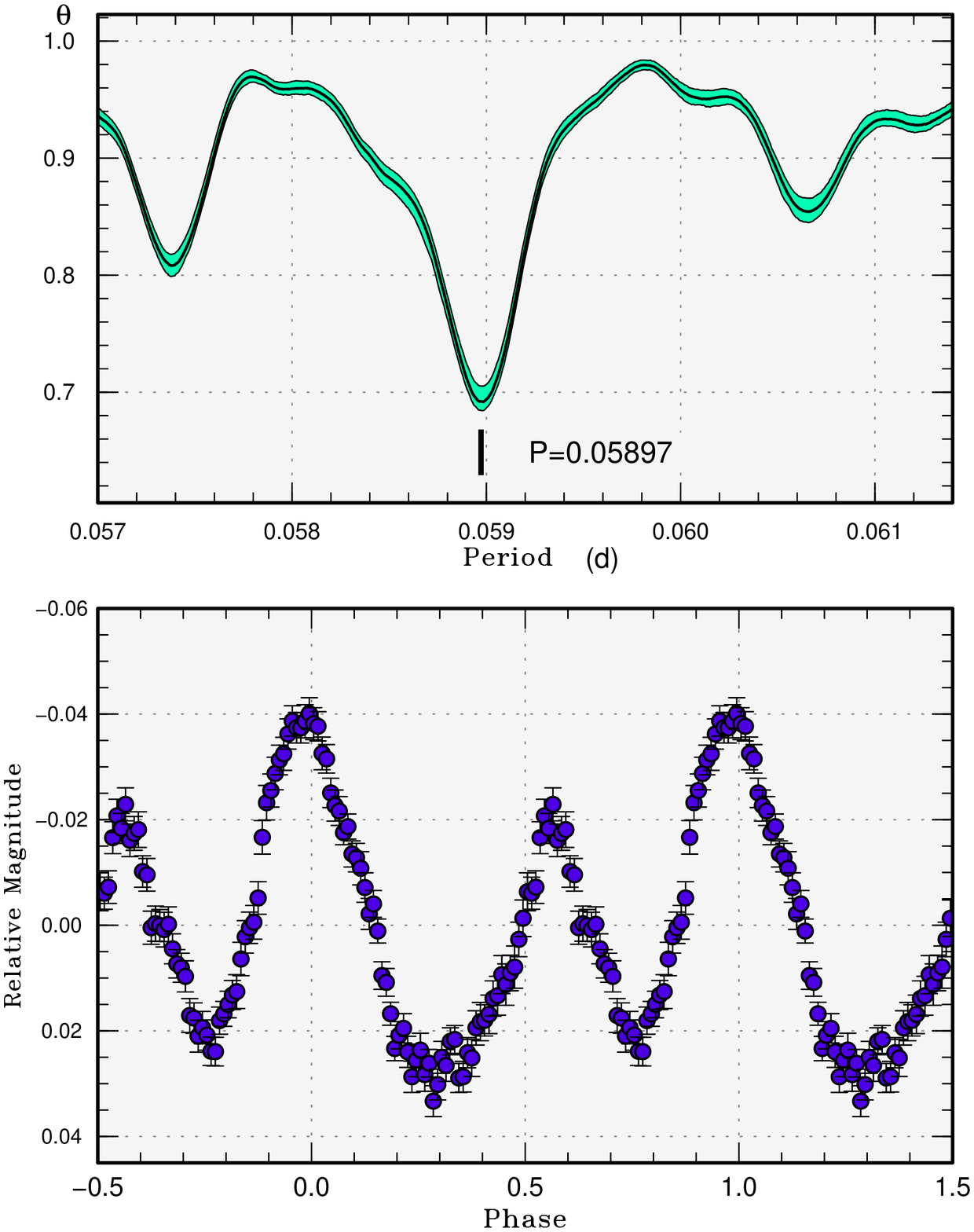}
  \end{center}
  \caption{Early superhumps in SDSS J0804 (2010). (Upper): PDM analysis.
     (Lower): Phase-averaged profile.}
  \label{fig:j0804eshpdm}
\end{figure}

\begin{figure}
  \begin{center}
    \FigureFile(88mm,110mm){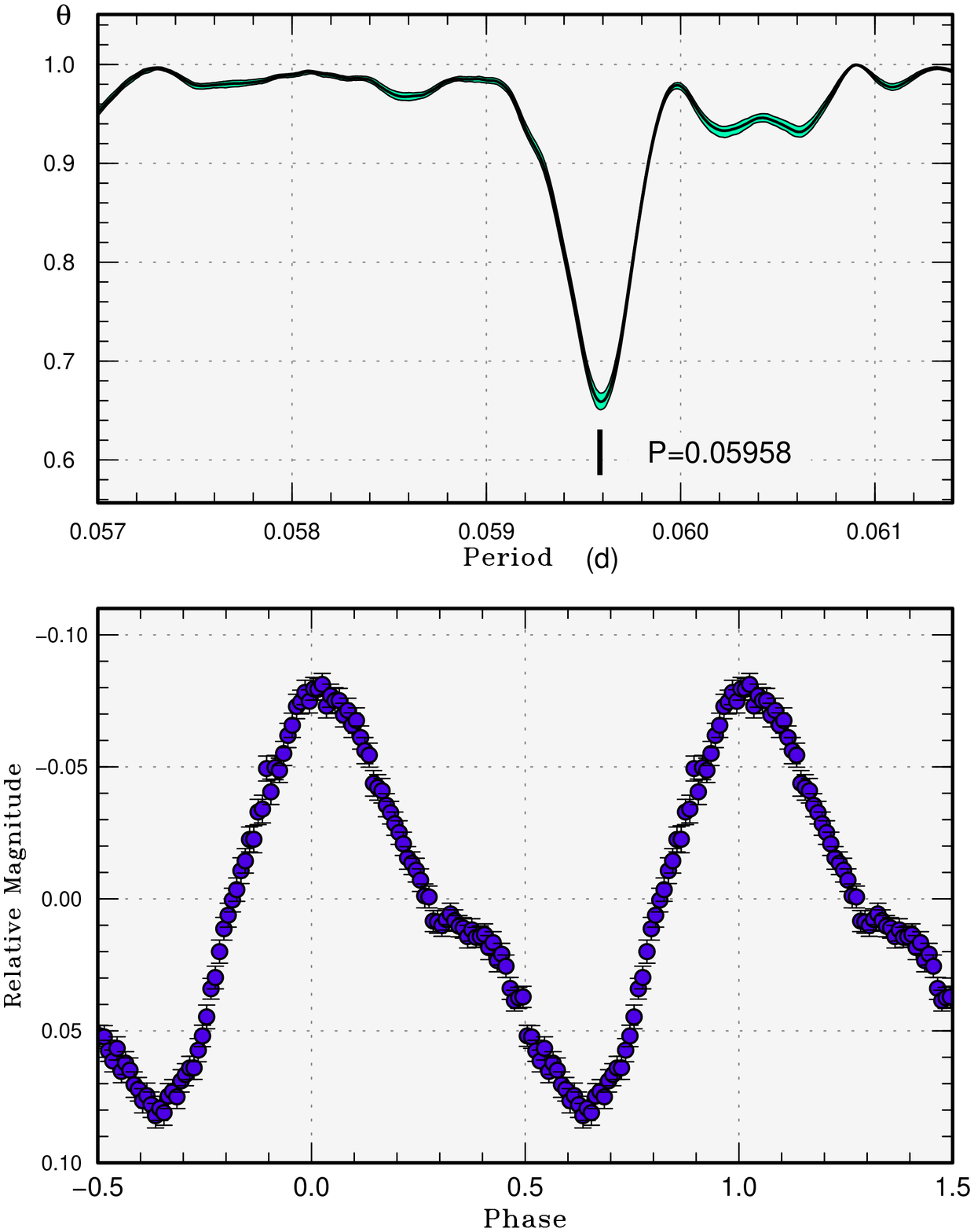}
  \end{center}
  \caption{Ordinary superhumps in SDSS J0804 (2010). (Upper): PDM analysis.
     (Lower): Phase-averaged profile.}
  \label{fig:j0804shpdm}
\end{figure}

\begin{figure}
  \begin{center}
    \FigureFile(88mm,110mm){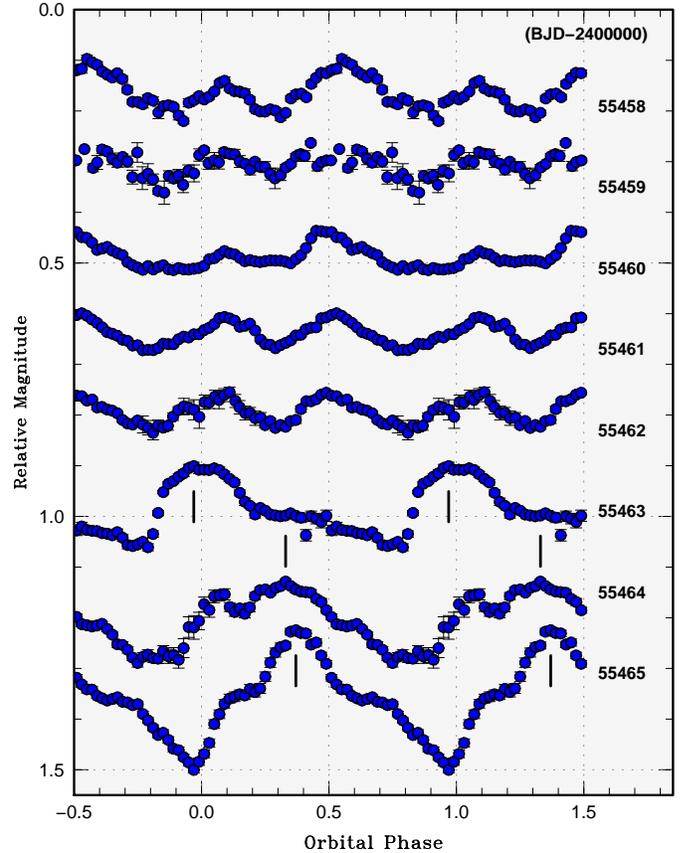}
  \end{center}
  \caption{Transition from early superhumps to ordinary superhumps in
  SDSS J0804 (2010).  Early superhumps were observed on the first five nights.
  Ordinary superhumps (ticks) seem to grow from the first (around orbital phase
  0.1) one of the double humps of early superhumps.
  The orbital phases were determined using the updated period of
  0.05900495 d after combination of quiescent analysis in \cite{kat09j0804}
  and post-superoutburst observations in 2010.  The phase zero is defined
  as the expected phase of eclipses.}
  \label{fig:j0804eshprof}
\end{figure}

   The mean period of early superhumps (before BJD 24455462)
was 0.058973(6) d, 0.05 \% shorter than $P_{\rm orb}$
(figure \ref{fig:j0804eshpdm}).  The mean period of ordinary
superhumps during the plateau phase (BJD 2455462--55472.5)
was 0.059584(2) d, 0.98 \% longer than  $P_{\rm orb}$
(figure \ref{fig:j0804shpdm}).

\begin{figure}
  \begin{center}
    \FigureFile(88mm,110mm){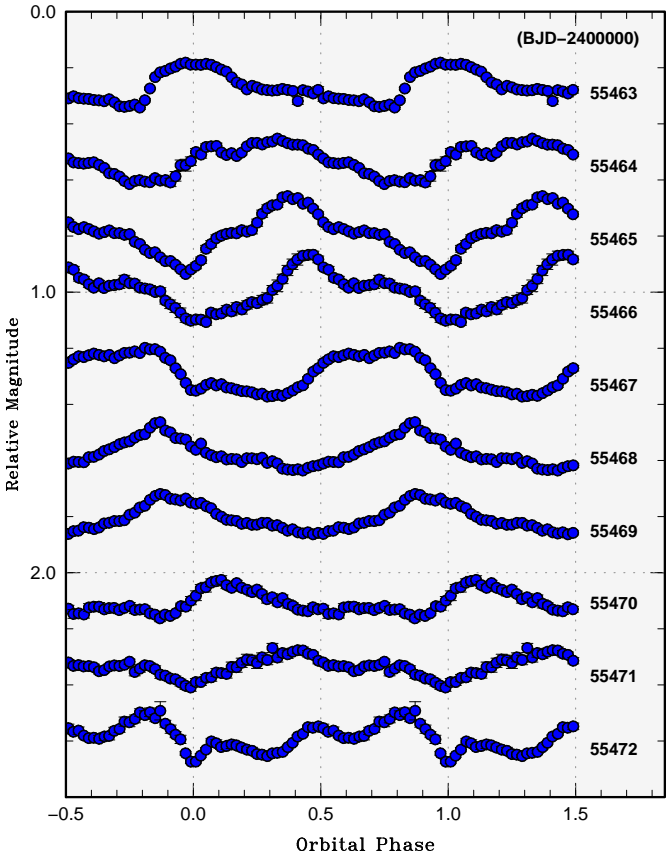}
  \end{center}
  \caption{Variation of profiles of ordinary superhumps in
  SDSS J0804 (2010).  The definition of the phase is as in figure
  \ref{fig:j0804eshprof}.  Shallow eclipses were seen on BJD
  2455464--2455467 and BJD 2455471--2455472.
  }
  \label{fig:j0804shprof}
\end{figure}

   The times of superhump maxima, determined after excluding the phases of
eclipses, during the plateau phase are listed in table \ref{tab:j0804oc2010}.
Since the contribution of orbital signature was small, we did not
subtract the mean orbital variation.
The stages A--C are clearly demonstrated.  The increase in the period
during the stage B was not uniform, and the major increase took place
during the final part of stage B.  This behavior in the period evolution
would explain the relatively long $P_{\rm SH}$ obtained during the
2006 superoutburst \citep{kat09j0804}.  In table \ref{tab:perlist}
we adopted $P_{\rm dot}$ = $+9.6(1.1) \times 10^{-5}$ ($31 \le E \le 159$).
This value is unusually large for a WZ Sge-type dwarf nova with
multiple rebrightenings.
If we neglect the final part with the major increase in the period
[cf. WZ Sge (2001) in \citet{Pdot}], we obtain
$P_{\rm dot}$ = $+3.5(0.9) \times 10^{-5}$ ($31 \le E \le 143$).
We will return to this issue in subsection \ref{sec:wzsgestat}.
In contrast to the expectation in \citet{kat09j0804}, shallow eclipses
were present during the entire course of the outburst.
The eclipses around the superoutburst maximum, however, were only observed
depending on superhump phases (figure \ref{fig:j0804shprof}).
Although stage B--C was apparently recorded, we did not determine
the period of stage C superhumps because the change in the $O-C$ diagram
was too large to be interpreted as a typical stage B--C transition.
This may have been a result of contamination of increasing signal of
orbital humps or a different component of superhumps as the system faded.

\begin{figure}
  \begin{center}
    \FigureFile(88mm,110mm){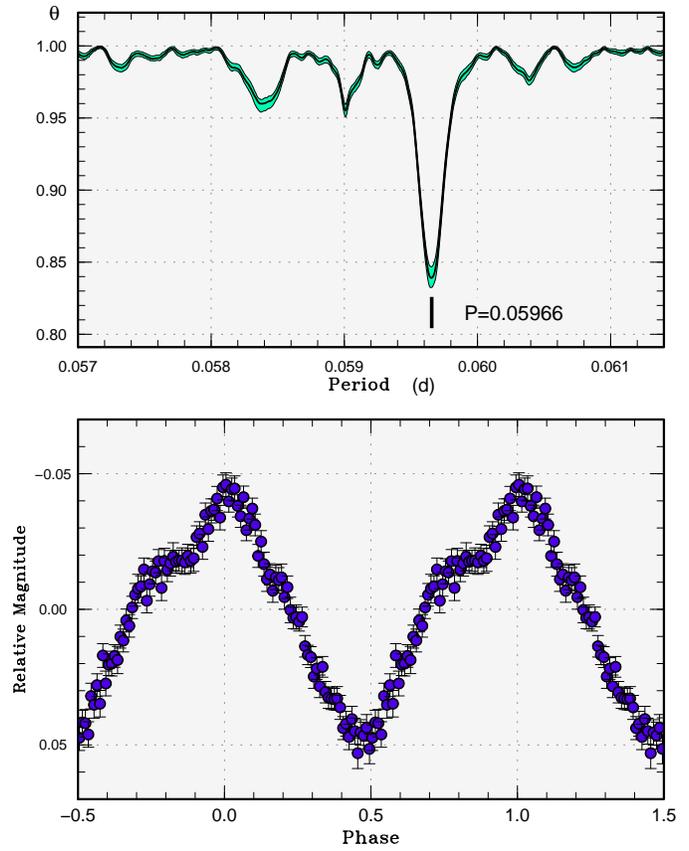}
  \end{center}
  \caption{Superhumps in SDSS J0804 (2010) during the rebrightening
     phase (BJD 2455472.5--2455494.0) after subtracting the global variations
     of rebrightenings, but not subtracting the orbital variations
     (Upper): PDM analysis.  A weaker signal at $P = 0.05901$ d
     corresponds to the orbital period.
     (Lower): Phase-averaged profile.}
  \label{fig:j0804rebpdm}
\end{figure}

\begin{figure}
  \begin{center}
    \FigureFile(88mm,110mm){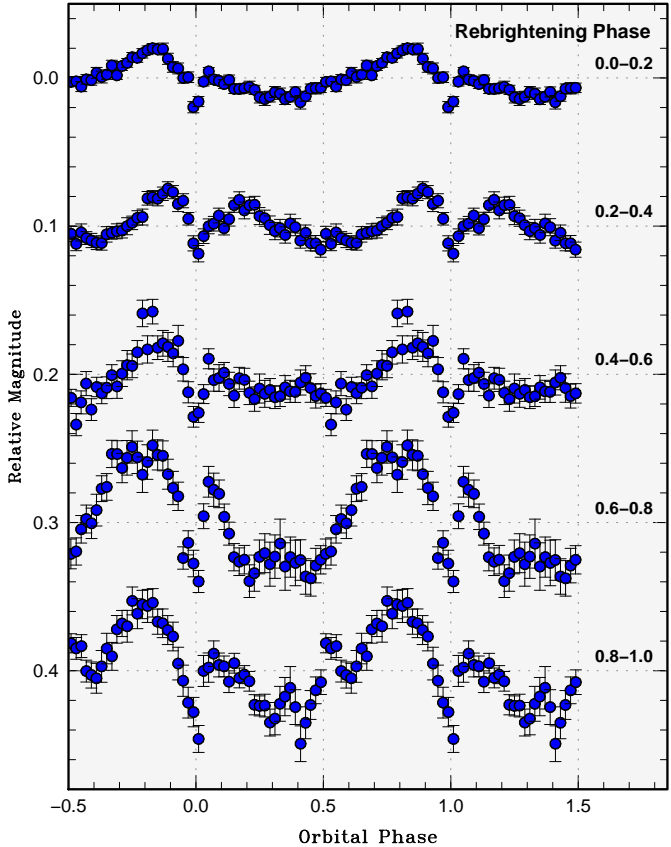}
  \end{center}
  \caption{Dependence of orbital variation of SDSS J0804 (2010)
     on the rebrightening phase.  The data are the same as in
     figure \ref{fig:j0804rebpdm}.  The zero phase of rebrightening
     is defined as the peak of each rebrightening.
     The amplitudes of orbital humps and depths of eclipses increased
     as the system faded from the peak of rebrightening.
     }
  \label{fig:j0804rebphprof}
\end{figure}

   During the six rebrightenings, persistent superhumps were observed
(figure \ref{fig:j0804rebpdm}).  The mean period was 0.059656(4) d,
confirming the slightly longer period observed during rebrightening
phase of the 2006 superoutburst \citep{kat09j0804}.
The rebrightenings following the current superoutburst was less numerous
than in 2006.  This may have resulted from the short interval
(for a WZ Sge-type dwarf nova) since the preceding superoutburst.
The amplitudes of orbital humps and depths of eclipses increased
by a factor of $\sim$2 as the system faded from the peak of rebrightening
(figure \ref{fig:j0804rebphprof}).  Since the typical amplitudes of
rebrightenings were $\sim$2 mag ($\sim$6 times in real scale),
this increase in the amplitudes cannot be simply explained by
a model assuming the orbital humps with a constant luminosity
and varying contribution from the accretion disk
(see also the discussion in \cite{kat09j0804}), and there was no
particular indication of an enhanced hot spot before the start of
rebrightenings (cf. \cite{osa03DNoutburst}).
The times of superhump maxima, determined after subtracting orbital
variations and excluding eclipses, during the rebrightening phase
are listed in table \ref{tab:j0804oc2010reb}.

\begin{figure}
  \begin{center}
    \FigureFile(88mm,110mm){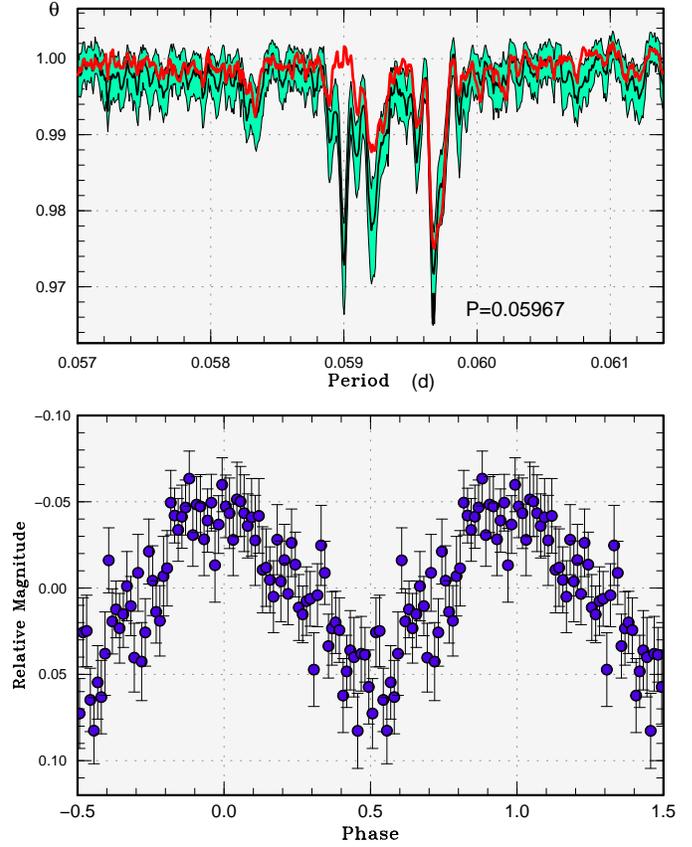}
  \end{center}
  \caption{Superhumps in SDSS J0804 (2010) during the post-rebrightening
     phase (after BJD 2455494.0) after subtracting the global variations
     of rebrightenings.
     (Upper): PDM analysis.  A signal at $P = 0.05901$ d
     corresponds to the orbital period.  The overlaid thick curve
     (red in the electronic version) is $\theta$ diagram after subtracting
     the orbital signal.
     (Lower): Phase-averaged profile.}
  \label{fig:j0804postpdm}
\end{figure}

   Even after rebrightenings, the superhump signal with a mean period
of 0.05967(2) d persisted (figure \ref{fig:j0804postpdm}).
The times of maxima determined after subtracting the orbital signal
are listed in table \ref{tab:j0804ocpostreb}.

\begin{figure*}
  \begin{center}
    \FigureFile(160mm,190mm){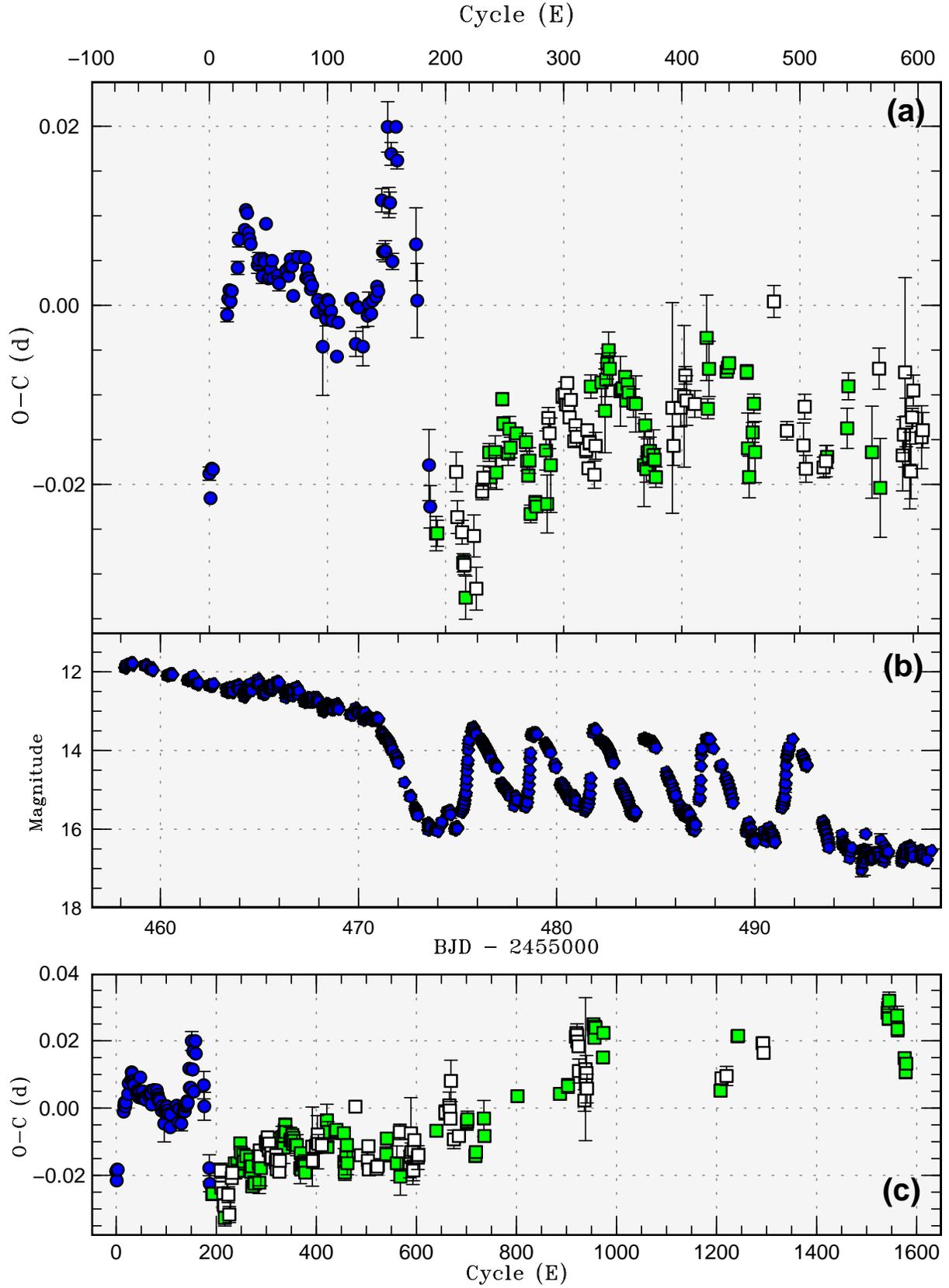}
  \end{center}
  \caption{$O-C$ diagram of superhumps in SDSS J0804 (2010).
      (a) $O-C$.
     Open squares indicate humps coinciding with the phase of orbital humps
     during the rebrightening phase.
     Filled squares are humps outside the phase of orbital humps and humps
     recorded during the plateau phase.
     We used a period of 0.05962 d for calculating the $O-C$'s.
     (b) Light curve.
     (c) $O-C$ diagram of the entire observation.
     The global evolution of the $O-C$ diagram is remarkably similar
     to those of GW Lib and V455 And \citep{Pdot} except the six rebrightenings
     after the major disturbance in the $O-C$ diagram.
  }
  \label{fig:j0804humpall}
\end{figure*}

   The $O-C$ diagram together with the light curve is shown in
figure \ref{fig:j0804humpall}.  
The global evolution of the $O-C$ diagram is remarkably similar
to those of GW Lib and V455 And \citep{Pdot} except the six rebrightenings
after the major disturbance in the $O-C$.  The late stage (during
rebrightenings and after rebrightenings) superhumps can be interpreted
as an extension on the stage B superhumps.

\begin{figure}
  \begin{center}
    \FigureFile(88mm,110mm){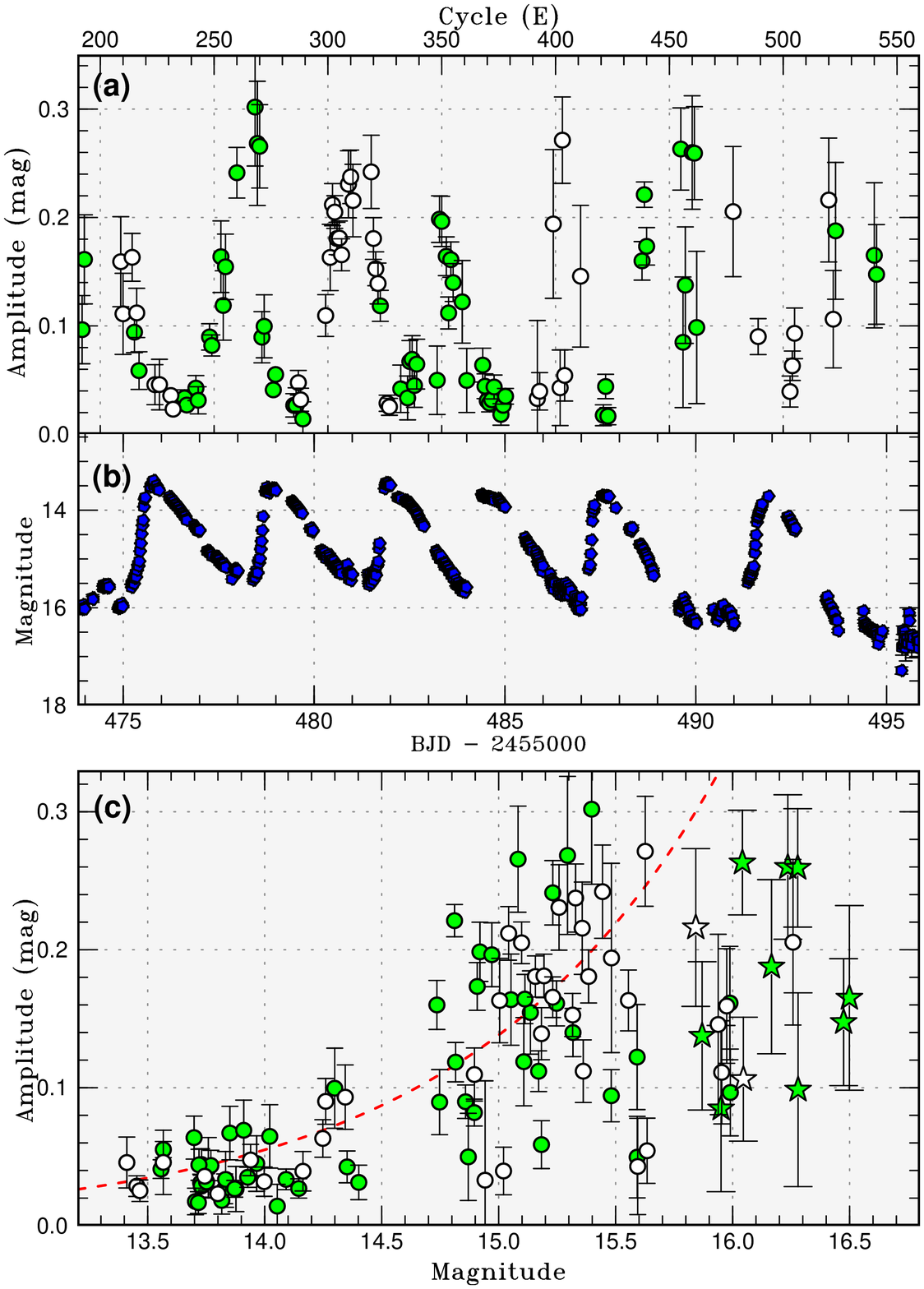}
  \end{center}
  \caption{Amplitudes of superhumps in SDSS J0804 (2010) during the
     rebrightening phase.
     (a) Amplitudes.  
     Open circles indicate humps coinciding with the phase of orbital humps
     during the rebrightening phase.
     Filled circles are humps outside the phase of orbital humps and humps
     recorded during the plateau phase.
     (b) Light curve.
     (c) Relation between amplitudes and system brightness.
     The circle symbols are as in the panel (a).  The star-shaped symbols
     represent humps during the quiescence immediately before and after
     the final rebrightening.
     The amplitudes increase when the system becomes fainter.
     The broken curve represents the relation assuming the constant
     luminosity of the superhump source and variable contribution
     from the accretion disk.
  }
  \label{fig:j0804humpamp}
\end{figure}

\begin{figure}
  \begin{center}
    \FigureFile(88mm,110mm){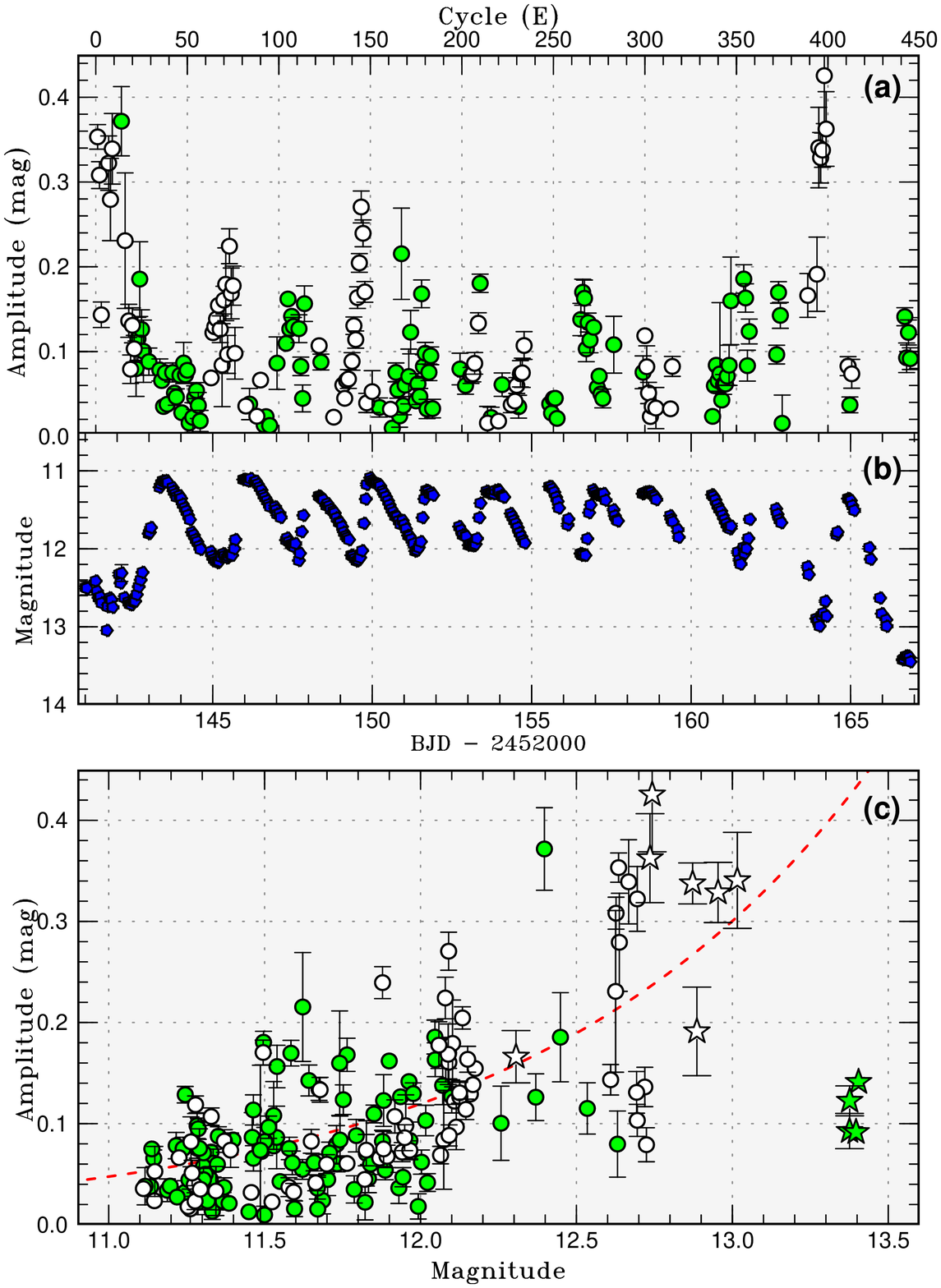}
  \end{center}
  \caption{Amplitudes of superhumps in WZ Sge (2001) during the
     rebrightening phase.
     (a) Amplitudes.  
     Open circles indicate humps coinciding with the phase of orbital humps
     during the rebrightening phase.
     Filled circles are humps outside the phase of orbital humps and humps
     recorded during the plateau phase.
     (b) Light curve.
     (c) Relation between amplitudes and system brightness.
     The circle symbols are as in the panel (a).  The star-shaped symbols
     represent humps during the quiescence immediately before and after
     the final rebrightening.
     The amplitudes increase when the system becomes fainter.
     The broken curve represents the relation assuming the constant
     luminosity of the superhump source and variable contribution
     from the accretion disk.  The amplitude of humps immediately
     preceding the final rebrightening may have been affected by
     the overlapping orbital humps.
  }
  \label{fig:wzsgerebhumpamp}
\end{figure}

   The amplitudes of superhumps (after subtracting for the orbital variation)
during the rebrightening systematically varied depending on the
brightness of the system (figure \ref{fig:j0804humpamp}).
The amplitudes increased when the system became fainter.
The tendency appears to be well explained by assuming the constant
luminosity of the superhump source and assuming variable contribution
from the accretion disk.
The general tendency in amplitudes of superhumps is similar to
what was observed in persistent negative superhumps in V503 Cyg
\citep{har95v503cyg} and MN Dra \citep{pav10mndraproc}, and superhumps
during the rebrightening phase of EG Cnc (\cite{pat98egcnc};
\cite{kat04egcnc}) and WZ Sge (cf. \cite{pat02wzsge};
reanalysis of the data in \cite{Pdot} is shown in figure
\ref{fig:wzsgerebhumpamp}).
The tendency showed a deviation from this relation around and after
the final rebrightening.  The decreased strength of superhumps, or tidal
dissipation, may have prevented further rebrightenings to occur.
The phenomenon (immediately following the rebrightening phase)
has been also confirmed in the 2001 superoutburst of WZ Sge
(figure \ref{fig:wzsgerebhumpamp}).

\begin{table}
\caption{Superhump maxima of SDSS J0804 (2010).}\label{tab:j0804oc2010}
\begin{center}
\begin{tabular}{cccccc}
\hline
$E$ & max\commenta & error & $O-C$\commentb & phase\commentc & $N$\commentd \\
\hline
0 & 55462.4512 & 0.0007 & $-$0.0188 & 0.53 & 59 \\
1 & 55462.5081 & 0.0006 & $-$0.0216 & 0.49 & 190 \\
2 & 55462.5710 & 0.0006 & $-$0.0183 & 0.56 & 198 \\
3 & 55462.6305 & 0.0006 & $-$0.0184 & 0.57 & 129 \\
15 & 55463.3632 & 0.0008 & $-$0.0014 & 0.99 & 174 \\
16 & 55463.4246 & 0.0001 & 0.0003 & 0.03 & 690 \\
17 & 55463.4852 & 0.0002 & 0.0013 & 0.06 & 529 \\
18 & 55463.5436 & 0.0002 & 0.0000 & 0.04 & 281 \\
19 & 55463.6044 & 0.0003 & 0.0011 & 0.07 & 185 \\
24 & 55463.9051 & 0.0007 & 0.0036 & 0.17 & 48 \\
25 & 55463.9678 & 0.0008 & 0.0067 & 0.23 & 48 \\
30 & 55464.2670 & 0.0003 & 0.0077 & 0.30 & 143 \\
31 & 55464.3289 & 0.0002 & 0.0099 & 0.35 & 744 \\
32 & 55464.3881 & 0.0001 & 0.0095 & 0.36 & 691 \\
33 & 55464.4456 & 0.0004 & 0.0073 & 0.33 & 180 \\
34 & 55464.5045 & 0.0002 & 0.0066 & 0.33 & 438 \\
35 & 55464.5635 & 0.0001 & 0.0060 & 0.33 & 871 \\
41 & 55464.9189 & 0.0009 & 0.0035 & 0.35 & 47 \\
42 & 55464.9792 & 0.0008 & 0.0041 & 0.37 & 44 \\
45 & 55465.1561 & 0.0008 & 0.0022 & 0.37 & 108 \\
46 & 55465.2177 & 0.0002 & 0.0041 & 0.42 & 521 \\
47 & 55465.2770 & 0.0002 & 0.0038 & 0.42 & 808 \\
48 & 55465.3409 & 0.0004 & 0.0080 & 0.50 & 266 \\
50 & 55465.4540 & 0.0007 & 0.0018 & 0.42 & 86 \\
51 & 55465.5137 & 0.0005 & 0.0019 & 0.43 & 151 \\
52 & 55465.5743 & 0.0003 & 0.0028 & 0.46 & 219 \\
53 & 55465.6348 & 0.0004 & 0.0038 & 0.49 & 147 \\
54 & 55465.6925 & 0.0004 & 0.0018 & 0.46 & 95 \\
58 & 55465.9313 & 0.0006 & 0.0020 & 0.51 & 56 \\
59 & 55465.9900 & 0.0008 & 0.0011 & 0.51 & 40 \\
65 & 55466.3492 & 0.0005 & 0.0024 & 0.59 & 49 \\
67 & 55466.4678 & 0.0005 & 0.0017 & 0.60 & 93 \\
68 & 55466.5285 & 0.0004 & 0.0027 & 0.63 & 200 \\
69 & 55466.5889 & 0.0004 & 0.0036 & 0.66 & 124 \\
70 & 55466.6478 & 0.0003 & 0.0028 & 0.65 & 159 \\
71 & 55466.7041 & 0.0004 & $-$0.0006 & 0.61 & 119 \\
75 & 55466.9469 & 0.0007 & 0.0037 & 0.72 & 47 \\
81 & 55467.3045 & 0.0007 & 0.0035 & 0.78 & 86 \\
82 & 55467.3619 & 0.0003 & 0.0012 & 0.76 & 275 \\
83 & 55467.4224 & 0.0002 & 0.0021 & 0.78 & 668 \\
84 & 55467.4811 & 0.0003 & 0.0011 & 0.78 & 181 \\
85 & 55467.5404 & 0.0003 & 0.0008 & 0.78 & 161 \\
86 & 55467.5991 & 0.0005 & $-$0.0002 & 0.78 & 55 \\
87 & 55467.6591 & 0.0003 & 0.0002 & 0.79 & 50 \\
91 & 55467.8947 & 0.0005 & $-$0.0029 & 0.78 & 64 \\
92 & 55467.9557 & 0.0003 & $-$0.0015 & 0.82 & 175 \\
96 & 55468.1889 & 0.0054 & $-$0.0068 & 0.77 & 26 \\
97 & 55468.2526 & 0.0005 & $-$0.0028 & 0.85 & 198 \\
98 & 55468.3127 & 0.0004 & $-$0.0023 & 0.87 & 681 \\
99 & 55468.3709 & 0.0008 & $-$0.0038 & 0.86 & 324 \\
100 & 55468.4326 & 0.0003 & $-$0.0017 & 0.90 & 306 \\
101 & 55468.4921 & 0.0002 & $-$0.0019 & 0.91 & 886 \\
102 & 55468.5504 & 0.0003 & $-$0.0031 & 0.90 & 289 \\
\hline
  \multicolumn{6}{l}{\commenta BJD$-$2400000.} \\
  \multicolumn{6}{l}{\commentb Against max $= 2455462.4700 + 0.059643 E$.} \\
  \multicolumn{6}{l}{\commentc Orbital phase.} \\
  \multicolumn{6}{l}{\commentd Number of points used to determine the maximum.} \\
\end{tabular}
\end{center}
\end{table}

\addtocounter{table}{-1}
\begin{table}
\caption{Superhump maxima of SDSS J0804 (2010) (continued).}
\begin{center}
\begin{tabular}{cccccc}
\hline
$E$ & max\commenta & error & $O-C$\commentb & phase\commentc & $N$\commentd \\
\hline
103 & 55468.6102 & 0.0003 & $-$0.0030 & 0.91 & 275 \\
104 & 55468.6687 & 0.0004 & $-$0.0041 & 0.90 & 90 \\
108 & 55468.9032 & 0.0006 & $-$0.0082 & 0.88 & 110 \\
109 & 55468.9667 & 0.0005 & $-$0.0044 & 0.95 & 174 \\
120 & 55469.6250 & 0.0004 & $-$0.0021 & 0.11 & 91 \\
121 & 55469.6848 & 0.0004 & $-$0.0020 & 0.12 & 92 \\
124 & 55469.8586 & 0.0014 & $-$0.0071 & 0.07 & 31 \\
125 & 55469.9223 & 0.0005 & $-$0.0030 & 0.15 & 169 \\
126 & 55469.9819 & 0.0005 & $-$0.0031 & 0.16 & 159 \\
130 & 55470.2160 & 0.0021 & $-$0.0076 & 0.13 & 62 \\
134 & 55470.4579 & 0.0012 & $-$0.0042 & 0.23 & 46 \\
135 & 55470.5189 & 0.0013 & $-$0.0029 & 0.26 & 31 \\
137 & 55470.6370 & 0.0004 & $-$0.0041 & 0.26 & 91 \\
138 & 55470.6981 & 0.0003 & $-$0.0027 & 0.30 & 89 \\
141 & 55470.8774 & 0.0004 & $-$0.0023 & 0.33 & 98 \\
142 & 55470.9381 & 0.0002 & $-$0.0011 & 0.36 & 218 \\
143 & 55470.9972 & 0.0003 & $-$0.0017 & 0.37 & 131 \\
146 & 55471.1862 & 0.0013 & 0.0084 & 0.57 & 189 \\
147 & 55471.2401 & 0.0009 & 0.0026 & 0.48 & 435 \\
148 & 55471.2997 & 0.0004 & 0.0026 & 0.49 & 479 \\
149 & 55471.3594 & 0.0012 & 0.0026 & 0.50 & 162 \\
151 & 55471.4925 & 0.0028 & 0.0165 & 0.76 & 26 \\
152 & 55471.5437 & 0.0017 & 0.0080 & 0.63 & 28 \\
153 & 55471.6033 & 0.0012 & 0.0080 & 0.64 & 111 \\
154 & 55471.6684 & 0.0013 & 0.0134 & 0.74 & 90 \\
155 & 55471.7160 & 0.0009 & 0.0014 & 0.55 & 70 \\
158 & 55471.9099 & 0.0006 & 0.0163 & 0.83 & 142 \\
159 & 55471.9658 & 0.0009 & 0.0126 & 0.78 & 175 \\
175 & 55472.9103 & 0.0041 & 0.0028 & 0.79 & 45 \\
176 & 55472.9637 & 0.0042 & $-$0.0035 & 0.69 & 45 \\
186 & 55473.5415 & 0.0040 & $-$0.0221 & 0.49 & 16 \\
187 & 55473.5964 & 0.0024 & $-$0.0268 & 0.42 & 16 \\
\hline
  \multicolumn{6}{l}{\commenta BJD$-$2400000.} \\
  \multicolumn{6}{l}{\commentb Against max $= 2455462.4701 + 0.059636 E$.} \\
  \multicolumn{6}{l}{\commentc Orbital phase.} \\
  \multicolumn{6}{l}{\commentd Number of points used to determine the maximum.} \\
\end{tabular}
\end{center}
\end{table}

\begin{table}
\caption{Superhump maxima of J0804 (2010) during the rebrightening phase.}\label{tab:j0804oc2010reb}
\begin{center}
\begin{tabular}{cccccc}
\hline
$E$ & max\commenta & error & $O-C$\commentb & phase\commentc & $N$\commentd \\
\hline
0 & 55473.8915 & 0.0019 & $-$0.0057 & 0.42 & 24 \\
1 & 55473.9512 & 0.0015 & $-$0.0056 & 0.43 & 23 \\
17 & 55474.9120 & 0.0022 & 0.0007 & 0.71 & 12 \\
18 & 55474.9665 & 0.0018 & $-$0.0044 & 0.64 & 23 \\
22 & 55475.2033 & 0.0013 & $-$0.0063 & 0.65 & 79 \\
23 & 55475.2595 & 0.0010 & $-$0.0097 & 0.60 & 126 \\
24 & 55475.3189 & 0.0011 & $-$0.0100 & 0.61 & 133 \\
25 & 55475.3749 & 0.0024 & $-$0.0137 & 0.56 & 46 \\
32 & 55475.7991 & 0.0023 & $-$0.0070 & 0.75 & 32 \\
34 & 55475.9125 & 0.0024 & $-$0.0130 & 0.67 & 21 \\
39 & 55476.2214 & 0.0009 & $-$0.0023 & 0.90 & 283 \\
40 & 55476.2826 & 0.0013 & $-$0.0008 & 0.94 & 342 \\
45 & 55476.5835 & 0.0010 & 0.0019 & 0.04 & 128 \\
46 & 55476.6403 & 0.0012 & $-$0.0010 & 0.00 & 140 \\
50 & 55476.8817 & 0.0018 & 0.0017 & 0.09 & 75 \\
51 & 55476.9390 & 0.0019 & $-$0.0006 & 0.07 & 90 \\
56 & 55477.2453 & 0.0007 & 0.0074 & 0.26 & 235 \\
57 & 55477.3022 & 0.0008 & 0.0047 & 0.22 & 269 \\
61 & 55477.5373 & 0.0013 & 0.0012 & 0.21 & 18 \\
62 & 55477.5997 & 0.0014 & 0.0039 & 0.26 & 20 \\
63 & 55477.6572 & 0.0012 & 0.0018 & 0.24 & 18 \\
68 & 55477.9569 & 0.0005 & 0.0032 & 0.32 & 77 \\
76 & 55478.4329 & 0.0010 & 0.0019 & 0.38 & 24 \\
77 & 55478.4904 & 0.0011 & $-$0.0002 & 0.36 & 22 \\
78 & 55478.5484 & 0.0007 & $-$0.0019 & 0.34 & 56 \\
79 & 55478.6097 & 0.0014 & $-$0.0002 & 0.38 & 60 \\
80 & 55478.6633 & 0.0010 & $-$0.0062 & 0.29 & 27 \\
84 & 55478.9030 & 0.0008 & $-$0.0051 & 0.35 & 81 \\
85 & 55478.9623 & 0.0005 & $-$0.0056 & 0.36 & 83 \\
93 & 55479.4455 & 0.0022 & 0.0004 & 0.55 & 23 \\
94 & 55479.4991 & 0.0032 & $-$0.0056 & 0.45 & 23 \\
95 & 55479.5684 & 0.0012 & 0.0040 & 0.63 & 22 \\
96 & 55479.6263 & 0.0017 & 0.0022 & 0.61 & 23 \\
97 & 55479.6823 & 0.0053 & $-$0.0014 & 0.56 & 15 \\
107 & 55480.2863 & 0.0008 & 0.0060 & 0.79 & 199 \\
109 & 55480.4057 & 0.0014 & 0.0061 & 0.82 & 14 \\
110 & 55480.4642 & 0.0005 & 0.0050 & 0.81 & 34 \\
111 & 55480.5262 & 0.0005 & 0.0073 & 0.86 & 84 \\
112 & 55480.5833 & 0.0005 & 0.0048 & 0.83 & 99 \\
113 & 55480.6416 & 0.0005 & 0.0034 & 0.82 & 102 \\
114 & 55480.7032 & 0.0006 & 0.0053 & 0.86 & 74 \\
117 & 55480.8774 & 0.0007 & 0.0006 & 0.81 & 45 \\
118 & 55480.9388 & 0.0007 & 0.0024 & 0.85 & 65 \\
119 & 55480.9971 & 0.0010 & 0.0010 & 0.84 & 55 \\
127 & 55481.4726 & 0.0009 & $-$0.0007 & 0.90 & 40 \\
128 & 55481.5345 & 0.0005 & 0.0015 & 0.95 & 126 \\
129 & 55481.5898 & 0.0007 & $-$0.0028 & 0.89 & 128 \\
130 & 55481.6523 & 0.0008 & 0.0000 & 0.95 & 108 \\
131 & 55481.7182 & 0.0013 & 0.0062 & 0.06 & 37 \\
134 & 55481.8872 & 0.0015 & $-$0.0037 & 0.93 & 66 \\
135 & 55481.9500 & 0.0015 & $-$0.0005 & 0.99 & 65 \\
\hline
  \multicolumn{6}{l}{\commenta BJD$-$2400000.} \\
  \multicolumn{6}{l}{\commentb Against max $= 2455473.8972 + 0.059655 E$.} \\
  \multicolumn{6}{l}{\commentc Orbital phase.} \\
  \multicolumn{6}{l}{\commentd Number of points used to determine the maximum.} \\
\end{tabular}
\end{center}
\end{table}

\addtocounter{table}{-1}
\begin{table}
\caption{Superhump maxima of J0804 (2010) during the rebrightening phase (continued).}
\begin{center}
\begin{tabular}{cccccc}
\hline
$E$ & max\commenta & error & $O-C$\commentb & phase\commentc & $N$\commentd \\
\hline
140 & 55482.2552 & 0.0032 & 0.0064 & 0.16 & 30 \\
143 & 55482.4309 & 0.0047 & 0.0031 & 0.14 & 62 \\
144 & 55482.4940 & 0.0019 & 0.0066 & 0.21 & 153 \\
145 & 55482.5554 & 0.0017 & 0.0083 & 0.25 & 198 \\
146 & 55482.6165 & 0.0020 & 0.0098 & 0.29 & 193 \\
147 & 55482.6741 & 0.0018 & 0.0077 & 0.26 & 149 \\
156 & 55483.2082 & 0.0039 & 0.0048 & 0.31 & 102 \\
157 & 55483.2680 & 0.0005 & 0.0050 & 0.33 & 170 \\
158 & 55483.3277 & 0.0006 & 0.0051 & 0.34 & 142 \\
160 & 55483.4483 & 0.0006 & 0.0063 & 0.38 & 152 \\
161 & 55483.5052 & 0.0007 & 0.0036 & 0.35 & 173 \\
162 & 55483.5667 & 0.0006 & 0.0054 & 0.39 & 178 \\
163 & 55483.6254 & 0.0007 & 0.0045 & 0.38 & 87 \\
167 & 55483.8627 & 0.0016 & 0.0031 & 0.41 & 34 \\
169 & 55483.9818 & 0.0031 & 0.0030 & 0.43 & 29 \\
176 & 55484.3923 & 0.0046 & $-$0.0041 & 0.38 & 11 \\
177 & 55484.4564 & 0.0013 & 0.0003 & 0.47 & 68 \\
178 & 55484.5111 & 0.0015 & $-$0.0046 & 0.40 & 88 \\
179 & 55484.5726 & 0.0015 & $-$0.0028 & 0.44 & 197 \\
180 & 55484.6318 & 0.0010 & $-$0.0032 & 0.44 & 195 \\
181 & 55484.6920 & 0.0015 & $-$0.0027 & 0.46 & 89 \\
184 & 55484.8704 & 0.0028 & $-$0.0033 & 0.48 & 124 \\
185 & 55484.9295 & 0.0012 & $-$0.0038 & 0.49 & 135 \\
186 & 55484.9872 & 0.0011 & $-$0.0058 & 0.46 & 139 \\
200 & 55485.8296 & 0.0118 & 0.0015 & 0.74 & 18 \\
201 & 55485.8850 & 0.0022 & $-$0.0028 & 0.68 & 38 \\
207 & 55486.2469 & 0.0021 & 0.0012 & 0.81 & 37 \\
210 & 55486.4271 & 0.0079 & 0.0024 & 0.87 & 17 \\
211 & 55486.4890 & 0.0010 & 0.0047 & 0.92 & 60 \\
212 & 55486.5459 & 0.0028 & 0.0019 & 0.88 & 96 \\
219 & 55486.9628 & 0.0015 & 0.0012 & 0.95 & 13 \\
229 & 55487.5664 & 0.0048 & 0.0083 & 0.18 & 64 \\
230 & 55487.6181 & 0.0011 & 0.0003 & 0.05 & 65 \\
231 & 55487.6822 & 0.0031 & 0.0047 & 0.14 & 63 \\
246 & 55488.5761 & 0.0005 & 0.0039 & 0.29 & 107 \\
247 & 55488.6362 & 0.0003 & 0.0043 & 0.31 & 111 \\
248 & 55488.6963 & 0.0005 & 0.0048 & 0.33 & 77 \\
263 & 55489.5897 & 0.0008 & 0.0033 & 0.47 & 64 \\
264 & 55489.6408 & 0.0039 & $-$0.0053 & 0.33 & 43 \\
265 & 55489.6972 & 0.0023 & $-$0.0085 & 0.29 & 43 \\
268 & 55489.8810 & 0.0011 & $-$0.0036 & 0.40 & 27 \\
269 & 55489.9438 & 0.0011 & $-$0.0005 & 0.47 & 23 \\
270 & 55489.9981 & 0.0034 & $-$0.0059 & 0.39 & 21 \\
286 & 55490.9688 & 0.0018 & 0.0104 & 0.84 & 18 \\
297 & 55491.6102 & 0.0010 & $-$0.0045 & 0.71 & 56 \\
311 & 55492.4432 & 0.0025 & $-$0.0066 & 0.83 & 74 \\
312 & 55492.5072 & 0.0014 & $-$0.0023 & 0.91 & 76 \\
313 & 55492.5598 & 0.0015 & $-$0.0093 & 0.80 & 37 \\
328 & 55493.4543 & 0.0012 & $-$0.0096 & 0.96 & 18 \\
330 & 55493.5742 & 0.0017 & $-$0.0090 & 1.00 & 71 \\
331 & 55493.6343 & 0.0013 & $-$0.0085 & 0.01 & 51 \\
\hline
  \multicolumn{6}{l}{\commenta BJD$-$2400000.} \\
  \multicolumn{6}{l}{\commentb Against max $= 2455473.8972 + 0.059655 E$.} \\
  \multicolumn{6}{l}{\commentc Orbital phase.} \\
  \multicolumn{6}{l}{\commentd Number of points used to determine the maximum.} \\
\end{tabular}
\end{center}
\end{table}

\begin{table}
\caption{Superhump maxima of SDSS J0804 (2010) during the post-rebrightening phase.}\label{tab:j0804ocpostreb}
\begin{center}
\begin{tabular}{cccccc}
\hline
$E$ & max\commenta & error & $O-C$\commentb & phase\commentc & $N$\commentd \\
\hline
0 & 55494.6511 & 0.0022 & $-$0.0052 & 0.25 & 39 \\
1 & 55494.7154 & 0.0015 & $-$0.0005 & 0.34 & 73 \\
21 & 55495.9004 & 0.0051 & $-$0.0086 & 0.42 & 21 \\
27 & 55496.2675 & 0.0023 & 0.0005 & 0.64 & 70 \\
28 & 55496.3138 & 0.0055 & $-$0.0128 & 0.42 & 94 \\
47 & 55497.4502 & 0.0040 & $-$0.0099 & 0.68 & 17 \\
48 & 55497.5121 & 0.0020 & $-$0.0076 & 0.73 & 17 \\
49 & 55497.5787 & 0.0106 & $-$0.0007 & 0.86 & 16 \\
50 & 55497.6327 & 0.0027 & $-$0.0063 & 0.78 & 17 \\
53 & 55497.8062 & 0.0043 & $-$0.0118 & 0.72 & 15 \\
54 & 55497.8657 & 0.0031 & $-$0.0120 & 0.73 & 23 \\
55 & 55497.9314 & 0.0010 & $-$0.0060 & 0.84 & 22 \\
56 & 55497.9940 & 0.0017 & $-$0.0030 & 0.90 & 21 \\
63 & 55498.4061 & 0.0035 & $-$0.0085 & 0.88 & 15 \\
64 & 55498.4665 & 0.0020 & $-$0.0077 & 0.91 & 16 \\
100 & 55500.6201 & 0.0009 & $-$0.0018 & 0.41 & 17 \\
118 & 55501.6987 & 0.0022 & 0.0029 & 0.69 & 28 \\
126 & 55502.1777 & 0.0039 & 0.0047 & 0.81 & 100 \\
127 & 55502.2358 & 0.0014 & 0.0032 & 0.79 & 228 \\
128 & 55502.2930 & 0.0014 & 0.0007 & 0.76 & 180 \\
129 & 55502.3639 & 0.0061 & 0.0119 & 0.96 & 116 \\
135 & 55502.7042 & 0.0028 & $-$0.0057 & 0.73 & 31 \\
145 & 55503.3014 & 0.0014 & $-$0.0051 & 0.85 & 86 \\
161 & 55504.2593 & 0.0023 & $-$0.0017 & 0.08 & 138 \\
162 & 55504.3199 & 0.0024 & $-$0.0007 & 0.11 & 102 \\
178 & 55505.2630 & 0.0020 & $-$0.0122 & 0.09 & 85 \\
179 & 55505.3237 & 0.0010 & $-$0.0111 & 0.12 & 132 \\
195 & 55506.2876 & 0.0054 & $-$0.0017 & 0.46 & 80 \\
196 & 55506.3420 & 0.0013 & $-$0.0070 & 0.38 & 87 \\
262 & 55510.2888 & 0.0015 & 0.0025 & 0.27 & 74 \\
347 & 55515.3572 & 0.0014 & 0.0000 & 0.17 & 60 \\
363 & 55516.3134 & 0.0022 & 0.0018 & 0.37 & 125 \\
379 & 55517.2820 & 0.0018 & 0.0158 & 0.79 & 25 \\
380 & 55517.3427 & 0.0015 & 0.0169 & 0.82 & 25 \\
381 & 55517.4017 & 0.0033 & 0.0162 & 0.82 & 25 \\
382 & 55517.4594 & 0.0013 & 0.0143 & 0.79 & 26 \\
384 & 55517.5772 & 0.0016 & 0.0128 & 0.79 & 26 \\
385 & 55517.6295 & 0.0036 & 0.0054 & 0.68 & 15 \\
396 & 55518.2766 & 0.0032 & $-$0.0038 & 0.64 & 57 \\
398 & 55518.4051 & 0.0213 & 0.0055 & 0.82 & 25 \\
399 & 55518.4603 & 0.0024 & 0.0010 & 0.76 & 25 \\
400 & 55518.5232 & 0.0052 & 0.0042 & 0.82 & 24 \\
401 & 55518.5782 & 0.0043 & $-$0.0004 & 0.76 & 26 \\
414 & 55519.3724 & 0.0015 & 0.0182 & 0.22 & 35 \\
415 & 55519.4303 & 0.0015 & 0.0165 & 0.20 & 25 \\
416 & 55519.4876 & 0.0008 & 0.0141 & 0.17 & 26 \\
417 & 55519.5505 & 0.0007 & 0.0174 & 0.23 & 26 \\
418 & 55519.6098 & 0.0008 & 0.0170 & 0.24 & 26 \\
433 & 55520.4953 & 0.0008 & 0.0077 & 0.25 & 26 \\
\hline
  \multicolumn{6}{l}{\commenta BJD$-$2400000.} \\
  \multicolumn{6}{l}{\commentb Against max $= 2455494.6562 + 0.059657 E$.} \\
  \multicolumn{6}{l}{\commentc Orbital phase.} \\
  \multicolumn{6}{l}{\commentd Number of points used to determine the maximum.} \\
\end{tabular}
\end{center}
\end{table}

\addtocounter{table}{-1}
\begin{table}
\caption{Superhump maxima of SDSS J0804 (2010) during the post-rebrightening phase (continued).}
\begin{center}
\begin{tabular}{cccccc}
\hline
$E$ & max\commenta & error & $O-C$\commentb & phase\commentc & $N$\commentd \\
\hline
434 & 55520.5622 & 0.0017 & 0.0149 & 0.38 & 24 \\
668 & 55534.4961 & 0.0006 & $-$0.0109 & 0.53 & 100 \\
670 & 55534.6191 & 0.0004 & $-$0.0073 & 0.61 & 142 \\
680 & 55535.2159 & 0.0029 & $-$0.0070 & 0.73 & 69 \\
702 & 55536.5394 & 0.0010 & 0.0040 & 0.16 & 15 \\
703 & 55536.5991 & 0.0008 & 0.0041 & 0.17 & 21 \\
752 & 55539.5183 & 0.0012 & 0.0001 & 0.64 & 19 \\
753 & 55539.5779 & 0.0013 & 0.0001 & 0.65 & 20 \\
754 & 55539.6347 & 0.0006 & $-$0.0028 & 0.62 & 19 \\
1002 & 55554.4324 & 0.0018 & $-$0.0001 & 0.40 & 15 \\
1003 & 55554.4934 & 0.0027 & 0.0013 & 0.44 & 20 \\
1004 & 55554.5499 & 0.0020 & $-$0.0018 & 0.39 & 20 \\
1005 & 55554.6148 & 0.0025 & 0.0034 & 0.49 & 21 \\
1021 & 55555.5644 & 0.0027 & $-$0.0016 & 0.59 & 21 \\
1022 & 55555.6199 & 0.0023 & $-$0.0057 & 0.53 & 21 \\
1036 & 55556.4460 & 0.0015 & $-$0.0148 & 0.53 & 20 \\
1038 & 55556.5611 & 0.0020 & $-$0.0189 & 0.48 & 21 \\
1039 & 55556.6232 & 0.0010 & $-$0.0166 & 0.53 & 18 \\
\hline
  \multicolumn{6}{l}{\commenta BJD$-$2400000.} \\
  \multicolumn{6}{l}{\commentb Against max $= 2455494.6562 + 0.059657 E$.} \\
  \multicolumn{6}{l}{\commentc Orbital phase.} \\
  \multicolumn{6}{l}{\commentd Number of points used to determine the maximum.} \\
\end{tabular}
\end{center}
\end{table}

\subsection{SDSS J081207.63$+$131824.4}\label{obj:j0812}

\begin{figure}
  \begin{center}
    \FigureFile(88mm,110mm){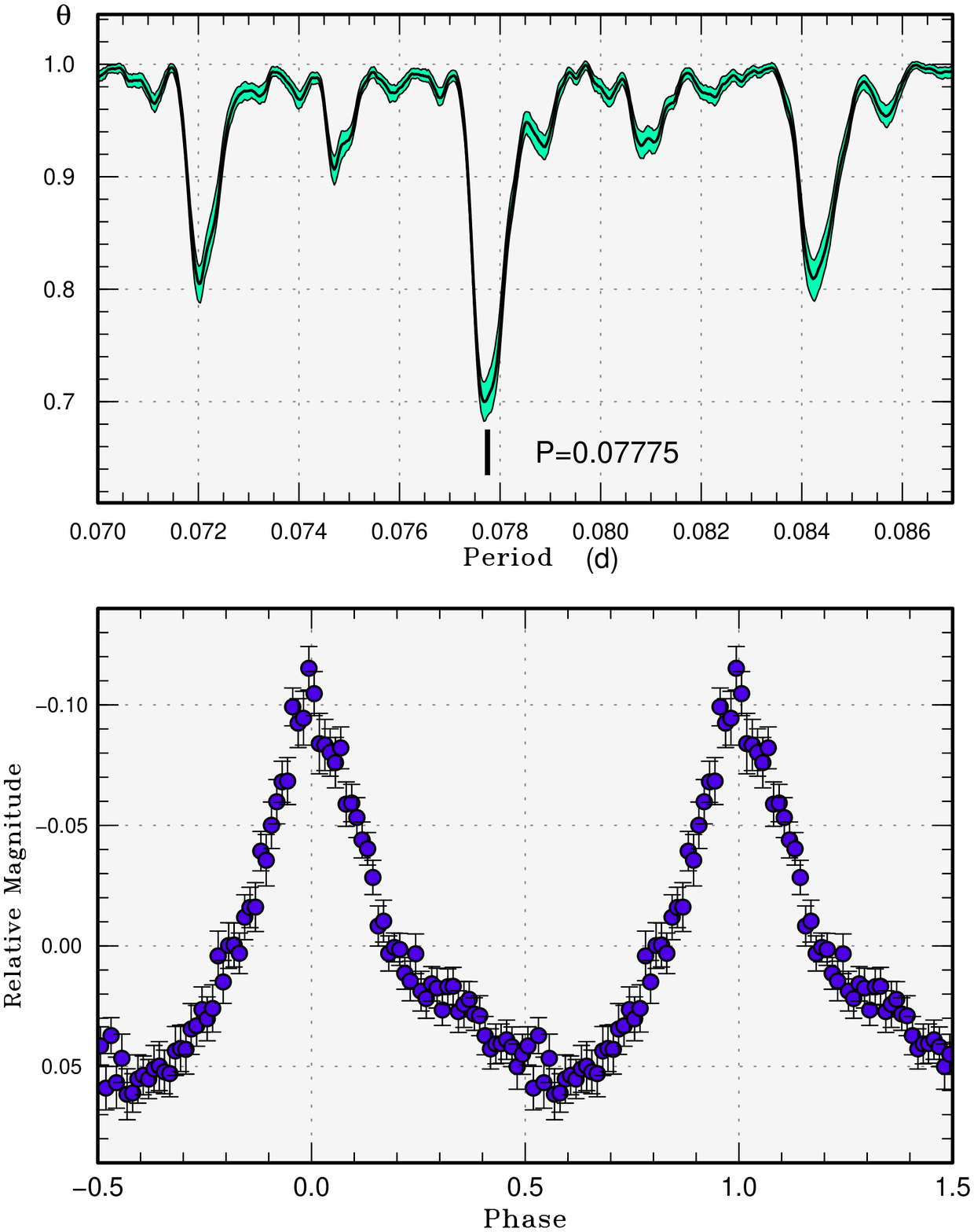}
  \end{center}
  \caption{Superhumps in SDSS J0812 (2011). (Upper): PDM analysis.
     (Lower): Phase-averaged profile.}
  \label{fig:j0812shpdm}
\end{figure}

   This object (hereafter SDSS J0812) was selected as a CV by
\citet{szk07SDSSCV6}.  The outburst detection by K. Itagaki in 2008
led to a classification as an SU UMa-type dwarf nova \citep{Pdot}.
The object was again in outburst in 2011 March (I. Miller, baavss-alert
2561, vsnet-alert 13019).  Subsequent observations detected superhumps
(vsnet-alert 13024, 13025).  Although \citet{Pdot} reported a period
of 0.08432(1) d, the new observations indicated that this period
was a wrong alias (vsnet-alert 13030; figure \ref{fig:j0812shpdm}).
The times of superhump maxima for 2008 (updated) and 2011 superoutbursts
are listed in tables \ref{tab:j0812oc2008} and \ref{tab:j0812oc2011},
respectively.

\begin{figure}
  \begin{center}
    \FigureFile(88mm,70mm){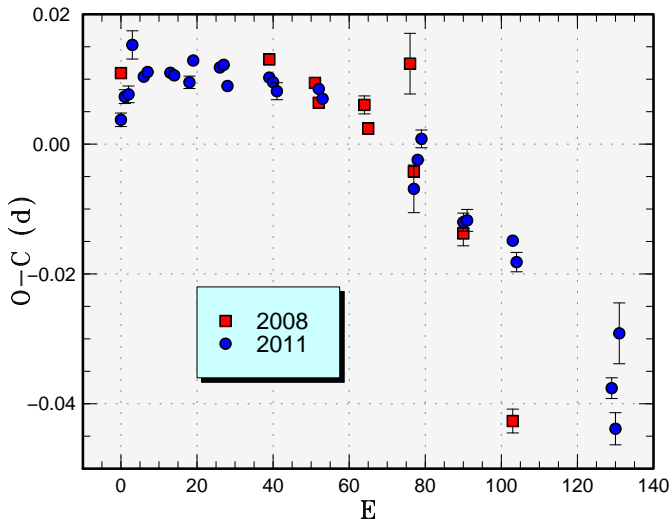}
  \end{center}
  \caption{Comparison of $O-C$ diagrams of SDSS J0812 between different
  superoutbursts.  A period of 0.07798 d was used to draw this figure.
  Cycle counts ($E$) after the detection of the outburst
  were used.  The true epochs of the starts of outbursts were unknown.
  }
  \label{fig:j0812comp}
\end{figure}

   A comparison of $O-C$ diagrams (figure \ref{fig:j0812comp}) suggests
that there is a short ($\sim$ 40 cycles or slightly longer) stage B with
a relatively constant $P_{\rm SH}$ and subsequent stage C common to
these two superoutbursts.  Since these stages were identifiable in the
2011 superoutburst, we gave values for these stages in table
\ref{tab:perlist}.  There was an apparent gap in stage B in the 2008
superoutburst and we gave a global value.  The large negative $P_{\rm dot}$
likely referred to a stage B--C transition.  The global behavior of
the $O-C$ diagram is not very different from that of HT Cas
(subsection \ref{obj:htcas}) with a similar $P_{\rm SH}$.

\begin{table}
\caption{Superhump maxima of SDSS J0812 (2008).}\label{tab:j0812oc2008}
\begin{center}
\begin{tabular}{ccccc}
\hline
$E$ & max\commenta & error & $O-C$\commentb & $N$\commentc \\
\hline
0 & 54751.2796 & 0.0002 & $-$0.0150 & 310 \\
39 & 54754.3230 & 0.0006 & 0.0035 & 148 \\
51 & 54755.2551 & 0.0004 & 0.0049 & 274 \\
52 & 54755.3300 & 0.0007 & 0.0023 & 179 \\
64 & 54756.2654 & 0.0014 & 0.0070 & 198 \\
65 & 54756.3398 & 0.0009 & 0.0038 & 113 \\
76 & 54757.2076 & 0.0047 & 0.0184 & 177 \\
77 & 54757.2689 & 0.0008 & 0.0022 & 295 \\
90 & 54758.2731 & 0.0019 & $-$0.0019 & 84 \\
103 & 54759.2580 & 0.0018 & $-$0.0253 & 102 \\
\hline
  \multicolumn{5}{l}{\commenta BJD$-$2400000.} \\
  \multicolumn{5}{l}{\commentb Against max $= 2454751.2946 + 0.077559 E$.} \\
  \multicolumn{5}{l}{\commentc Number of points used to determine the maximum.} \\
\end{tabular}
\end{center}
\end{table}

\begin{table}
\caption{Superhump maxima of SDSS J0812 (2011).}\label{tab:j0812oc2011}
\begin{center}
\begin{tabular}{ccccc}
\hline
$E$ & max\commenta & error & $O-C$\commentb & $N$\commentc \\
\hline
0 & 55644.9450 & 0.0010 & $-$0.0130 & 144 \\
1 & 55645.0265 & 0.0011 & $-$0.0091 & 212 \\
2 & 55645.1049 & 0.0013 & $-$0.0084 & 111 \\
3 & 55645.1905 & 0.0022 & $-$0.0005 & 45 \\
6 & 55645.4195 & 0.0003 & $-$0.0044 & 83 \\
7 & 55645.4982 & 0.0003 & $-$0.0033 & 73 \\
13 & 55645.9660 & 0.0003 & $-$0.0014 & 96 \\
14 & 55646.0435 & 0.0003 & $-$0.0015 & 102 \\
18 & 55646.3544 & 0.0010 & $-$0.0012 & 46 \\
19 & 55646.4357 & 0.0005 & 0.0025 & 48 \\
26 & 55646.9805 & 0.0002 & 0.0038 & 405 \\
27 & 55647.0589 & 0.0003 & 0.0045 & 350 \\
28 & 55647.1336 & 0.0008 & 0.0016 & 85 \\
39 & 55647.9927 & 0.0004 & 0.0066 & 307 \\
40 & 55648.0700 & 0.0006 & 0.0062 & 178 \\
41 & 55648.1466 & 0.0013 & 0.0051 & 68 \\
52 & 55649.0047 & 0.0004 & 0.0092 & 164 \\
53 & 55649.0812 & 0.0005 & 0.0080 & 165 \\
77 & 55650.9388 & 0.0037 & 0.0021 & 118 \\
78 & 55651.0212 & 0.0006 & 0.0069 & 363 \\
79 & 55651.1025 & 0.0014 & 0.0105 & 200 \\
90 & 55651.9474 & 0.0014 & 0.0014 & 59 \\
91 & 55652.0257 & 0.0017 & 0.0020 & 83 \\
103 & 55652.9583 & 0.0008 & 0.0029 & 303 \\
104 & 55653.0330 & 0.0015 & $-$0.0001 & 264 \\
129 & 55654.9630 & 0.0016 & $-$0.0111 & 110 \\
130 & 55655.0348 & 0.0025 & $-$0.0170 & 112 \\
131 & 55655.1275 & 0.0047 & $-$0.0020 & 86 \\
\hline
  \multicolumn{5}{l}{\commenta BJD$-$2400000.} \\
  \multicolumn{5}{l}{\commentb Against max $= 2455644.9580 + 0.077645 E$.} \\
  \multicolumn{5}{l}{\commentc Number of points used to determine the maximum.} \\
\end{tabular}
\end{center}
\end{table}

\subsection{SDSS J093249.57$+$472523.0}\label{obj:j0932}

   This object (hereafter SDSS J0932) was selected as a short-period
($P_{\rm orb} = 1.7$ hr) CV by \citet{szk04SDSSCV3}.
\citet{hom06j0932j1023} clarified its eclipsing nature and suspected that
it is an intermediate polar based on X-ray observations.

\begin{figure}
  \begin{center}
    \FigureFile(88mm,110mm){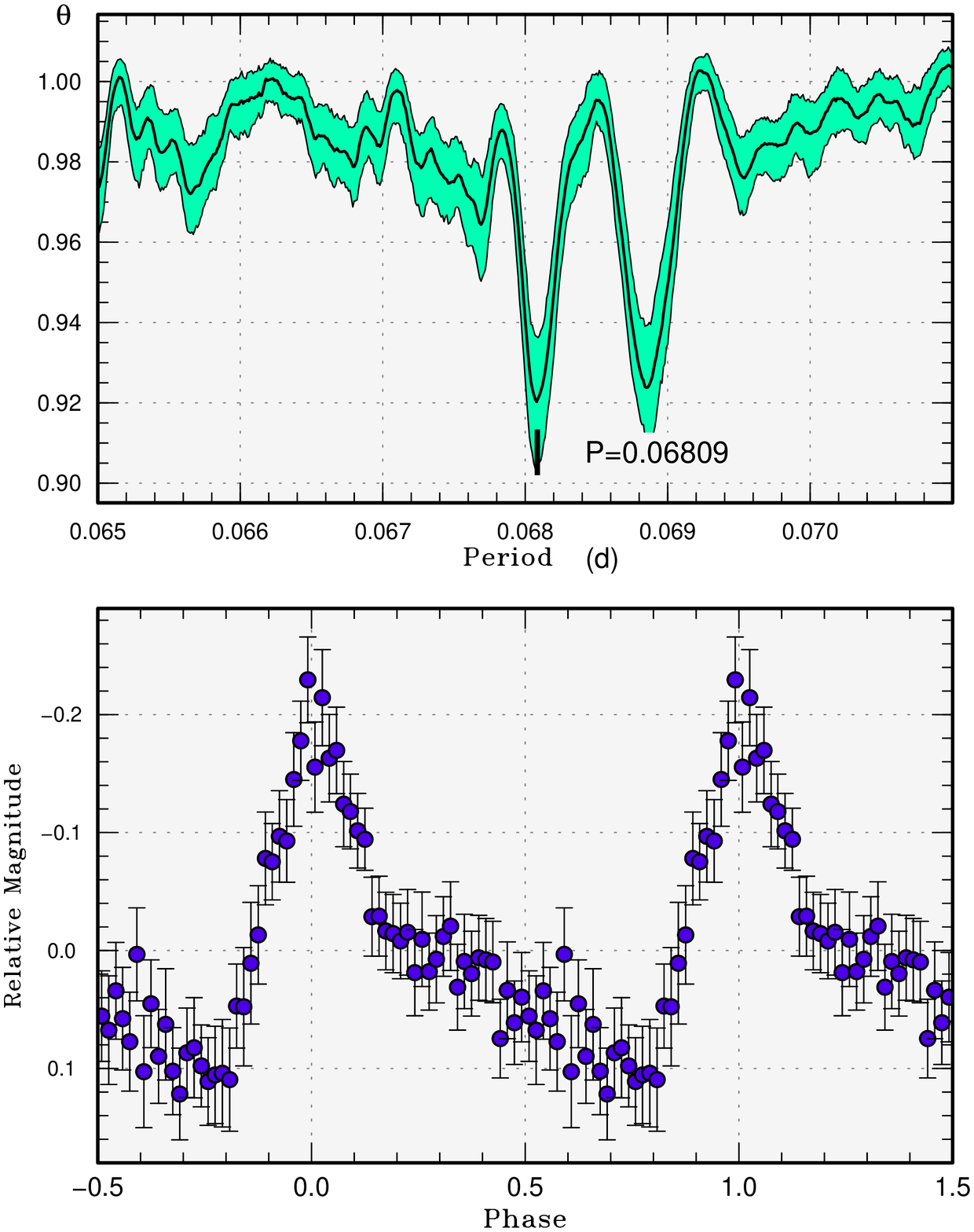}
  \end{center}
  \caption{Superhumps in SDSS J0932 (2011). (Upper): PDM analysis
     outside the eclipses.
     (Lower): Phase-averaged profile.}
  \label{fig:j0932shpdm}
\end{figure}

   This object was for the first time detected in outburst on 2011
March 18 at an unfiltered CCD magnitude of 14.9 (J. Shears, baavss-alert
2537).  Subsequent observations immediately detected superhumps
(vsnet-alert 13006, 13020, 13034, 13042; figure \ref{fig:j0932shpdm}).
The observed eclipse minima during the current outburst confirmed
the shorter alias for $P_{\rm orb}$ given by \citet{hom06j0932j1023}.
The times of recorded eclipses, determined with the Kwee and van Woerden
(KW) method \citep{KWmethod}, after removing linearly approximated trends
around eclipses in order to minimize the effect of superhumps,
are summarized in table \ref{tab:j0932ecl}.
The resultant refined ephemeris is as follows:

\begin{equation}
{\rm Min(BJD)} = 2453106.6834(2) + 0.066303547(6) E
\label{equ:j0932ecl}.
\end{equation}

   The times of superhump maxima determined from observations outside
the eclipses are listed in table \ref{tab:j0932oc2011}.
There was no significant global trend of period variation.
It is likely that we only observed stage C superhumps.
The values given in table \ref{tab:perlist} follows this identification.
The fractional superhump excess was 2.7 \%.
We also obtained post-superoutburst observations.  Due to the faintness
of the object and low signal of superhumps, we only obtained a mean period
of 0.06814(4) d with the PDM method for the interval of
BJD 2455650.3--2455656.2.
This period is practically identical with the period in stage C
and these superhumps are likely persistent superhumps in post-superoutburst
stage of short-$P_{\rm orb}$ systems (e.g. \cite{ohs11qzvir}).

\begin{table}
\caption{Eclipse minima of SDSS J0932.}\label{tab:j0932ecl}
\begin{center}
\begin{tabular}{ccccc}
\hline
$E$ & Minimum\commenta & error & $O-C$\commentb & Source\commentc \\
\hline
0 & 53106.68364 & 0.00045 & 0.00023 & 1 \\
1 & 53106.74963 & 0.00020 & -0.00008 & 1 \\
2 & 53106.81609 & 0.00028 & 0.00008 & 1 \\
467 & 53137.64733 & 0.00016 & 0.00017 & 1 \\
468 & 53137.71340 & 0.00014 & -0.00007 & 1 \\
469 & 53137.77931 & 0.00023 & -0.00046 & 1 \\
470 & 53137.84620 & 0.00030 & 0.00013 & 1 \\
38202 & 55639.61176 & 0.00005 & 0.00026 & 2 \\
38203 & 55639.67801 & 0.00006 & 0.00020 & 2 \\
38209 & 55640.07542 & 0.00007 & -0.00021 & 2 \\
38210 & 55640.14193 & 0.00006 & 0.00000 & 2 \\
38211 & 55640.20853 & 0.00007 & 0.00029 & 2 \\
38214 & 55640.40617 & 0.00010 & -0.00097 & 2 \\
38215 & 55640.47469 & 0.00011 & 0.00124 & 2 \\
38216 & 55640.53892 & 0.00006 & -0.00083 & 2 \\
38217 & 55640.60456 & 0.00006 & -0.00149 & 2 \\
38218 & 55640.67125 & 0.00005 & -0.00111 & 2 \\
38220 & 55640.80509 & 0.00017 & 0.00013 & 2 \\
38227 & 55641.26942 & 0.00007 & 0.00033 & 2 \\
38228 & 55641.33602 & 0.00010 & 0.00063 & 2 \\
38229 & 55641.40250 & 0.00005 & 0.00080 & 2 \\
38232 & 55641.60086 & 0.00004 & 0.00025 & 2 \\
38244 & 55642.39646 & 0.00007 & 0.00022 & 2 \\
38304 & 55646.37486 & 0.00005 & 0.00039 & 2 \\
38305 & 55646.44106 & 0.00004 & 0.00030 & 2 \\
38314 & 55647.03754 & 0.00009 & 0.00004 & 2 \\
38315 & 55647.10351 & 0.00007 & -0.00029 & 2 \\
38319 & 55647.36886 & 0.00004 & -0.00015 & 2 \\
38320 & 55647.43547 & 0.00005 & 0.00015 & 2 \\
38328 & 55647.96488 & 0.00006 & -0.00087 & 2 \\
38329 & 55648.03180 & 0.00008 & -0.00025 & 2 \\
38330 & 55648.09873 & 0.00007 & 0.00038 & 2 \\
38364 & 55650.35295 & 0.00006 & 0.00028 & 2 \\
38365 & 55650.41921 & 0.00004 & 0.00023 & 2 \\
38366 & 55650.48497 & 0.00006 & -0.00031 & 2 \\
38367 & 55650.55199 & 0.00008 & 0.00041 & 2 \\
38410 & 55653.40198 & 0.00005 & -0.00066 & 2 \\
38440 & 55655.39143 & 0.00010 & -0.00032 & 2 \\
38442 & 55655.52489 & 0.00005 & 0.00054 & 2 \\
\hline
  \multicolumn{5}{l}{\commenta BJD$-$2400000.} \\
  \multicolumn{5}{l}{\commentb Against equation \ref{equ:j0932ecl}.} \\
  \multicolumn{5}{l}{\commentc 1: \citet{hom06j0932j1023}, 2: this work.} \\
\end{tabular}
\end{center}
\end{table}

\begin{table}
\caption{Superhump maxima of SDSS J0932 (2011).}\label{tab:j0932oc2011}
\begin{center}
\begin{tabular}{cccccc}
\hline
$E$ & max\commenta & error & $O-C$\commentb & phase\commentc & $N$\commentd \\
\hline
0 & 55639.5874 & 0.0013 & $-$0.0089 & 0.64 & 42 \\
1 & 55639.6605 & 0.0007 & $-$0.0038 & 0.74 & 55 \\
7 & 55640.0662 & 0.0016 & $-$0.0068 & 0.86 & 56 \\
8 & 55640.1376 & 0.0016 & $-$0.0035 & 0.94 & 66 \\
9 & 55640.2064 & 0.0008 & $-$0.0028 & 0.97 & 43 \\
11 & 55640.3556 & 0.0017 & 0.0101 & 0.22 & 41 \\
12 & 55640.4179 & 0.0014 & 0.0044 & 0.16 & 56 \\
13 & 55640.4874 & 0.0019 & 0.0058 & 0.21 & 57 \\
14 & 55640.5540 & 0.0012 & 0.0042 & 0.22 & 57 \\
16 & 55640.6848 & 0.0034 & $-$0.0012 & 0.19 & 49 \\
15 & 55640.6167 & 0.0018 & $-$0.0012 & 0.16 & 64 \\
17 & 55640.7487 & 0.0027 & $-$0.0054 & 0.15 & 52 \\
18 & 55640.8208 & 0.0044 & $-$0.0014 & 0.24 & 50 \\
19 & 55640.8884 & 0.0036 & $-$0.0019 & 0.26 & 28 \\
20 & 55640.9532 & 0.0068 & $-$0.0053 & 0.24 & 29 \\
25 & 55641.2991 & 0.0007 & 0.0001 & 0.45 & 72 \\
26 & 55641.3651 & 0.0008 & $-$0.0020 & 0.45 & 139 \\
30 & 55641.6353 & 0.0015 & $-$0.0043 & 0.52 & 71 \\
41 & 55642.3853 & 0.0019 & $-$0.0034 & 0.83 & 56 \\
56 & 55643.4156 & 0.0043 & 0.0052 & 0.37 & 33 \\
70 & 55644.3637 & 0.0007 & $-$0.0002 & 0.67 & 29 \\
71 & 55644.4333 & 0.0008 & 0.0013 & 0.72 & 30 \\
72 & 55644.5006 & 0.0020 & 0.0005 & 0.74 & 35 \\
81 & 55645.1174 & 0.0032 & 0.0043 & 0.04 & 22 \\
85 & 55645.3914 & 0.0019 & 0.0058 & 0.17 & 31 \\
86 & 55645.4577 & 0.0016 & 0.0040 & 0.17 & 35 \\
87 & 55645.5217 & 0.0024 & $-$0.0001 & 0.14 & 35 \\
88 & 55645.5918 & 0.0017 & 0.0019 & 0.20 & 35 \\
99 & 55646.3441 & 0.0012 & 0.0050 & 0.54 & 90 \\
100 & 55646.4122 & 0.0008 & 0.0050 & 0.57 & 110 \\
101 & 55646.4783 & 0.0006 & 0.0030 & 0.57 & 61 \\
102 & 55646.5480 & 0.0011 & 0.0046 & 0.62 & 37 \\
103 & 55646.6142 & 0.0008 & 0.0027 & 0.62 & 34 \\
109 & 55647.0182 & 0.0022 & $-$0.0020 & 0.71 & 23 \\
110 & 55647.0894 & 0.0027 & 0.0011 & 0.78 & 37 \\
111 & 55647.1597 & 0.0027 & 0.0033 & 0.84 & 43 \\
114 & 55647.3625 & 0.0009 & 0.0018 & 0.90 & 61 \\
115 & 55647.4297 & 0.0009 & 0.0008 & 0.92 & 72 \\
123 & 55647.9826 & 0.0018 & 0.0088 & 0.25 & 47 \\
124 & 55648.0505 & 0.0019 & 0.0087 & 0.28 & 53 \\
158 & 55650.3542 & 0.0024 & $-$0.0034 & 0.02 & 36 \\
159 & 55650.4175 & 0.0016 & $-$0.0082 & 0.98 & 36 \\
160 & 55650.4818 & 0.0015 & $-$0.0120 & 0.95 & 37 \\
161 & 55650.5473 & 0.0022 & $-$0.0146 & 0.94 & 36 \\
\hline
  \multicolumn{6}{l}{\commenta BJD$-$2400000.} \\
  \multicolumn{6}{l}{\commentb Against max $= 2455639.5962 + 0.068110 E$.} \\
  \multicolumn{6}{l}{\commentc Orbital phase.} \\
  \multicolumn{6}{l}{\commentd Number of points used to determine the maximum.} \\
\end{tabular}
\end{center}
\end{table}

\subsection{SDSS J112003.40$+$663632.4}\label{obj:j1120}

\begin{figure}
  \begin{center}
    \FigureFile(88mm,110mm){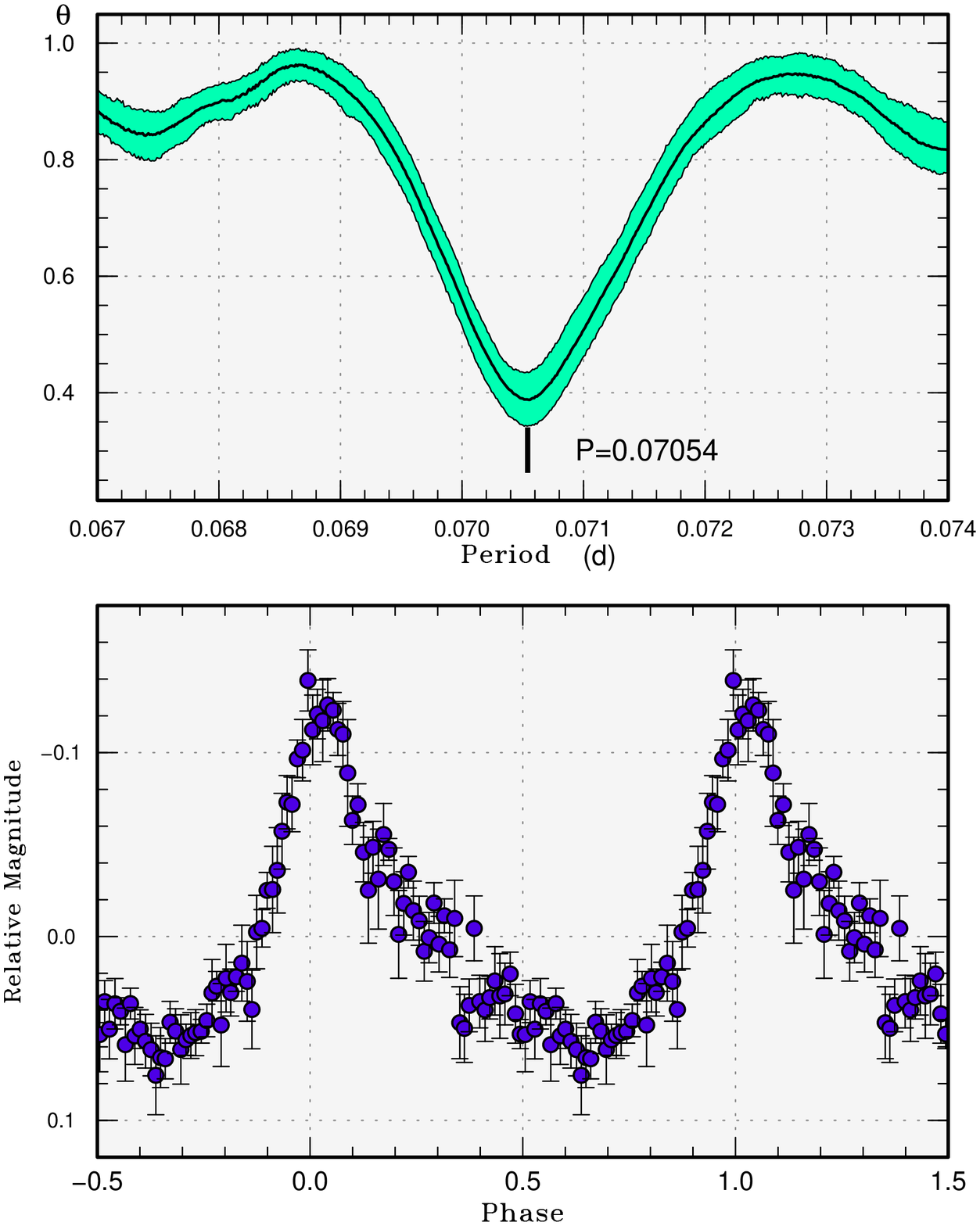}
  \end{center}
  \caption{Superhumps in SDSS J1120 (2011). (Upper): PDM analysis.
     (Lower): Phase-averaged profile.}
  \label{fig:j1120shpdm}
\end{figure}

   This object (hereafter SDSS J1120) was initially selected as
a dwarf nova by \citet{wil10newCVs}.
The 2011 outburst was detected on February 14.838 UT at an unfiltered
CCD magnitude of 15.7 by J. Shears (baavss-alert 2513).
Superhump were detected by S. Brady (baavss-alert 2514).
We analyzed the available data and obtained the times of superhump
maxima (table \ref{tab:j1120oc2011}).  It was not possible to
identify the stage of superhump evolution.  The result of PDM analysis
and mean profile are shown in figure \ref{fig:j1120shpdm}.

\begin{table}
\caption{Superhump maxima of SDSS J1120.}\label{tab:j1120oc2011}
\begin{center}
\begin{tabular}{ccccc}
\hline
$E$ & max\commenta & error & $O-C$\commentb & $N$\commentc \\
\hline
0 & 55607.4060 & 0.0008 & $-$0.0016 & 95 \\
1 & 55607.4797 & 0.0011 & 0.0016 & 94 \\
7 & 55607.9019 & 0.0004 & 0.0005 & 73 \\
8 & 55607.9716 & 0.0004 & $-$0.0004 & 73 \\
21 & 55608.8893 & 0.0005 & 0.0001 & 72 \\
22 & 55608.9594 & 0.0005 & $-$0.0002 & 73 \\
\hline
  \multicolumn{5}{l}{\commenta BJD$-$2400000.} \\
  \multicolumn{5}{l}{\commentb Against max $= 2455607.4076 + 0.070550 E$.} \\
  \multicolumn{5}{l}{\commentc Number of points used to determine the maximum.} \\
\end{tabular}
\end{center}
\end{table}

\subsection{SDSS J114628.80$+$675907.7}\label{obj:j1146}

\begin{figure}
  \begin{center}
    \FigureFile(88mm,110mm){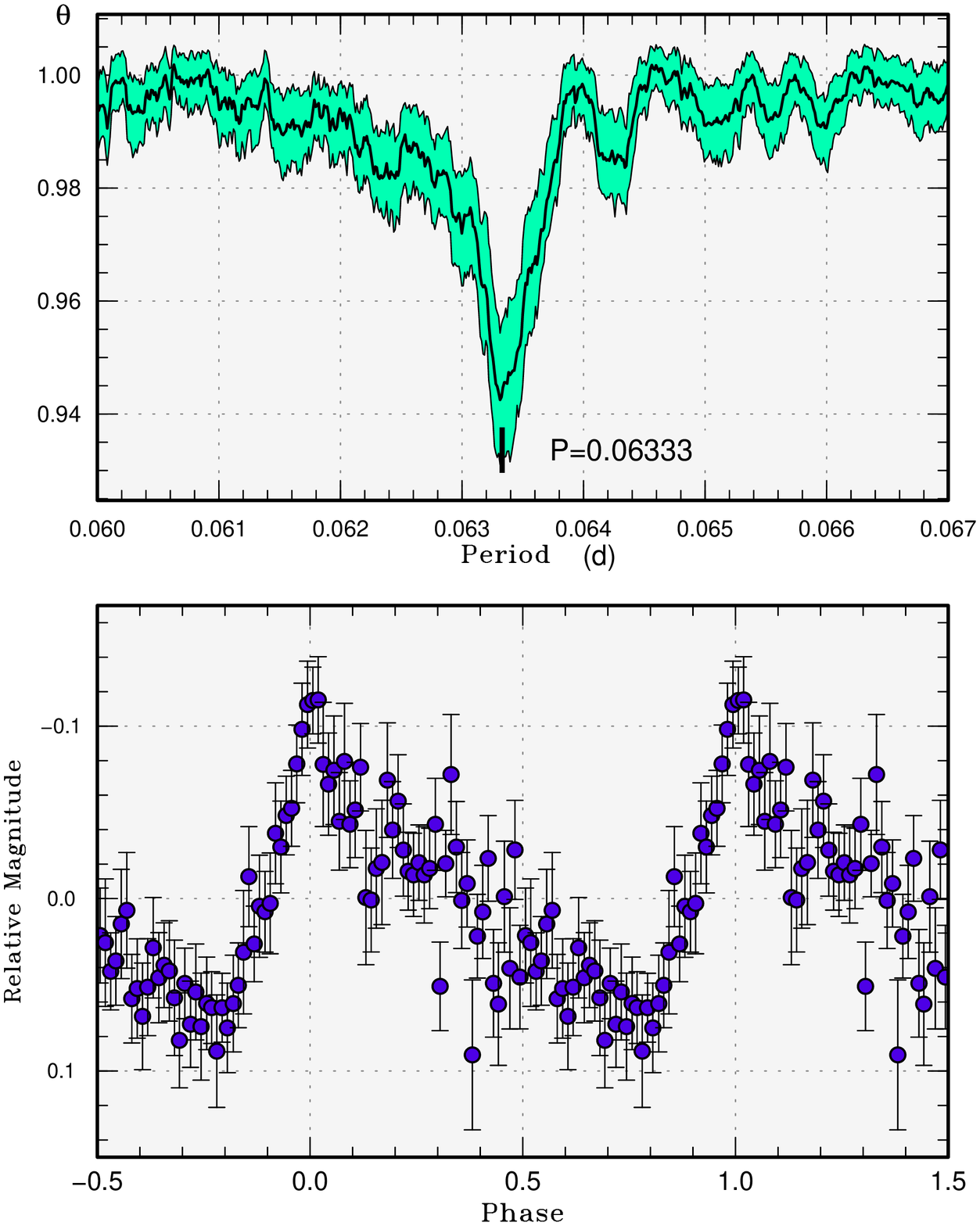}
  \end{center}
  \caption{Superhumps in SDSS J1146 (2011). (Upper): PDM analysis.
     (Lower): Phase-averaged profile.}
  \label{fig:j1146shpdm}
\end{figure}

   This object (hereafter SDSS J1146) is a CV selected during the course of
the SDSS \citep{szk03SDSSCV2}.  \citet{szk03SDSSCV2} reported an orbital
period of 1.6 hr based on radial-velocity study.  Although the spectrum
suggested a dwarf nova in quiescence, no major outburst had been recorded.
A small outburst was recorded on 2009 April 17 at an unfiltered CCD
magnitude of 16.9 (J. Shears, cvnet-outburst 3062).
J. Shears also detected an outburst on 2011 January 3 during its
rising phase (baavss-alert 2459).  Subsequent observations recorded
growing superhumps (vsnet-alert 12567, 12576, 12579;
figure \ref{fig:j1146shpdm}).

   The times of superhump maxima are listed in table \ref{tab:j1146oc2011}.
The values given in table \ref{tab:perlist} are somewhat uncertain
because the middle of stage B was not very well observed and
only the initial part of stage C was observed.

\begin{table}
\caption{Superhump maxima of SDSS J1146 (2011).}\label{tab:j1146oc2011}
\begin{center}
\begin{tabular}{ccccc}
\hline
$E$ & max\commenta & error & $O-C$\commentb & $N$\commentc \\
\hline
0 & 55565.6927 & 0.0043 & $-$0.0242 & 30 \\
1 & 55565.7744 & 0.0016 & $-$0.0060 & 29 \\
2 & 55565.8387 & 0.0015 & $-$0.0051 & 32 \\
3 & 55565.9048 & 0.0016 & $-$0.0024 & 19 \\
9 & 55566.2962 & 0.0018 & 0.0085 & 111 \\
10 & 55566.3593 & 0.0015 & 0.0081 & 126 \\
14 & 55566.6131 & 0.0004 & 0.0082 & 65 \\
15 & 55566.6766 & 0.0005 & 0.0083 & 51 \\
16 & 55566.7379 & 0.0004 & 0.0063 & 32 \\
42 & 55568.3795 & 0.0006 & $-$0.0011 & 123 \\
43 & 55568.4464 & 0.0005 & 0.0024 & 105 \\
72 & 55570.2859 & 0.0010 & 0.0026 & 249 \\
73 & 55570.3475 & 0.0009 & 0.0009 & 180 \\
75 & 55570.4714 & 0.0012 & $-$0.0021 & 42 \\
76 & 55570.5387 & 0.0010 & 0.0018 & 66 \\
77 & 55570.6011 & 0.0007 & 0.0008 & 68 \\
78 & 55570.6651 & 0.0010 & 0.0014 & 68 \\
79 & 55570.7292 & 0.0008 & 0.0021 & 46 \\
80 & 55570.7937 & 0.0016 & 0.0032 & 23 \\
81 & 55570.8526 & 0.0008 & $-$0.0014 & 46 \\
82 & 55570.9195 & 0.0007 & 0.0021 & 49 \\
83 & 55570.9829 & 0.0013 & 0.0021 & 39 \\
88 & 55571.2960 & 0.0018 & $-$0.0020 & 38 \\
89 & 55571.3605 & 0.0007 & $-$0.0009 & 100 \\
90 & 55571.4255 & 0.0009 & 0.0007 & 96 \\
91 & 55571.4889 & 0.0008 & 0.0007 & 94 \\
92 & 55571.5482 & 0.0019 & $-$0.0035 & 96 \\
93 & 55571.6120 & 0.0009 & $-$0.0031 & 99 \\
94 & 55571.6763 & 0.0012 & $-$0.0022 & 72 \\
105 & 55572.3698 & 0.0009 & $-$0.0063 & 33 \\
\hline
  \multicolumn{5}{l}{\commenta BJD$-$2400000.} \\
  \multicolumn{5}{l}{\commentb Against max $= 2455565.7170 + 0.063420 E$.} \\
  \multicolumn{5}{l}{\commentc Number of points used to determine the maximum.} \\
\end{tabular}
\end{center}
\end{table}

\subsection{SDSS J122740.83$+$513925.0}\label{obj:j1227}

   This object (hereafter SDSS J1227) is an eclipsing SU UMa-type
dwarf nova whose superoutburst was first recorded in 2007.
This 2007 superoutburst was analyzed by \citet{she08j1227} and
\citet{Pdot}.  Another superoutburst was observed in 2011.
By using eclipses observed during the 2011 observations
(table \ref{tab:j1227ecl}), it has become evident that the alias
selection of the orbital ephemeris in \citet{Pdot} was incorrect.
We have revised the eclipse ephemeris as follows:

\begin{table}
\caption{Eclipse Minima of SDSS J1227 (2011).}\label{tab:j1227ecl}
\begin{center}
\begin{tabular}{cccc}
\hline
$E$ & Minimum\commenta & error & $O-C$\commentb \\
\hline
29202 & 55634.51978 & 0.00002 & 0.00014 \\
29203 & 55634.58272 & 0.00002 & 0.00012 \\
29215 & 55635.33819 & 0.00004 & 0.00019 \\
29216 & 55635.40145 & 0.00014 & 0.00050 \\
29217 & 55635.46423 & 0.00002 & 0.00033 \\
29218 & 55635.52706 & 0.00003 & 0.00021 \\
29230 & 55636.28281 & 0.00009 & 0.00056 \\
29231 & 55636.34561 & 0.00007 & 0.00040 \\
29232 & 55636.40864 & 0.00005 & 0.00048 \\
29246 & 55637.28967 & 0.00005 & 0.00021 \\
29247 & 55637.35270 & 0.00003 & 0.00028 \\
29248 & 55637.41573 & 0.00007 & 0.00036 \\
29284 & 55639.68236 & 0.00008 & 0.00078 \\
29300 & 55640.68954 & 0.00003 & 0.00075 \\
29391 & 55646.41865 & 0.00008 & 0.00138 \\
\hline
  \multicolumn{4}{l}{\commenta BJD$-$2400000.} \\
  \multicolumn{4}{l}{\commentb Against equation \ref{equ:j1227ecl}.} \\
\end{tabular}
\end{center}
\end{table}

\begin{equation}
{\rm Min(BJD)} = 2453796.2417(6) + 0.06295041(3) E
\label{equ:j1227ecl}.
\end{equation}

   We have updated the results of the 2007 superoutburst because
the significant addition to the data published in \citet{Pdot}.
The times of superhump maxima are listed in table \ref{tab:j1227oc2007}.
The data now clearly show all stages A--C and a positive $P_{\rm dot}$
for stage B.  The updated parameters are listed in table \ref{tab:perlist}.

\begin{figure}
  \begin{center}
    \FigureFile(88mm,70mm){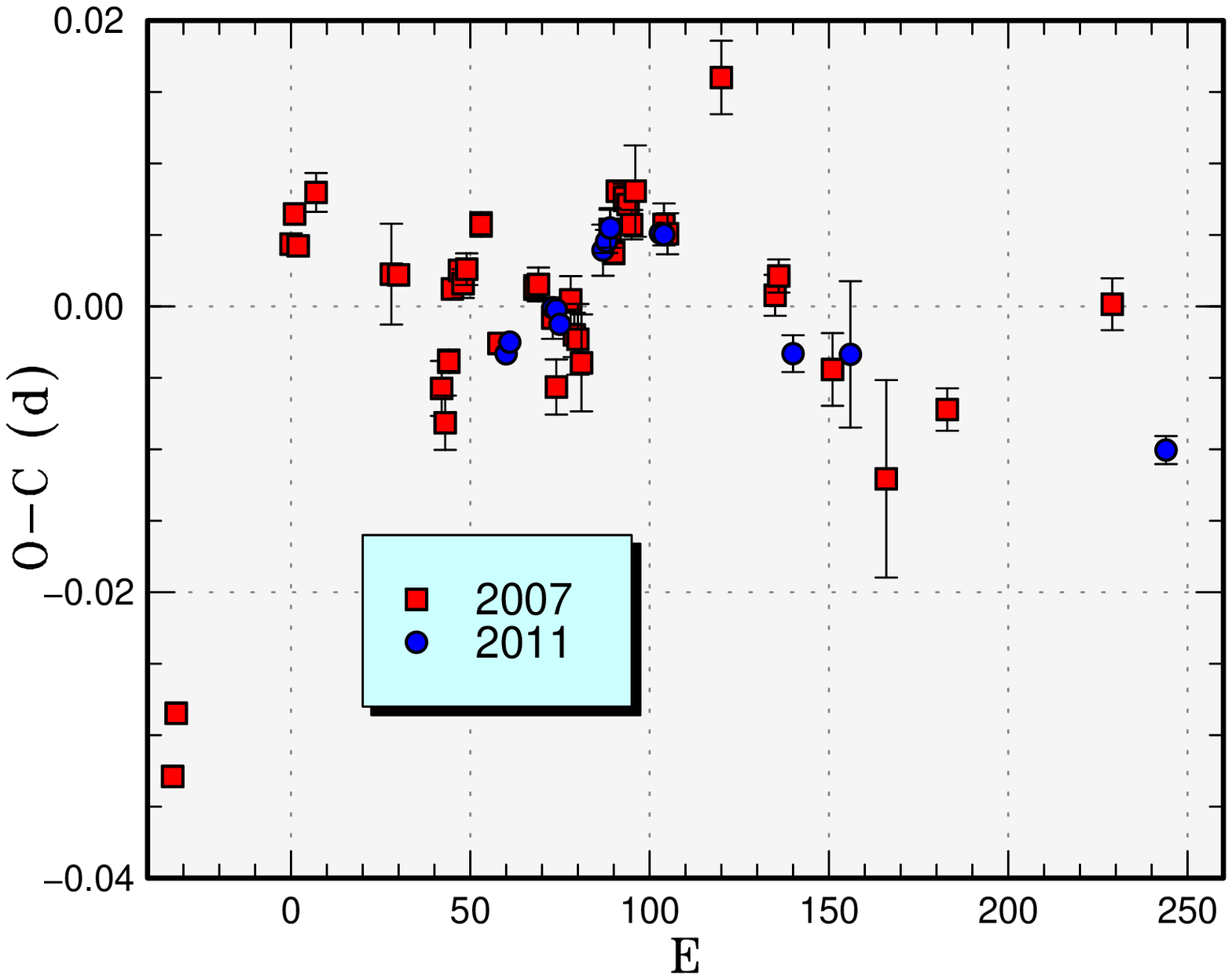}
  \end{center}
  \caption{Comparison of $O-C$ diagrams of SDSS J1227 between different
  superoutbursts.  A period of 0.06460 d was used to draw this figure.
  Approximate cycle counts ($E$) after the start of the stage B
  were used.  Since the start of the 2011 superoutburst
  was not well constrained, we shifted the $O-C$ diagrams
  to best fit the best-recorded 2007 one.
  }
  \label{fig:j1227comp}
\end{figure}

   The 2011 outburst was detected by CRTS (= CSS110312:122741$+$513925).
The outburst was apparently observed during its middle-to-late stage.
The times of superhump maxima outside the eclipses
are listed in table \ref{tab:j1227oc2011}.
A stage B--C transition was recorded.  A combined $O-C$ diagram is
shown in figure \ref{fig:j1227comp}.

\begin{table}
\caption{Superhump maxima of SDSS J1227 (2007).}\label{tab:j1227oc2007}
\begin{center}
\begin{tabular}{cccccc}
\hline
$E$ & max\commenta & error & $O-C$\commentb & phase\commentc & $N$\commentd \\
\hline
0 & 54256.4379 & 0.0004 & $-$0.0286 & 0.45 & -- \\
1 & 54256.5069 & 0.0006 & $-$0.0242 & 0.55 & -- \\
33 & 54258.6069 & 0.0003 & 0.0073 & 0.91 & 122 \\
34 & 54258.6736 & 0.0002 & 0.0094 & 0.97 & 120 \\
35 & 54258.7360 & 0.0005 & 0.0071 & 0.96 & 118 \\
40 & 54259.0627 & 0.0014 & 0.0107 & 0.15 & 107 \\
61 & 54260.4136 & 0.0035 & 0.0041 & 0.61 & 106 \\
63 & 54260.5428 & 0.0003 & 0.0040 & 0.66 & 192 \\
75 & 54261.3100 & 0.0019 & $-$0.0045 & 0.85 & 9 \\
76 & 54261.3722 & 0.0019 & $-$0.0069 & 0.84 & 18 \\
77 & 54261.4411 & 0.0008 & $-$0.0026 & 0.93 & 206 \\
78 & 54261.5108 & 0.0006 & 0.0024 & 0.04 & 196 \\
80 & 54261.6413 & 0.0006 & 0.0036 & 0.11 & 31 \\
81 & 54261.7049 & 0.0010 & 0.0026 & 0.13 & 35 \\
82 & 54261.7706 & 0.0011 & 0.0036 & 0.17 & 35 \\
86 & 54262.0321 & 0.0009 & 0.0066 & 0.32 & 49 \\
91 & 54262.3468 & 0.0005 & $-$0.0020 & 0.32 & 15 \\
101 & 54262.9967 & 0.0009 & 0.0015 & 0.64 & 82 \\
102 & 54263.0615 & 0.0012 & 0.0017 & 0.67 & 114 \\
106 & 54263.3175 & 0.0014 & $-$0.0008 & 0.74 & 15 \\
107 & 54263.3773 & 0.0019 & $-$0.0057 & 0.69 & 28 \\
111 & 54263.6418 & 0.0017 & 0.0003 & 0.89 & 35 \\
112 & 54263.7040 & 0.0015 & $-$0.0022 & 0.88 & 35 \\
113 & 54263.7683 & 0.0025 & $-$0.0026 & 0.90 & 35 \\
114 & 54263.8312 & 0.0034 & $-$0.0043 & 0.90 & 35 \\
122 & 54264.3573 & 0.0014 & 0.0048 & 0.26 & 28 \\
123 & 54264.4203 & 0.0008 & 0.0031 & 0.26 & 157 \\
124 & 54264.4892 & 0.0005 & 0.0073 & 0.35 & 140 \\
126 & 54264.6180 & 0.0010 & 0.0068 & 0.40 & 41 \\
127 & 54264.6821 & 0.0013 & 0.0063 & 0.42 & 39 \\
128 & 54264.7453 & 0.0010 & 0.0048 & 0.42 & 40 \\
129 & 54264.8122 & 0.0032 & 0.0072 & 0.49 & 39 \\
137 & 54265.3267 & 0.0015 & 0.0045 & 0.66 & 28 \\
138 & 54265.3906 & 0.0014 & 0.0038 & 0.67 & 29 \\
153 & 54266.3706 & 0.0026 & 0.0141 & 0.24 & 29 \\
168 & 54267.3243 & 0.0014 & $-$0.0017 & 0.39 & 24 \\
169 & 54267.3903 & 0.0012 & $-$0.0004 & 0.44 & 23 \\
184 & 54268.3527 & 0.0025 & $-$0.0076 & 0.73 & 29 \\
199 & 54269.3141 & 0.0069 & $-$0.0158 & 0.00 & 18 \\
216 & 54270.4171 & 0.0015 & $-$0.0117 & 0.52 & 17 \\
262 & 54273.3961 & 0.0018 & $-$0.0062 & 0.85 & 25 \\
\hline
  \multicolumn{6}{l}{\commenta BJD$-$2400000.} \\
  \multicolumn{6}{l}{\commentb Against max $= 2454256.4664 + 0.064641 E$.} \\
  \multicolumn{6}{l}{\commentc Orbital phase.} \\
  \multicolumn{6}{l}{\commentd Number of points used to determine the maximum.} \\
\end{tabular}
\end{center}
\end{table}

\begin{table}
\caption{Superhump maxima of SDSS J1227 (2011).}\label{tab:j1227oc2011}
\begin{center}
\begin{tabular}{cccccc}
\hline
$E$ & max\commenta & error & $O-C$\commentb & phase\commentc & $N$\commentd \\
\hline
0 & 55634.5319 & 0.0003 & $-$0.0056 & 0.20 & 88 \\
1 & 55634.5973 & 0.0005 & $-$0.0048 & 0.24 & 63 \\
13 & 55635.3749 & 0.0006 & $-$0.0017 & 0.59 & 56 \\
14 & 55635.4394 & 0.0004 & $-$0.0018 & 0.61 & 190 \\
15 & 55635.5029 & 0.0004 & $-$0.0028 & 0.62 & 121 \\
27 & 55636.2833 & 0.0018 & 0.0030 & 0.02 & 41 \\
28 & 55636.3485 & 0.0008 & 0.0037 & 0.06 & 50 \\
29 & 55636.4141 & 0.0014 & 0.0047 & 0.10 & 30 \\
43 & 55637.3181 & 0.0007 & 0.0051 & 0.46 & 63 \\
44 & 55637.3826 & 0.0007 & 0.0050 & 0.48 & 65 \\
80 & 55639.6999 & 0.0013 & $-$0.0014 & 0.29 & 53 \\
96 & 55640.7334 & 0.0051 & $-$0.0006 & 0.71 & 38 \\
184 & 55646.4115 & 0.0010 & $-$0.0027 & 0.91 & 59 \\
\hline
  \multicolumn{6}{l}{\commenta BJD$-$2400000.} \\
  \multicolumn{6}{l}{\commentb Against max $= 2455634.537 + 0.064547 E$.} \\
  \multicolumn{6}{l}{\commentc Orbital phase.} \\
  \multicolumn{6}{l}{\commentd Number of points used to determine the maximum.} \\
\end{tabular}
\end{center}
\end{table}

\subsection{SDSS J125023.85$+$665525.5}\label{obj:j1250}

   We have observed the 2011 superoutburst of this eclipsing SU UMa-type
dwarf nova (hereafter SDSS J1250).  In analyzing the data, we used
the refined ephemeris of eclipses (equation \ref{equ:j1250ecl}).

\begin{equation}
{\rm Min(BJD)} = 2453407.5595(3) + 0.058735696(9) E
\label{equ:j1250ecl}.
\end{equation}

   The times of superhump maxima are listed in table \ref{tab:j1250oc2011}.
Although we likely observed stages B and C, we only measured the mean
period due to the gap in the observation.

\begin{table}
\caption{Superhump maxima of SDSS J1250 (2011).}\label{tab:j1250oc2011}
\begin{center}
\begin{tabular}{cccccc}
\hline
$E$ & max\commenta & error & $O-C$\commentb & phase\commentc & $N$\commentd \\
\hline
0 & 55665.0937 & 0.0005 & 0.0003 & 0.47 & 124 \\
40 & 55667.5096 & 0.0044 & 0.0058 & 0.60 & 62 \\
41 & 55667.5603 & 0.0005 & $-$0.0038 & 0.47 & 133 \\
42 & 55667.6206 & 0.0006 & $-$0.0038 & 0.49 & 134 \\
52 & 55668.2272 & 0.0013 & 0.0003 & 0.82 & 106 \\
53 & 55668.2868 & 0.0015 & $-$0.0005 & 0.84 & 104 \\
55 & 55668.4099 & 0.0012 & 0.0022 & 0.93 & 48 \\
56 & 55668.4674 & 0.0014 & $-$0.0006 & 0.91 & 55 \\
58 & 55668.5902 & 0.0014 & 0.0017 & 0.00 & 112 \\
59 & 55668.6474 & 0.0023 & $-$0.0014 & 0.98 & 71 \\
\hline
  \multicolumn{6}{l}{\commenta BJD$-$2400000.} \\
  \multicolumn{6}{l}{\commentb Against max $= 2455665.0934 + 0.060261 E$.} \\
  \multicolumn{6}{l}{\commentc Orbital phase.} \\
  \multicolumn{6}{l}{\commentd Number of points used to determine the maximum.} \\
\end{tabular}
\end{center}
\end{table}

\subsection{SDSS J133941.11$+$484727.5}\label{obj:j1339}

   SDSS J133941.11$+$484727.5 (hereafter SDSS J1339) is a CV selected
during the course of the SDSS \citep{szk06SDSSCV5}.
\citet{szk06SDSSCV5} reported a possible orbital period of 1.7 hr
and the presence of features of a white dwarf in the spectrum,
suggesting a WZ Sge-type dwarf nova.
\citet{gan06j1339} obtained an orbital period of 0.05731(2) d
and showed the presence of ZZ Cet-type pulsation of the white dwarf.
\citet{gan06j1339} also found a long-period photometric variation,
similar to the ones observed in GW Lib \citep{wou02gwlib} and FS Aur
\citep{neu02fsaur}, whose origin is yet unknown.

   The 2011 outburst, the first known outburst in its history,
was detected by J. Shears on February 7.919 at a visual magnitude of 10.4
(baavss-alert 2501).  P. Schmeer also made an independent detection
of the outburst (vsnet-alert 12806).

   Early observations detected low-amplitude ($\sim$0.015--0.025 mag)
variations which can be attributed to early superhumps (vsnet-alert 12817,
12819, 12820).
Early spectroscopy by A. Arai detected H$\alpha$ in emission and
other Balmer lines in absorption.  Although He\textsc{II} emission was
absent, C\textsc{III}/N\textsc{III} emission lines were possibly
present (vsnet-alert 12822).  The spectroscopic feature was less
striking compared to the 2001 superoutburst of WZ Sge
\citep{bab02wzsgeletter}.

   Ordinary superhumps emerged four days after the outburst detection
(vsnet-alert 12835, 12836, 12837).  After 13 d, the object started to
fade rapidly.  In contrast to many WZ Sge-type dwarf novae, the object
did not show post-superoutburst rebrightenings.

\begin{figure}
  \begin{center}
    \FigureFile(88mm,110mm){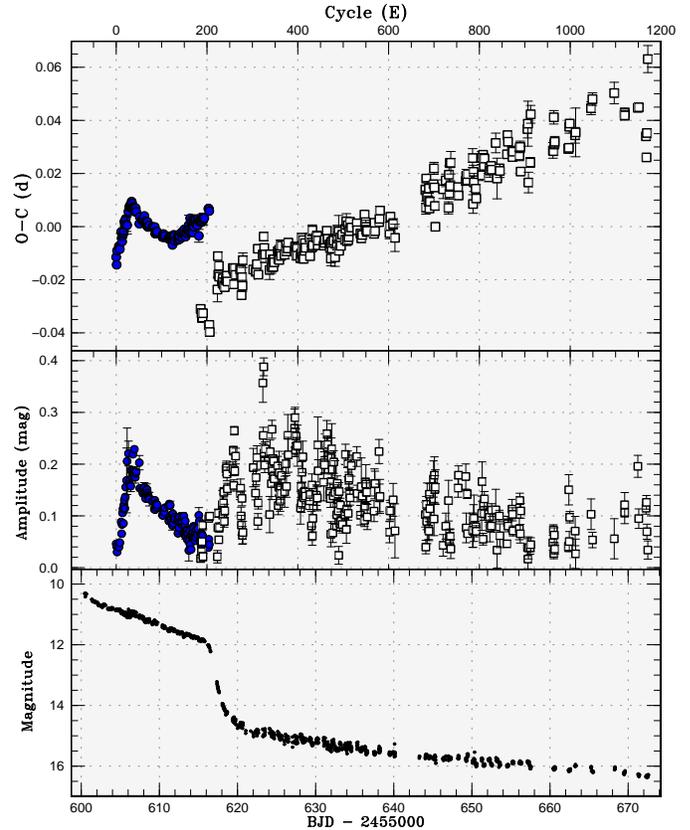}
  \end{center}
  \caption{$O-C$ diagram of superhumps in SDSS J1339 (2011).
     (Upper): $O-C$.
     Filled circles and open squares represent superhumps during the
     plateau phase and post-superoutburst superhumps, respectively.
     We used a period of 0.058112 d for calculating the $O-C$'s.
     (Middle): Amplitudes.  The symbols are common to the upper panel.
     (Lower): Light curve.
  }
  \label{fig:j1339humpall}
\end{figure}

   The times of ordinary superhumps are listed in table \ref{tab:j1339oc2011}.
The stages A and B are clearly recognized (cf. figure \ref{fig:j1339humpall}).
During the rapid fading phase, the phase of superhumps were apparently
reversed and the reversed phase continued during the entire post-superoutburst
state.

\begin{figure}
  \begin{center}
    \FigureFile(88mm,110mm){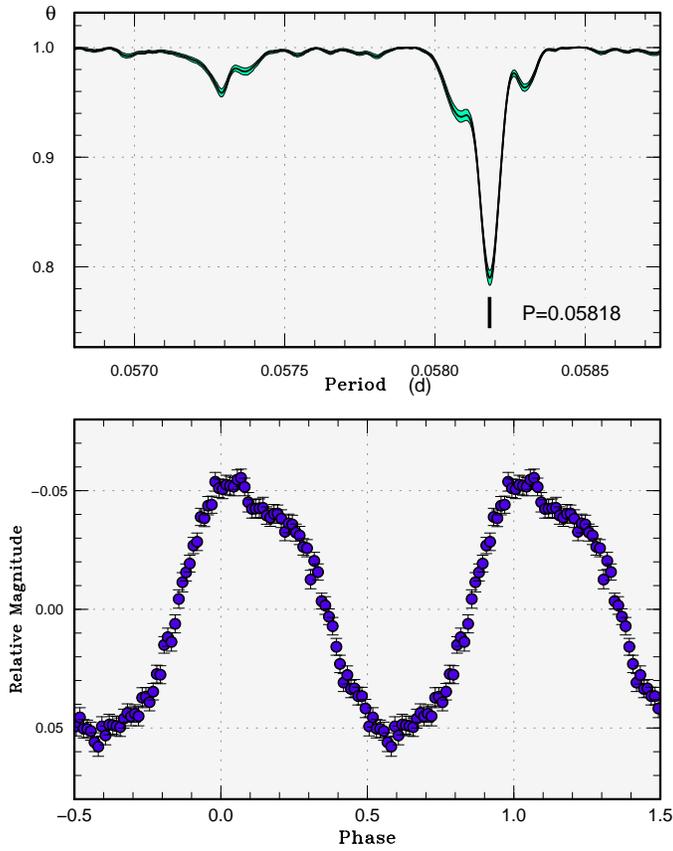}
  \end{center}
  \caption{Superhumps during the post-superoutburst stage in SDSS J1339 (2011).
     (Upper): PDM analysis.
     The signal at 0.057289(1) d is the orbital period.
     The tick is located at the orbital period.
     (Lower): Phase-averaged profile.}
  \label{fig:j1339lateshpdm}
\end{figure}

\begin{figure}
  \begin{center}
    \FigureFile(88mm,110mm){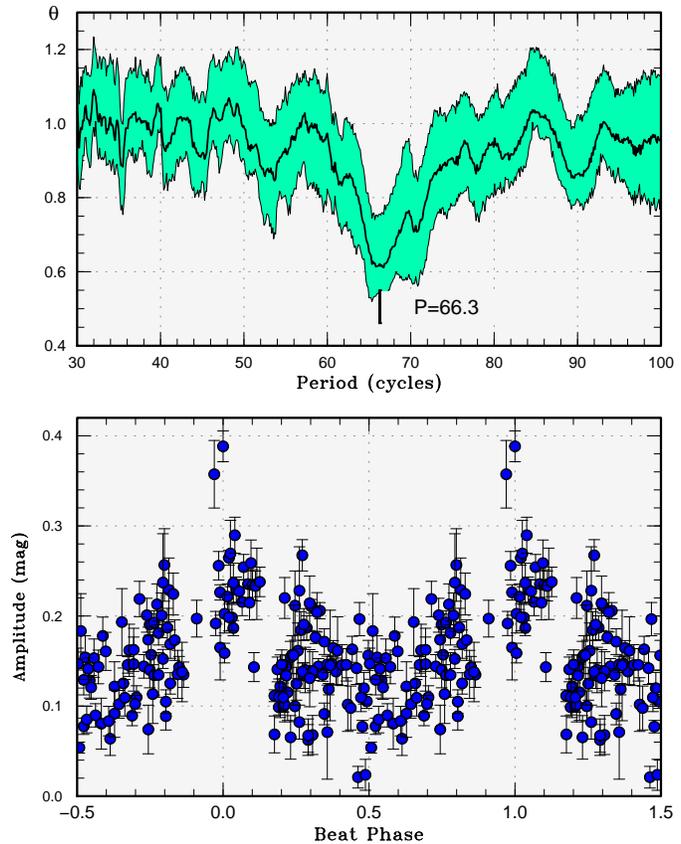}
  \end{center}
  \caption{Beat phenomenon during the post-superoutburst phase in SDSS J1339 (2011).
     (Upper): PDM analysis.
     Superhump maxima for $18 \le E \le 428$ were used.
     The obtained beat period of 66.3(3) cycles is close to the expected
     beat period of 64.7 cycles based on the orbital and superhump periods.
     (Lower): Phase-averaged profile using the measured beat period.}
  \label{fig:j1339latebeat}
\end{figure}

   During the post-superoutburst phase, a beat phenomenon between
persisting superhumps and the orbital signal was strongly present
(vsnet-alert 12943; figure \ref{fig:j1339latebeat}).
We have obtained the orbital period of
0.057289(1) d from these observations (cf. figure \ref{fig:j1339lateshpdm}).
This period is in good agreement with the period determined from
radial-velocity study \citep{gan06j1339}.
The times of superhumps during the post-superoutburst stage were determined
after subtracting the averaged orbital variations
(table \ref{tab:j1339oc2011late}).  The tables includes the initial
eight epochs which were observed as the secondary maxima during the
final stage of the plateau (cf. subsection \ref{obj:kepv344lyr}).
These maxima were measured using phases $-$0.2 to 0.2 of the humps
in order to reduce the contamination from the primary maxima.
The mean periods of late-stage superhumps were
0.058175(3) d ($18 \le E \le 428$) and 0.058183(5) d ($495 \le E \le 985$).

\begin{figure}
  \begin{center}
    \FigureFile(88mm,110mm){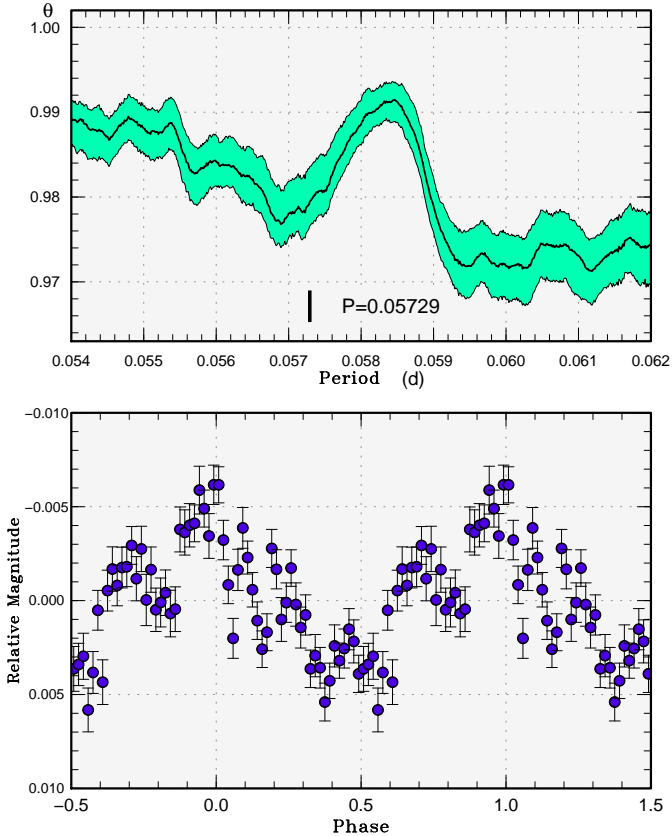}
  \end{center}
  \caption{Possible early superhumps in SDSS J1339 (2011).
     (Upper): PDM analysis.
     The tick is located at the orbital period.
     (Lower): Phase-averaged profile using the orbital period.}
  \label{fig:j1339eshpdm}
\end{figure}

   Since the amplitudes of early superhumps were very low, and since some of
observations were made under non-ideal conditions (due to the sparse
field, the comparison star was much fainter than the variable in some
observations), we have analyzed a selected high-quality subset selected from
the data before BJD 2455604.2.
A PDM analysis of the data yielded a signal of 0.05689(2) d
in addition to the likely signal of emerging ordinary superhumps
(signals around 0.059--0.062 d, upper panel of figure \ref{fig:j1339eshpdm}).
Since this signal is very weak and the baseline of the observation
was not sufficiently long, we used the orbital period as an
approximation of the period of early superhumps
and extracted the averaged profile.  The resultant profile had
double-wave modulations characteristic to early superhumps
\citep{kat02wzsgeESH}, although the full amplitude was only 0.014 mag
(lower panel of figure \ref{fig:j1339eshpdm}).

   The relatively large $P_{\rm dot}$ of $+5.4(0.2) \times 10^{-5}$
during stage B in combination with the lack of post-superoutburst
rebrightenings is compatible with the type-D classification of
WZ Sge-type outbursts (\cite{Pdot}, subsection 5.3).
The delay times of appearance of ordinary superhumps in these type-D
outbursts, however, tend to be around $\sim$10 d.  This delay, corresponding
to the phase with early superhumps, was observed only for 4 d in SDSS J1339.
Although this short duration more resembled those of type-B outbursts
(outbursts with multiple rebrightenings), it may have been a result
of the lack of observations around the true maximum.  Indeed, S. Ueda
reported retrospective positive detections 1.7 d before Miller's detection
at unfiltered CCD magnitudes of 9.9 (vsnet-alert 12859).
The last negative observations by S. Ueda were reported on January 29,
9 d before detection of the outburst.  This possible failure to record
the earliest stage of the outburst might explain the lack of highly
excited emission lines in the spectrum and low amplitudes of early
superhumps.  If the true maximum was 6 d earlier (assuming a typical
delay time of 10 d), the object should have reached a magnitude of 9.5
or brighter, giving an outburst amplitude of $\sim$8.0 mag.  The object
is thus one of the brightest members of WZ Sge-type dwarf novae
and deserves further detailed investigation.

\begin{table}
\caption{Superhump maxima of SDSS J1339 (2011).}\label{tab:j1339oc2011}
\begin{center}
\begin{tabular}{ccccc}
\hline
$E$ & max\commenta & error & $O-C$\commentb & $N$\commentc \\
\hline
0 & 55604.4595 & 0.0005 & $-$0.0113 & 152 \\
1 & 55604.5148 & 0.0009 & $-$0.0141 & 83 \\
2 & 55604.5782 & 0.0006 & $-$0.0088 & 66 \\
6 & 55604.8123 & 0.0008 & $-$0.0071 & 72 \\
7 & 55604.8699 & 0.0006 & $-$0.0077 & 71 \\
8 & 55604.9274 & 0.0008 & $-$0.0083 & 71 \\
11 & 55605.1082 & 0.0004 & $-$0.0018 & 64 \\
\end{tabular}
\end{center}
\end{table}

\addtocounter{table}{-1}
\begin{table}
\caption{Superhump maxima of SDSS J1339 (2011) (continued).}
\begin{center}
\begin{tabular}{ccccc}
\hline
$E$ & max\commenta & error & $O-C$\commentb & $N$\commentc \\
\hline
84 & 55609.3521 & 0.0003 & $-$0.0001 & 361 \\
85 & 55609.4104 & 0.0003 & 0.0001 & 536 \\
86 & 55609.4660 & 0.0002 & $-$0.0024 & 480 \\
87 & 55609.5248 & 0.0003 & $-$0.0017 & 538 \\
88 & 55609.5829 & 0.0003 & $-$0.0017 & 170 \\
89 & 55609.6405 & 0.0003 & $-$0.0022 & 67 \\
90 & 55609.6992 & 0.0004 & $-$0.0017 & 45 \\
\end{tabular}
\end{center}
\end{table}

\addtocounter{table}{-1}
\begin{table}
\caption{Superhump maxima of SDSS J1339 (2011) (continued).}
\begin{center}
\begin{tabular}{ccccc}
\hline
$E$ & max\commenta & error & $O-C$\commentb & $N$\commentc \\
\hline
166 & 55614.1193 & 0.0005 & 0.0020 & 63 \\
167 & 55614.1762 & 0.0005 & 0.0007 & 61 \\
168 & 55614.2339 & 0.0004 & 0.0003 & 231 \\
169 & 55614.2897 & 0.0004 & $-$0.0020 & 239 \\
170 & 55614.3529 & 0.0004 & 0.0031 & 153 \\
171 & 55614.4081 & 0.0004 & 0.0002 & 140 \\
172 & 55614.4680 & 0.0005 & 0.0020 & 172 \\
173 & 55614.5272 & 0.0006 & 0.0031 & 132 \\
174 & 55614.5842 & 0.0004 & 0.0019 & 115 \\
175 & 55614.6433 & 0.0004 & 0.0029 & 116 \\
176 & 55614.6996 & 0.0003 & 0.0011 & 87 \\
178 & 55614.8163 & 0.0014 & 0.0016 & 17 \\
179 & 55614.8735 & 0.0007 & 0.0007 & 29 \\
180 & 55614.9341 & 0.0010 & 0.0032 & 16 \\
182 & 55615.0439 & 0.0024 & $-$0.0032 & 47 \\
183 & 55615.1075 & 0.0007 & 0.0023 & 31 \\
185 & 55615.2240 & 0.0004 & 0.0025 & 223 \\
186 & 55615.2827 & 0.0003 & 0.0031 & 206 \\
187 & 55615.3404 & 0.0007 & 0.0027 & 174 \\
188 & 55615.3977 & 0.0007 & 0.0019 & 64 \\
189 & 55615.4575 & 0.0005 & 0.0036 & 147 \\
190 & 55615.5150 & 0.0005 & 0.0030 & 144 \\
191 & 55615.5725 & 0.0006 & 0.0024 & 115 \\
192 & 55615.6313 & 0.0023 & 0.0030 & 102 \\
194 & 55615.7479 & 0.0007 & 0.0034 & 48 \\
203 & 55616.2745 & 0.0008 & 0.0070 & 63 \\
204 & 55616.3326 & 0.0005 & 0.0070 & 83 \\
205 & 55616.3898 & 0.0009 & 0.0061 & 56 \\
\hline
  \multicolumn{5}{l}{\commenta BJD$-$2400000.} \\
  \multicolumn{5}{l}{\commentb Against max $= 2455604.4708 + 0.058112 E$.} \\
  \multicolumn{5}{l}{\commentc Number of points used to determine the maximum.} \\
\end{tabular}
\end{center}
\end{table}

\begin{table}
\caption{Late stage superhumps in SDSS J1339 (2011).}\label{tab:j1339oc2011late}
\begin{center}
\begin{tabular}{ccccc}
\hline
$E$ & max\commenta & error & $O-C$\commentb & $N$\commentc \\
\hline
0 & 55615.2489 & 0.0014 & $-$0.0055 & 120 \\
2 & 55615.3635 & 0.0016 & $-$0.0072 & 94 \\
3 & 55615.4198 & 0.0005 & $-$0.0091 & 28 \\
4 & 55615.4781 & 0.0005 & $-$0.0090 & 87 \\
5 & 55615.5377 & 0.0016 & $-$0.0075 & 56 \\
18 & 55616.2883 & 0.0005 & $-$0.0134 & 30 \\
19 & 55616.3470 & 0.0006 & $-$0.0129 & 28 \\
\end{tabular}
\end{center}
\end{table}

\addtocounter{table}{-1}
\begin{table}
\caption{Late stage superhumps in SDSS J1339 (2011) (continued).}
\begin{center}
\begin{tabular}{ccccc}
\hline
$E$ & max\commenta & error & $O-C$\commentb & $N$\commentc \\
\hline
156 & 55624.3306 & 0.0029 & $-$0.0008 & 41 \\
157 & 55624.3881 & 0.0005 & $-$0.0015 & 147 \\
158 & 55624.4474 & 0.0005 & $-$0.0003 & 163 \\
159 & 55624.5073 & 0.0003 & 0.0014 & 284 \\
160 & 55624.5626 & 0.0003 & $-$0.0015 & 274 \\
161 & 55624.6236 & 0.0003 & 0.0013 & 271 \\
162 & 55624.6805 & 0.0008 & 0.0001 & 134 \\
\end{tabular}
\end{center}
\end{table}

\addtocounter{table}{-1}
\begin{table}
\caption{Late stage superhumps in SDSS J1339 (2011) (continued).}
\begin{center}
\begin{tabular}{ccccc}
\hline
$E$ & max\commenta & error & $O-C$\commentb & $N$\commentc \\
\hline
258 & 55630.2657 & 0.0009 & $-$0.0006 & 45 \\
259 & 55630.3256 & 0.0008 & 0.0011 & 70 \\
271 & 55631.0210 & 0.0014 & $-$0.0017 & 60 \\
272 & 55631.0806 & 0.0007 & $-$0.0003 & 103 \\
273 & 55631.1383 & 0.0005 & $-$0.0008 & 181 \\
274 & 55631.1976 & 0.0007 & 0.0003 & 46 \\
278 & 55631.4334 & 0.0006 & 0.0033 & 42 \\
\end{tabular}
\end{center}
\end{table}

\addtocounter{table}{-1}
\begin{table}
\caption{Late stage superhumps in SDSS J1339 (2011) (continued).}
\begin{center}
\begin{tabular}{ccccc}
\hline
$E$ & max\commenta & error & $O-C$\commentb & $N$\commentc \\
\hline
365 & 55636.4895 & 0.0012 & $-$0.0028 & 40 \\
366 & 55636.5468 & 0.0009 & $-$0.0036 & 38 \\
367 & 55636.6053 & 0.0020 & $-$0.0033 & 38 \\
382 & 55637.4759 & 0.0014 & $-$0.0055 & 38 \\
383 & 55637.5382 & 0.0011 & $-$0.0014 & 38 \\
384 & 55637.5931 & 0.0011 & $-$0.0047 & 40 \\
393 & 55638.1195 & 0.0007 & $-$0.0019 & 64 \\
\end{tabular}
\end{center}
\end{table}

\addtocounter{table}{-1}
\begin{table}
\caption{Late stage superhumps in SDSS J1339 (2011) (continued).}
\begin{center}
\begin{tabular}{ccccc}
\hline
$E$ & max\commenta & error & $O-C$\commentb & $N$\commentc \\
\hline
622 & 55651.4525 & 0.0015 & 0.0065 & 61 \\
623 & 55651.5095 & 0.0007 & 0.0053 & 65 \\
624 & 55651.5646 & 0.0009 & 0.0022 & 62 \\
625 & 55651.6254 & 0.0010 & 0.0048 & 40 \\
634 & 55652.1414 & 0.0011 & $-$0.0029 & 120 \\
636 & 55652.2611 & 0.0012 & 0.0004 & 122 \\
637 & 55652.3183 & 0.0025 & $-$0.0006 & 109 \\
\end{tabular}
\end{center}
\end{table}

\subsection{SDSS J160501.35$+$203056.9}\label{obj:j1605}

   This transient was detected by K. Itagaki on 2010 November 21.36782 UT
at an unfiltered CCD magnitude of 12.0 (vsnet-alert 12414).
A. Drake informed that this object was spectroscopically recorded
as an evident CV in SDSS and there was a possible faint outburst in 2005
(vsnet-alert 12413).  Based on this information, we refer to this
object by the SDSS designation (hereafter SDSS J1605).
Although the outburst of the amplitude was sufficiently suggestive of
an outburst of a WZ Sge-type dwarf nova, the short visibility in the
evening and morning sky made observations very difficult.
The rapid fading from the main superoutburst took place on December 24
(vsnet-alert 12525), 33 d after the initial detection.
The object still remained bright on 2011 January 21 (61 d after
the initial detection).

\begin{figure}
  \begin{center}
    \FigureFile(88mm,110mm){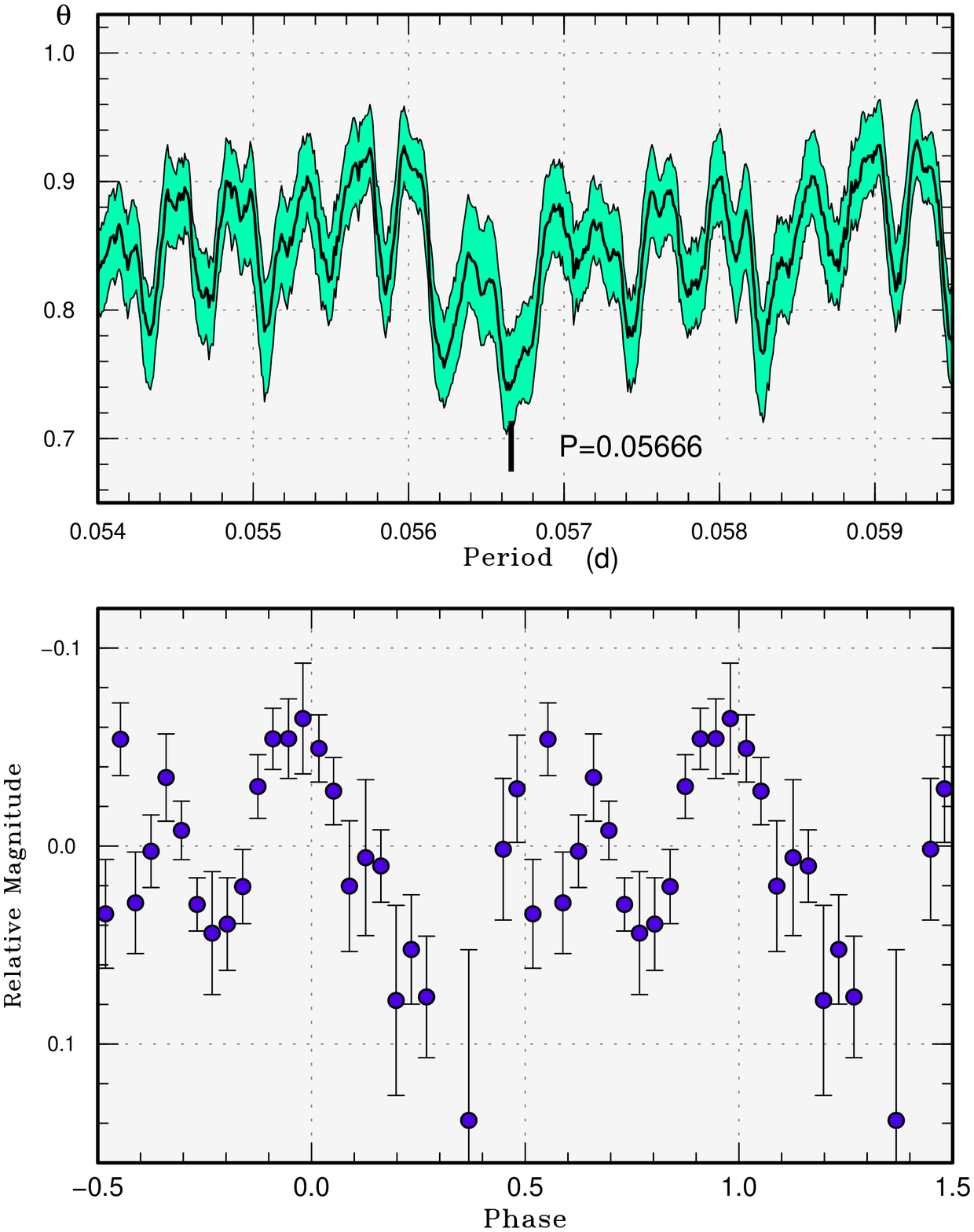}
  \end{center}
  \caption{Possible early superhumps in SDSS J1605 (2010).
     (Upper): PDM analysis.
     The rejection rate for bootstrapping was reduced to 0.2 for
     better visualization.
     (Lower): Phase-averaged profile.}
  \label{fig:j1605eshpdm}
\end{figure}

   Since the later stage of the outburst was poorly observed and
the state of the object (either continuation of the plateau phase or
some kind of rebrightening phenomenon) was unknown, we restrict our
analysis to the earlier part of the observation.
Since some of the observations were recorded at high airmasses
(larger than 3), secondary corrections for atmospheric extinction
were applied to these observations.  The early data showed some hint
of double-wave modulations, which are possibly early superhumps
(figure \ref{fig:j1605eshpdm}).  Due to the short visibility,
it was impossible to uniquely determine the period, we selected the
most likely period based on our knowledge in periods of known WZ Sge-type
dwarf novae with long plateau phases (\cite{Pdot}, subsection 5.3).
This period should be tested by future observations.

   During the later stage of the outburst, there was a signal
around 0.05718 d (0.9 \% longer than the period of possible early
superhumps) attributable to ordinary superhumps, but we did not adopt
this value due to very poor statistics.

\subsection{SDSS J210014.12$+$004446.0}\label{obj:j2100}

   We reported on the 2007 superoutburst of this object (hereafter
SDSS J2100) in \citet{Pdot}.  We further obtained observations on
three consecutive nights during the 2010 superoutburst
(table \ref{tab:j2100oc2010}).  The $O-C$ values suggest that this
object is either a long-$P_{\rm SH}$ system with a large negative
$P_{\rm dot}$ or we incidentally spotted an epoch of stage transition.
The global period with the PDM method was 0.08743(2) d, close
to the value of 0.08746(8) d in \citet{tra05j2100}.

\begin{table}
\caption{Superhump maxima of SDSS J2100 (2010).}\label{tab:j2100oc2010}
\begin{center}
\begin{tabular}{ccccc}
\hline
$E$ & max\commenta & error & $O-C$\commentb & $N$\commentc \\
\hline
0 & 55530.2355 & 0.0021 & $-$0.0030 & 41 \\
1 & 55530.3224 & 0.0003 & $-$0.0034 & 82 \\
35 & 55533.3065 & 0.0009 & 0.0127 & 81 \\
69 & 55536.2554 & 0.0012 & $-$0.0064 & 77 \\
\hline
  \multicolumn{5}{l}{\commenta BJD$-$2400000.} \\
  \multicolumn{5}{l}{\commentb Against max $= 2455530.2384 + 0.087295 E$.} \\
  \multicolumn{5}{l}{\commentc Number of points used to determine the maximum.} \\
\end{tabular}
\end{center}
\end{table}

\subsection{OT J000938.3$-$121017}\label{obj:j0009}

\begin{figure}
  \begin{center}
    \FigureFile(88mm,110mm){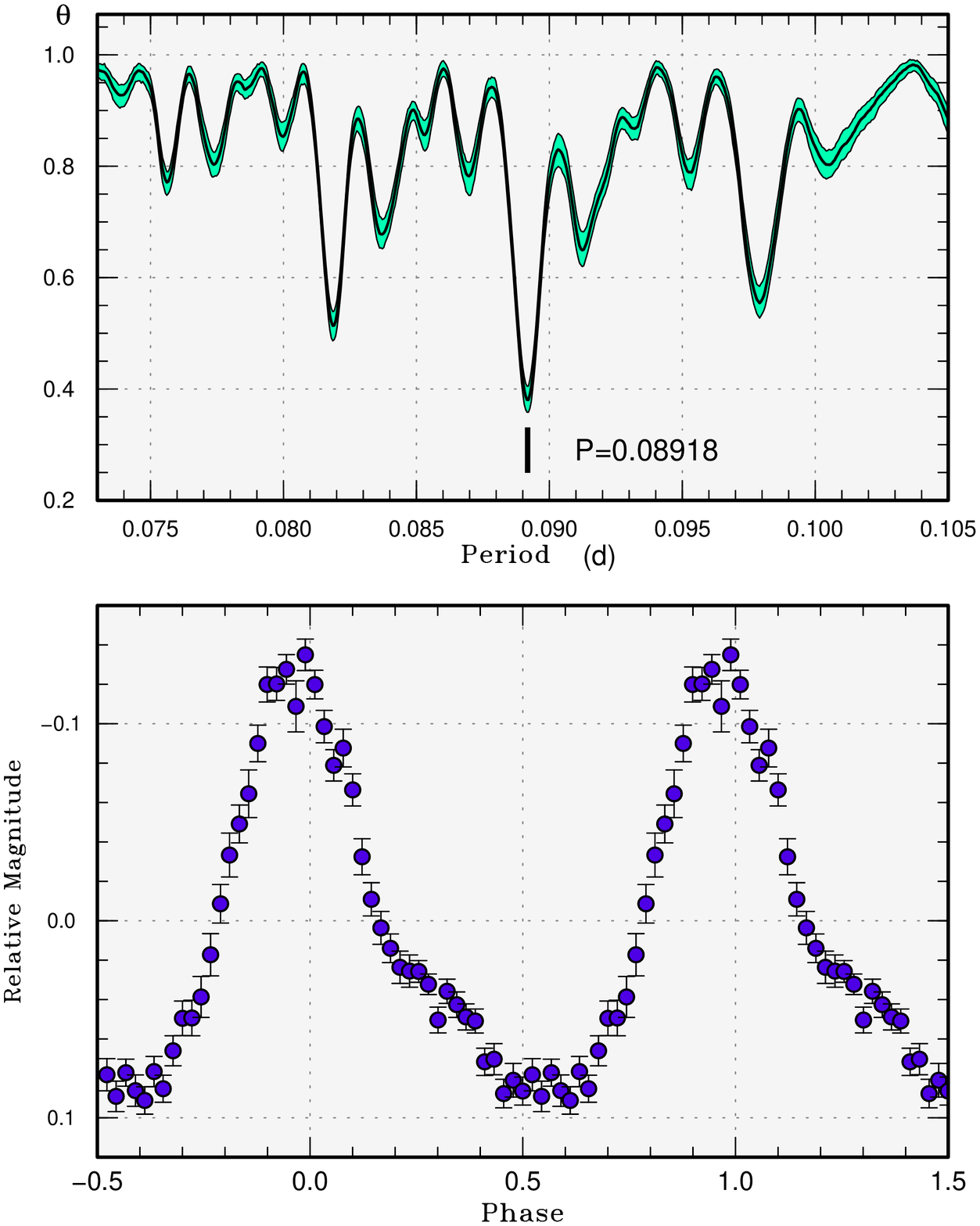}
  \end{center}
  \caption{Superhumps in OT J0009 (2010). (Upper): PDM analysis.
     (Lower): Phase-averaged profile.}
  \label{fig:j0009shpdm}
\end{figure}

\begin{figure}
  \begin{center}
    \FigureFile(88mm,110mm){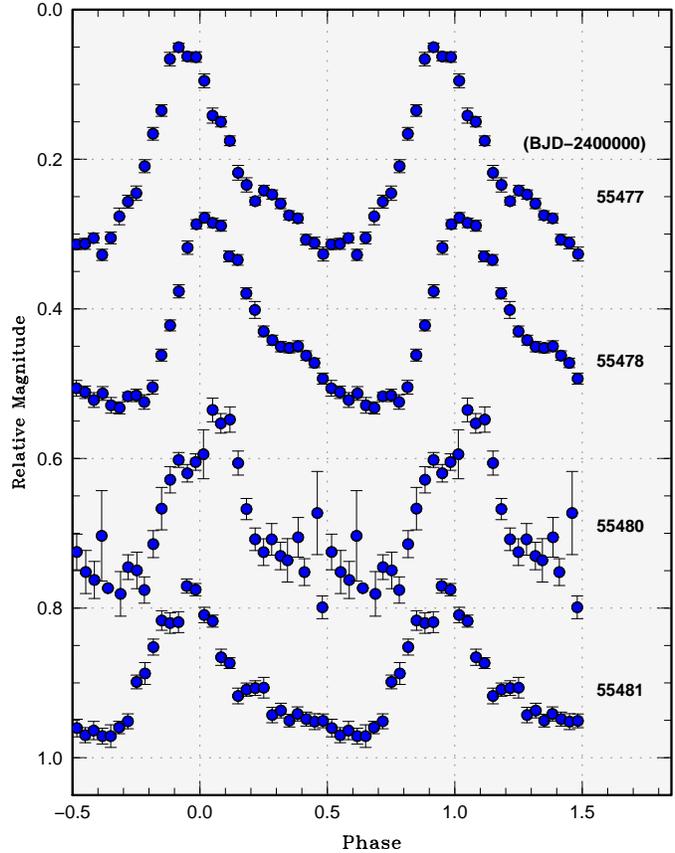}
  \end{center}
  \caption{Superhump profiles of OT J0009 (2010).
     The profiles of superhumps strongly varied between nights.
     The figure was drawn against a mean period of 0.089117 d.}
  \label{fig:j0009prof}
\end{figure}

   This object (= CSS101007:000938$-$121017, hereafter OT J0009) was
discovered by the CRTS on 2010 October 10.  The object has a blue
counterpart on the Digitized Sky Survey (DSS).  Following the initial
detection, an inspection of the backlog by the CRTS team clarified another
outburst in 2005 (cf. vsnet-alert 12235).
Subsequent observations confirmed superhumps (vsnet-alert 12234,
figure \ref{fig:j0009shpdm}),
identifying the object a dwarf nova near the period gap
(vsnet-alert 12238, 12241, 12244).
There was a distinct shortening of the superhump period after $E=14$,
which we attribute to a stage B-C transition (table \ref{tab:j0009oc2010}).
Although we list our tentative identifications of periods in table
\ref{tab:perlist}, this variation in the period may be interpreted
as a continuous change with a global $P_{\rm dot}$ of $-51(4) \times 10^{-5}$
(see also figure \ref{fig:j0009prof}).
Although the object resembles EF Peg in its long $P_{\rm SH}$, large
outburst amplitude and rare outbursts, the large variation in the superhump
period resembles those of more frequently outbursting SU UMa-type
dwarf novae (cf. \cite{Pdot}, subsection 4.10).
As proposed for V342 Cam (=1RXS J042332.8+745300) \citep{she10j0423},
the appearance of orbital signature in some stage of the outburst may have
contributed this change of the period.
Further detailed observations during future superoutbursts and in quiescence
are desired.

\begin{table}
\caption{Superhump maxima of OT J0009 (2010).}\label{tab:j0009oc2010}
\begin{center}
\begin{tabular}{ccccc}
\hline
$E$ & max\commenta & error & $O-C$\commentb & $N$\commentc \\
\hline
0 & 55477.0085 & 0.0006 & $-$0.0048 & 65 \\
1 & 55477.0983 & 0.0003 & $-$0.0041 & 227 \\
2 & 55477.1877 & 0.0004 & $-$0.0039 & 331 \\
3 & 55477.2770 & 0.0004 & $-$0.0037 & 94 \\
12 & 55478.0893 & 0.0004 & 0.0065 & 93 \\
13 & 55478.1778 & 0.0004 & 0.0059 & 94 \\
14 & 55478.2674 & 0.0004 & 0.0065 & 94 \\
34 & 55480.0493 & 0.0008 & 0.0060 & 111 \\
35 & 55480.1349 & 0.0006 & 0.0025 & 222 \\
45 & 55481.0229 & 0.0024 & $-$0.0007 & 50 \\
46 & 55481.1100 & 0.0006 & $-$0.0028 & 94 \\
47 & 55481.1989 & 0.0008 & $-$0.0030 & 94 \\
48 & 55481.2864 & 0.0009 & $-$0.0045 & 94 \\
\hline
  \multicolumn{5}{l}{\commenta BJD$-$2400000.} \\
  \multicolumn{5}{l}{\commentb Against max $= 2455477.0133 + 0.089117 E$.} \\
  \multicolumn{5}{l}{\commentc Number of points used to determine the maximum.} \\
\end{tabular}
\end{center}
\end{table}

\subsection{OT J012059.6$+$325545}\label{obj:j0120}

\begin{figure}
  \begin{center}
    \FigureFile(88mm,110mm){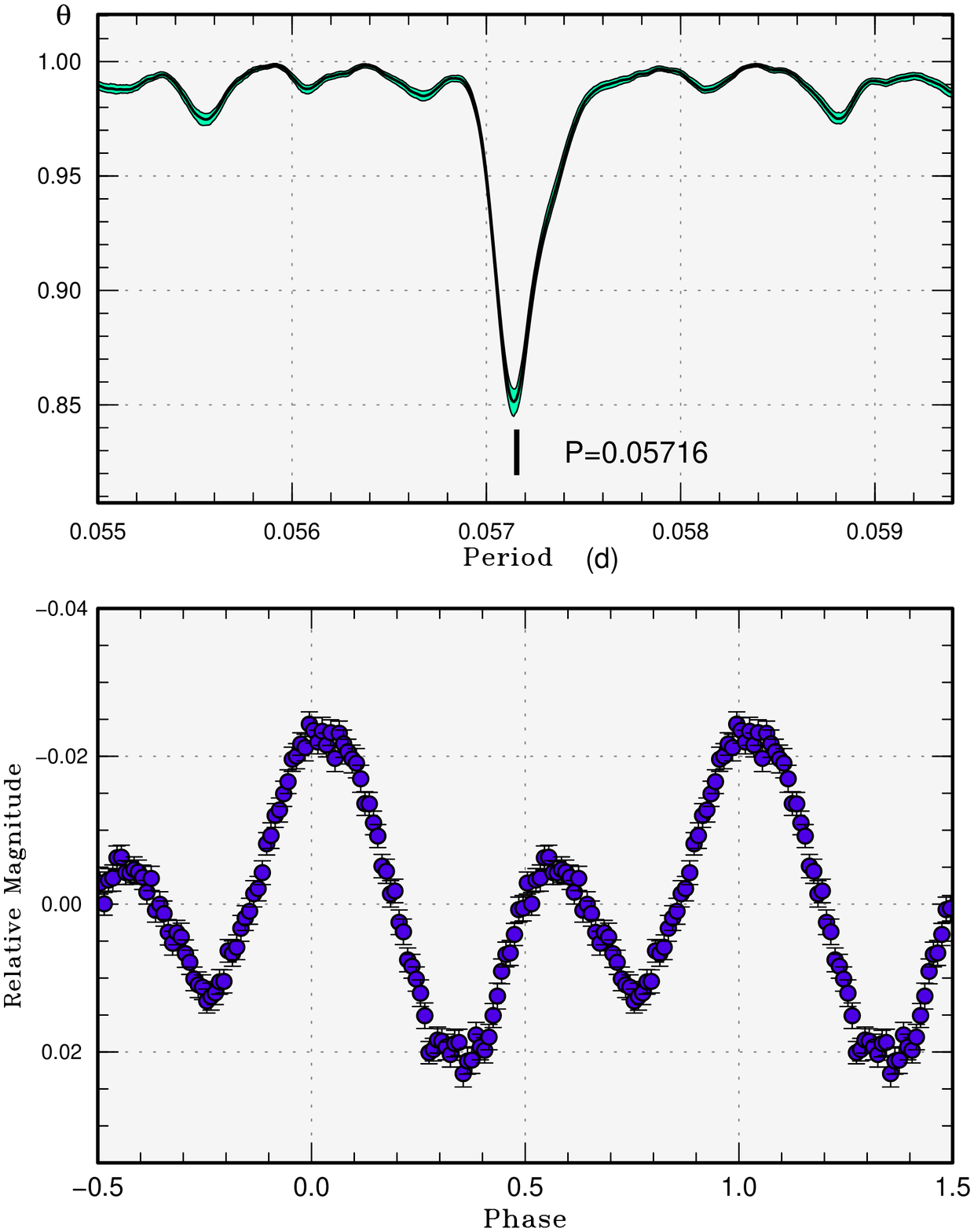}
  \end{center}
  \caption{Early superhumps in OT J0120 (2010). (Upper): PDM analysis.
     (Lower): Phase-averaged profile.}
  \label{fig:j0120eshpdm}
\end{figure}

\begin{figure}
  \begin{center}
    \FigureFile(88mm,110mm){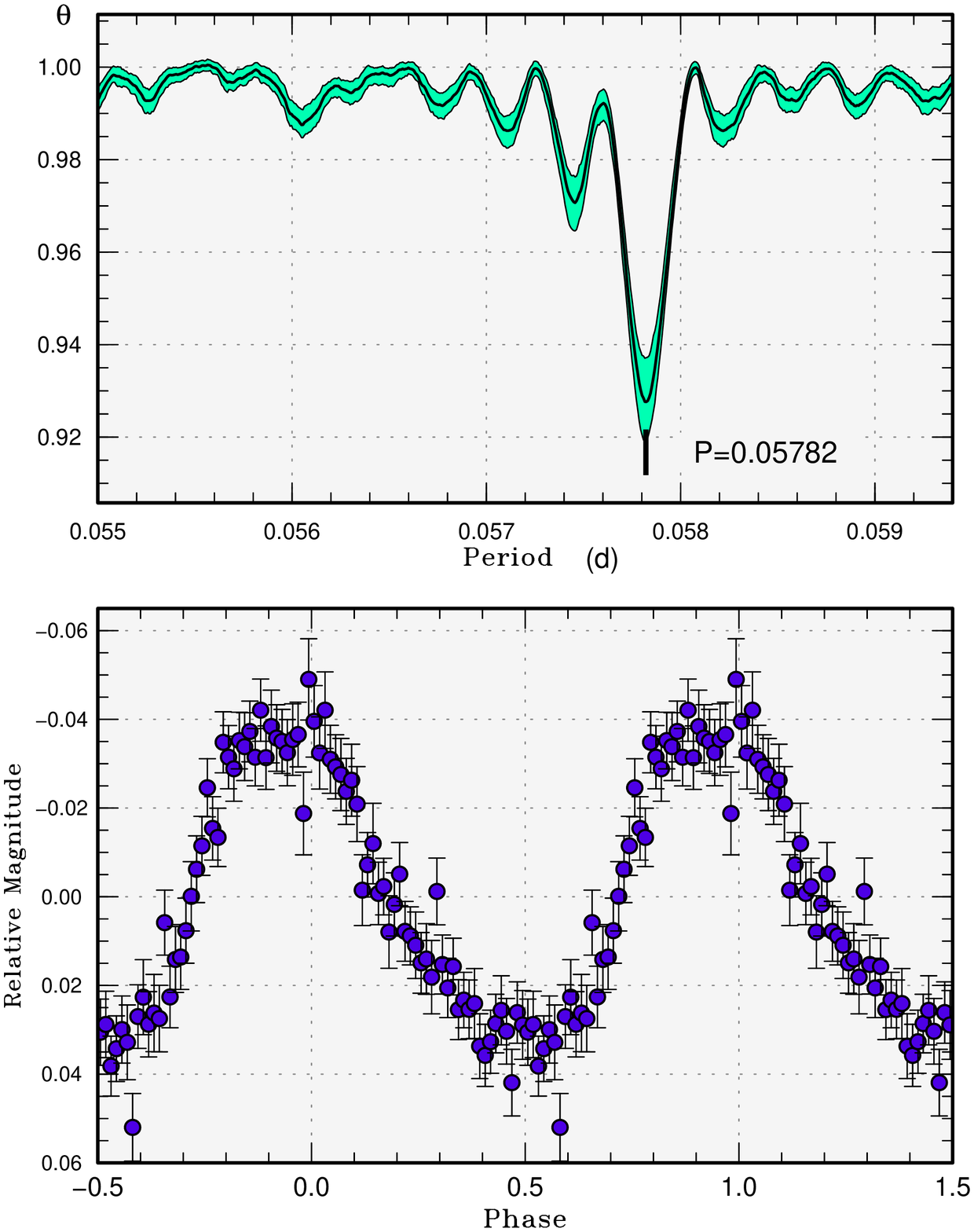}
  \end{center}
  \caption{Ordinary superhumps in OT J0120 (2010). (Upper): PDM analysis.
     (Lower): Phase-averaged profile.}
  \label{fig:j0120shpdm}
\end{figure}

   This object (hereafter OT J0120) was discovered as a large-amplitude
transient by K. Itagaki on 2010 November 30.50663 at an unfiltered
CCD magnitude of 12.3 (vsnet-alert 12431).
Immediate follow-up observations clearly indicated the presence of
early superhumps (vsnet-alert 12436; figure \ref{fig:j0120eshpdm}).
The object started to show ordinary superhumps 11 d after the
discovery (vsnet-alert 12491, 12496; figure \ref{fig:j0120shpdm}).
Since some of the observations were recorded at high airmasses
(larger than 3), secondary corrections for atmospheric extinction
were applied to these observations.

   The times of superhump maxima are listed in table \ref{tab:j0120oc2010}.
Due to the gap in the observations, stage A superhumps were not recorded.
Following the stage B with a significant increase in the period
[$P_{\rm dot} = +4.3(0.5) \times 10^{-5}$], the period quickly shortened.
The mean period for the interval BJD 2455554.4--2455556.3 was 
0.05672(8) d (PDM method), which is 0.8 \% shorter than the period of
early superhumps (close to $P_{\rm orb}$).  This rapid excursion to
a shorter period is similar to the case in SDSS J0804 (subsection
\ref{obj:j0804}), and it might be a common property in WZ Sge-type
dwarf novae.  Whether this period is related to negative superhumps or
not requires further detailed examination.

\begin{figure}
  \begin{center}
    \FigureFile(88mm,110mm){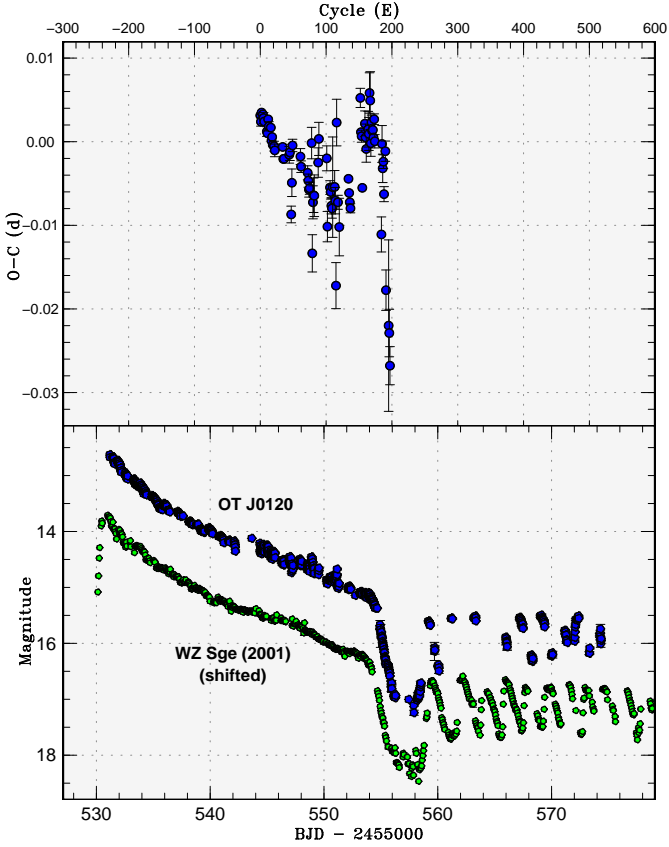}
  \end{center}
  \caption{$O-C$ diagram of superhumps in OT J0120 (2010).
     Upper: $O-C$ diagram.
     We used a period of 0.05784 d for calculating the $O-C$'s.
     Lower: Light curve and comparison with WZ Sge (2001).
     After a temporary ``dip" following the rapid fading from the
     superoutburst, the object underwent multiple rebrightenings.
     The behavior was almost exactly same to the one in WZ Sge (2001).
  }
  \label{fig:j0120humpall}
\end{figure}

\begin{figure}
  \begin{center}
    \FigureFile(88mm,110mm){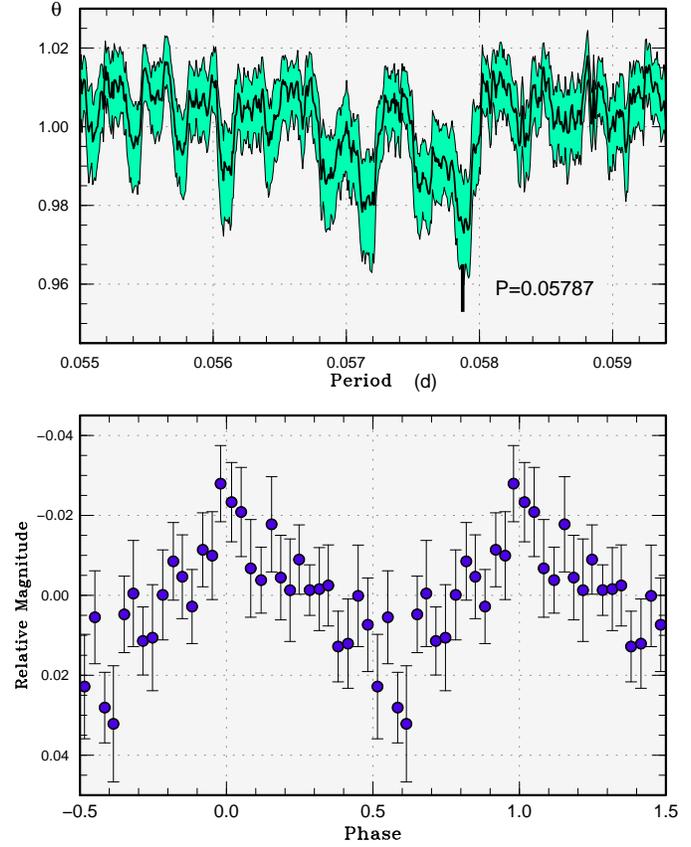}
  \end{center}
  \caption{Superhumps in OT J0120 during rebrightening phase (2010).
     (Upper): PDM analysis of the CRI and Shu data.
     The rejection rate for bootstrapping was reduced to 0.2 for
     better visualization.
     (Lower): Phase-averaged profile.}
  \label{fig:j0120rebpdm}
\end{figure}

   The object underwent multiple rebrightenings (vsnet-alert 12538,
12562, 12577, 12596, 12602, 12610, 12674).  Although the exact number
of rebrightenings was unclear, the shortest intervals of rebrightenings
were $\sim$2 d, close to those of WZ Sge (2001; \cite{pat02wzsge};
\cite{ish02wzsgeletter}; \cite{Pdot}) and AL Com (2007; \cite{uem08alcom}),
a group of rebrightening classified as type-A WZ Sge-type superoutbursts
in \citet{ima06tss0222} and \citet{Pdot}.  A comparison with WZ Sge (2001)
(figure \ref{fig:j0120humpall}) suggests at least nine rebrightenings
during the observable period.
An analysis of selected high signal-to-noise observations during the
rebrightening phase supports the presence of persistent superhumps with
a period of 0.05787(3) d, which is close to $P_{\rm SH}$ during
the main superoutburst \ref{fig:j0120rebpdm}.
The observed amplitude of these modulations
were small because the observations were relatively restricted to
maxima of rebrightening (see a discussion in SDSS J0804, subsection
\ref{obj:j0804}).  These modulations apparently persisted, even with
larger amplitudes, 18 d after the final observation of the rebrightening
phase (vsnet-alert 12762).

\begin{table}
\caption{Superhump maxima of OT J0120 (2010).}\label{tab:j0120oc2010}
\begin{center}
\begin{tabular}{ccccc}
\hline
$E$ & max\commenta & error & $O-C$\commentb & $N$\commentc \\
\hline
0 & 55544.4060 & 0.0003 & 0.0031 & 189 \\
1 & 55544.4631 & 0.0005 & 0.0024 & 100 \\
2 & 55544.5221 & 0.0003 & 0.0036 & 55 \\
4 & 55544.6375 & 0.0004 & 0.0034 & 53 \\
5 & 55544.6949 & 0.0006 & 0.0030 & 47 \\
6 & 55544.7524 & 0.0004 & 0.0026 & 50 \\
10 & 55544.9825 & 0.0006 & 0.0015 & 77 \\
11 & 55545.0401 & 0.0004 & 0.0013 & 95 \\
12 & 55545.0997 & 0.0005 & 0.0031 & 94 \\
13 & 55545.1567 & 0.0006 & 0.0022 & 95 \\
16 & 55545.3300 & 0.0004 & 0.0022 & 56 \\
17 & 55545.3863 & 0.0003 & 0.0007 & 77 \\
18 & 55545.4446 & 0.0004 & 0.0011 & 79 \\
19 & 55545.5015 & 0.0005 & 0.0002 & 51 \\
21 & 55545.6170 & 0.0006 & 0.0001 & 35 \\
22 & 55545.6743 & 0.0008 & $-$0.0003 & 48 \\
34 & 55546.3688 & 0.0005 & 0.0004 & 86 \\
35 & 55546.4252 & 0.0002 & $-$0.0010 & 109 \\
36 & 55546.4831 & 0.0004 & $-$0.0009 & 73 \\
44 & 55546.9463 & 0.0010 & $-$0.0002 & 127 \\
45 & 55547.0044 & 0.0010 & 0.0001 & 83 \\
47 & 55547.1127 & 0.0010 & $-$0.0072 & 216 \\
48 & 55547.1743 & 0.0016 & $-$0.0034 & 42 \\
49 & 55547.2366 & 0.0007 & 0.0011 & 29 \\
61 & 55547.9294 & 0.0010 & 0.0001 & 104 \\
62 & 55547.9860 & 0.0007 & $-$0.0010 & 134 \\
72 & 55548.5637 & 0.0008 & $-$0.0015 & 45 \\
73 & 55548.6206 & 0.0008 & $-$0.0024 & 47 \\
74 & 55548.6773 & 0.0009 & $-$0.0034 & 47 \\
75 & 55548.7353 & 0.0007 & $-$0.0032 & 52 \\
78 & 55548.9143 & 0.0019 & 0.0023 & 105 \\
79 & 55548.9589 & 0.0022 & $-$0.0109 & 109 \\
80 & 55549.0229 & 0.0020 & $-$0.0047 & 108 \\
81 & 55549.0815 & 0.0021 & $-$0.0039 & 115 \\
82 & 55549.1393 & 0.0025 & $-$0.0039 & 93 \\
88 & 55549.4903 & 0.0018 & 0.0003 & 30 \\
89 & 55549.5510 & 0.0020 & 0.0031 & 32 \\
101 & 55550.2428 & 0.0015 & 0.0012 & 13 \\
102 & 55550.2924 & 0.0018 & $-$0.0070 & 17 \\
106 & 55550.5285 & 0.0008 & $-$0.0022 & 46 \\
107 & 55550.5858 & 0.0008 & $-$0.0027 & 39 \\
108 & 55550.6420 & 0.0010 & $-$0.0043 & 53 \\
109 & 55550.6995 & 0.0012 & $-$0.0046 & 48 \\
110 & 55550.7599 & 0.0060 & $-$0.0020 & 28 \\
113 & 55550.9334 & 0.0020 & $-$0.0019 & 169 \\
114 & 55550.9896 & 0.0016 & $-$0.0035 & 158 \\
115 & 55551.0373 & 0.0028 & $-$0.0136 & 169 \\
116 & 55551.1146 & 0.0028 & 0.0059 & 165 \\
118 & 55551.2208 & 0.0008 & $-$0.0035 & 31 \\
120 & 55551.3335 & 0.0034 & $-$0.0064 & 10 \\
134 & 55552.1490 & 0.0003 & $-$0.0002 & 239 \\
\hline
  \multicolumn{5}{l}{\commenta BJD$-$2400000.} \\
  \multicolumn{5}{l}{\commentb Against max $= 2455544.4029 + 0.057809 E$.} \\
  \multicolumn{5}{l}{\commentc Number of points used to determine the maximum.} \\
\end{tabular}
\end{center}
\end{table}

\addtocounter{table}{-1}
\begin{table}
\caption{Superhump maxima of OT J0120 (2010) (continued).}
\begin{center}
\begin{tabular}{ccccc}
\hline
$E$ & max\commenta & error & $O-C$\commentb & $N$\commentc \\
\hline
135 & 55552.2052 & 0.0003 & $-$0.0019 & 180 \\
136 & 55552.2619 & 0.0003 & $-$0.0030 & 253 \\
137 & 55552.3190 & 0.0005 & $-$0.0037 & 204 \\
152 & 55553.1998 & 0.0011 & 0.0100 & 255 \\
153 & 55553.2536 & 0.0003 & 0.0059 & 118 \\
154 & 55553.3109 & 0.0007 & 0.0055 & 253 \\
155 & 55553.3626 & 0.0005 & $-$0.0007 & 188 \\
159 & 55553.6016 & 0.0015 & 0.0071 & 47 \\
160 & 55553.6578 & 0.0016 & 0.0055 & 40 \\
161 & 55553.7143 & 0.0016 & 0.0042 & 44 \\
164 & 55553.8902 & 0.0021 & 0.0066 & 79 \\
165 & 55553.9474 & 0.0020 & 0.0061 & 104 \\
166 & 55554.0102 & 0.0024 & 0.0110 & 105 \\
167 & 55554.0671 & 0.0035 & 0.0102 & 219 \\
168 & 55554.1198 & 0.0016 & 0.0051 & 216 \\
169 & 55554.1802 & 0.0009 & 0.0076 & 226 \\
170 & 55554.2370 & 0.0003 & 0.0066 & 174 \\
171 & 55554.2950 & 0.0004 & 0.0068 & 204 \\
172 & 55554.3519 & 0.0003 & 0.0059 & 254 \\
173 & 55554.4119 & 0.0006 & 0.0081 & 99 \\
174 & 55554.4671 & 0.0008 & 0.0055 & 62 \\
184 & 55555.0344 & 0.0021 & $-$0.0053 & 168 \\
185 & 55555.1030 & 0.0022 & 0.0055 & 213 \\
186 & 55555.1580 & 0.0017 & 0.0026 & 285 \\
187 & 55555.2166 & 0.0008 & 0.0035 & 111 \\
188 & 55555.2705 & 0.0009 & $-$0.0004 & 186 \\
190 & 55555.3914 & 0.0012 & 0.0048 & 192 \\
191 & 55555.4326 & 0.0024 & $-$0.0118 & 136 \\
195 & 55555.6597 & 0.0103 & $-$0.0159 & 62 \\
196 & 55555.7167 & 0.0028 & $-$0.0167 & 61 \\
197 & 55555.7706 & 0.0023 & $-$0.0206 & 61 \\
\hline
  \multicolumn{5}{l}{\commenta BJD$-$2400000.} \\
  \multicolumn{5}{l}{\commentb Against max $= 2455544.4029 + 0.057809 E$.} \\
  \multicolumn{5}{l}{\commentc Number of points used to determine the maximum.} \\
\end{tabular}
\end{center}
\end{table}

\subsection{OT J014150.4$+$090822}\label{obj:j0141}

\begin{figure}
  \begin{center}
    \FigureFile(88mm,110mm){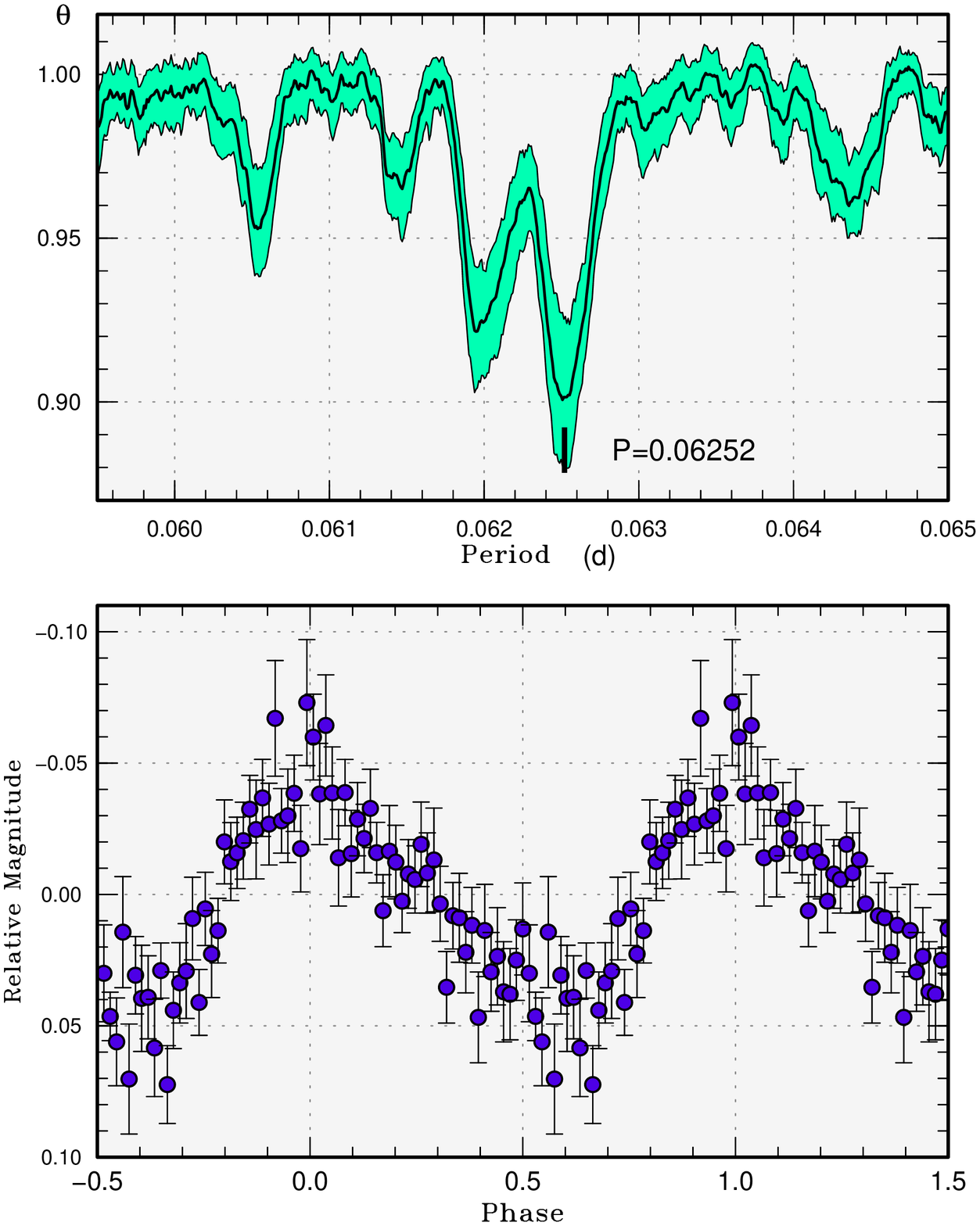}
  \end{center}
  \caption{Superhumps in OT J0141 (2010). (Upper): PDM analysis.
     (Lower): Phase-averaged profile.}
  \label{fig:j0141shpdm}
\end{figure}

   This object (= CSS101127:014150$+$090822), hereafter OT J0141) was
discovered by the CRTS on 2010 November 27.  No previous outburst was
known.  Soon after the discovery, superhumps were detected (vsnet-alert
12422; figure \ref{fig:j0141shpdm}).
The times of superhump maxima are listed in table \ref{tab:j0141oc2010}.
Stage B with a $P_{\rm dot}$ of $+10.1(1.6) \times 10^{-5}$
and the subsequent stage C were apparently observed,
and we listed parameters in table \ref{tab:perlist} based on this
identification.

\begin{table}
\caption{Superhump maxima of OT J0141 (2010).}\label{tab:j0141oc2010}
\begin{center}
\begin{tabular}{ccccc}
\hline
$E$ & max\commenta & error & $O-C$\commentb & $N$\commentc \\
\hline
0 & 55527.9421 & 0.0010 & 0.0025 & 126 \\
1 & 55528.0072 & 0.0015 & 0.0051 & 125 \\
2 & 55528.0677 & 0.0008 & 0.0032 & 128 \\
19 & 55529.1230 & 0.0007 & $-$0.0037 & 128 \\
34 & 55530.0618 & 0.0022 & $-$0.0021 & 96 \\
35 & 55530.1254 & 0.0010 & $-$0.0010 & 131 \\
67 & 55532.1189 & 0.0027 & $-$0.0067 & 79 \\
68 & 55532.1811 & 0.0033 & $-$0.0070 & 114 \\
114 & 55535.0687 & 0.0033 & 0.0065 & 128 \\
115 & 55535.1298 & 0.0041 & 0.0051 & 93 \\
130 & 55536.0598 & 0.0026 & $-$0.0020 & 128 \\
131 & 55536.1245 & 0.0020 & 0.0001 & 115 \\
\hline
  \multicolumn{5}{l}{\commenta BJD$-$2400000.} \\
  \multicolumn{5}{l}{\commentb Against max $= 2455527.9396 + 0.062479 E$.} \\
  \multicolumn{5}{l}{\commentc Number of points used to determine the maximum.} \\
\end{tabular}
\end{center}
\end{table}

\subsection{OT J041350.0$+$094515}\label{obj:j0413}

\begin{figure}
  \begin{center}
    \FigureFile(88mm,110mm){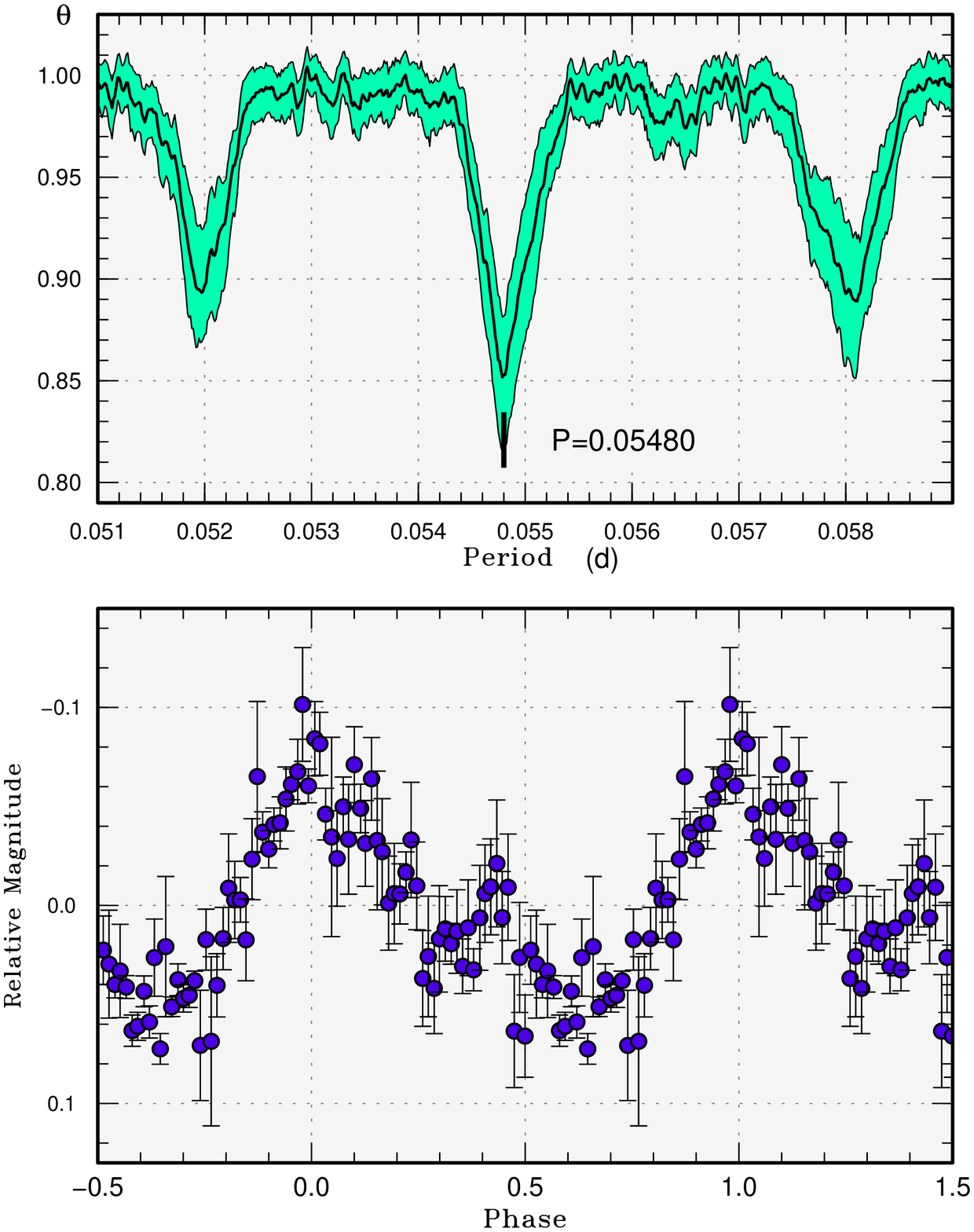}
  \end{center}
  \caption{Superhumps in OT J0413 (2011). (Upper): PDM analysis.
     (Lower): Phase-averaged profile.}
  \label{fig:j0413shpdm}
\end{figure}

   This object (= CSS110125:041350$+$094515), hereafter OT J0413) was
discovered by the CRTS on 2011 January 25.  Short-period superhumps
were immediately detected (vsnet-alert 12713, 12719; figure
\ref{fig:j0413shpdm}).

   The times of superhump maxima are listed in table \ref{tab:j0413oc2011}.
A relatively small $P_{\rm dot} = +4.2(0.8) \times 10^{-5}$ was detected.
The superhump period resembles those of WZ Sge-type dwarf novae,
such as GW Lib and UW Tri.  Although the early superhumps were not observed,
this object may have been detected well after the true maximum
(the last preceding CRTS negative observation was on 2011 January 9).
Although the lack of past outbursts might also support the interpretation
as a potential WZ Sge-type dwarf nova, the object might belong to
group ``X'' in \citet{uem10j0557}, a small group of short-$P_{\rm orb}$
SU UMa-type dwarf novae without noticeable WZ Sge-type characteristics.
Observations of future outbursts are desired.

\begin{table}
\caption{Superhump maxima of OT J0413 (2011).}\label{tab:j0413oc2011}
\begin{center}
\begin{tabular}{ccccc}
\hline
$E$ & max\commenta & error & $O-C$\commentb & $N$\commentc \\
\hline
0 & 55587.6103 & 0.0005 & 0.0007 & 54 \\
1 & 55587.6645 & 0.0006 & 0.0001 & 54 \\
2 & 55587.7205 & 0.0004 & 0.0013 & 54 \\
3 & 55587.7736 & 0.0006 & $-$0.0004 & 54 \\
18 & 55588.5962 & 0.0005 & $-$0.0001 & 54 \\
19 & 55588.6502 & 0.0006 & $-$0.0008 & 54 \\
20 & 55588.7055 & 0.0007 & $-$0.0004 & 54 \\
21 & 55588.7597 & 0.0006 & $-$0.0010 & 53 \\
36 & 55589.5831 & 0.0007 & 0.0002 & 54 \\
37 & 55589.6360 & 0.0006 & $-$0.0018 & 54 \\
38 & 55589.6918 & 0.0007 & $-$0.0008 & 53 \\
39 & 55589.7477 & 0.0007 & 0.0003 & 53 \\
54 & 55590.5699 & 0.0012 & 0.0002 & 27 \\
55 & 55590.6239 & 0.0008 & $-$0.0005 & 38 \\
56 & 55590.6796 & 0.0007 & 0.0003 & 38 \\
57 & 55590.7340 & 0.0008 & $-$0.0001 & 37 \\
58 & 55590.7890 & 0.0012 & 0.0001 & 22 \\
73 & 55591.6125 & 0.0009 & 0.0014 & 36 \\
74 & 55591.6676 & 0.0010 & 0.0017 & 38 \\
75 & 55591.7226 & 0.0009 & 0.0018 & 37 \\
76 & 55591.7766 & 0.0012 & 0.0010 & 29 \\
100 & 55593.0881 & 0.0024 & $-$0.0031 & 87 \\
\hline
  \multicolumn{5}{l}{\commenta BJD$-$2400000.} \\
  \multicolumn{5}{l}{\commentb Against max $= 2455587.6096 + 0.054816 E$.} \\
  \multicolumn{5}{l}{\commentc Number of points used to determine the maximum.} \\
\end{tabular}
\end{center}
\end{table}

\subsection{OT J043112.5$-$031452}\label{obj:j0431}

\begin{figure}
  \begin{center}
    \FigureFile(88mm,110mm){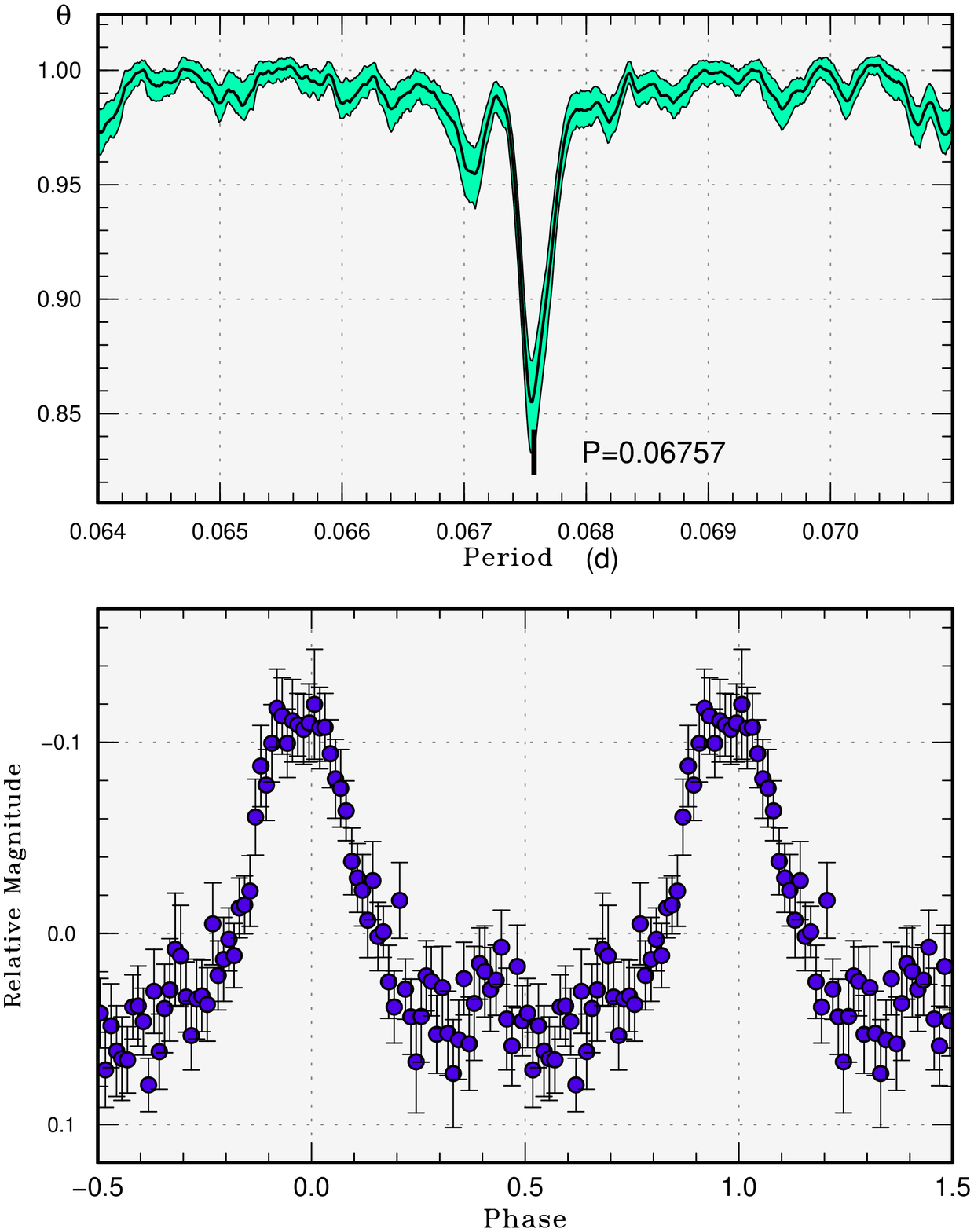}
  \end{center}
  \caption{Superhumps in OT J0431 (2011). (Upper): PDM analysis
     for the data before BJD 2455589.  The data within orbital phases
     of 0.09 were removed before the analysis.
     (Lower): Phase-averaged profile.}
  \label{fig:j0431shpdm}
\end{figure}

   This object (= CSS110113:043112$-$031452), hereafter OT J0431) was
discovered by the CRTS on 2011 January 13.  Subsequent observations
detected both superhumps and eclipses (vsnet-alert 12612, 12613, 12615;
figure \ref{fig:j0431shpdm}).
The times of recorded eclipses, determined with the KW method,
after removing linearly approximated trends around eclipses in order to
minimize the effect of superhumps, are summarized in table \ref{tab:j0431ecl}.
We obtained an ephemeris of

\begin{equation}
{\rm Min(BJD)} = 2455575.27211(6) + 0.0660495(5) E
\label{equ:j0431ecl}.
\end{equation}

\begin{figure}
  \begin{center}
    \FigureFile(88mm,110mm){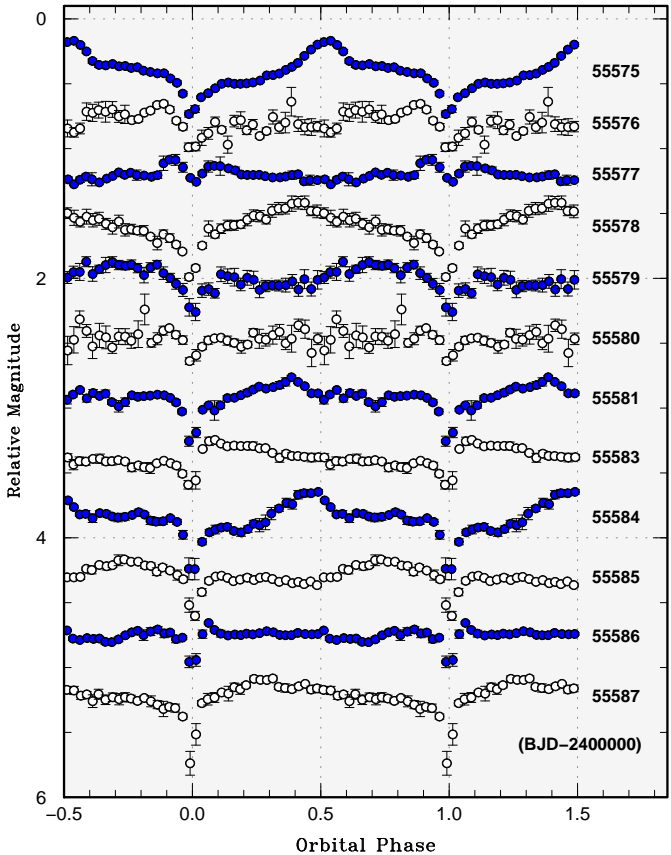}
  \end{center}
  \caption{Variation of superhumps and eclipses in OT J0431 (2011).
  The data for the first 12 d are shown.  A strong beat phenomenon
  was observed.  The symbols are interchangeably used every other day
  for easier visualization.
  }
  \label{fig:j0431shprof}
\end{figure}

   The times of superhump maxima outside the eclipses
are listed in table \ref{tab:j0431oc2011}.
Well-refined stage B and C superhumps were recorded.
The $P_{\rm dot}$ and fractional excess of stage B superhumps are
$+8.4(1.2) \times 10^{-5}$ and 2.3 \%, respectively, and
are usual values for this $P_{\rm orb}$.  It is noteworthy that
this object is the first eclipsing SU UMa-type dwarf nova with
a very distinct positive $P_{\rm dot}$ and with distinct stages B and C
[the best observed such an object dates back to XZ Eri \citep{uem04xzeri},
with poorer statistics].  As in HT Cas (subsection \ref{obj:htcas}),
a strong beat phenomenon was observed (figure \ref{fig:j0431shprof}).

\begin{table}
\caption{Eclipse Minima of OT J0431 (2011).}\label{tab:j0431ecl}
\begin{center}
\begin{tabular}{cccc}
\hline
$E$ & Minimum\commenta & error & $O-C$\commentb \\
\hline
0 & 55575.27191 & 0.00012 & -0.00020 \\
1 & 55575.33844 & 0.00018 & 0.00028 \\
2 & 55575.40389 & 0.00007 & -0.00032 \\
15 & 55576.26287 & 0.00007 & 0.00002 \\
16 & 55576.32920 & 0.00030 & 0.00030 \\
23 & 55576.79243 & 0.00005 & 0.00118 \\
37 & 55577.71577 & 0.00006 & -0.00017 \\
38 & 55577.78178 & 0.00004 & -0.00021 \\
39 & 55577.84786 & 0.00006 & -0.00018 \\
52 & 55578.70661 & 0.00005 & -0.00007 \\
53 & 55578.77274 & 0.00005 & 0.00001 \\
58 & 55579.10194 & 0.00010 & -0.00103 \\
67 & 55579.69737 & 0.00005 & -0.00005 \\
68 & 55579.76367 & 0.00009 & 0.00020 \\
69 & 55579.82847 & 0.00006 & -0.00106 \\
82 & 55580.68795 & 0.00005 & -0.00022 \\
83 & 55580.75397 & 0.00005 & -0.00025 \\
84 & 55580.81997 & 0.00006 & -0.00029 \\
113 & 55582.73495 & 0.00009 & -0.00075 \\
114 & 55582.80158 & 0.00006 & -0.00017 \\
128 & 55583.72630 & 0.00005 & -0.00014 \\
129 & 55583.79240 & 0.00006 & -0.00009 \\
143 & 55584.71737 & 0.00004 & 0.00019 \\
144 & 55584.78362 & 0.00006 & 0.00038 \\
158 & 55585.70787 & 0.00006 & -0.00005 \\
159 & 55585.77401 & 0.00008 & 0.00004 \\
173 & 55586.69850 & 0.00003 & -0.00017 \\
174 & 55586.76447 & 0.00003 & -0.00024 \\
187 & 55587.62349 & 0.00004 & 0.00013 \\
188 & 55587.68957 & 0.00006 & 0.00016 \\
189 & 55587.75559 & 0.00006 & 0.00013 \\
203 & 55588.68015 & 0.00004 & -0.00000 \\
204 & 55588.74617 & 0.00002 & -0.00003 \\
205 & 55588.81211 & 0.00003 & -0.00014 \\
\hline
  \multicolumn{4}{l}{\commenta BJD$-$2400000.} \\
  \multicolumn{4}{l}{\commentb Against equation \ref{equ:j0431ecl}.} \\
\end{tabular}
\end{center}
\end{table}

\begin{table}
\caption{Superhump maxima of OT J0431 (2011).}\label{tab:j0431oc2011}
\begin{center}
\begin{tabular}{cccccc}
\hline
$E$ & max\commenta & error & $O-C$\commentb & phase\commentc & $N$\commentd \\
\hline
0 & 55575.2417 & 0.0006 & 0.0021 & 0.54 & 51 \\
1 & 55575.3078 & 0.0004 & 0.0007 & 0.54 & 66 \\
2 & 55575.3750 & 0.0004 & 0.0004 & 0.56 & 59 \\
3 & 55575.4422 & 0.0005 & $-$0.0001 & 0.57 & 62 \\
15 & 55576.2565 & 0.0011 & 0.0039 & 0.90 & 30 \\
16 & 55576.3209 & 0.0009 & 0.0007 & 0.88 & 54 \\
23 & 55576.7924 & 0.0011 & $-$0.0005 & 0.02 & 50 \\
37 & 55577.7350 & 0.0007 & $-$0.0034 & 0.29 & 56 \\
38 & 55577.8011 & 0.0006 & $-$0.0048 & 0.29 & 56 \\
40 & 55577.9387 & 0.0012 & $-$0.0023 & 0.37 & 102 \\
45 & 55578.2738 & 0.0006 & $-$0.0048 & 0.45 & 45 \\
46 & 55578.3437 & 0.0006 & $-$0.0025 & 0.50 & 42 \\
47 & 55578.4088 & 0.0006 & $-$0.0050 & 0.49 & 48 \\
52 & 55578.7490 & 0.0008 & $-$0.0024 & 0.64 & 59 \\
56 & 55579.0109 & 0.0028 & $-$0.0107 & 0.61 & 26 \\
57 & 55579.0860 & 0.0029 & $-$0.0031 & 0.74 & 77 \\
66 & 55579.6945 & 0.0020 & $-$0.0023 & 0.96 & 32 \\
67 & 55579.7617 & 0.0019 & $-$0.0027 & 0.97 & 55 \\
81 & 55580.7120 & 0.0006 & 0.0021 & 0.36 & 59 \\
82 & 55580.7798 & 0.0006 & 0.0024 & 0.39 & 62 \\
83 & 55580.8459 & 0.0011 & 0.0010 & 0.39 & 56 \\
110 & 55582.6780 & 0.0024 & 0.0097 & 0.13 & 41 \\
111 & 55582.7444 & 0.0014 & 0.0085 & 0.13 & 55 \\
112 & 55582.8100 & 0.0012 & 0.0066 & 0.13 & 55 \\
125 & 55583.6898 & 0.0011 & 0.0085 & 0.45 & 49 \\
126 & 55583.7572 & 0.0005 & 0.0083 & 0.47 & 68 \\
127 & 55583.8255 & 0.0009 & 0.0091 & 0.50 & 66 \\
140 & 55584.7010 & 0.0005 & 0.0066 & 0.76 & 55 \\
141 & 55584.7681 & 0.0007 & 0.0062 & 0.77 & 56 \\
155 & 55585.7060 & 0.0008 & $-$0.0014 & 0.97 & 57 \\
156 & 55585.7705 & 0.0010 & $-$0.0045 & 0.95 & 56 \\
170 & 55586.7186 & 0.0009 & $-$0.0018 & 0.30 & 56 \\
171 & 55586.7854 & 0.0009 & $-$0.0025 & 0.31 & 55 \\
183 & 55587.6027 & 0.0026 & 0.0044 & 0.69 & 64 \\
184 & 55587.6666 & 0.0012 & 0.0007 & 0.65 & 66 \\
185 & 55587.7316 & 0.0010 & $-$0.0018 & 0.64 & 70 \\
199 & 55588.6665 & 0.0028 & $-$0.0124 & 0.79 & 55 \\
200 & 55588.7333 & 0.0044 & $-$0.0131 & 0.80 & 56 \\
\hline
  \multicolumn{6}{l}{\commenta BJD$-$2400000.} \\
  \multicolumn{6}{l}{\commentb Against max $= 2455575.2396 + 0.067534 E$.} \\
  \multicolumn{6}{l}{\commentc Orbital phase.} \\
  \multicolumn{6}{l}{\commentd Number of points used to determine the maximum.} \\
\end{tabular}
\end{center}
\end{table}

\subsection{OT J044216.0$-$002334}\label{obj:j0442}

\begin{figure}
  \begin{center}
    \FigureFile(88mm,110mm){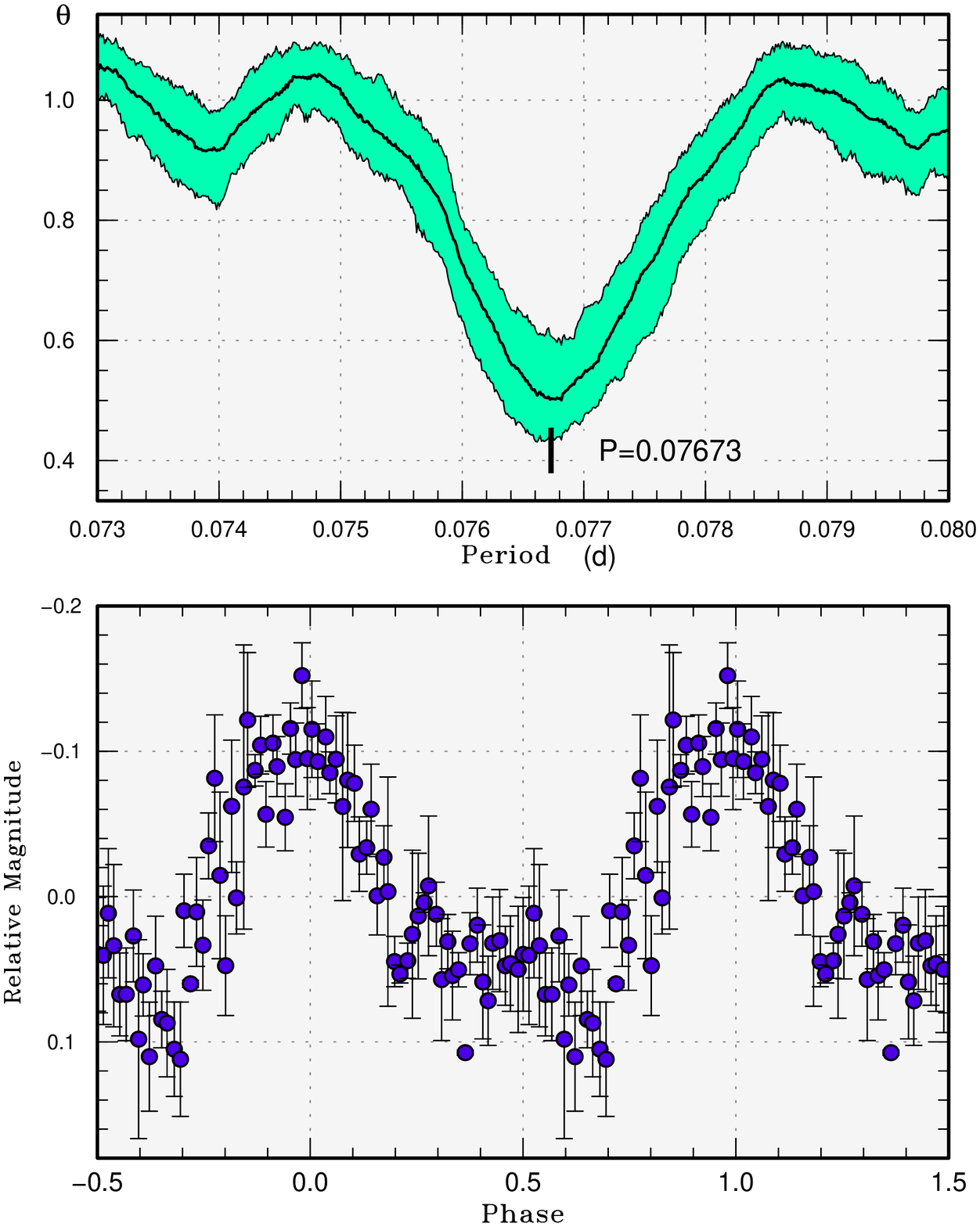}
  \end{center}
  \caption{Superhumps in OT J0442 (2011). (Upper): PDM analysis
     (Lower): Phase-averaged profile.}
  \label{fig:j0442shpdm}
\end{figure}

   This object (= CSS071115:044216$-$002334), hereafter OT J0442) was
discovered by the CRTS in 2007 January.  Three outburst have been known,
and the 2011 January one was the brightest.  Superhumps were subsequently
detected (vsnet-alert 12784, 12785, 12786; figure \ref{fig:j0442shpdm}).
Since the observations were obtained at high airmasses,
we corrected observations by using a second-order atmospheric extinction.
The times of superhump maxima are listed in table \ref{tab:j0442oc2011}.

\begin{table}
\caption{Superhump maxima of OT J0442 (2011).}\label{tab:j0442oc2011}
\begin{center}
\begin{tabular}{ccccc}
\hline
$E$ & max\commenta & error & $O-C$\commentb & $N$\commentc \\
\hline
0 & 55596.3030 & 0.0030 & $-$0.0009 & 23 \\
1 & 55596.3821 & 0.0011 & 0.0015 & 41 \\
13 & 55597.3014 & 0.0023 & $-$0.0003 & 24 \\
14 & 55597.3777 & 0.0013 & $-$0.0008 & 40 \\
26 & 55598.2999 & 0.0017 & 0.0002 & 33 \\
27 & 55598.3767 & 0.0019 & 0.0003 & 34 \\
\hline
  \multicolumn{5}{l}{\commenta BJD$-$2400000.} \\
  \multicolumn{5}{l}{\commentb Against max $= 2455596.3039 + 0.076761 E$.} \\
  \multicolumn{5}{l}{\commentc Number of points used to determine the maximum.} \\
\end{tabular}
\end{center}
\end{table}

\subsection{OT J064804.5$+$414702}\label{obj:j0648}

\begin{figure}
  \begin{center}
    \FigureFile(88mm,110mm){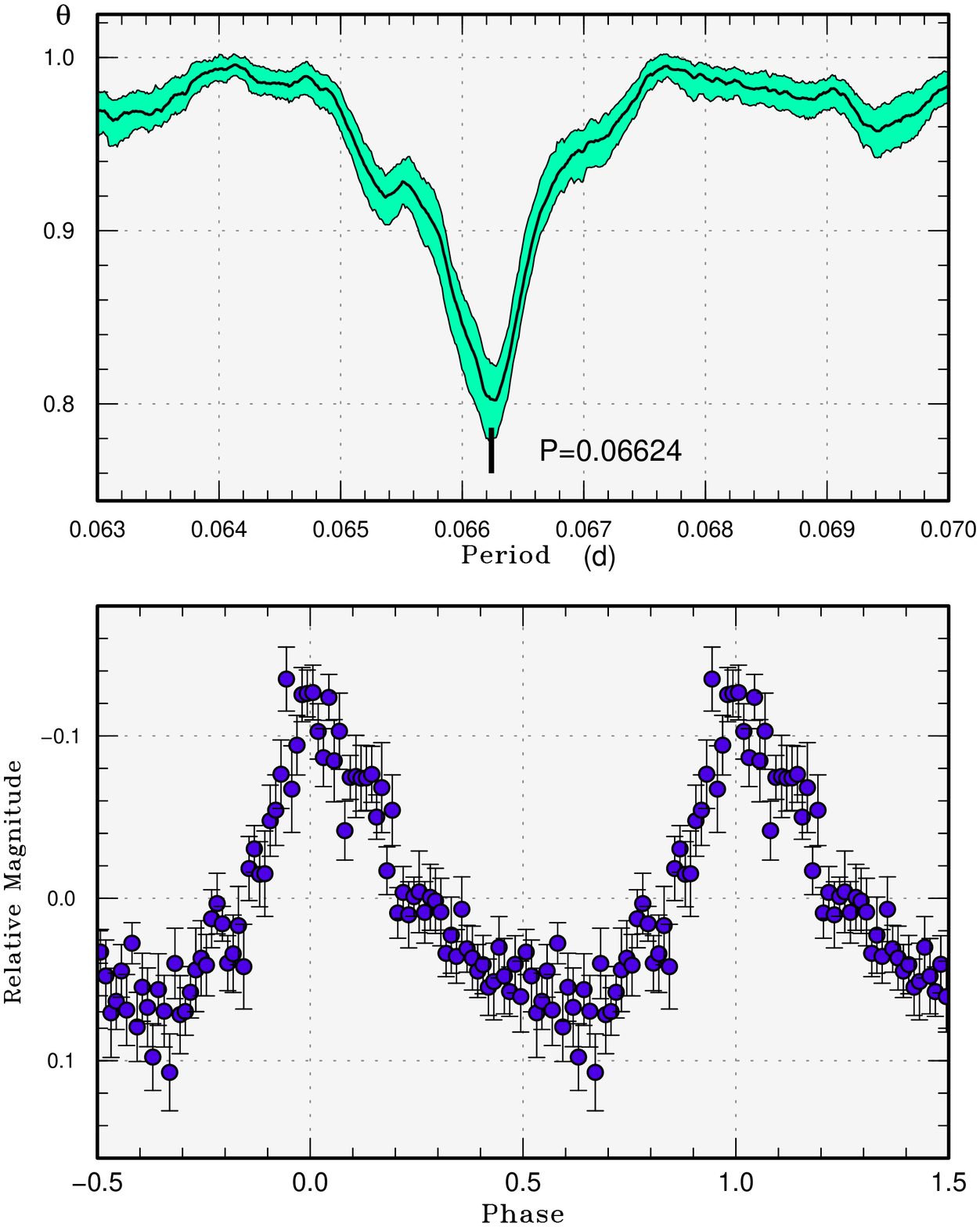}
  \end{center}
  \caption{Superhumps in OT J0648 (2011). (Upper): PDM analysis.
     (Lower): Phase-averaged profile.}
  \label{fig:j0648shpdm}
\end{figure}

   This object (= CSS091026:064805$+$414702, hereafter OT J0648) was
discovered in outburst by the CRTS in 2009 October.  During its
2011 January outburst, superhumps were detected (vsnet-alert 12735, 12737;
figure \ref{fig:j0648shpdm}).
The times of superhump maxima are listed in table \ref{tab:j0648oc2011}.
The outburst was apparently already in stage B, and a transition
to stage C superhumps was recorded five days later.

\begin{table}
\caption{Superhump maxima of OT J0648 (2011).}\label{tab:j0648oc2011}
\begin{center}
\begin{tabular}{ccccc}
\hline
$E$ & max\commenta & error & $O-C$\commentb & $N$\commentc \\
\hline
0 & 55590.5903 & 0.0006 & 0.0011 & 67 \\
1 & 55590.6583 & 0.0002 & 0.0027 & 126 \\
2 & 55590.7236 & 0.0003 & 0.0017 & 126 \\
3 & 55590.7902 & 0.0003 & 0.0020 & 126 \\
4 & 55590.8552 & 0.0004 & 0.0007 & 126 \\
5 & 55590.9218 & 0.0004 & 0.0011 & 125 \\
10 & 55591.2531 & 0.0011 & 0.0008 & 133 \\
12 & 55591.3849 & 0.0008 & 0.0001 & 76 \\
13 & 55591.4514 & 0.0007 & 0.0003 & 88 \\
14 & 55591.5168 & 0.0009 & $-$0.0006 & 68 \\
23 & 55592.1114 & 0.0013 & $-$0.0026 & 105 \\
24 & 55592.1804 & 0.0013 & 0.0001 & 129 \\
25 & 55592.2461 & 0.0018 & $-$0.0005 & 119 \\
26 & 55592.3094 & 0.0021 & $-$0.0035 & 82 \\
27 & 55592.3749 & 0.0013 & $-$0.0043 & 60 \\
28 & 55592.4402 & 0.0023 & $-$0.0053 & 66 \\
29 & 55592.5112 & 0.0013 & $-$0.0006 & 58 \\
42 & 55593.3715 & 0.0022 & $-$0.0021 & 63 \\
43 & 55593.4412 & 0.0017 & 0.0013 & 23 \\
44 & 55593.5072 & 0.0023 & 0.0010 & 68 \\
69 & 55595.1644 & 0.0056 & 0.0008 & 134 \\
70 & 55595.2399 & 0.0102 & 0.0100 & 81 \\
83 & 55596.0875 & 0.0028 & $-$0.0042 & 122 \\
\hline
  \multicolumn{5}{l}{\commenta BJD$-$2400000.} \\
  \multicolumn{5}{l}{\commentb Against max $= 2455590.5893 + 0.066295 E$.} \\
  \multicolumn{5}{l}{\commentc Number of points used to determine the maximum.} \\
\end{tabular}
\end{center}
\end{table}

\subsection{OT J075414.5$+$313216}\label{obj:j0754}

\begin{figure}
  \begin{center}
    \FigureFile(88mm,110mm){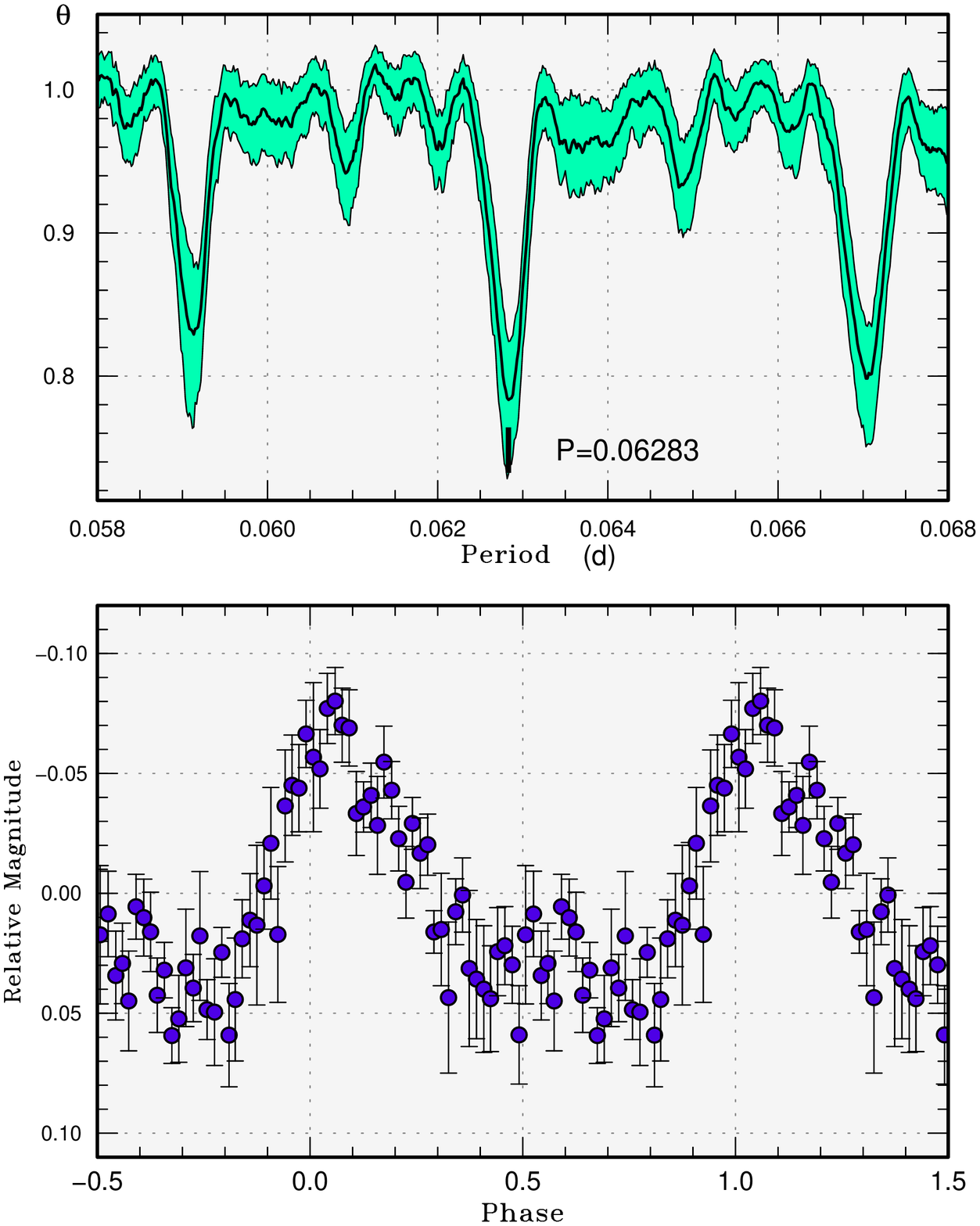}
  \end{center}
  \caption{Superhumps in OT J0754 (2011). (Upper): PDM analysis.
     (Lower): Phase-averaged profile.}
  \label{fig:j0754shpdm}
\end{figure}

   This object (= CSS110414:075414$+$313216, hereafter OT J0754) was
discovered by the CRTS on 2011 April 14.  Two earlier outbursts
(2007 February and 2008 November) were also detected by the CRTS,
and the 2008 one reached a maximum of 14.2 mag.
Subsequent observations detected superhumps (vsnet-alert 13166,
13186, 13202).  Due to the limited nightly baselines, the possibility
of one-day alias is not completely excluded (figure \ref{fig:j0754shpdm}).
The times of superhump maxima are listed in table \ref{tab:j0754oc2011}.
Although both the overall light curve and, $O-C$ variation and amplitudes
of superhumps (cf. subsection \ref{sec:humpamp}) showed a typical course
of a stage B--C transition (vsnet-alert 13219),
the baseline was not sufficient to determine $P_{\rm dot}$ for stage B.

\begin{table}
\caption{Superhump maxima of OT J0754 (2011).}\label{tab:j0754oc2011}
\begin{center}
\begin{tabular}{ccccc}
\hline
$E$ & max\commenta & error & $O-C$\commentb & $N$\commentc \\
\hline
0 & 55666.3975 & 0.0012 & $-$0.0050 & 65 \\
1 & 55666.4626 & 0.0015 & $-$0.0027 & 55 \\
16 & 55667.4058 & 0.0009 & $-$0.0019 & 64 \\
17 & 55667.4713 & 0.0016 & 0.0008 & 43 \\
31 & 55668.3503 & 0.0011 & 0.0002 & 47 \\
32 & 55668.4154 & 0.0009 & 0.0025 & 65 \\
33 & 55668.4844 & 0.0023 & 0.0086 & 38 \\
63 & 55670.3587 & 0.0015 & $-$0.0018 & 33 \\
64 & 55670.4252 & 0.0018 & 0.0019 & 34 \\
65 & 55670.4858 & 0.0022 & $-$0.0004 & 29 \\
79 & 55671.3661 & 0.0015 & 0.0004 & 31 \\
80 & 55671.4291 & 0.0026 & 0.0005 & 34 \\
81 & 55671.4993 & 0.0038 & 0.0079 & 18 \\
95 & 55672.3662 & 0.0050 & $-$0.0048 & 34 \\
96 & 55672.4276 & 0.0045 & $-$0.0061 & 34 \\
\hline
  \multicolumn{5}{l}{\commenta BJD$-$2400000.} \\
  \multicolumn{5}{l}{\commentb Against max $= 2455666.4025 + 0.062826 E$.} \\
  \multicolumn{5}{l}{\commentc Number of points used to determine the maximum.} \\
\end{tabular}
\end{center}
\end{table}

\subsection{OT J102616.0$+$192045}\label{obj:j102616}

\begin{figure}
  \begin{center}
    \FigureFile(88mm,110mm){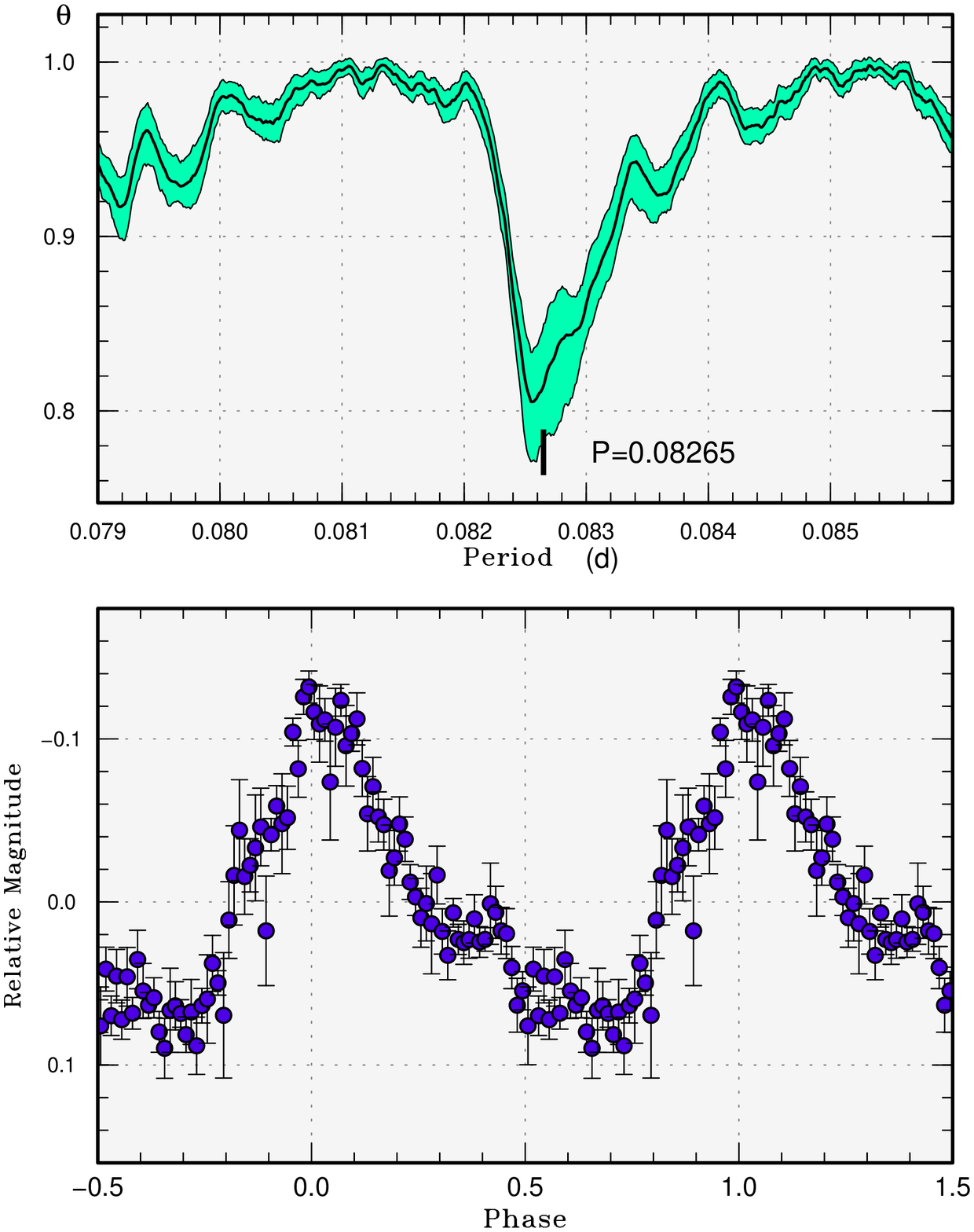}
  \end{center}
  \caption{Superhumps in OT J102616 (2010). (Upper): PDM analysis.
     The indicated period was determined from a PDM analysis of the
     entire data, and is slightly different from the bootstrap median.
     This difference probably arises from the relatively large variation
     in the period.
     (Lower): Phase-averaged profile.}
  \label{fig:j102616shpdm}
\end{figure}

   This object (= CSS101130:102616$+$192045, hereafter OT J102616) was
discovered by the CRTS on 2010 November 30.  Two earlier outbursts
(2006 April and 2007 April) were also detected by the CRTS.
Subsequent observations immediately clarified the presence of
superhumps (vsnet-alert 12447, 12452, 12459; figure \ref{fig:j102616shpdm}).
The large amplitudes ($\sim$0.3 mag) of superhumps at the start of the
observation suggests that the observation succeeded in recording the
relatively early stage of the superoutburst.

The times of superhump maxima are listed in table \ref{tab:j102616oc2010}.
There were a large systematic decrease in the superhump period,
amounting to a global $P_{\rm dot} = -8.6(1.5) \times 10^{-5}$
and a mean global $P_{\rm SH} = 0.08267(4)$ d from timing analysis.
There appeared to have been a discontinuous decrease of the period
at $E=36$ and early epochs ($E \le 1$) may have resulted from
stage A superhumps.  Although we list values in table \ref{tab:perlist}
based on this interpretation, there remains a possibility that the
entire observation should be interpreted as stage B, as discussed
in \citet{Pdot} (subsection 4.10).  It is noteworthy that a relatively
large negative $P_{\rm dot}$ was recorded in a system with a long-$P_{\rm orb}$
and with relatively infrequent outbursts.  This object should require
further attention.

\begin{table}
\caption{Superhump maxima of OT J102616 (2010).}\label{tab:j102616oc2010}
\begin{center}
\begin{tabular}{ccccc}
\hline
$E$ & max\commenta & error & $O-C$\commentb & $N$\commentc \\
\hline
0 & 55533.4881 & 0.0006 & $-$0.0094 & 131 \\
1 & 55533.5744 & 0.0001 & $-$0.0058 & 245 \\
13 & 55534.5738 & 0.0004 & 0.0016 & 74 \\
22 & 55535.3196 & 0.0003 & 0.0034 & 321 \\
25 & 55535.5677 & 0.0002 & 0.0035 & 291 \\
36 & 55536.4789 & 0.0003 & 0.0053 & 127 \\
47 & 55537.3859 & 0.0005 & 0.0029 & 92 \\
48 & 55537.4685 & 0.0003 & 0.0029 & 131 \\
61 & 55538.5434 & 0.0004 & 0.0030 & 109 \\
62 & 55538.6257 & 0.0006 & 0.0026 & 73 \\
70 & 55539.2847 & 0.0008 & 0.0003 & 105 \\
82 & 55540.2738 & 0.0006 & $-$0.0026 & 100 \\
95 & 55541.3495 & 0.0023 & $-$0.0017 & 100 \\
107 & 55542.3373 & 0.0012 & $-$0.0059 & 185 \\
\hline
  \multicolumn{5}{l}{\commenta BJD$-$2400000.} \\
  \multicolumn{5}{l}{\commentb Against max $= 2455533.4975 + 0.082670 E$.} \\
  \multicolumn{5}{l}{\commentc Number of points used to determine the maximum.} \\
\end{tabular}
\end{center}
\end{table}

\subsection{OT J102705.8$-$434341}\label{obj:j1027}

\begin{figure}
  \begin{center}
    \FigureFile(88mm,110mm){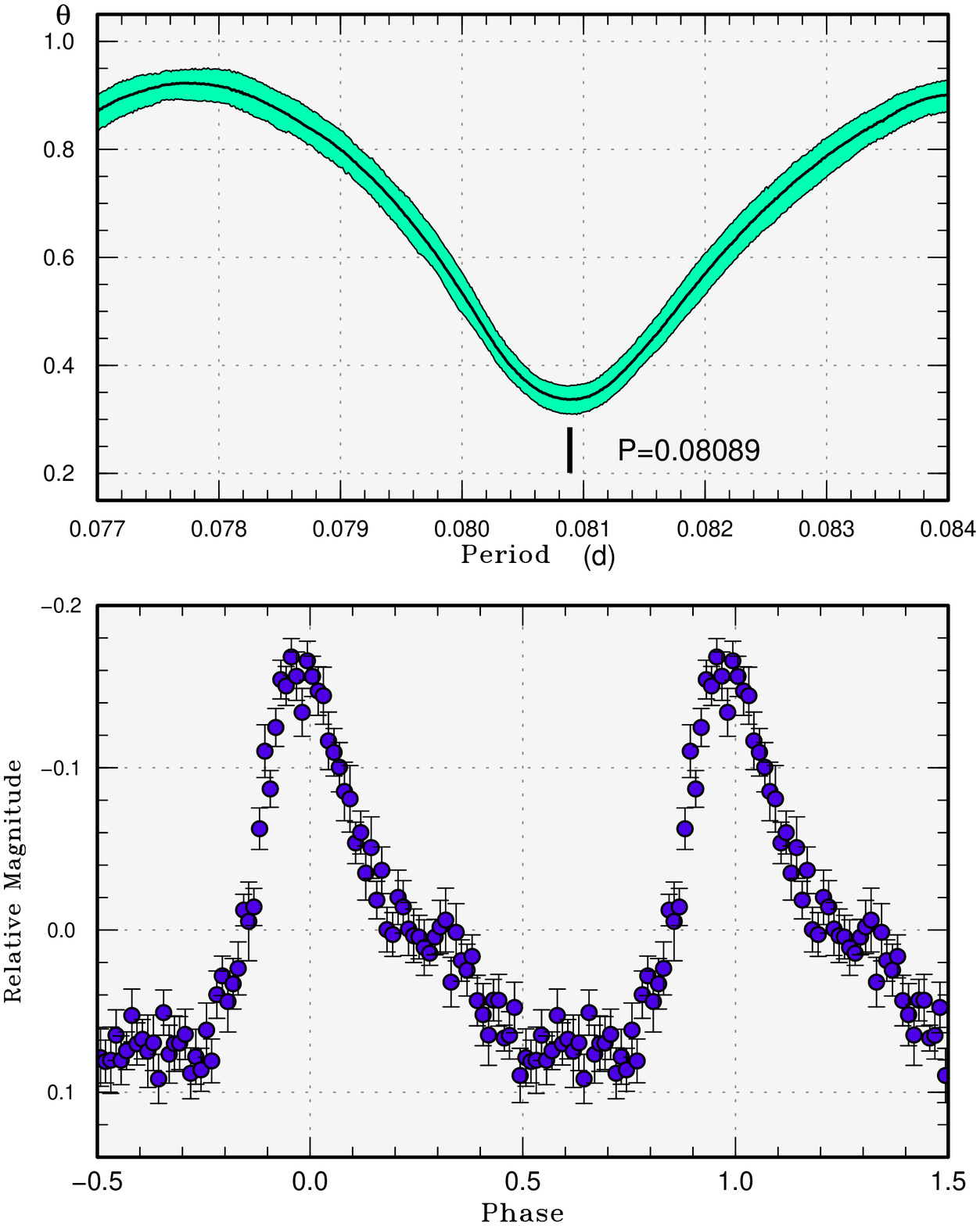}
  \end{center}
  \caption{Superhumps in OT J1027 (2011). (Upper): PDM analysis.
     (Lower): Phase-averaged profile.}
  \label{fig:j1027shpdm}
\end{figure}

   This object (= SSS110314:102706$-$434341, hereafter OT J1027) was
discovered by the CRTS SSS on 2011 March 14.
There were four past outbursts in CRTS SSS data, and has an X-ray
counterpart 1RXP J102706$-$4343.5.
Subsequent observations detected superhumps (vsnet-alert 12997, 13009;
figure \ref{fig:j1027shpdm}).
The times of superhump maxima are listed in table \ref{tab:j1027oc2011}.
The period given in table \ref{tab:perlist} was determined by the
PDM method.

\begin{table}
\caption{Superhump maxima of OT J1027 (2011).}\label{tab:j1027oc2011}
\begin{center}
\begin{tabular}{ccccc}
\hline
$E$ & max\commenta & error & $O-C$\commentb & $N$\commentc \\
\hline
0 & 55637.0723 & 0.0007 & $-$0.0001 & 109 \\
1 & 55637.1532 & 0.0006 & $-$0.0001 & 109 \\
2 & 55637.2351 & 0.0006 & 0.0009 & 109 \\
3 & 55637.3142 & 0.0007 & $-$0.0009 & 110 \\
12 & 55638.0444 & 0.0008 & 0.0013 & 79 \\
13 & 55638.1242 & 0.0009 & 0.0003 & 84 \\
14 & 55638.2024 & 0.0010 & $-$0.0024 & 84 \\
15 & 55638.2867 & 0.0011 & 0.0010 & 85 \\
\hline
  \multicolumn{5}{l}{\commenta BJD$-$2400000.} \\
  \multicolumn{5}{l}{\commentb Against max $= 2455637.0724 + 0.080885 E$.} \\
  \multicolumn{5}{l}{\commentc Number of points used to determine the maximum.} \\
\end{tabular}
\end{center}
\end{table}

\subsection{OT J120052.9$-$152620}\label{obj:j1200}

\begin{figure}
  \begin{center}
    \FigureFile(88mm,110mm){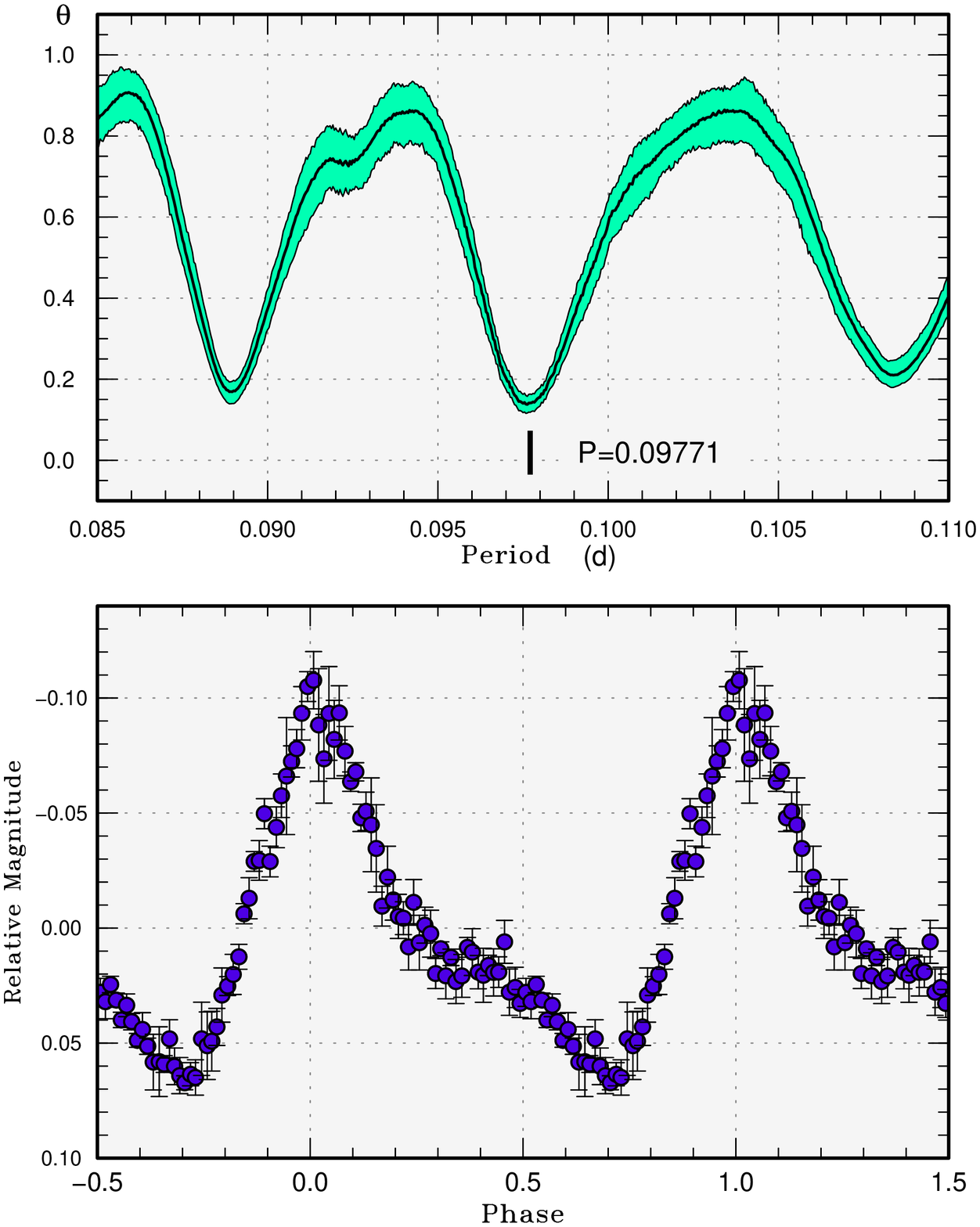}
  \end{center}
  \caption{Superhumps in OT J1200 (2011). (Upper): PDM analysis.
     (Lower): Phase-averaged profile.}
  \label{fig:j1200shpdm}
\end{figure}

   This object (= CSS110205:120053$-$152620, hereafter OT J1200) was
discovered by the CRTS on 2011 February 5.  Both CRTS and ASAS-3
\citep{ASAS3} recorded numerous past outbursts, and the brightest
outburst reaching $V=13.39$ on 2001 February 14 recorded by ASAS-3
was indicative of a superoutburst (vsnet-alert 12792).  Upon this information,
follow-up observations indeed clarified the presence of superhumps
(vsnet-alert 12803, 12815; figure \ref{fig:j1200shpdm}).
The times of superhump maxima are listed in table \ref{tab:j1200oc2011}.
Due to the short baseline of observations, we adopted the most likely
one-day alias.  Other aliases still remain viable.
We adopted the period determined by the PDM method in table \ref{tab:perlist}.

\begin{table}
\caption{Superhump maxima of OT J1200 (2011).}\label{tab:j1200oc2011}
\begin{center}
\begin{tabular}{ccccc}
\hline
$E$ & max\commenta & error & $O-C$\commentb & $N$\commentc \\
\hline
0 & 55599.6371 & 0.0005 & $-$0.0004 & 87 \\
9 & 55600.5242 & 0.0011 & 0.0041 & 62 \\
10 & 55600.6145 & 0.0004 & $-$0.0037 & 97 \\
\hline
  \multicolumn{5}{l}{\commenta BJD$-$2400000.} \\
  \multicolumn{5}{l}{\commentb Against max $= 2455599.6375 + 0.098066 E$.} \\
  \multicolumn{5}{l}{\commentc Number of points used to determine the maximum.} \\
\end{tabular}
\end{center}
\end{table}

\subsection{OT J132900.9$-$365859}\label{obj:j1329}

\begin{figure}
  \begin{center}
    \FigureFile(88mm,110mm){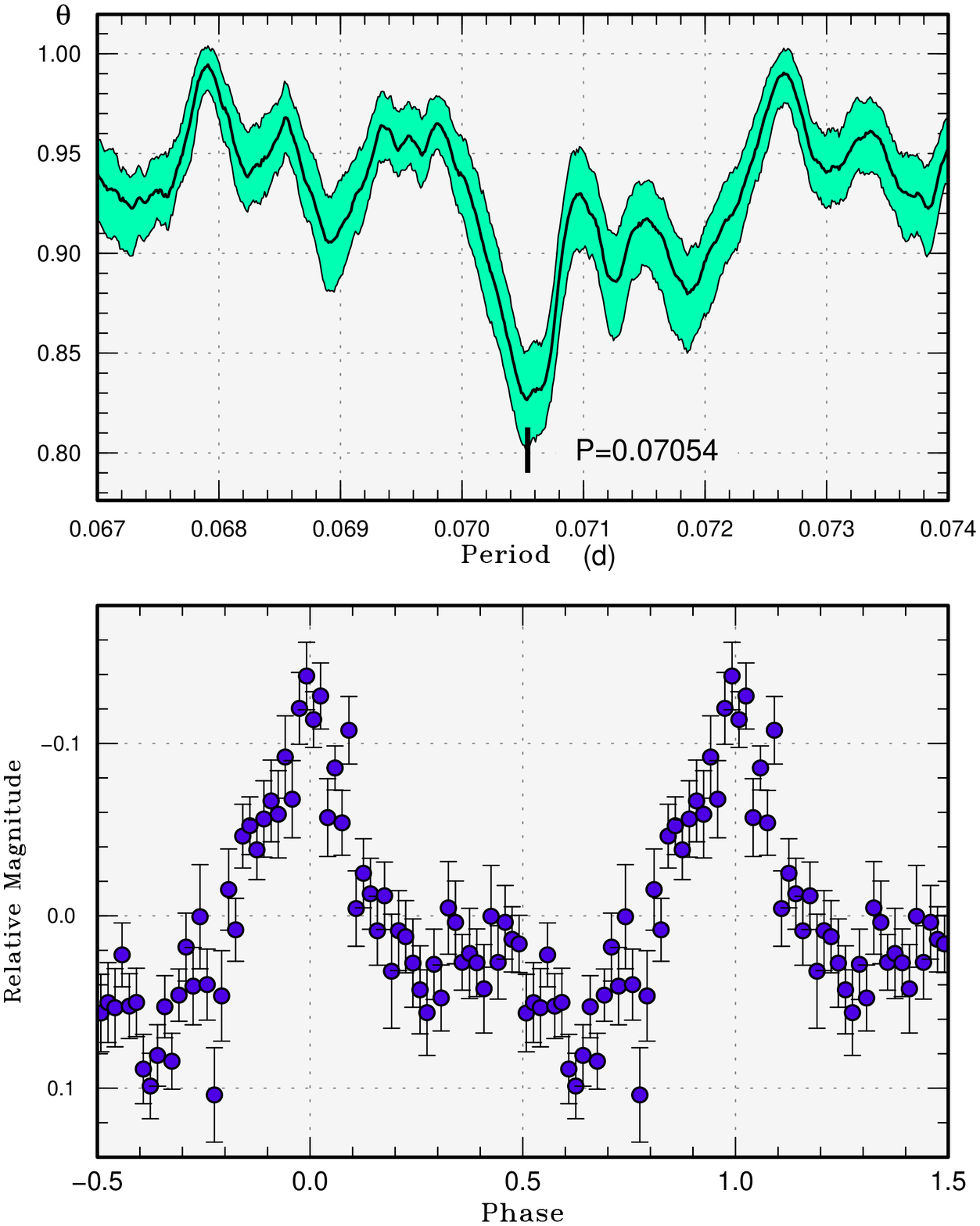}
  \end{center}
  \caption{Superhumps in OT J1329 (2011). (Upper): PDM analysis.
     (Lower): Phase-averaged profile.}
  \label{fig:j1329shpdm}
\end{figure}

   This object (= SSS110403:132901$-$365859), hereafter OT J1329) was
discovered by the CRTS SSS on 2011 April 3.
Subsequent observations detected apparent superhumps (vsnet-alert 13105;
figure \ref{fig:j1329shpdm}).
The times of superhump maxima are listed in table \ref{tab:j1329oc2011}.
There appears to have been a stage transition between $E=16$ and $E=59$,
and we listed a period from $E \le 16$ for the period of stage B superhumps
in table \ref{tab:perlist}.
Since there was no earlier record of outbursts in CRTS SSS, the object
seems to have a relatively low outburst frequency.

\begin{table}
\caption{Superhump maxima of OT J1329 (2011).}\label{tab:j1329oc2011}
\begin{center}
\begin{tabular}{ccccc}
\hline
$E$ & max\commenta & error & $O-C$\commentb & $N$\commentc \\
\hline
0 & 55656.4074 & 0.0011 & $-$0.0050 & 118 \\
1 & 55656.4796 & 0.0014 & $-$0.0033 & 162 \\
13 & 55657.3296 & 0.0011 & 0.0007 & 162 \\
14 & 55657.4017 & 0.0010 & 0.0023 & 161 \\
15 & 55657.4743 & 0.0009 & 0.0045 & 162 \\
16 & 55657.5439 & 0.0020 & 0.0035 & 155 \\
59 & 55660.5723 & 0.0020 & 0.0007 & 162 \\
60 & 55660.6386 & 0.0020 & $-$0.0034 & 159 \\
\hline
  \multicolumn{5}{l}{\commenta BJD$-$2400000.} \\
  \multicolumn{5}{l}{\commentb Against max $= 2455656.4125 + 0.070493 E$.} \\
  \multicolumn{5}{l}{\commentc Number of points used to determine the maximum.} \\
\end{tabular}
\end{center}
\end{table}

\subsection{OT J154544.9$+$442830}\label{obj:j1545}

\begin{figure}
  \begin{center}
    \FigureFile(88mm,110mm){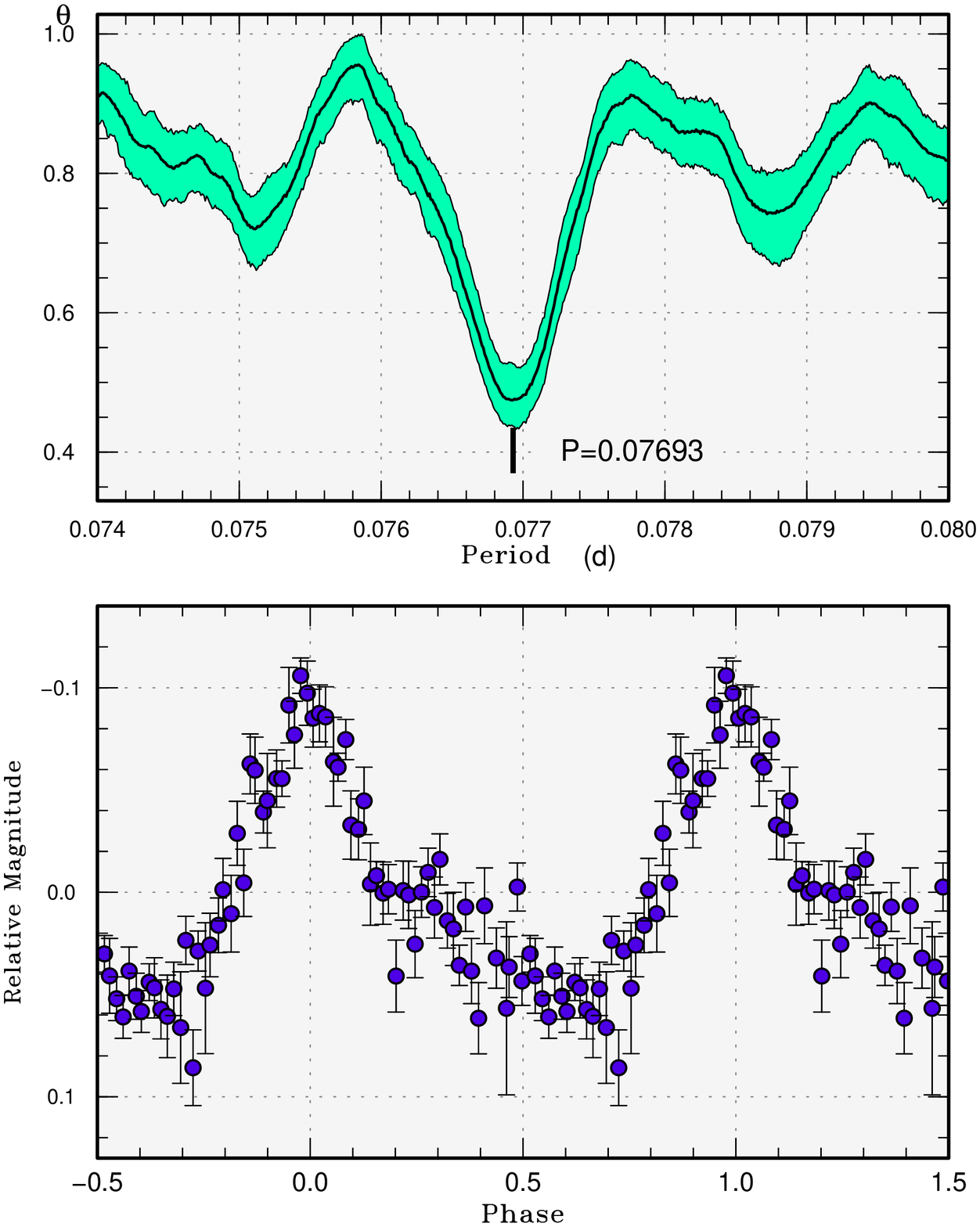}
  \end{center}
  \caption{Superhumps in OT J2234 (2011). (Upper): PDM analysis.
     (Lower): Phase-averaged profile.}
  \label{fig:j1545shpdm}
\end{figure}

   This object (= CSS110428:154545+442830), hereafter OT J1545) was
discovered by the CRTS on 2011 April 28.  Subsequent observations
successfully detected superhumps
(vsnet-alert 13275; figure \ref{fig:j1545shpdm}).
The times of superhump maxima are listed in table \ref{tab:j1545oc2011}.
The object started rapid fading only three days after the initial
observation, and the outburst was observed only during its final phase.
We identified the superhumps as being stage C superhumps.

\begin{table}
\caption{Superhump maxima of OT J1545 (2011).}\label{tab:j1545oc2011}
\begin{center}
\begin{tabular}{ccccc}
\hline
$E$ & max\commenta & error & $O-C$\commentb & $N$\commentc \\
\hline
0 & 55681.1443 & 0.0050 & $-$0.0045 & 83 \\
1 & 55681.2289 & 0.0007 & 0.0030 & 157 \\
17 & 55682.4600 & 0.0037 & 0.0020 & 18 \\
18 & 55682.5354 & 0.0021 & 0.0004 & 25 \\
29 & 55683.3812 & 0.0020 & $-$0.0008 & 22 \\
30 & 55683.4587 & 0.0019 & $-$0.0003 & 21 \\
31 & 55683.5344 & 0.0017 & $-$0.0017 & 22 \\
32 & 55683.6176 & 0.0021 & 0.0045 & 14 \\
42 & 55684.3821 & 0.0011 & $-$0.0011 & 25 \\
43 & 55684.4567 & 0.0018 & $-$0.0035 & 23 \\
44 & 55684.5342 & 0.0018 & $-$0.0029 & 25 \\
45 & 55684.6191 & 0.0040 & 0.0049 & 15 \\
\hline
  \multicolumn{5}{l}{\commenta BJD$-$2400000.} \\
  \multicolumn{5}{l}{\commentb Against max $= 2455681.1488 + 0.077008 E$.} \\
  \multicolumn{5}{l}{\commentc Number of points used to determine the maximum.} \\
\end{tabular}
\end{center}
\end{table}

\subsection{OT J223418.5$-$035530}\label{obj:j2234}

\begin{figure}
  \begin{center}
    \FigureFile(88mm,110mm){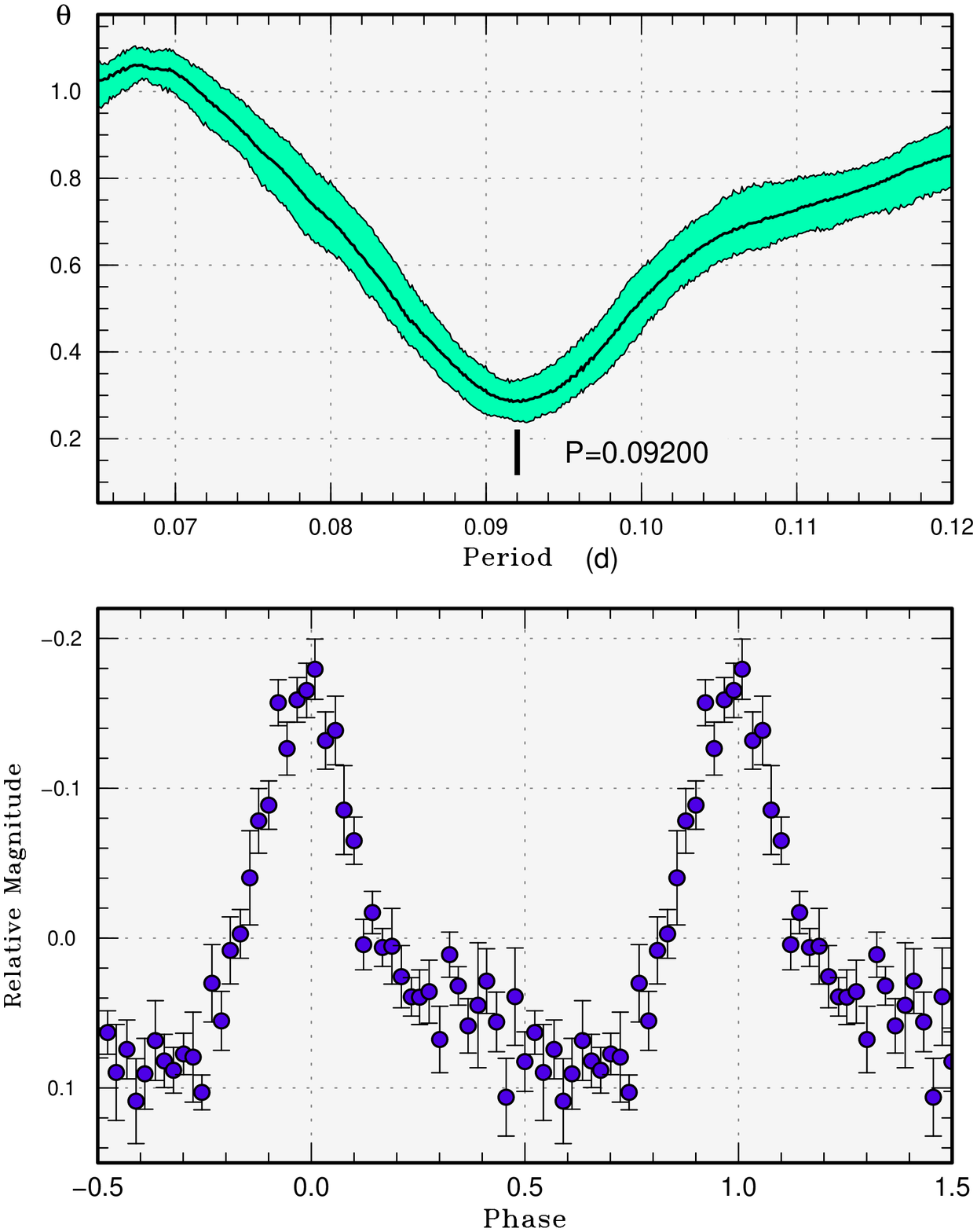}
  \end{center}
  \caption{Superhumps in OT J2234 (2010). (Upper): PDM analysis.
     (Lower): Phase-averaged profile.}
  \label{fig:j2234shpdm}
\end{figure}

   This object (= CSS090910:223418$-$035530), hereafter OT J2234) was
discovered by the CRTS on 2009 September 10.
We observed this object during two superoutbursts in 2009 and 2010.
Although the existence of superhumps were already apparent in 2009
(cf. vsnet-alert 11480), the short observations were unable to
select a unique period.  Using the 2010 observations, we have selected
a unique period, and the $O-C$'s referring to this period
are listed in tables \ref{tab:j2234oc2009} and \ref{tab:j2234oc2010}.
Since the amplitudes of superhumps were low ($\sim$ 0.03 mag) and
they grew to 0.18 mag six days later, the initial observations
were apparently performed during the growing stage of superhumps.
In table \ref{tab:perlist}, we listed averaged periods for the entire
runs since observations were too insufficient for determining stages
and period changes.  The superhumps recorded during the 2010
superoutburst is shown in figure \ref{fig:j2234shpdm}.

\begin{table}
\caption{Superhump maxima of OT J2234 (2009).}\label{tab:j2234oc2009}
\begin{center}
\begin{tabular}{ccccc}
\hline
$E$ & max\commenta & error & $O-C$\commentb & $N$\commentc \\
\hline
0 & 55084.9749 & 0.0184 & 0.0014 & 134 \\
1 & 55085.0638 & 0.0052 & $-$0.0018 & 194 \\
66 & 55091.0591 & 0.0012 & 0.0010 & 194 \\
109 & 55095.0217 & 0.0026 & $-$0.0006 & 186 \\
\hline
  \multicolumn{5}{l}{\commenta BJD$-$2400000.} \\
  \multicolumn{5}{l}{\commentb Against max $= 2455084.9734 + 0.092192 E$.} \\
  \multicolumn{5}{l}{\commentc Number of points used to determine the maximum.} \\
\end{tabular}
\end{center}
\end{table}

\begin{table}
\caption{Superhump maxima of OT J2234 (2010).}\label{tab:j2234oc2010}
\begin{center}
\begin{tabular}{ccccc}
\hline
$E$ & max\commenta & error & $O-C$\commentb & $N$\commentc \\
\hline
0 & 55482.0178 & 0.0008 & 0.0001 & 93 \\
1 & 55482.1094 & 0.0007 & $-$0.0002 & 95 \\
2 & 55482.2017 & 0.0011 & 0.0001 & 78 \\
\hline
  \multicolumn{5}{l}{\commenta BJD$-$2400000.} \\
  \multicolumn{5}{l}{\commentb Against max $= 2455482.0177 + 0.091914 E$.} \\
  \multicolumn{5}{l}{\commentc Number of points used to determine the maximum.} \\
\end{tabular}
\end{center}
\end{table}

\subsection{OT J230425.8$+$062546}\label{obj:j2304}

\begin{figure}
  \begin{center}
    \FigureFile(88mm,110mm){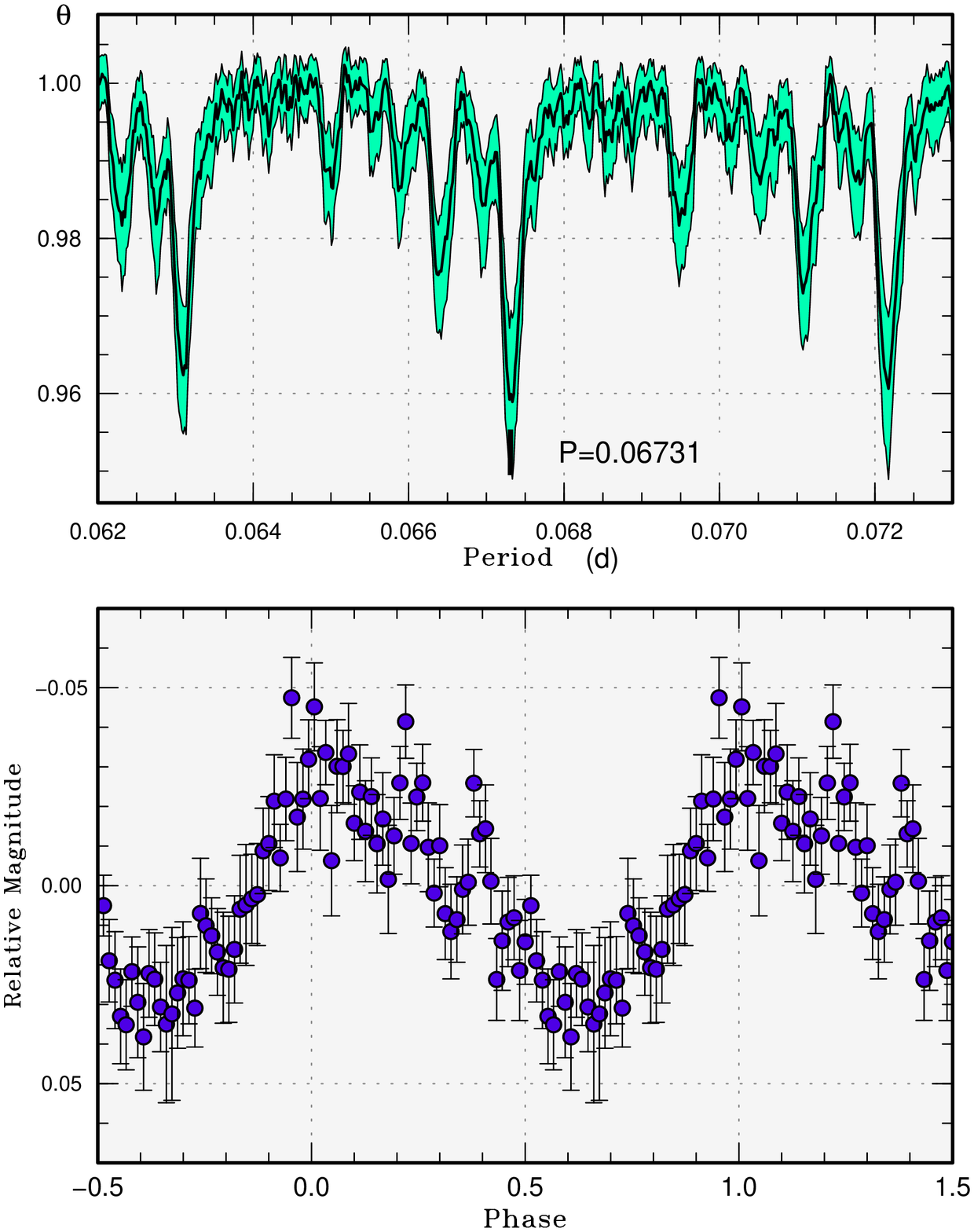}
  \end{center}
  \caption{Superhumps in OT J2304 (2011). (Upper): PDM analysis.
     (Lower): Phase-averaged profile.}
  \label{fig:j2304shpdm}
\end{figure}

   This transient was originally reported as a possible nova
\citep{nak11j2304cbet2616}\footnote{
  According to the information from the discoverer, the large difference
in magnitudes between the discovery and confirmatory observations
was probably caused by transformation of image formats in the digital
camera, and probably did not reflect true variation. 
}.  Soon after the announcement, the color of
the quiescent counterpart inferred a dwarf-nova outburst (vsnet-alert 12548).
Initial time-resolved photometry recorded relatively small variations
(A. Arai, vsnet-alert 12558).  Subsequent observations indicated the
presence of superhumps with amplitude of 0.06 mag
(A. Arai, vsnet-alert 12563).  This finding was confirmed by subsequent
observations (vsnet-alert 12564, 12578, 12633; figure \ref{fig:j2304shpdm}).

\begin{figure}
  \begin{center}
    \FigureFile(88mm,110mm){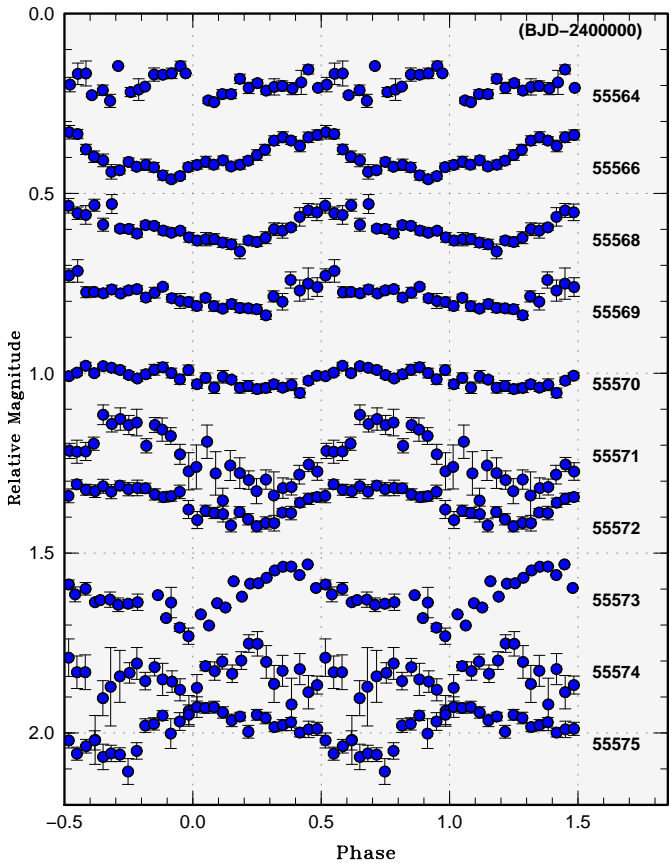}
  \end{center}
  \caption{Superhump profiles of OT J2304 (2011) for the first 10 d.
     The amplitudes of superhumps gradually grew until BJD 2455571.
     The period apparently shortened after this.  The profiles of
     superhumps were rather irregular and the amplitudes were generally
     small.  The figure was drawn against a period of 0.06705 d.}
  \label{fig:j2304prof}
\end{figure}

   Since the observations were obtained at high airmasses,
we corrected observations by using a second-order atmospheric extinction.
The resultant times of superhump maxima are listed in table
\ref{tab:j2304oc2011}.  Since the observing windows were limited to
the low evening sky, there remained an ambiguity in selecting the
alias (cf. vsnet-alert 12578).  The superhumps queerly
grew since the detection until $E \sim120$, when the period and amplitudes
suddenly decreased (figure \ref{fig:j2304prof}).
Since the fractional decrease (1.4 \%) is exceptionally
large for a stage B-C transition \citep{Pdot}, we leave the interpretation
of the later stage open though we tentatively assign the values in table
\ref{tab:perlist}.  Although the results may have been affected by
adverse observing conditions, the unusual development of superhumps
in this system should warrant further study during future outbursts.

\begin{table}
\caption{Superhump maxima of OT J2304 (2011).}\label{tab:j2304oc2011}
\begin{center}
\begin{tabular}{ccccc}
\hline
$E$ & max\commenta & error & $O-C$\commentb & $N$\commentc \\
\hline
0 & 55563.9791 & 0.0028 & $-$0.0354 & 108 \\
29 & 55565.9214 & 0.0011 & $-$0.0273 & 274 \\
30 & 55565.9956 & 0.0023 & $-$0.0198 & 64 \\
59 & 55567.9479 & 0.0017 & $-$0.0019 & 104 \\
60 & 55568.0077 & 0.0015 & $-$0.0087 & 70 \\
74 & 55568.9571 & 0.0016 & 0.0069 & 215 \\
88 & 55569.8972 & 0.0014 & 0.0132 & 347 \\
89 & 55569.9608 & 0.0018 & 0.0101 & 427 \\
103 & 55570.9046 & 0.0013 & 0.0201 & 116 \\
104 & 55570.9694 & 0.0011 & 0.0182 & 107 \\
118 & 55571.9022 & 0.0009 & 0.0172 & 217 \\
119 & 55571.9743 & 0.0014 & 0.0226 & 211 \\
122 & 55572.1788 & 0.0022 & 0.0271 & 15 \\
123 & 55572.2380 & 0.0014 & 0.0195 & 12 \\
137 & 55573.1684 & 0.0071 & 0.0162 & 16 \\
149 & 55573.9517 & 0.0014 & $-$0.0009 & 148 \\
164 & 55574.9499 & 0.0013 & $-$0.0032 & 107 \\
179 & 55575.9383 & 0.0022 & $-$0.0153 & 150 \\
224 & 55578.9187 & 0.0043 & $-$0.0364 & 237 \\
254 & 55580.9340 & 0.0014 & $-$0.0221 & 192 \\
\hline
  \multicolumn{5}{l}{\commenta BJD$-$2400000.} \\
  \multicolumn{5}{l}{\commentb Against max $= 2455564.0145 + 0.066699 E$.} \\
  \multicolumn{5}{l}{\commentc Number of points used to determine the maximum.} \\
\end{tabular}
\end{center}
\end{table}

\section{Discussion}

\subsection{Re-calibration of $P_{\rm SH}$--$P_{\rm orb}$ Relation}\label{sec:pshporb}

   The empirical relation by \citet{sto84tumen} between $P_{\rm SH}$
and $P_{\rm orb}$ have been widely used in estimating $P_{\rm orb}$
from $P_{\rm SH}$ (e.g. \cite{RitterCV7}).  Since we now have a plenty
of objects and know the difference between stage B and C superhumps,
we provide an updated calibration.  Since it was already evident that
$P_{\rm orb}$--$\epsilon$ relation deviates from a linear one
(figure 15 in \cite{Pdot}), we have introduced a
$1/(P_{\rm orb}-{\rm const})$-type dependence instead of the linear
one in \citet{sto84tumen}.
Using all samples with known $P_{\rm orb}$ and well-defined stage B
superhumps (following the criteria used in \cite{Pdot}), we have derived
relations for objects below the period gap:

\begin{equation}
\epsilon = 0.000346(36)/(0.043-P_{\rm orb}) + 0.0443(21)
\label{equ:p1porb}
\end{equation}

for the mean period of stage B superhumps, and

\begin{equation}
\epsilon = 0.000273(24)/(0.044-P_{\rm orb}) + 0.0381(13)
\label{equ:p2porb}
\end{equation}

for the mean period of stage C superhumps.

\begin{figure}
  \begin{center}
    \FigureFile(88mm,77mm){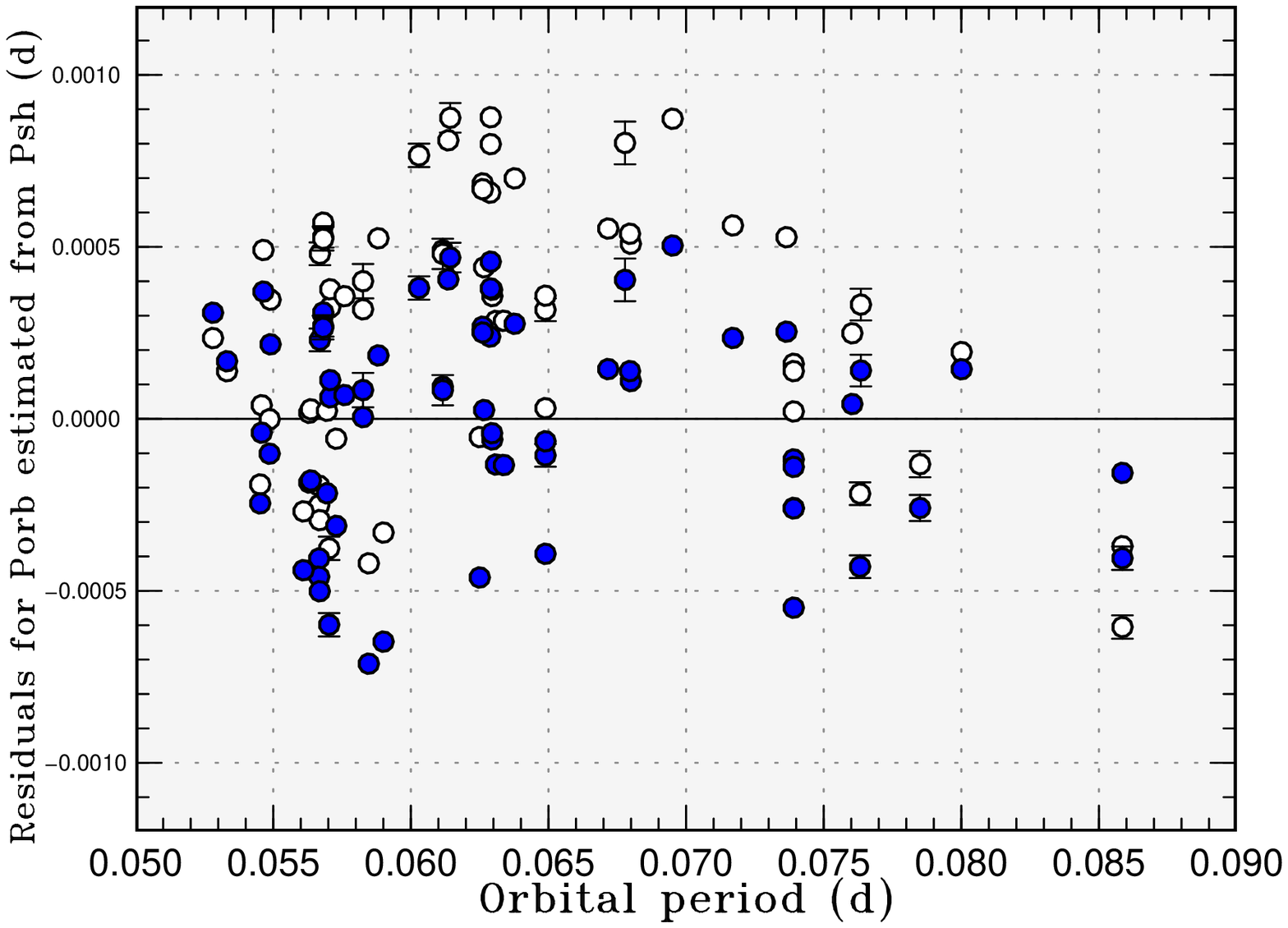}
  \end{center}
  \caption{Residuals for estimated $P_{\rm orb}$ from mean $P_{\rm SH}$
  of stage B superhumps.
  Filled and open circles represent the present relation
  (equation \ref{equ:p1porb}) and \citet{sto84tumen}, respectively.
  The relation by \citet{sto84tumen} systematically gives longer $P_{\rm orb}$.
  }
  \label{fig:p1porbres}
\end{figure}

   Figure \ref{fig:p1porbres} represents the residuals for estimated
$P_{\rm orb}$ from mean $P_{\rm SH}$ of stage B superhumps.
The relation by \citet{sto84tumen} systematically (0.0003 d in average)
gives longer $P_{\rm orb}$.  The 1-$\sigma$ error for the estimated
$P_{\rm orb}$ from the updated relation is 0.0003 d.

\begin{figure}
  \begin{center}
    \FigureFile(88mm,77mm){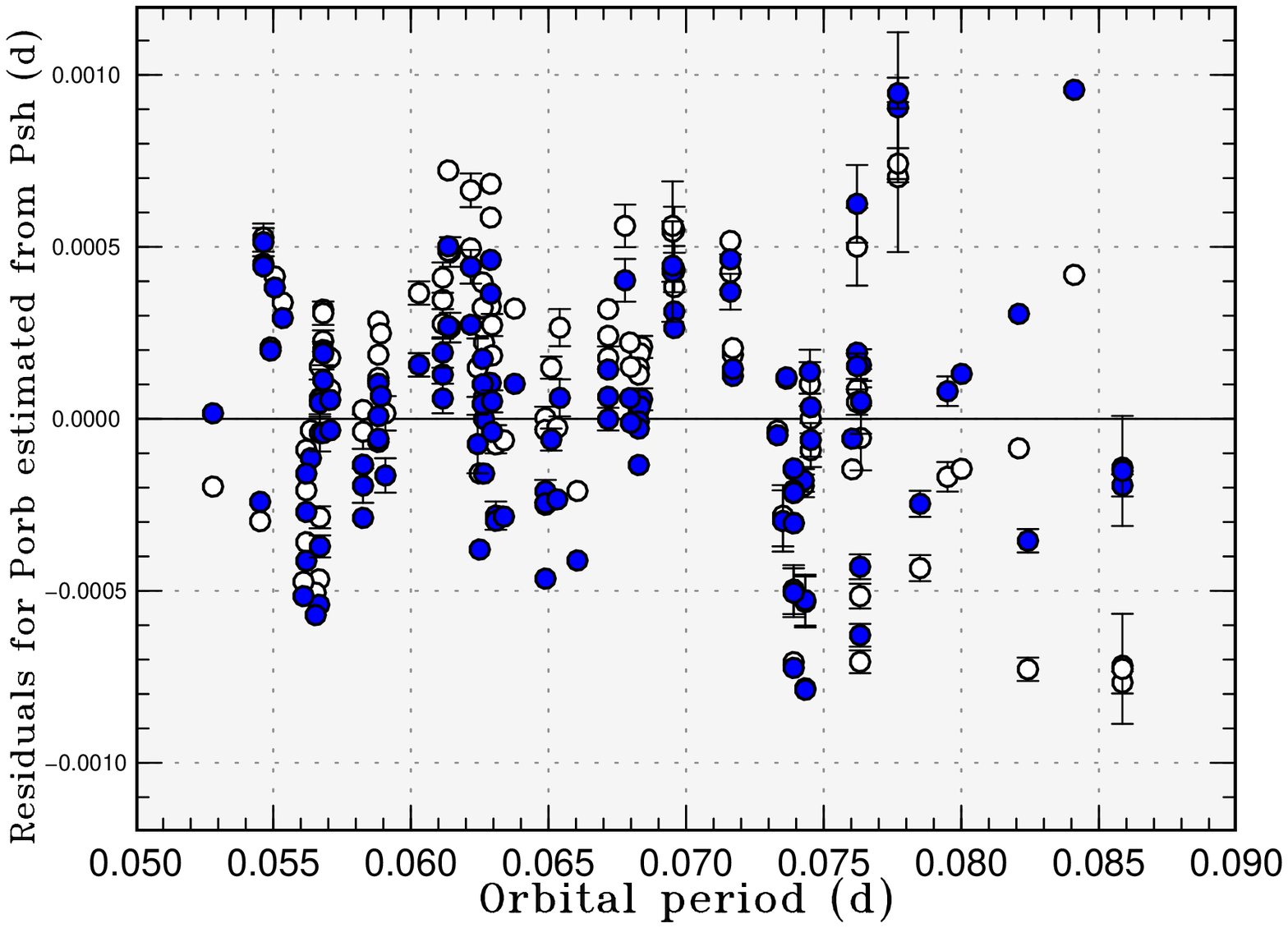}
  \end{center}
  \caption{Residuals for estimated $P_{\rm orb}$ from mean $P_{\rm SH}$
  of stage C superhumps.
  Filled and open circles represent the present relation
  (equation \ref{equ:p1porb}) and \citet{sto84tumen}, respectively.
  Although the relation by \citet{sto84tumen} generally well reproduces
  the real $P_{\rm orb}$, there remains a small systematic trend.
  }
  \label{fig:p2porbres}
\end{figure}

   Figure \ref{fig:p2porbres} represents the residuals for estimated
$P_{\rm orb}$ from mean $P_{\rm SH}$ of stage C superhumps.
Although the relation by \citet{sto84tumen} generally well reproduces
the real $P_{\rm orb}$, there remains a small systematic trend
(shorter values for short-$P_{\rm orb}$ and long-$P_{\rm orb}$ systems
and longer values for intermediate-$P_{\rm orb}$ systems.
We conclude that the relation by \citet{sto84tumen} appears to be generally
useful for stage C superhumps, and suggest to use our updated relation
for stage B superhumps especially when only early-stage observations and
mean superhump periods are available.  The 1-$\sigma$ error for the
estimated $P_{\rm orb}$ from the updated relation is 0.0003 d.

   In the following subsections, orbital periods were estimated using these
relations for objects without known orbital periods.

\subsection{Period Derivatives during Stage B}

\begin{figure}
  \begin{center}
    \FigureFile(88mm,77mm){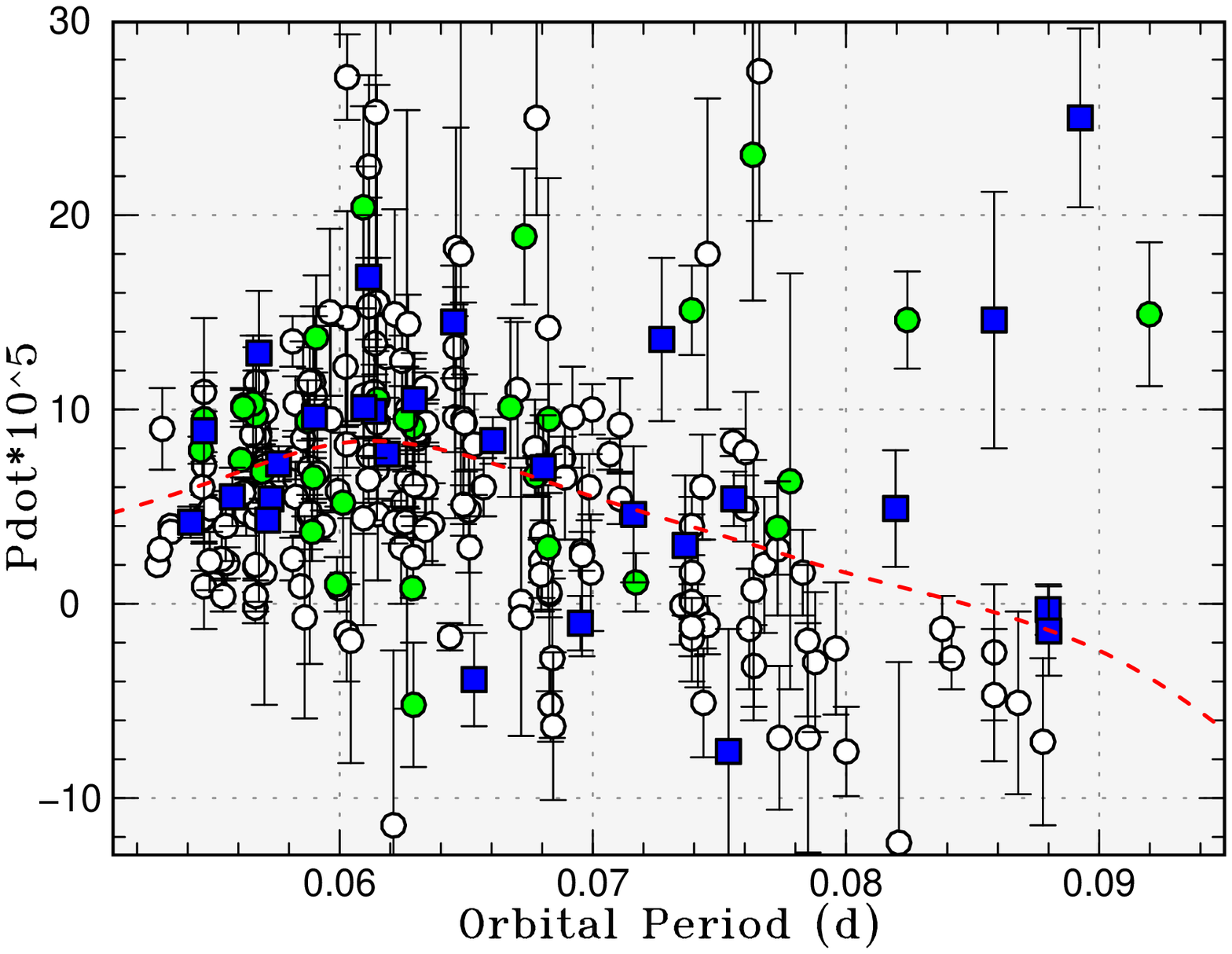}
  \end{center}
  \caption{$P_{\rm dot}$ for stage B versus $P_{\rm orb}$.
  Open circles, filled circles and filled squares represent samples in
  \citet{Pdot}, \citet{Pdot2} and this paper, respectively.
  The curve represents the spline-smoothed global trend.
  }
  \label{fig:pdotporb3}
\end{figure}

   Figure \ref{fig:pdotporb3} represents the relation between $P_{\rm orb}$
and $P_{\rm dot}$ during stage B (note that we use $P_{\rm orb}$
instead of $P_{\rm SH}$ used in earlier papers).
The enlarged figure (corresponding to figure 10 in \cite{Pdot})
is only shown here.  The new data generally confirmed the predominance
of positive period derivatives during stage B in systems with orbital
periods shorter than 0.07 d, in agreement with the tendency reported
in \citet{Pdot} and \citet{Pdot2}.  Although longer period systems
($P_{\rm orb} > 0.08$ d) tend to show zero or slightly negative $P_{\rm dot}$,
the new study suggest the presence of positive-$P_{\rm orb}$ systems
in this period regime.  The most striking example is GX Cas (2010,
subsection \ref{obj:gxcas}).

\subsection{Periods of Stage A Superhumps}

\begin{figure}
  \begin{center}
    \FigureFile(88mm,77mm){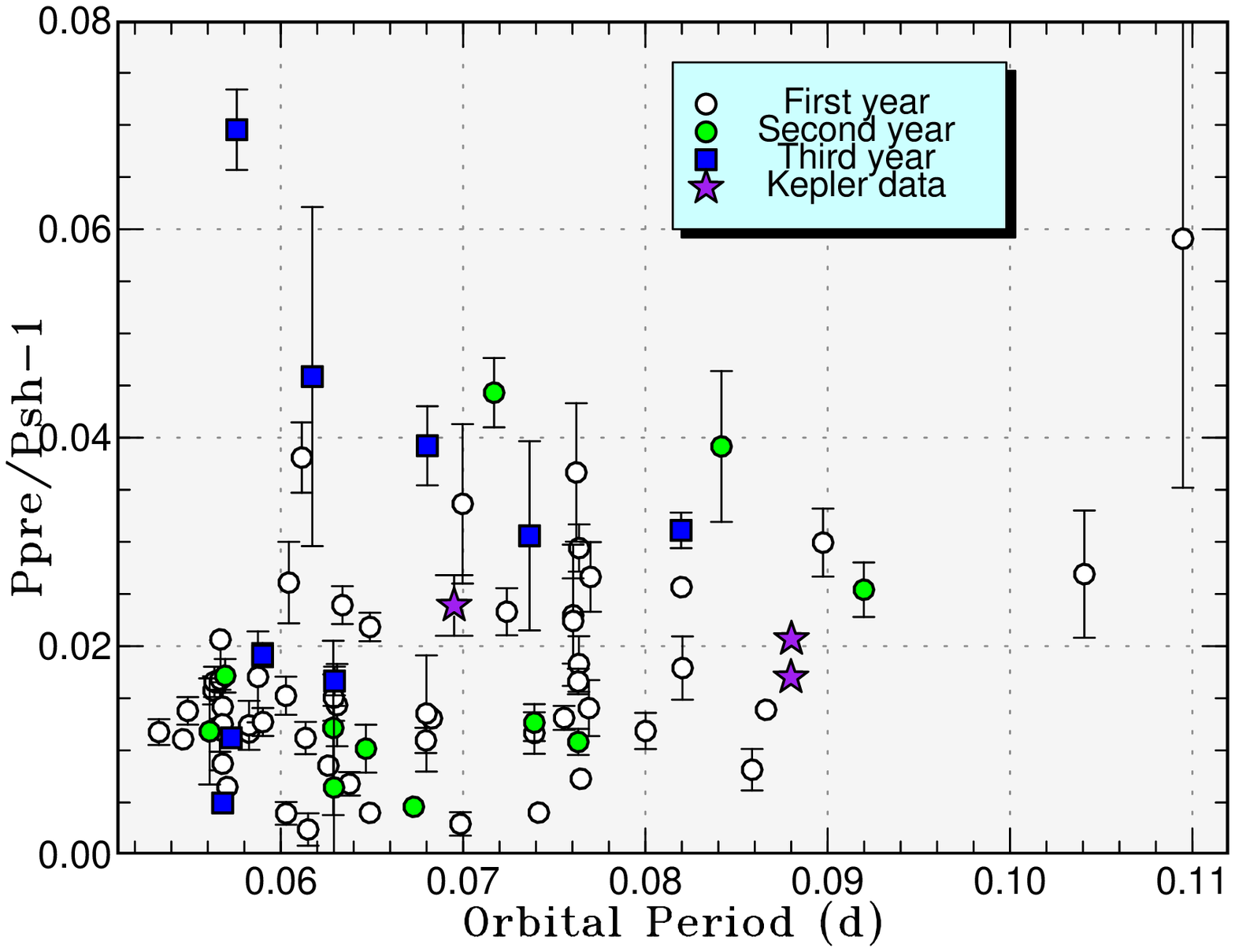}
  \end{center}
  \caption{Superhump periods during the stage A.  Superhumps in this stage
  has a period typically 1.0--1.5 \% longer than the one during the stage B.
  There is a slight tendency of increasing fractional period excess for
  longer-$P_{\rm orb}$ systems.  The symbols for first, second and third
  years represent data in \citet{Pdot}, \citet{Pdot2} and this paper.
  }
  \label{fig:ppre2}
\end{figure}

   Since this table was missing in \citet{Pdot2}, we list values for
the second year and this (third) year (table \ref{tab:pera}) following
the manner in \citet{Pdot}.  The combined figure is presented in
figure \ref{fig:ppre2}.  The fractional period excess (against periods
of stage B superhumps) has a tendency to increase as $P_{\rm orb}$ increases.
The result for high-accuracy Kepler data (subsection \ref{sec:kepler})
agrees with other ground-based observations.

   As shown in subsection \ref{obj:mmhya}, MM Hya likely had an unusually
large fractional period excess for stage A superhumps.  This needs to be
confirmed by future observations.

   We discussed the origin of stage A superhumps in \citet{Pdot} and
presented two possibilities: larger precession rate for larger disks
or longer periods during the growing stage suggested by smoothed particle
hydrodynamics (SPH) simulations.  There have been growing evidence
favoring the latter with SPH simulations (\cite{mur98SH}; \cite{smi07SHsimul})
and most recently in \citet{kle08SH}, which first successfully produced
the growing eccentric mode with a grid-based simulation.

\begin{table}
\caption{Superhump Periods during Stage A}\label{tab:pera}
\begin{center}
\begin{tabular}{cccc}
\hline
Object\commenta & Year & period (d) & err \\
\hline
NN Cam & 2009 & 0.07755 & 0.00025 \\
V592 Her & 2010 & 0.05728 & 0.00029 \\
CT Hya & 2010 & 0.06718 & 0.00015 \\
EF Peg & 2009 & 0.09077 & 0.00063 \\
EK TrA & 2009 & 0.06562 & 0.00054 \\
IY UMa & 2009 & 0.07717 & 0.00014 \\
1RXS J0423 & 2010 & 0.07930 & 0.00010 \\
SDSS J1610 & 2009 & 0.05881 & 0.00009 \\
SDSS J1625 & 2010 & 0.09849 & 0.00025 \\
OT J0506 & 2009 & 0.06964 & 0.00005 \\
OT J1044 & 2010 & 0.06084 & 0.00003 \\
OT J1440 & 2009 & 0.06503 & 0.00051 \\
\hline
BG Ari & 2010 & 0.08754 & 0.00015 \\
HT Cas & 2010 & 0.07868 & 0.00069 \\
V1504 Cyg & 2009b & 0.07395 & 0.00021 \\
MM Hya & 2011 & 0.06295 & 0.00023 \\
V344 Lyr & 2009 & 0.09314 & 0.00009 \\
V344 Lyr & 2009b & 0.09351 & 0.00011 \\
V1212 Tau & 2011 & 0.07286 & 0.00027 \\
SW UMa & 2010 & 0.05850 & 0.00002 \\
SDSS J0804 & 2010 & 0.06077 & 0.00007 \\
SDSS J1146 & 2011 & 0.06623 & 0.00103 \\
SDSS J1227 & 2007 & 0.06568 & 0.00009 \\
SDSS J1339 & 2011 & 0.05874 & 0.00003 \\
\hline
  \multicolumn{4}{l}{\commenta Abbreviations for \citet{Pdot2} objects:} \\
  \multicolumn{4}{l}{1RXS J0423: 1RXS J042332$+$745300} \\
  \multicolumn{4}{l}{SDSS J1610: SDSS J161027.61$+$090738.4} \\
  \multicolumn{4}{l}{SDSS J1625: SDSS J162520.29$+$120308.7} \\
  \multicolumn{4}{l}{OT J0506: OT J050617.4$+$354738} \\
  \multicolumn{4}{l}{OT J1044: OT J104411.4$+$211307} \\
  \multicolumn{4}{l}{OT J1440: OT J144011.0$+$494734} \\
\end{tabular}
\end{center}
\end{table}

\subsection{WZ Sge-Type Stars: Statistics}\label{sec:wzsgestat}

   Table \ref{tab:wztab} summarizes the properties of new outbursts of
WZ Sge-type and related objects reported in \citet{Pdot2} and this paper
(the selection criteria are the same as in \cite{Pdot}).  Although many
of recently discovered WZ Sge-type objects were not sufficiently observed
during the earliest stages of their outbursts, the fractional superhump
excesses were relatively well determined.

\begin{figure}
  \begin{center}
    \FigureFile(88mm,70mm){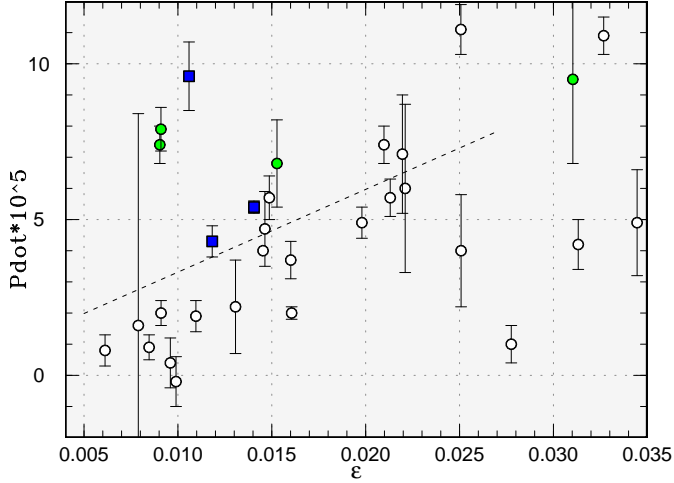}
  \end{center}
  \caption{$P_{\rm dot}$ versus $\epsilon$ for WZ Sge-type
  dwarf novae.  Open circles, filled circles and filled squares represents
  outbursts reported in \citet{Pdot}, \citet{Pdot2} and this paper,
  respectively.
  The dashed line represents a linear regression for points with
  $\epsilon < 0.026$ as in \citet{Pdot} figure 36.
  }
  \label{fig:wzpdoteps3}
\end{figure}

   Figure \ref{fig:wzpdoteps3} shows the relation between $P_{\rm dot}$
versus $\epsilon$ for WZ Sge-type dwarf novae studied in these series
of papers.  In \citet{Pdot2} and this paper, higher values $P_{\rm dot}$
were recorded for short-$P_{\rm orb}$ systems, and the relation described
in \citet{Pdot} became less clear.  These outbursts are V592 Her (2010),
OT J2138 (2010) and SDSS J0804 (2010).  As already described in
subsection \ref{obj:j0804}, the large $P_{\rm dot}$ in SDSS J0804 (2010)
is an outlier among objects with multiple rebrightenings.
The large $P_{\rm dot}$ in this system may have been affected by
a large ``disturbance'' in the $O-C$ diagram near the end of the
superoutburst plateau, and the $P_{\rm dot}$ during the earlier part
of the stage B superhumps was much smaller
[$P_{\rm dot}$ = $+3.5(0.9) \times 10^{-5}$].  The same phenomenon is
sometimes obvious in other WZ Sge-type dwarf novae [V455 And (2007),
figure 44 and the secondary component in WZ Sge (2001), figure 127
in \citet{Pdot}] and all of them are eclipsing objects.
The high inclination may be somehow responsible for this phenomenon
(see also a discussion in subsection \ref{sec:humpamp}), and we might
better exclude the ``disturbance'' part of the $O-C$ diagram in determining
the representative $P_{\rm dot}$ in such systems.
The interpretation of this phenomenon needs to be investigated further.

\begin{figure}
  \begin{center}
    \FigureFile(88mm,70mm){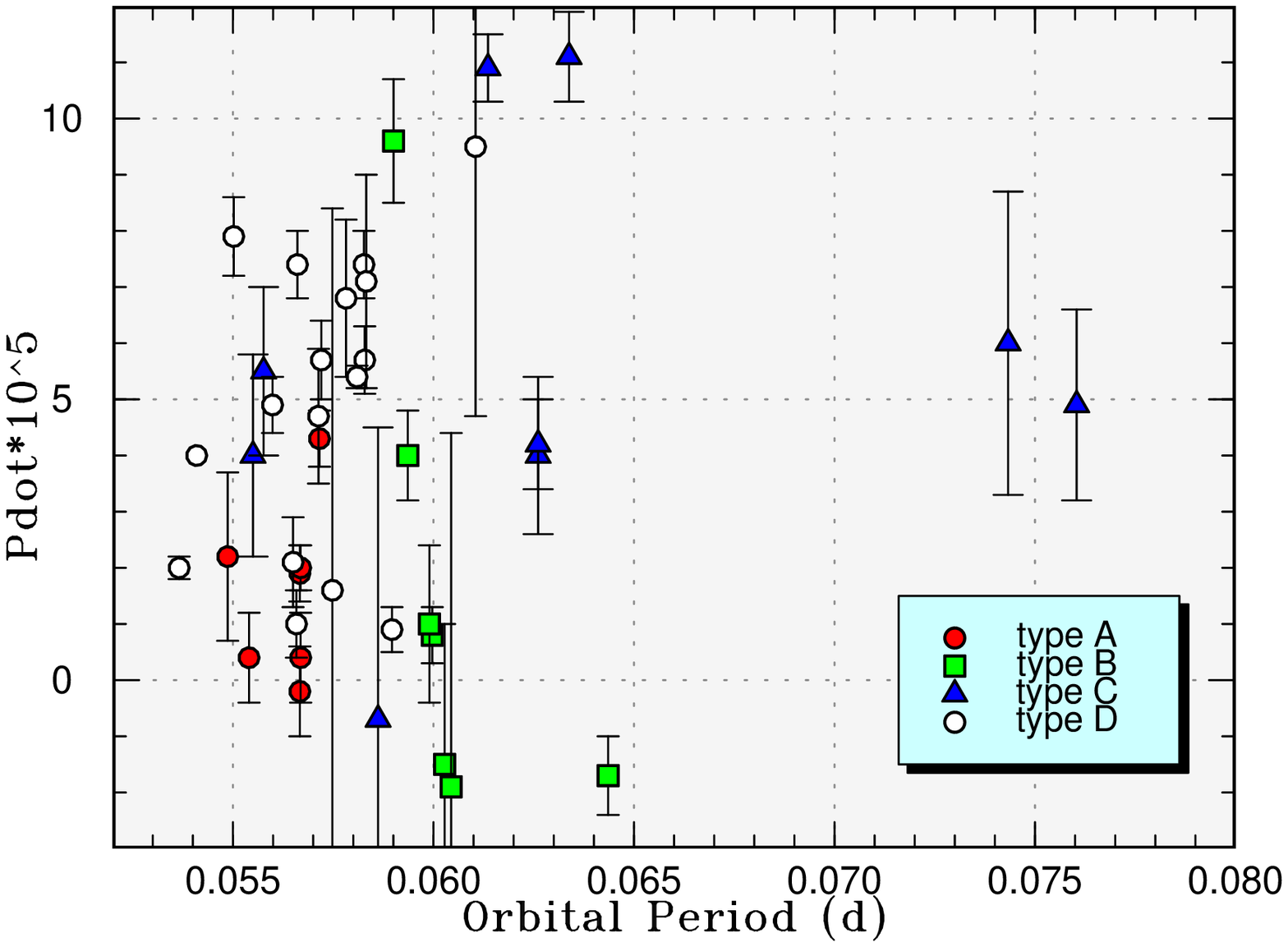}
  \end{center}
  \caption{$P_{\rm dot}$ versus $P_{\rm SH}$ for WZ Sge-type
  dwarf novae.  Symbols represent the type (cf. \cite{Pdot}) of outburst:
  type-A (filled circles), type-B (filled squares),
  type-C (filled triangles), type-D (open circles).
  }
  \label{fig:wzpdottype3}
\end{figure}

   WZ Sge-type dwarf novae show various activities during the
post-superoutburst stage.  In \citet{Pdot}, we introduced a classification:
type-A (long-lasting post-outburst rebrightening), type-B (multiple
discrete rebrightenings), type-C (single rebrightening) and type-D
(no rebrightening).  It is known that there is a dependence of the types
of outbursts on $P_{\rm orb}$ and $\epsilon$ (figure 37 in \cite{Pdot};
note $P_{\rm SH}$ was instead used).  The updated statistics
(figure \ref{fig:wzpdottype3}) basically strengthens
the earlier finding: type-B objects mostly occupy low-$\epsilon$ and
middle-$P_{\rm orb}$ region, type-A and type-D objects occupy
middle-$\epsilon$ and short-$P_{\rm orb}$ region, and type-C objects
are more scattered.  The type-D outlier is SDSS J0804 as described above.

\begin{table*}
\caption{Parameters of WZ Sge-type superoutbursts.}\label{tab:wztab}
\begin{center}
\begin{tabular}{cccccccccccc}
\hline
Object & Year & $P_{\rm SH}$ & $P_{\rm orb}$ & $P_{\rm dot}$\commenta & err\commenta & $\epsilon$ & Type\commentb & $N_{\rm reb}$\commentc & delay\commentd & Max & Min \\
\hline
VX For & 2009 & 0.061327 & -- & 1.0 & 1.4 & -- & B & 5 & -- & ]12.6 & 20.6 \\
V592 Her & 2010 & 0.056607 & 0.0561 & 7.4 & 0.6 & 0.009 & D & 0 & $\geq$4 & ]14.2 & 21.3 \\
BC UMa & 2009 & 0.064553 & 0.06261 & 9.5 & 2.7 & 0.031 & -- & -- & -- & ]11.6 & 18.6 \\
SDSS J1610 & 2009 & 0.057820 & 0.05695 & 6.8 & 1.4 & 0.015 & D? & 0? & $\geq$2 & ]13.9 & 20.1 \\
OT J1044 & 2010 & 0.060236 & 0.05909 & -- & -- & 0.019 & C & 1 & $\geq$5 & ]12.6 & 19.3 \\
OT J2138 & 2010 & 0.055019 & 0.05450 & 7.9 & 0.7 & 0.010 & D & 0 & 6 & 8.7 & [15.4 \\
OT J2230 & 2009 & -- & 0.05841 & -- & -- & -- & -- & -- & $\geq$4 & ]14.4 & 21.0 \\
\hline
ASAS J1025 & 2011 & 0.063538 & 0.06136 & 9.9 & 2.4 & 0.035 & -- & -- & -- & ]12.9 & 19.3 \\
MisV 1443 & 2011 & 0.056725 & -- & 5.5 & 1.5 & -- & C? & 1? & -- & ]12.8 & 19.7 \\
SDSS J0804 & 2010 & 0.059630 & 0.059005 & 9.6 & 1.1 & 0.011 & B & 6 & 4 & 12.0 & 17.8 \\
SDSS J1339 & 2011 & 0.058094 & 0.057289 & 5.4 & 0.2 & 0.014 & D & 0 & $\geq$6 & ]9.9 & 17.7 \\
SDSS J1605 & 2010 & -- & 0.05666 & -- & -- & -- & A? & ? & $\geq$3 & ]12.0 & 19.9 \\
OT J0120 & 2010 & 0.057833 & 0.057157 & 4.3 & 0.5 & 0.012 & A & $\geq$9 & $\geq$11 & ]12.3 & 20.0 \\
\hline
  \multicolumn{12}{l}{\commenta Unit $10^{-5}$.} \\
  \multicolumn{12}{l}{\commentb A: long-lasting rebrightening; B: multiple rebegitehnings; C: single rebrightening; D: no rebrightening.} \\
  \multicolumn{12}{l}{\commentc Number of rebrightenings.} \\
  \multicolumn{12}{l}{\commentd Days before ordinary superhumps appeared.} \\
\end{tabular}
\end{center}
\end{table*}

\subsection{Analysis of Kepler Observations}\label{sec:kepler}

   The data for two SU UMa-type dwarf novae are available in Kepler
\citep{Kepler} public release data.  One of them (V344 Lyr) has already
been published \citep{sti10v344lyr}.
\citet{can10v344lyr} compared the Kepler data with model simulations
based on thermal-tidal disk instability.
We examined Kepler observations from the viewpoint of the current study.
The magnitudes were calculated from the corrected aperture-integrated 
flux obtained from the Kepler pipeline, with an arbitrary photometric
zero-point.

\subsubsection{Superhumps in V1504 Cygni}\label{obj:kepv1504cyg}

\begin{figure*}
  \begin{center}
    \FigureFile(170mm,90mm){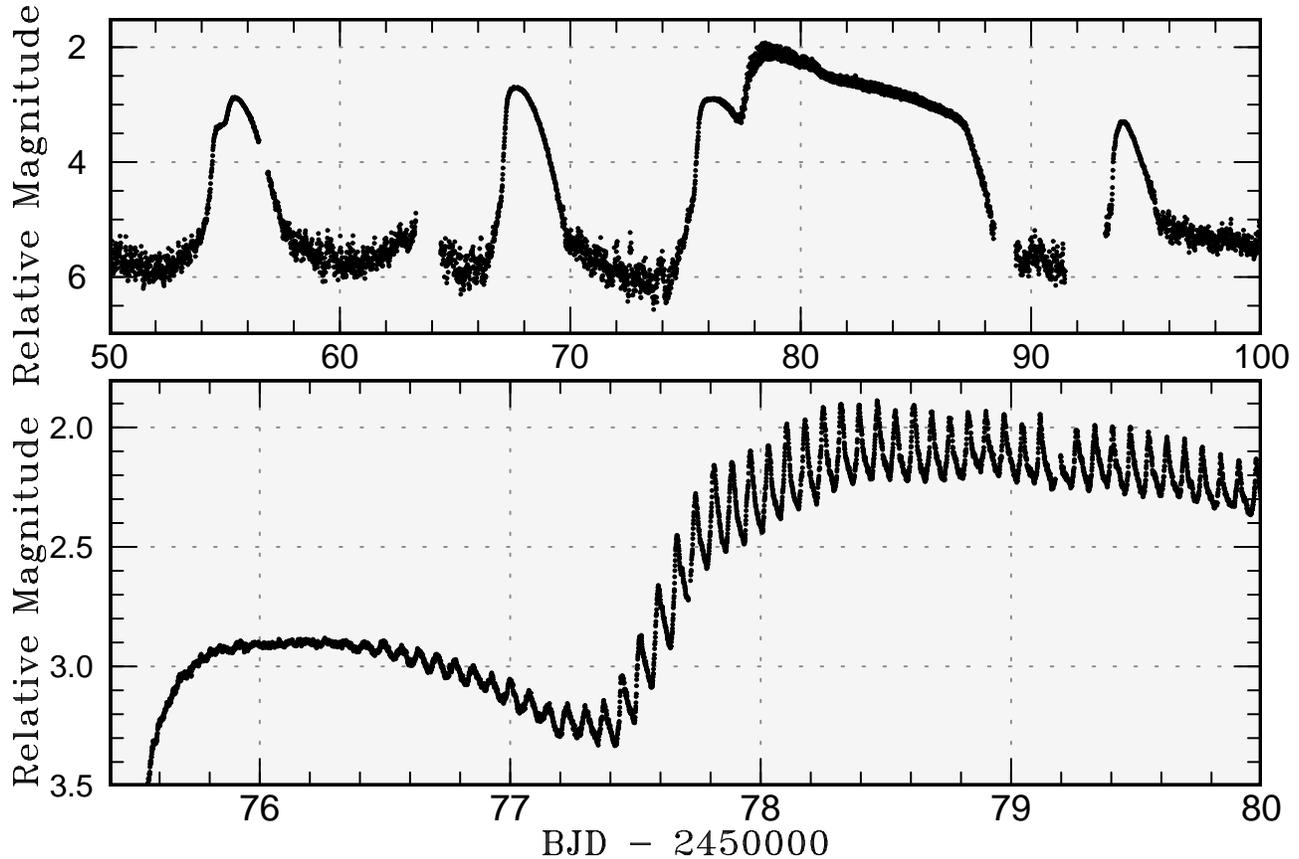}
  \end{center}
  \caption{Kepler light curve of V1504 Cyg (2009b).
     (Upper): light curve around the superoutburst in 2009 September.
     Note the presence of precursor outbursts (shoulders) in
     the first (normal) outburst and the third (super) outburst.
     The relative magnitudes are given against an arbitrary zero point.
     (Lower): enlargement of the initial part of the superoutburst.
     Following a distinct precursor, superhumps developed into
     a full superoutburst.
  }
  \label{fig:v1504kep}
\end{figure*}

   Kepler recorded the 2009 September superoutburst of this object
(see subsection \ref{obj:v1504cyg} for detailed data).
This outburst was composed of a precursor outburst lasting for
$\sim$2 d, and superhumps started to develop around the maximum of this
precursor.  As the superhumps developed, the object stopped fading and
entered the plateau phase of the superoutburst (figure \ref{fig:v1504kep}).
The plateau phase lasted for $\sim$9 d, followed by a steeper decline.
During the entire course of the plateau phase and decline phase,
ordinary superhumps were present.  The amplitudes of superhumps were
large around the maximum, initially decayed relatively rapidly for 3.2 d.
After reaching a minimum, the amplitudes slightly increased,
but again decayed in $\sim$ 2 d.

   The $O-C$ diagram indicates the presence of fine structures
rather than our discrete three stages.  The transitions between
stages were smoother than in many objects in our present and past
studies.  In table \ref{tab:perlist}, we assigned stage A before
the epoch of maximum amplitude, and stages B and C were divided
by a kink in the $O-C$ diagram.
The relatively smooth variation
in the $O-C$ diagram and the lack of segment with a positive
$P_{\rm dot}$ might be attributable to the notably high frequency of
outbursts in this system (\cite{raj87v1504cyg}; \cite{nog97v1504cyg}),
suggesting a high mass-transfer rate.
Although the tendency that objects with high outburst frequencies tend
to show smoother $O-C$ variation was originally proposed for
long-$P_{\rm orb}$ systems (\cite{Pdot}, subsection 4.10), it might
also be extended to shorter-$P_{\rm orb}$ systems as in V1504 Cyg.

   Yet another interesting feature is in the fine structure during stage A.
During the first 8 superhump cycles, the period of superhumps once
shortened while the amplitudes stayed low.  Following this stage of
initial development, the amplitudes quickly grew and the period
stayed at a relatively constant value.  This later stage has likely
been recognized as stage A superhumps in ground-based observations.

   During the growing stage, small-amplitude ($\sim$0.01 mag)
quasi-periodic oscillations (QPOs) with time-scales of $\sim$10 min were
superimposed on superhumps while they disappeared when superhumps quickly grew.
These modulations resemble so-called ``super-QPOs'' in some SU UMa-type
dwarf novae during the evolutionary stage of superhumps
(cf. \cite{kat92swumasuperQPO}; \cite{kat02efpeg}).
These QPOs must be somehow related to the evolutionary process of superhumps.

   Although \citet{ant05v1504cyg} reported humps with a period close to
the orbital period on the first day of the superoutburst, there was no
special indication of such humps in Kepler data.

\begin{figure}
  \begin{center}
    \FigureFile(88mm,110mm){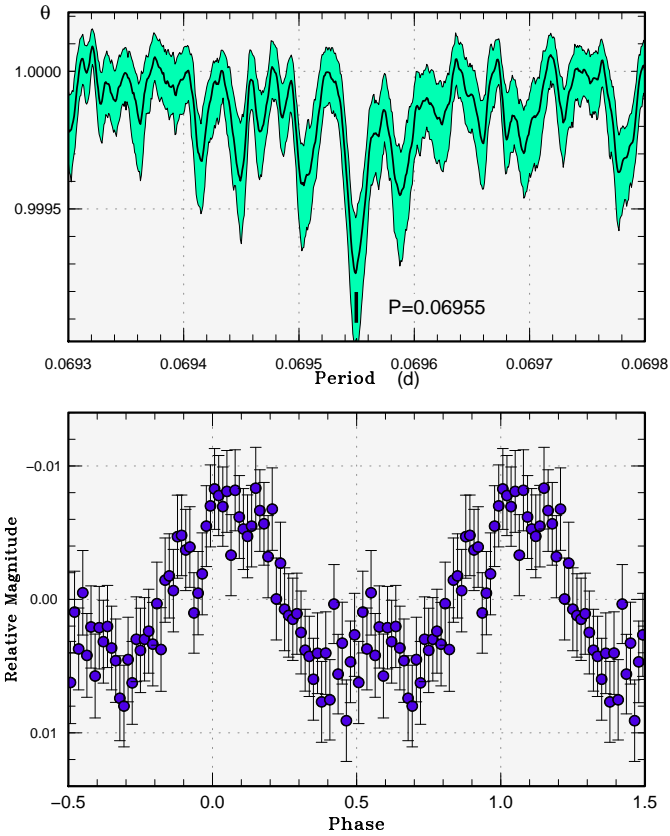}
  \end{center}
  \caption{Period analysis of Kepler V1504 Cyg data in quiescence.
     (Upper): PDM analysis.
     (Lower): Phase-averaged profile.}
  \label{fig:v1504cygquipdm}
\end{figure}

   Using quiescent segments of Kepler observations, we have obtained a
refined orbital period of 0.069549(2) d with a mean amplitude of
orbital modulation of 0.016 mag (figure \ref{fig:v1504cygquipdm}).

\subsubsection{Transient Superhumps during Normal Outbursts of V1504 Cygni}\label{obj:kepv1504cygtrans}

\begin{figure}
  \begin{center}
    \FigureFile(88mm,110mm){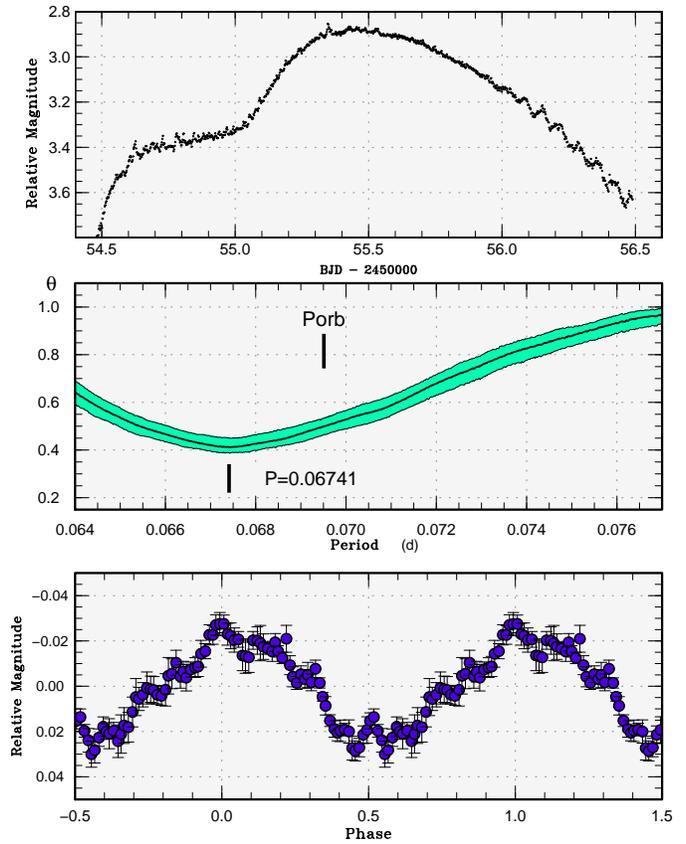}
  \end{center}
  \caption{Negative superhumps emerging during a normal outburst of
     V1504 Cyg.
     (Upper): Light curve.  This outburst is the first outburst in
     figure \ref{fig:v1504kep}.  Low-amplitude modulations appeared
     during its fading phase.
     (Middle): PDM analysis.
     (Lower): Phase-averaged profile.}
  \label{fig:v1504prec1}
\end{figure}

   During the normal outburst starting on 2009 August 10 (two outburst cycles
before this superoutburst), a prominent ``precursor'' (or a shoulder)
preceding this outburst was recorded just as in the superoutburst.
During the fading stage of its second maxima, negative superhumps
with a period of 0.0677(2) d [2.6(3) \% shorter than $P_{\rm orb}$]
appeared (figure \ref{fig:v1504prec1}).
This outburst may have been a failed superoutburst in which
tidal instability was excited but it was not strong enough to
trigger a full superoutburst.  It is very suggestive that negative superhumps,
instead of ordinary (positive) superhumps, were excited.
This suggests that ordinary superhumps and negative superhumps can be
excited under similar conditions (or have the common underlying mechanism)
within a relatively short time.  In ER UMa, \citet{ohs11eruma} reported
the presence of negative superhumps during its 2011 January superoutburst
and the state with strong negative superhumps appeared to prevent
a long-lasting superoutburst.  \citet{can10v344lyr} also discussed the
prevention of disk-instability in the state with strong negative superhumps
in V344 Lyr.  During this normal outburst of V1504 Cyg,
excitation of negative superhumps by chance may have prevented an evolution
of a full superoutburst.

\begin{figure}
  \begin{center}
    \FigureFile(88mm,110mm){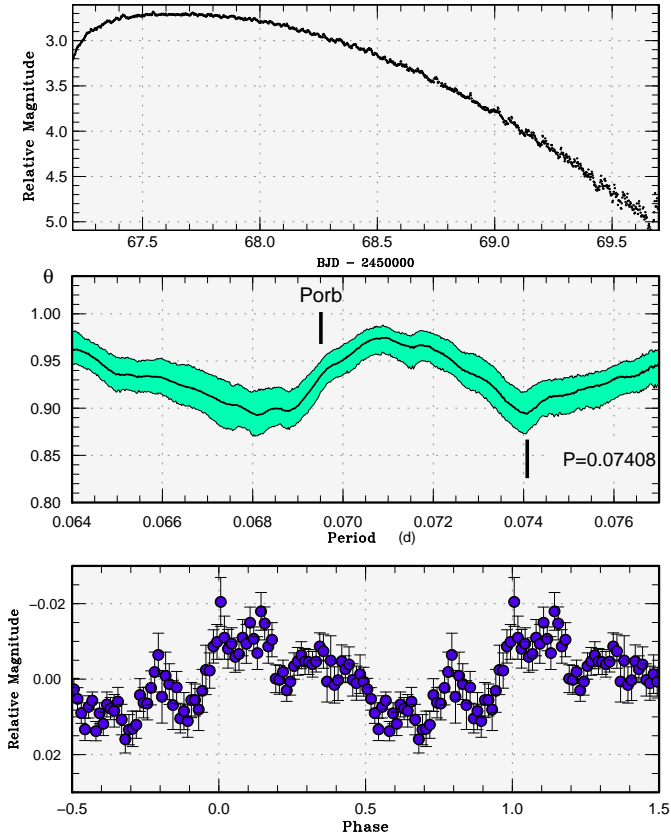}
  \end{center}
  \caption{Positive superhumps emerging during a normal outburst of
     V1504 Cyg.
     (Upper): Light curve.  This outburst is the second outburst in
     figure \ref{fig:v1504kep}.  Low-amplitude modulations appeared
     during its fading phase.
     (Middle): PDM analysis.  In addition to negative superhumps,
     there is a signal of positive superhumps.
     (Lower): Phase-averaged profile.}
  \label{fig:v1504prec2}
\end{figure}

   During the next normal outburst starting on August 23 (outburst just
prior to the superoutburst), transient low-amplitude superhumps
with a period of 0.07402(7) d appeared during its fading branch
(figure \ref{fig:v1504prec2}).
This period is very close to the period [0.0739(2) d] of stage A
superhumps, and the excitation of the superhumps apparently occurred
around the maximum of this outburst.  This outburst appears to be
phenomenologically similar to the precursor part of the main superoutburst,
but the strength of the tidal instability may have not been sufficient
to trigger a full superoutburst.

   \citet{pav02v1504cygproc} also reported the presence of short outbursts
with durations of less than 1 d and maximum brightness not exceeding
15.2 mag.  \citet{pav02v1504cygproc} referred to these outburst as
``third type'' of outbursts.  These outbursts are known to occur during
the early phase of the supercycle.  They also noted the light variations
close to superhump period we detected during one normal outburst
(immediately following a superoutburst).
These short outbursts may be somehow related to ``failed'' outbursts
observed by Kepler, and it is expected that further Kepler observations
can detect these outbursts.

\subsubsection{Superhumps in V344 Lyrae}\label{obj:kepv344lyr}

   The Kepler observations of V344 Lyr were analyzed by \citet{sti10v344lyr},
who reported the presence of (most likely) negative superhumps in quiescence
and during normal outbursts.  During superoutbursts, postive superhumps
were the predominant signal.  We analyzed Kepler observations between
2009 June 20 and December 17.  During this period, two superoutbursts
were recorded (2009 August and 2009 November--December).

   Both superoutbursts showed prominent precursors as in V1504 Cyg.
\citet{can10v344lyr} compared calculations based on the thermal-tidal
instability model and the enhanced mass-transfer model, and indicated that
the thermal-tidal instability model generally well reproduces the precursor
although some differences remained between observations and
model calculations.

\begin{figure}
  \begin{center}
    \FigureFile(88mm,110mm){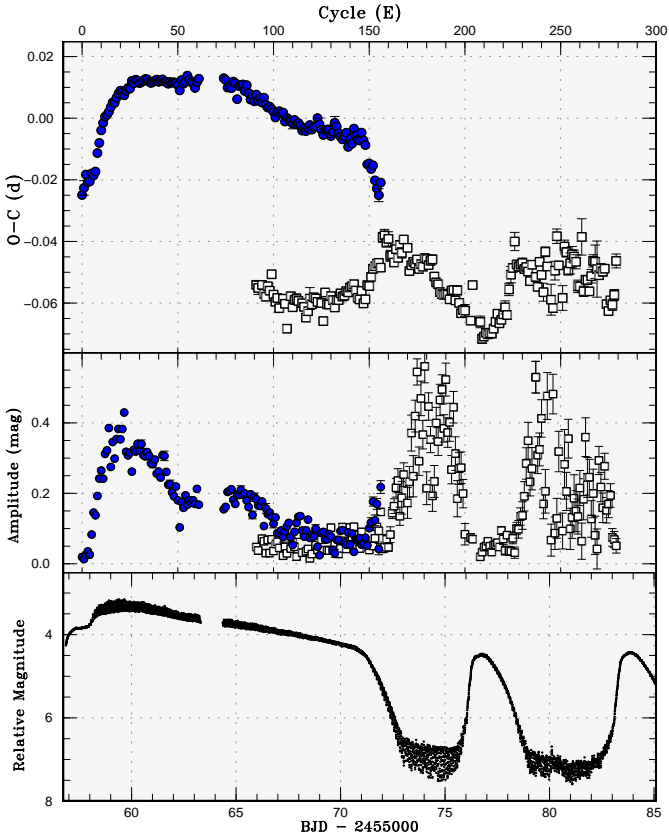}
  \end{center}
  \caption{$O-C$ diagram of superhumps in V344 Lyr (2009).
     (Upper): $O-C$ diagram.  The filled circles represent primary
     maxima of superhumps.  The open squares represent secondary maxima
     of superhumps and persisting superhumps.
     (Middle): Amplitudes of superhumps.  The symbols are common to
     the upper panel.
     (Lower): Light curve.}
  \label{fig:v344lyrhumpall}
\end{figure}

\begin{figure}
  \begin{center}
    \FigureFile(88mm,110mm){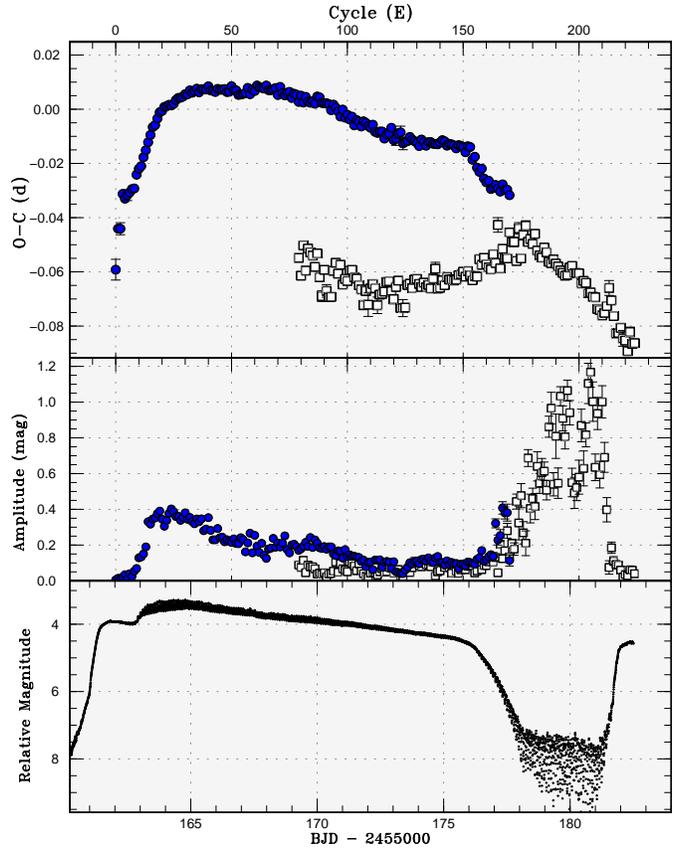}
  \end{center}
  \caption{$O-C$ diagram of superhumps in V344 Lyr (2009b).
     (Upper): $O-C$ diagram.  The filled circles represent primary
     maxima of superhumps.  The open squares represent secondary maxima
     of superhumps and persisting superhumps.
     (Middle): Amplitudes of superhumps.  The symbols are common to
     the upper panel.
     (Lower): Light curve.}
  \label{fig:v344lyrhumpall2}
\end{figure}

   We applied our methods to these Kepler observations.  In contrast to
V1504 Cyg, the superhumps had secondary maxima (a bump around superhump
phase of $\sim$ 0.4) which became stronger
in later stages.  The $O-C$ diagrams for these two superoutbursts
are similar (figures \ref{fig:v344lyrhumpall} and \ref{fig:v344lyrhumpall2}).
The $O-C$ diagrams of primary maxima showed stages A and B, although
the transitions between them were not sharp, as in V1504 Cyg.
These relatively smooth transitions may have been caused by a high
mass-transfer rate, as in V1504 Cyg, which enables short recurrence
times of normal outbursts and superoutbursts (\cite{kat02v344lyr};
\cite{can10v344lyr}; the interval of superoutbursts was 104 d
in Kepler observations).

   Stages A were less structured compared to V1504 Cyg, and there were
smooth variations in the periods.
Stages B with relatively constant superhump periods apparently lasted
for $\sim$50 cycles, then the periods switched to slightly shorter ones.
As in other SU UMa-type dwarf novae, this change in period can be
attributed to a stage B--C transition.
After this transition, the period gradually increased for approximately
$120 \le E \le 150$, then suddenly decreased.  Since the periods after
$E$ = 150 were similar to the periods seen in the early stage C,
we interpreted that there were temporary ($120 \le E \le 150$) excursions
to more positive $O-C$ in addition to the basic period of stage C.
We therefore adopted mean periods of entire stage C in table
\ref{tab:perlist}.

   After this stage B--C transition, the secondary maxima of superhumps
became stronger, and this appearance of secondary maxima may be somehow
related to the origin of stage B--C transition.

   The periods derived from times of secondary maxima were initially
close to superhump periods of stage B, despite that the periods
of primary superhumps had already decreased after stage B--C transitions.
Due to the difference in periods, the primary and secondary maxima
of superhumps approached and became indistinguishable approximately
after $E$ = 160--170.  The period then suddenly shortened.
The interference between these two components of superhump maxima
appeared to cause temporary excursions of primary maxima toward
larger $O-C$.

   The merged two components of maxima then formed single-humped
superhumps which lasted for two cycles of normal outbursts
in 2009 August, and at least for one cycle in 2009 November--December.
Although individual times of maxima became inapparent, the signal
can be traced in phase-averaged light curves for one more cycle
of normal outbursts.

   The origin of the secondary maxima is not evident.  Since the periods
of the secondary component of maxima were longer than the periods of
the primary component, the precession rate should be larger.
Considering that the periods of the secondary component are close to
those of stage B superhumps (primary maxima), it would be possible
to assume that the secondary component was formed in the outermost
region of the accretion disk, which had already cooled, and the
hot region of the accretion disk had shrunken to a smaller radius,
resulting shorter superhump periods (primary maxima).
This picture would naturally explain why the secondary component
was not apparent before the period of superhumps (primary maxima)
shortened, i.e. stage B--C transition.
The mass-transfer from the secondary might be responsible for producing
a light source in the outermost region of the disk
(cf. \cite{min88superhump}), although it will be premature to directly
associate the source with an ordinary stream-impact hot spot.\footnote{
   See also a discussion in \citet{hes92lateSH}, who reported that the
   traditional model of late superhumps by \citet{vog83lateSH} did not
   trivially explain the observed eclipse depths in OY Car and suggested
   that the light source of late-stage superhumps is more extended.
}
The behavior of amplitudes (middle panels in figures \ref{fig:v344lyrhumpall}
and \ref{fig:v344lyrhumpall2}) was similar to those of persisting
superhumps in SDSS J0804 and WZ Sge (subsection \ref{obj:j0804}),
and the luminosity from this source can be estimated to be approximately
constant.  It would be noteworthy that secondary maxima were temporarily
present in V1504 Cyg, another high mass-transfer ($\dot{M}$) system, but did not
evolve further as in V344 Lyr.  The behavior in V1504 Cyg seems to be
difficult to explain by a simple model based on stream-impact
hot spot.\footnote{
   During the refereeing period, \citet{woo11v344lyr} independently
   reported an analysis of Kepler observations of V344 Lyr.
   \citet{woo11v344lyr} interpreted that the secondary maxima
   are generated as the accretion stream bright spot sweeps around
   the rim of the non-axisymmetric disk, and attempted to reproduce
   the observed profile by assuming an enhanced mass-transfer from
   the secondary.  This interpretation corresponds to the traditional
   interpretation of traditional late superhumps discussed in the text.
}

   As a possibility alternative to the emergence of stream-impact hot spot,
we propose the (1,3)-mode wave (the $(k,l)$ mode designation follows
\cite{lub91SHa}), which is a one-armed wave traveling with an angular
velocity third times larger than that of the secondary.  This wave meets
the tidal stress of the secondary twice in a single binary rotation, and
might produce observable double-wave modulations.  \citet{lub91SHb} has shown
that this (1,3)-mode wave tends to grow and persist at later epochs.
It is known that short-$P_{\rm orb}$ systems with distinct stage B-C
transitions tend to show low-amplitude double-wave modulations prior to
the stage B-C transition
(e.g. \cite{uem10j0557}; \cite{kat03hodel}; \cite{bab00v1028cyg}).
Such a phenomenon would be understood if the (1,0)-mode wave, the main
superhump component, weakens and the other low-amplitudes modes become
observable.  These objects are likely low-$\dot{M}$ systems and
it would be more difficult to attribute the secondary maxima to stream-impact
hot spot than in V344 Lyr.  The interpretation based on the (1,3)-mode wave
might have a potentiality in explaining the double-wave superhump
modulations in wide range of objects, and should be further investigated.

\begin{figure}
  \begin{center}
    \FigureFile(88mm,110mm){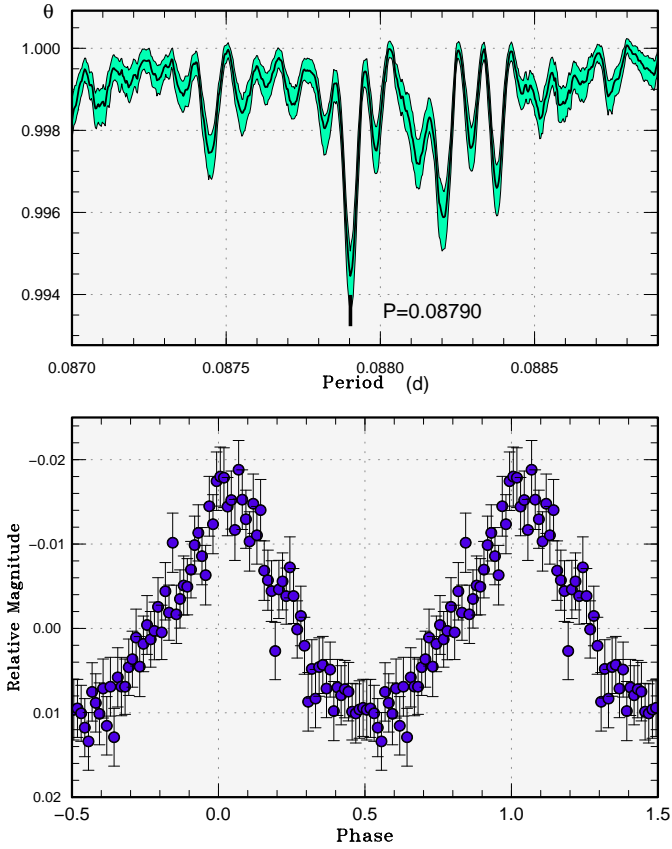}
  \end{center}
  \caption{Period analysis of Kepler V344 Lyr data in quiescence.
     (Upper): PDM analysis.
     (Lower): Phase-averaged profile.}
  \label{fig:v344lyrquipdm}
\end{figure}

   After selecting quiescent segments without strong negative
superhumps, we searched a signal of the orbital period with a help
of estimated orbital period (0.0877--0.0880 d) from superhump periods
(using the relation of \cite{sto84tumen}).
We have found a candidate period 0.087903(2) d with a mean amplitude
of 0.026 mag (figure \ref{fig:v344lyrquipdm}).
Assuming that this is the orbital period, the fractional
excesses for positive (stage B) and negative superhumps are 4.2 \%
and $-$2.1 \%, respectively.

\subsubsection{Comparison between Kepler observations and other samples}

   Kepler obtained highest precision, uninterrupted observations of
two SU UMa-type dwarf novae.  Some of the features, including
substructures in stage A in V1504 Cyg and complex behavior in late
stages in V344 Lyr, were new and were outside descriptions derived
from our large set of ground-based samples.  Apart from them,
the $O-C$ behavior of Kepler observations generally fits the descriptions
obtained from ground-based samples.
The two systems observed by Kepler were either long-$P_{\rm orb}$ or
high-$\dot{M}$, and neither of them were short-$P_{\rm orb}$
and low-$\dot{M}$ systems which show distinct stage A--C
and long-persisting superhumps.

   Since there were not sufficient ground-based observations for
long-$P_{\rm orb}$ or high-$\dot{M}$ systems particularly after
the termination of superoutbursts, it is not clear whether
persisting superhumps in V1504 Cyg and V344 Lyr have the origin
common to persistent stage C superhumps in short-$P_{\rm orb}$
and low-$\dot{M}$ systems.
Future key Kepler observations should include these short-$P_{\rm orb}$
and low-$\dot{M}$ systems.

\subsection{Late-Stage Superhumps in WZ Sge Stars}\label{sec:latehumps}

   In well-observed WZ Sge-type dwarf novae, late-stage superhumps
having periods longer than those observed during the main superoutburst
have been reported.  Although there was already a suggestion in EG Cnc
\citep{pat98egcnc}, \citet{kat08wzsgelateSH} established the prevalence
of such superhumps in GW Lib, V455 And and WZ Sge.  \citet{Pdot}
further reported further detections in FL Psc (=ASAS J002511$+$1217.2),
ASAS J153616$-$0839.1, SDSS J0804 and OT J074727.6$+$065050.
In OT J213806.6$+$261957 (hereafter OT J2138), however,
a slightly shorter period was detected \citep{Pdot2}.
In this paper, we further detected long-period
late-stage superhumps in a new superoutburst of SDSS J0804 and in
SDSS J1339.  Since the variation in SDSS J1339 looks somewhat unique,
we first examine the phenomenon in this object, and present a survey
of the existing data of WZ Sge-type dwarf novae and similar systems
for detecting weaker periodicities other than the reported late-stage
superhumps.

\subsubsection{Late-Stage Superhumps in SDSS J1339}\label{sec:j1339latehump}

   The late-stage superhumps in SDSS J1339 were unique in that they
apparently showed a phase reversal, as in ``traditional" late superhumps
(\cite{sch80vwhyi}; \cite{vog83lateSH}), despite the fact that most of
late-stage superhumps (stage C superhumps) in ordinary SU UMa-type
dwarf novae do not show a phase reversal \citep{Pdot}.
The overall tendency of the $O-C$ diagram, however, resembles those of
GW Lib \citep{Pdot} and SDSS J0804 (subsection \ref{obj:j0804}),
and the $O-C$ variation could be understood without a phase jump
if the phase of rapid period decrease immediately following the end of
stage B was not observed due to a gap in observation.

   In interpreting the structure of these $O-C$ diagrams in \citet{Pdot},
we did not assume a phase reversal because the phases of superhumps
were observationally continuous.  Since this continuity may have been
superficial and there could have been multiple components or periods
as in V344 Lyr\footnote{
   An attempt to introduce a second component in interpreting the
   $O-C$ diagram in ER UMa was presented in \citet{Pdot}.
}, we made an alternative analysis following the method used in
V344 Lyr.  Since there were sometimes small secondary superhump maxima
during the plateau phase, we also measured the times of these secondary maxima
(the first eight epochs in table \ref{tab:j1339oc2011late}).
As seen in figure \ref{fig:j1339humpall}, these maxima do not seem
to be on a smooth extension of late-stage superhumps unless we assume
a discontinuous period change.  The situation appears to be different
from the case of V344 Lyr.

   The long period of persisting superhumps, however, is analogous
to that of persistent superhumps in V344 Lyr, and these superhumps
are expected to arise from the outermost region of the accretion disk.

   As in the case of Kepler observations of V344 Lyr, it would be
controversial whether these late-stage superhumps are ``traditional"
late-superhumps arising from the stream impact point or superhumps
arising from varying tidal dissipation.  Being a WZ Sge-type dwarf nova,
the system parameters, including the mass-transfer rate, are expected
to be different from V344 Lyr, and the light from the stream impact point
would be expected to be much less in V344 Lyr.  Although one might assume
an irradiation-induced enhanced mass-transfer, the lack of post-outburst
rebrightening suggests that the mass-transfer rate was not sufficient to
produce a normal outburst in contrast to V344 Lyr.  The situation may be
closer to low-$\dot{M}$ systems described in subsection
\ref{obj:kepv344lyr}.  We might speculate that the (1,0)-mode wave temporarily
vanished as the cooling front passed the disk, and the other weaker modes
remained and coupled with others to excite the (1,0)-mode, which was
by chance formed in the opposite direction to the original (1,0)-mode.

   The presence of the beat phenomenon might also provide
a clue to the origin of late-stage superhumps. In SDSS J1339, the singly
peaked orbital signal was clearly present during the post-superoutburst stage.
The actually observed light curve was not explained by a simple addition
of the orbital modulation to the superhumps with constant amplitudes,
and the amplitudes (subtracted for orbital modulations) of the superhumps
cyclically varied with a period of the beat period.\footnote{
   The same phenomenon appears to have been recorded in the SU UMa-type
   dwarf nova OY Car \citep{sch86oycar} during the first two nights
   of the post-superoutburst state.  Since OY Car is an eclipsing system,
   the condition may be different from the case of SDSS J1339.
}

   As a working hypothesis, we propose
that these cyclic variations were produced by the cyclically variable
eccentricity of the accretion disk precessing with the beat period
(cf. \cite{hir90SHexcess}):
when the disk becomes more circular, the amplitudes become smaller,
and when the disk becomes more eccentric, the amplitudes becomes larger.
This explanation would apply either to ``traditional" late-superhumps-type
explanation, assuming the varying release of the gravitational energy,
and or to the cyclic variation of the superhump light source itself
from varying tidal dissipation.

\subsubsection{Late-Stage Superhumps in Other Systems}

\begin{figure}
  \begin{center}
    \FigureFile(88mm,110mm){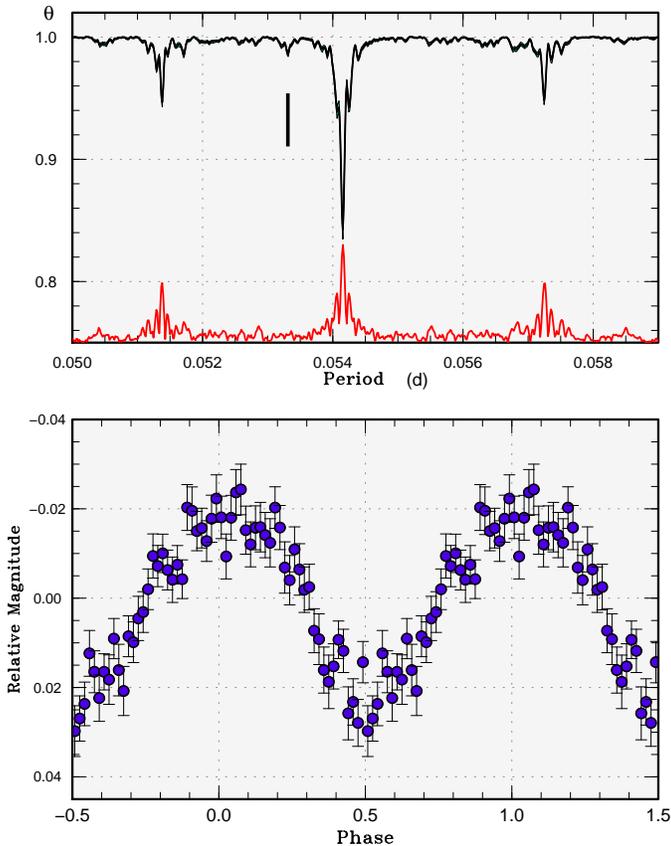}
  \end{center}
  \caption{Post-superoutburst superhumps in GW Lib (2007).
     (Upper): PDM analysis.  The strong signal at 0.054153(8) d is
     the post-superoutburst superhumps.
     The signal at 0.053310(2) d is the orbital period.
     The curve drawn in the bottom of the figure indicates the window
     function.
     (Lower): Phase-averaged profile of the orbital modulations.}
  \label{fig:gwlibpdmlate}
\end{figure}

   We have analyzed post-superoutburst observations of GW Lib
\citep{kat08wzsgelateSH}.  Although it was less striking than in SDSS J1339,
GW Lib also showed the orbital signal with an amplitude of 0.045 mag
and a period of 0.053310(2) d (figure \ref{fig:gwlibpdmlate}),
which is in agreement with the radial-velocity study [0.05332(2) d,
\cite{tho02gwlibv844herdiuma}].
A beat phenomenon in the superhump amplitudes was also recorded
with a smaller degree than in SDSS J1339.  The detection of orbital
signal is surprising since the inclination of the system is suggested
to be very low (11$^\circ$, \cite{tho02gwlibv844herdiuma}), and none
of quiescent observations detected orbital modulations
(\cite{vanzyl00gwlib}; \cite{wou02gwlib}; \cite{sch10gwlib}).
It appears that GW Lib and SDSS J1339 are very similar in their
outburst behavior, evolution of superhumps and the post-superoutburst
superhumps and beat phenomenon.

\begin{figure}
  \begin{center}
    \FigureFile(88mm,110mm){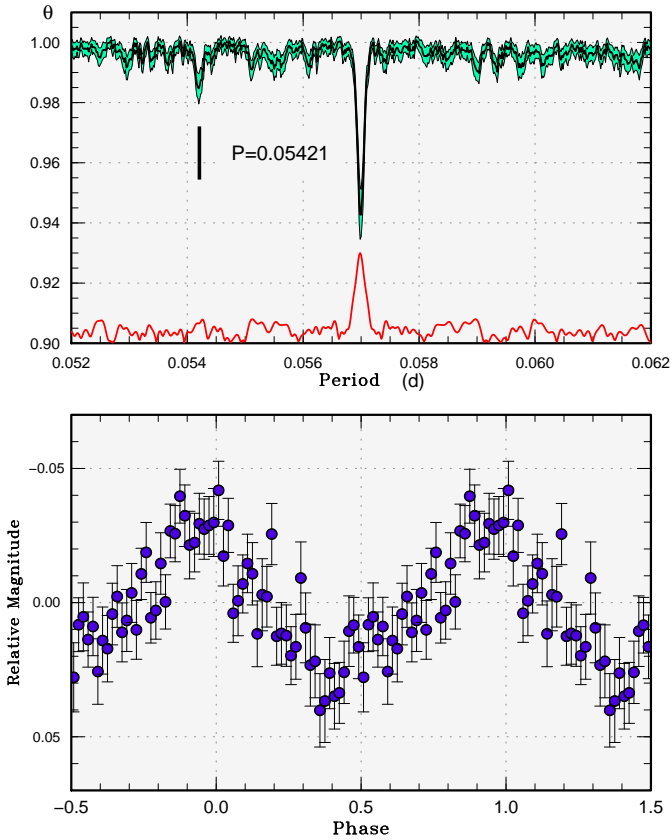}
  \end{center}
  \caption{Post-superoutburst superhumps in FL Psc (2004).
     (Upper): PDM analysis.  The strong signal at 0.05700(3) d is
     the post-superoutburst (positive) superhumps.
     The signal at 0.054210(8) d is likely negative superhumps.
     The curve drawn in the bottom of the figure indicates the window
     function.  The signal at 0.054210 d is not an alias period of
     the (positive) superhumps.
     (Lower): Phase-averaged profile by a period of 0.054210 d.}
  \label{fig:flpscpdmlate}
\end{figure}

   An analysis of post-superoutburst observations of FL Psc \citep{Pdot}
yielded a strong periodicity at 0.054210(8) d with a mean amplitude
of 0.08 mag (figure \ref{fig:flpscpdmlate}).
The period is too short for an orbital period (5 \% shorter than the
superhump period during the main outburst), and is most likely identified
as negative superhumps. This is the first indication of negative superhumps
during the post-superoutburst stage of WZ Sge-type dwarf novae.

   As discussed in subsection \ref{obj:kepv1504cygtrans}, the state
with negative superhumps is considered to have a tendency to suppress
the disk instability to occur.  The appearance of negative superhumps
might explain the reason why multiple rebrightenings did not occur.
There has been a suggestion that there is a minimum mass-transfer
rate to generate a tilt, which has been proposed as the mechanism for
negative superhumps \citep{mon10disktilt}.  Although the quiescent
mass-transfer rates in WZ Sge-type dwarf novae generally are lower than
this limit, the mass-transfer rate may have exceeded this limit in outburst
and generated a tilt and negative superhumps.  The generation of tilts
in WZ Sge-type dwarf novae may affect the dynamics of superoutbursts
than had been supposed and requires further investigation.
Further systematic, sufficiently long, observations during the
post-superoutburst states of WZ Sge-type or SU UMa-type dwarf novae
are needed.

   We also propose a candidate orbital period of 0.056096(5) d, with
a mean amplitude of 0.06 mag.  This candidate period is well in agreement
with the predicted orbital period of 0.05609 d from stage B superhumps
(subsection \ref{sec:pshporb}).  Assuming this orbital period, fractional
superhump excesses for positive (stage B) and negative superhumps are
1.8 \% and $-$3.4 \%, respectively.  It is known that fractional excesses
(in absolute values) for positive superhumps are approximately twice larger
in long-$P_{\rm orb}$ systems (cf. \cite{pat97v603aql};
table 2 of \cite{mon10negSH}), and the larger fractional excess for
negative superhumps in FL Psc is unusual.  It has been established, however,
the fractional excesses (in absolute values) for positive and negative
superhumps in ER UMa are almost equal \citep{ohs11eruma}, and there was
a suggestion of a large negative fractional excess for the supposed
negative superhumps in V1159 Ori \citep{pat95v1159ori}.
The negative superhumps in very short-$P_{\rm orb}$ systems may have
different properties from those in traditional negative superhumps
in long-$P_{\rm orb}$ systems.

\begin{figure}
  \begin{center}
    \FigureFile(88mm,110mm){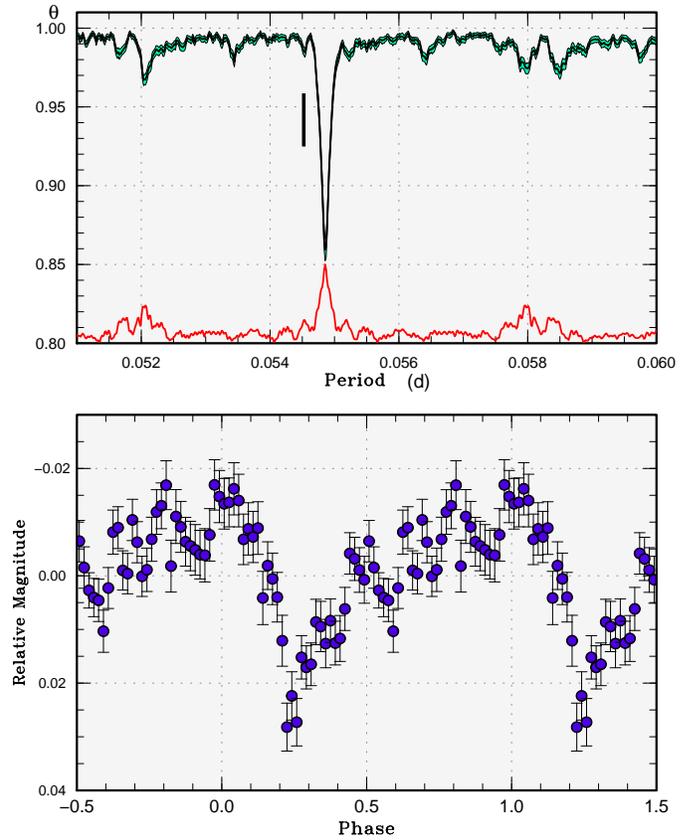}
  \end{center}
  \caption{Post-superoutburst superhumps in OT J2138 (2010).
     (Upper): PDM analysis.  The strong signal at 0.054853(6) d is
     the post-superoutburst superhumps.
     The signal at 0.054523(4) d is identified with the orbital period.
     The curve drawn in the bottom of the figure indicates the window
     function.
     (Lower): Phase-averaged profile by a period of 0.054523 d.}
  \label{fig:j2138pdmlate}
\end{figure}

   An analysis of the post-superoutburst observations of OT J2138 (2010,
BJD after 2455348) yielded two periodicities (figure \ref{fig:j2138pdmlate}):
persistent superhumps with a period of 0.054853(6) d and
an amplitude of 0.11 mag, the orbital period of 0.054523(4) d and
an amplitude of 0.05 mag.  The other signals were one-day or two-day
aliases of the superhumps.
The orbital period determined from late-stage observations
is consistent with the period of early superhumps.
The fractional excess of stage B superhumps has been refined to be 0.91 \%.

   The origin of orbital signals in these systems during the
post-superoutburst stage is unclear, although they were detected
in several WZ Sge-type dwarf novae.
Although it might have originated from the irradiated secondary
by the heated white dwarf, the low inclination in GW Lib would make
this interpretation difficult.  The assymmetric profile in OT J2138
resembles an combination of a shallow eclipse and a orbital hump.
Although the amplitudes were too small to be regarded as true eclipses,
geometrical effect can be responsible for systems with intermediate
inclinations.

\subsection{Amplitudes of Superhumps}\label{sec:humpamp}

   \citet{sma10SHamp} recently raised a question against the widely
believed descriptions about amplitudes of superhumps.  \citet{sma10SHamp}
summarized as follows:

\begin{enumerate}

\item Superhumps first appear around superoutburst maximum
and reach their highest amplitude either at maximum or 1-2
days later.
\item Superhump amplitudes decrease during superoutburst; this effect,
however, has been well documented only for relatively few cases.
\item At their full development the superhumps have a range of 0.3-0.4 mag,
and are equally prominent in all SU UMa stars independent of inclination
(\cite{war95book}, p.194).
\item There is no strong modulation of the superhump profile at the
beat period (\cite{war85suuma}, p.372).

\end{enumerate}

   The claim by \citet{sma10SHamp} was that the points 3 and 4 are wrong.
These observational findings were, however, derived from observations
in the 1970's and 1980's, when most of the knowledges in superhumps
came from regularly outbursting bright southern SU UMa-type dwarf novae,
i.e. relatively long-$P_{\rm orb}$ objects with higher mass-transfer
rates.

   Since the modern accumulated data cover much wider range of objects,
we first present statistical descriptions of superhump amplitudes,
and examine the items individually.  Note that we already
presented a typical variation of superhump amplitudes in figure 3
of \citet{Pdot}.

   In this paper, we present a statistical analysis and treat only a pilot
study of representative objects.  We are going to devote ourselves in
a separate work
to more comprehensive survey of the amplitudes of superhumps as well as
detailed data on individual objects.

\subsubsection{Amplitudes of Superhumps: Samples and General Behavior in
   Non-Eclipsing Systems}\label{sec:shampgeneral}

   The samples are observations in \citet{Pdot}, \citet{Pdot2} and
this paper.  We restricted analysis to superoutbursts in which
the starts of stage B were determined.  Additional criterion for choosing
samples was that more than 20 epochs of superhumps maxima were measured.
The resultant samples are listed in table \ref{tab:ampoutlist}.
The amplitudes of superhumps
were measured using the template fitting method described in \citet{Pdot},
and the derived amplitudes are not literally full minimum-to-maximum
amplitudes in usual sense but more reflect a kind of first-order moment.
It has become evident that some (less than 1 \%) of recorded maxima
tabulated in \citet{Pdot} were inadequate for measuring amplitudes because
they were only observed around superhump maxima.  We rejected these maxima
in the analysis.

\begin{table}
\caption{Superoutbursts used in analyzing superhump amplitudes.}\label{tab:ampoutlist}
\begin{center}
\begin{tabular}{lcclcc}
\hline
Object\commenta & Year & $P_{\rm SH}$ & Object\commenta & Year & $P_{\rm SH}$ \\
\hline
V455 And & 2007 & 0.0571 & WZ Sge & 2001 & 0.0572 \\ 
V466 And & 2008 & 0.0572 & V551 Sgr & 2003 & 0.0676 \\ 
DH Aql & 2002 & 0.0800 & V1212 Tau & 2011 & 0.0701 \\ 
VY Aqr & 2008 & 0.0647 & FL TrA & 2005 & 0.0599 \\ 
EG Aqr & 2006 & 0.0790 & SU UMa & 1999 & 0.0791 \\ 
BG Ari & 2010 & 0.0849 & SW UMa & 2000 & 0.0583 \\ 
TT Boo & 2004 & 0.0781 & SW UMa & 2006 & 0.0582 \\ 
NN Cam & 2009 & 0.0743 & SW UMa & 2010 & 0.0582 \\ 
HT Cas & 2010 & 0.0763 & BC UMa & 2003 & 0.0646 \\ 
V1040 Cen & 2002 & 0.0622 & BZ UMa & 2007 & 0.0702 \\ 
WX Cet & 1998 & 0.0595 & DV UMa & 1997 & 0.0888 \\ 
GO Com & 2003 & 0.0631 & DV UMa & 2007 & 0.0885 \\ 
V632 Cyg & 2008 & 0.0658 & IY UMa & 2000 & 0.0758 \\ 
V1028 Cyg & 1995 & 0.0617 & IY UMa & 2006 & 0.0761 \\ 
V1504 Cyg & 2009b & 0.0722 & IY UMa & 2009 & 0.0762 \\ 
HO Del & 2008 & 0.0644 & KS UMa & 2003 & 0.0702 \\ 
KV Dra & 2009 & 0.0601 & HV Vir & 2002 & 0.0583 \\ 
XZ Eri & 2007 & 0.0628 & HV Vir & 2008 & 0.0583 \\ 
XZ Eri & 2008 & 0.0628 & 1RXS J0423 & 2008 & 0.0784 \\ 
UV Gem & 2003 & 0.0935 & 1RXS J0423 & 2010 & 0.0785 \\ 
V592 Her & 2010 & 0.0566 & ASAS J1025 & 2006 & 0.0634 \\ 
V1108 Her & 2004 & 0.0575 & ASAS J1536 & 2004 & 0.0646 \\ 
RU Hor & 2008 & 0.0710 & ASAS J1600 & 2005 & 0.0650 \\ 
MM Hya & 2011 & 0.0589 & Mis V1443 & 2011 & 0.0567 \\ 
RZ Leo & 2000 & 0.0787 & SDSS J0804 & 2010 & 0.0596 \\ 
GW Lib & 2007 & 0.0541 & SDSS J0812 & 2011 & 0.0779 \\ 
V344 Lyr & 2009 & 0.0916 & SDSS J1146 & 2011 & 0.0633 \\ 
V344 Lyr & 2009b & 0.0916 & SDSS J1227 & 2007 & 0.0646 \\ 
V585 Lyr & 2003 & 0.0604 & SDSS J1339 & 2011 & 0.0581 \\ 
DT Oct & 2003 & 0.0748 & SDSS J1627 & 2008 & 0.1097 \\ 
V2527 Oph & 2004 & 0.0721 & OT J0238 & 2008 & 0.0537 \\ 
UV Per & 2000 & 0.0666 & OT J1443 & 2009 & 0.0722 \\ 
UV Per & 2003 & 0.0667 & OT J1610 & 2009 & 0.0578 \\ 
UV Per & 2007 & 0.0663 & OT J1625 & 2010 & 0.0961 \\ 
QY Per & 1999 & 0.0786 & OT J2138 & 2010 & 0.0550 \\ 
\hline
  \multicolumn{6}{l}{\commenta Abbreviations for \citet{Pdot} objects:} \\
  \multicolumn{6}{l}{ASAS J1536: ASAS J153616$-$0839.1} \\
  \multicolumn{6}{l}{ASAS J1600: ASAS J160048$-$4846.2} \\
  \multicolumn{6}{l}{SDSS J1627: SDSS J162718.39$+$120435.0} \\
  \multicolumn{6}{l}{OT J0238: OT J023839.1$+$355648 = CT Tri} \\
  \multicolumn{6}{l}{OT J1443: OT J144341.9$-$175550} \\
\end{tabular}
\end{center}
\end{table}

\begin{figure}
  \begin{center}
    \FigureFile(88mm,70mm){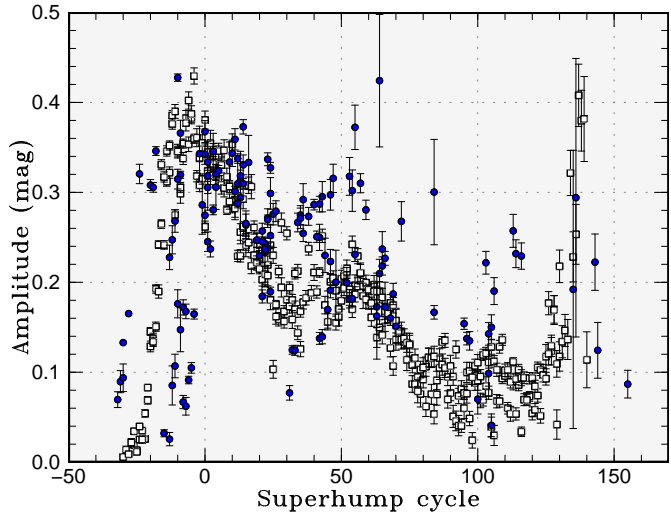}
  \end{center}
  \caption{Variation of superhump amplitudes in systems with $P_{\rm orb}$
    longer than 0.08 d.  Open squares represent Kepler observations of
    V344 Lyr.}
  \label{fig:humpamp1}
\end{figure}

\begin{figure}
  \begin{center}
    \FigureFile(88mm,70mm){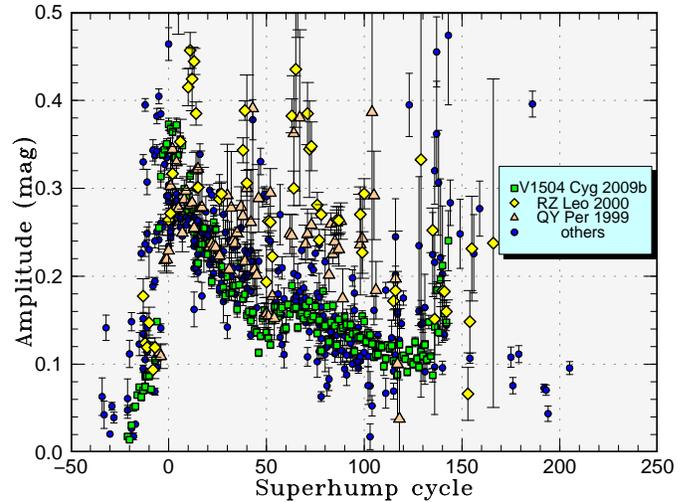}
  \end{center}
  \caption{Variation of superhump amplitudes in systems with
    $0.07 < P_{\rm orb} {\rm (d)} \le 0.08$.
    Filled squares represent Kepler observations of
    V1504 Cyg.  Two objects with large-amplitude superhumps (RZ Leo
    and QY Per) are shown in different symbols.}
  \label{fig:humpamp2}
\end{figure}

\begin{figure}
  \begin{center}
    \FigureFile(88mm,70mm){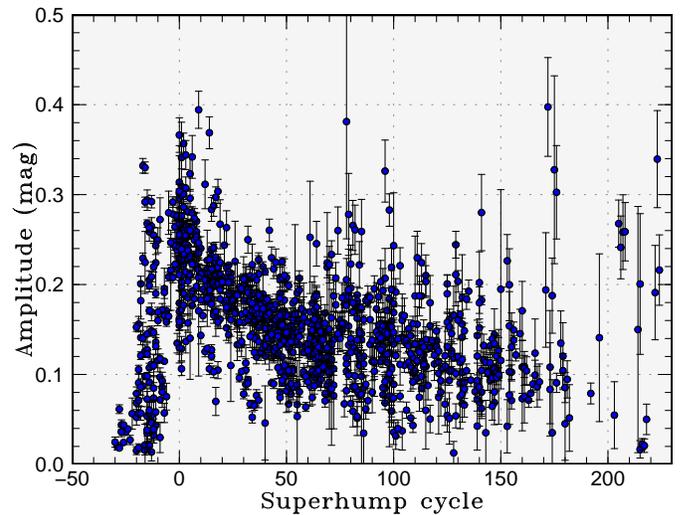}
  \end{center}
  \caption{Variation of superhump amplitudes in systems with
    $0.06 < P_{\rm orb} {\rm (d)} \le 0.07$.}
  \label{fig:humpamp3}
\end{figure}

\begin{figure}
  \begin{center}
    \FigureFile(88mm,70mm){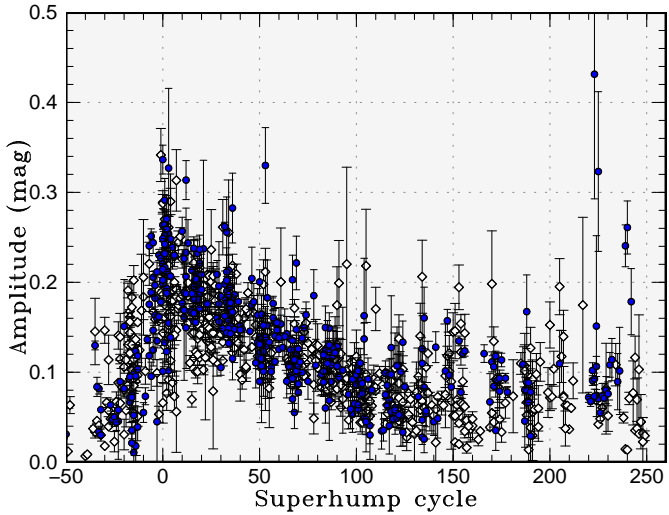}
  \end{center}
  \caption{Variation of superhump amplitudes in systems with
    $P_{\rm orb} {\rm (d)} \le 0.06$.
    Filled circles and open diamonds represent
    ordinary SU UMa-type dwarf novae and extreme WZ Sge-type dwarf novae,
    respectively.}
  \label{fig:humpamp4}
\end{figure}

   The variation of superhump amplitudes in non-eclipsing systems
are shown in figures \ref{fig:humpamp1}--\ref{fig:humpamp4}.
The superhump amplitudes are plotted against cycle numbers after
the start of stage B.  Since the starts of stage B
were measured from $O-C$ variations (shown in tables in \cite{Pdot},
\cite{Pdot2} and in this paper), they can be different from the maxima
of superhump amplitudes.  There is also a tendency of 1-d periodicities
in the density of data, resulting from uneven distribution of
observations against longitudes.  There should have been typically
this degree (10--15 cycles) of uncertainty in determining the starts
of stage B.
Since the quality of the data were sometimes low, there are sometimes
a number of scattered points in large-amplitude
regions.  They should be better considered as a result of a selection bias
when the errors in the original data were large.  Although there must
have equally been scattered points in low-amplitude regions, they
were not detected either due to poor convergence in numerical fitting
or because the resultant amplitudes were negative.  Nevertheless,
the majority of the measured data are in relatively narrow strips
in each figure, and we can regard these strips as typical behavior of
amplitude variation.
The large increase in the amplitude at late stages (e.g. $E > 130$ in
figure \ref{fig:humpamp1}) corresponds to the final fading of superoutbursts.

\begin{figure}
  \begin{center}
    \FigureFile(88mm,70mm){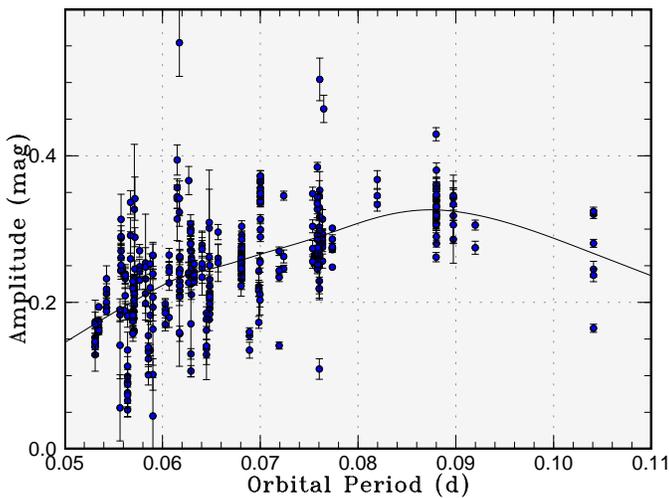}
  \end{center}
  \caption{Dependence of superhump amplitudes on orbital period.
    The superoutburst samples are described in subsection
    \ref{sec:shampgeneral}.  We selected epochs $-5 < E < 10$ to illustrate
    the maximum superhump amplitudes.  The curve indicates a spline-smoothed
    interpolation.}
  \label{fig:humpampporb}
\end{figure}

   Several features can be seen in these figures:

\begin{enumerate}

\item The amplitudes of superhumps usually reach a maximum around the
stage A--B transition.  This means that the time before the system reaches
the peak of superhump amplitude is variable between objects or between
superoutbursts.

\item Although the amplitudes of superhumps globally decrease with time,
secondary peaks of amplitudes are usually seen $\sim$50 cycles after
the main peak in long-$P_{\rm orb}$ systems (figure \ref{fig:humpamp1}) and
$\sim$70--80 cycles after the main peak in intermediate- or
short-$P_{\rm orb}$ systems (figures \ref{fig:humpamp2}, \ref{fig:humpamp3}).
In short-$P_{\rm orb}$ systems, the secondary peaks become less evident
(figure \ref{fig:humpamp4}).  The epochs of the secondary peaks
are close to the epochs of stage B--C transitions.  In some cases,
the secondary peaks appear later than stage B--C transitions.

\item The amplitude of superhumps are well-correlated with $P_{\rm orb}$:
longer-$P_{\rm orb}$ systems show larger amplitudes.
The relation for maximum amplitudes of superhumps is shown in
figure \ref{fig:humpampporb}.  Figures for other stages of superoutburst
basically show the same trend.  Warner's point 3 (0.3--0.4 mag for
maximum amplitudes) apparently came from the long-$P_{\rm orb}$ samples
in the older times.  The maximum amplitudes seem to decrease for
longer-$P_{\rm orb}$ ($P_{\rm orb} > 0.09$ d) objects.

\item Two systems (RZ Leo and QY Per) show systematically larger
amplitudes of superhumps.  These systems have long recurrence times
similar to WZ Sge-type dwarf novae, but with longer $P_{\rm orb}$.

\item The amplitudes of superhumps in extreme WZ Sge-type (we refer to
systems with recurrence times longer than $\sim$10 yr or systems with
multiple rebrightenings) are generally smaller than in ordinary
SU UMa-type dwarf novae.

\end{enumerate}

\subsubsection{Amplitudes of Superhumps: Dependence on Orbital Inclination}\label{sec:shampinc}

\begin{figure}
  \begin{center}
    \FigureFile(88mm,70mm){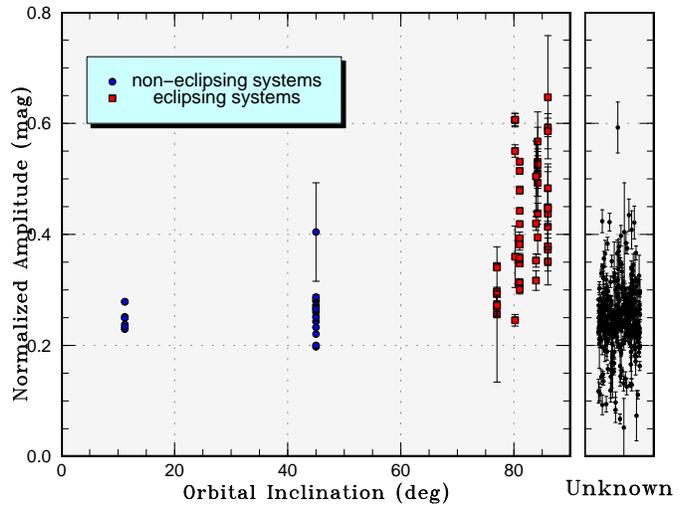}
  \end{center}
  \caption{Dependence of superhump amplitudes on orbital inclination.
    The superoutburst samples are described in subsection
    \ref{sec:shampgeneral}.  We selected epochs $-5 < E < 10$ to illustrate
    the maximum superhump amplitudes.  The amplitudes were normalized
    by using the dependence on superhump periods.  The right panel indicates
    amplitudes for objects without known orbital inclinations.}
  \label{fig:humpampinc}
\end{figure}

   We further studied the dependence of superhump amplitudes on orbital
inclination ($i$).  As in subsection \ref{sec:shampgeneral}, we only used
epochs $-5 < E < 10$ to illustrate the maximum superhump amplitudes.
Due to the strong dependence of superhump amplitudes
on $P_{\rm orb}$, we first removed this effect by using a relation
derived in subsection \ref{sec:shampgeneral} (curve in figure
\ref{fig:humpampporb}).  The normalization was taken so that the average
of amplitudes agrees with the pre-normalized value of the sample (0.25 mag).
We used $i$ values in the on-line version 7.15 of \citet{RitterCV7}.
As seen in figure \ref{fig:humpampinc}, the normalized maximum amplitudes
do not strongly depend on $i$ for systems with $i < 80^{\circ}$,
including an eclipsing system WZ Sge ($i = 77^{\circ}$).
The right panel of the figure shows the amplitudes for non-eclipsing
SU UMa-type dwarf novae without known inclinations.  The 90 \% of
the normalized amplitudes without known inclinations fall in the range
of 0.14--0.35 mag, well in agreement with $i < 80^{\circ}$ samples.
This result seems to confirm that Warner's point 3 is basically valid for
non-eclipsing systems.  It is worth noting that there is no indication
of smaller amplitudes in systems with very low inclinations.

\subsubsection{Amplitudes of Superhumps: Beat Phenomenon}\label{sec:shampbeat}

   As we have already shown, the beat phenomenon in amplitudes of superhumps
are clearly present in eclipsing objects (e.g. HT Cas, subsection
\ref{obj:htcas} and SDSS J0804 subsection \ref{obj:j0804}).
While it is not clear what the term ``profile'' exactly mentioned
in Warner's point 4, the present observations seem to support the
presence of a beat phenomenon (at least in amplitudes) in eclipsing systems.

\begin{figure}
  \begin{center}
    \FigureFile(88mm,70mm){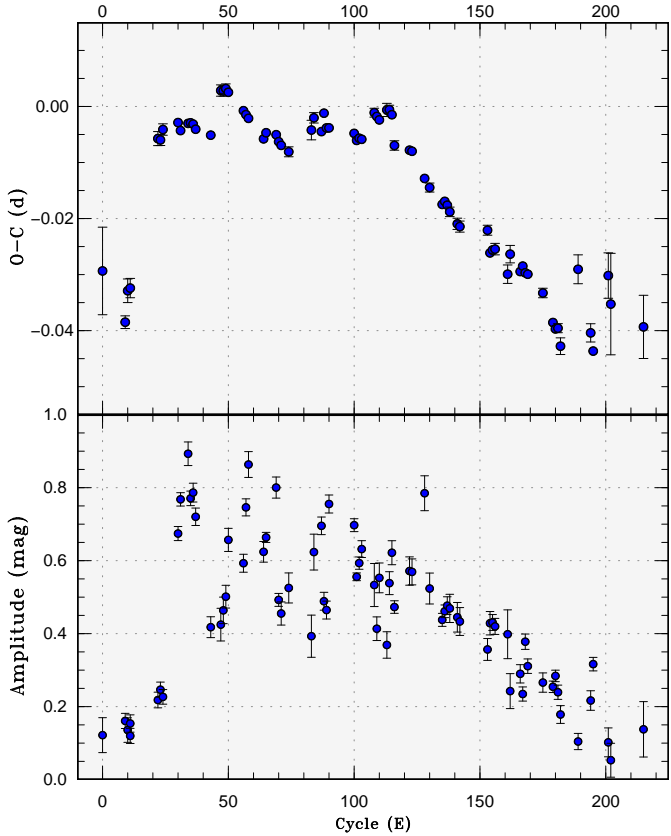}
  \end{center}
  \caption{Beat phenomenon in IY UMa (2009). (Upper): $O-C$ diagram.
     (Lower): Amplitudes of superhumps.  Although the beat phenomenon was
     strongly seen during stage B, it became weaker during stage C.}
  \label{fig:iyuma2009amp}
\end{figure}

   In figure \ref{fig:iyuma2009amp}, we show variations of superhump
amplitudes in IY UMa (2009), which was reported in \citet{Pdot2}.
Although the beat phenomenon was strongly seen during stage B,
it became weaker with time and almost disappeared during stage C.
In HT Cas (2010, figure \ref{fig:htcasbeat}), however, the beat phenomenon
persisted during its entire plateau phase.  It would be worth noting
that double-wave modulation in amplitude mentioned in \citet{sma10SHamp}
($A_2$ in his designation) is not apparent even in the best observed
case of HT Cas.

\begin{figure}
  \begin{center}
    \FigureFile(88mm,70mm){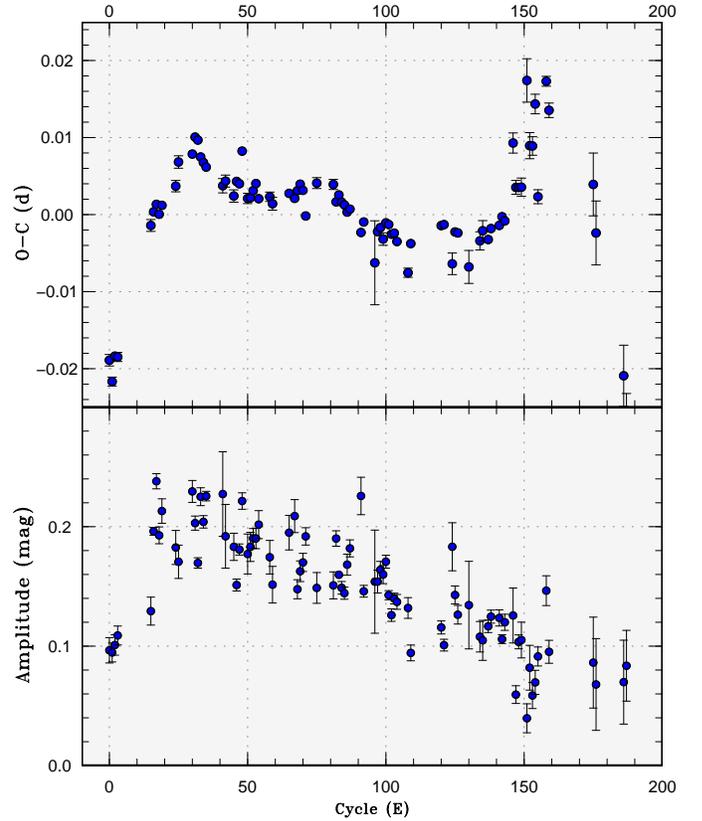}
  \end{center}
  \caption{Beat phenomenon in SDSS J0804 (2010). (Upper): $O-C$ diagram.
     (Lower): Amplitudes of superhumps.  The beat phenomenon is much
     weaker than in IY UMa (2009), and became inapparent during the
     late plateau phase.}
  \label{fig:j0804beat}
\end{figure}

   The beat phenomenon is much weaker in SDSS J0804 (2010, figure
\ref{fig:j0804beat}) than in IY UMa (2009), and became inapparent during
the late plateau phase.

   Among non-eclipsing systems, RZ Leo, likely a high-inclination
system (\cite{men01rzleo}; \cite{ish01rzleo}) showed the presence of
the beat phenomenon [modulations with a 30.6(1)-cycle period is already
apparent in figure \ref{fig:humpamp2}].  The mean amplitude of modulations
($A_1$ in \cite{sma10SHamp}) was 0.07(2) mag.  QY Per, another outlier in
figure \ref{fig:humpamp2}, has a candidate weak beat phenomenon with
a period of 28.5(2) cycles and $A_1$ of 0.02(1) mag.  Since the signal
was weak and the true orbital period is unknown, this needs to be
confirmed by further studies.  We could not confirm the presence of
the beat phenomenon in TT Boo (2004).  In KS UMa (2003), we found
a weak beat phenomenon with $A_1$ of 0.02(1) mag whose beat period
is consistent with the known $P_{\rm orb}$.

   The degree of the beat phenomenon during the superoutburst plateau
appears to be related to the maximum amplitudes of superhumps, and appears
to be smaller in WZ Sge-type dwarf novae.

   The beat phenomenon in non-eclipsing systems, especially in
short-$P_{\rm orb}$ systems, was undetected or close to the detection limit.
This finding seems to be consistent with numerical modeling of
inclination effect on superhumps \citep{sim98SHincl}, in which only
systems with $i$ larger than 45--60$^{\circ}$ show significant
orbital modulation of superhumps.

\subsubsection{Amplitudes of Superhumps: Discussion}\label{sec:shampint}

   Although \citet{sma10SHamp} discussed a new interpretation on
the origin of superhumps in combination with his earlier findings
(e.g. \cite{sma09zcha}; \cite{sma09SH}), we restrict our discussion
within the traditional framework (cf. \cite{osa89suuma};
\cite{hir90SHexcess}; \cite{osa96review})
since we do not have sufficient materials to compare with the analysis
by earlier works by Smak.\footnote{
   This particularly refers to a best-observed southern eclipsing
   SU UMa-type dwarf novae Z Cha and OY Car, for which we have virtually
   no new data.
   The new data in HT Cas, however, seem to provide some degree of
   negative evidence (subsection \ref{obj:htcas}) to Smak's idea.
   Since the difference in the presence of a hot spot during superoutburst
   or the irradiated secondary might be dependent on object [for example,
   in HT Cas, the frequency of outbursts is unusually low,
   and orbital humps in quiescence is often inapparent
   (\cite{woo92htcas}; \cite{fel05gycncircomhtcas}).
   HT Cas therefore might not be an ideal object representing the entire
   SU UMa-type population], and we should
   leave the problem as a future work.  Detailed modeling of the eclipse
   and superhump light curve assuming a vertically extended disk is
   indispensable for further discussion, which is clearly outside the scope
   of this paper.
}

   In non-eclipsing systems, there is evidence that our normalized
amplitudes of superhumps were fairly constant over a wide range of
orbital inclinations ($i \le 80^{\circ}$).\footnote{
   This way of normalization of amplitudes is different from the one
   in \citet{sma10SHamp}.  We employed a normalization for
   $P_{\rm orb}$-dependence while \citet{sma10SHamp} introduced
   a concept of normalization to the average luminosity of the disk.
   This concept by \citet{sma10SHamp} implicitly assumes that
   the observed light from the disk
   suffers from the $i$-dependent projection and obscuration effects,
   while the emission from the superhump light source is isotropic,
   or observable independent of $i$.
}
This can be understood
if the superhump light source has almost the same $i$-dependence
as the background disk light.  For geometrically thin, and optically thick
accretion disks, the projection effect ($\cos i$) plays the strongest role
in low-to-intermediate $i$.  This would imply that the $i$-dependence
of the superhump light source can be explained as well, such as
in a form of the (superhump phase-dependent) variations of
the surface brightness (or the temperature) on a geometrically thin disk.
The apparent lack of deviations from this
relation up to $i \sim 80^{\circ}$ suggest that the vertical structure
in the superhump light source is smaller than $10^{\circ}$ in slope
against the remaining disk.  This value is not much greater than
the flaring angle of 3--6$^{\circ}$ inferred for a normal outburst
of Z Cha \citep{rob99zcha}, and may not require a special mechanism
other than additional tidal heating around the superhump light source.

   In deeply eclipsing systems, the amplitudes are systematically higher
than the above simple assumption of the projection effect.
Although this phenomenon can likely be understood as a consequence
of the vertical expansion around the superhump light source,
even a simple treatment should require numerical estimation of different
contribution from a vertically extended and structured disk,
self-obscuration and eclipse by the secondary, and will be
an elaborate future work.

   The dependence of amplitudes on $P_{\rm orb}$ may be a combined
effect of the strength of the tidal torque (larger in higher-$q$
systems) and the strength of the 3:1 resonance (which will be expected
to be smaller in higher-$q$ systems around the border of the instability
condition for tidal instabilities).  This would explain a maximum
of amplitudes around $P_{\rm orb} = 0.09$ d.
Although not included in our samples, the low ($\sim$0.10--0.15 mag)
maximum amplitudes in RZ LMi (\cite{nog95rzlmi}; \cite{rob95eruma};
\cite{ole08rzlmi}) -- would be an outlier for a $P_{\rm SH}$ = 0.0594 d
object -- may be similarly explained by a supposed lower tidal torque
than in other SU UMa-type dwarf novae as proposed in
\citet{osa95rzlmi} and \citet{hel01eruma}.

   There is an indication that amplitudes of superhumps are higher
in systems with longer intervals of outbursts (RZ Leo and QY Per)
among long-$P_{\rm orb}$ systems.  These systems may have stored
larger disk mass, either as a result of a lower quiescent viscosity or
by other mechanisms, and the superhump light source or the vertical
extent of the disk may be more pronounced at the time of the outburst.
The higher vertical extent might also explain the very strong beat
phenomenon in HT Cas.

   In extreme WZ Sge-type dwarf novae, the superhump amplitudes tend
to be lower and the degree of the beat phenomenon is weaker than
in other SU UMa-type dwarf novae.  This may be explained if the
most of the disk mass outside the 3:1 resonance is efficiently accreted
during the stage of early superhumps [the development of the 2:1
resonance is expected to suppress the 3:1 resonance (\cite{lub91SHa};
\cite{osa95wzsge})] and there is relatively little amount of mass
at the onset of development of ordinary superhumps.
Following this interpretation, the maximum amplitudes of superhumps
in WZ Sge-type dwarf novae would be a discriminating feature in
estimating the strength of the 2:1 resonance prior to the development
of ordinary superhumps.

   The relatively weak beat phenomenon in stage C superhumps
(subsection \ref{sec:shampbeat}) would also deserve attention.
The weaker signal of the beat phenomenon would suggest a lower
vertical extent for stage C superhumps.  An examination whether
the lower temperature of the light source of stage C superhumps
or its location, can explain the phenomenon could shed light on
clarifying the nature of still poorly understood stage C superhumps.

\section{Summary}

   We studied the characteristics of superhumps for 51 SU UMa-type
dwarf novae whose superoutbursts were mainly observed during the 2010--2011
season.  In addition to the purpose of our previous surveys focusing on
systematic variation of superhump periods, we extended our analysis to
post-superoutburst variations and a survey of variations of superhump
amplitudes for selected objects.  We also analyzed public Kepler data
for V1504 Cyg and V344 Lyr.
Most of the new data for systems with short superhump periods basically
confirmed the earlier findings.
We recorded early superhumps for four WZ Sge-type dwarf novae
(SDSS J0804, SDSS J1339, OT J0120, and likely SDSS J1605).
We also refined the ephemerides of eclipsing systems HT Cas, SDSS J0932,
SDSS J1227 and OT J0431, and provided firm estimates of fractional
superhump excesses for these objects.
We have also updated statistics in relations
$P_{\rm SH}$--$P_{\rm orb}$ for stage B and C superhumps,
$P_{\rm orb}$--$P_{\rm dot}$ and 
$P_{\rm orb}$--$\epsilon$ for stage B superhumps.
The traditional \citet{sto84tumen} relation gives systematically longer
$P_{\rm orb}$ estimates for stage B superhumps.

In addition to them, we found:

\begin{itemize}

\item The spread of period derivative is broader in long-$P_{\rm orb}$
systems.  In particular, GX Cas unexpectedly showed a large positive
$P_{\rm dot}$.

\item We recorded the long-waited superoutburst of
the eclipsing SU UMa-type dwarf nova HT Cas, and described the full
evolution of superhumps during its earliest stage to post-superoutburst
stage.  A very strong beat phenomenon was recorded particularly when
superhumps reached the full amplitudes.

\item There was no particularly indication of the reflection effect
and conspicuous appearance of the hot spot in HT Cas as judged from
the light curves outside the eclipses.  These results do not favor
an interpretation assuming an irradiation-induced enhanced mass-transfer.

\item In SDSS J0804, we for the first time recorded full evolution of
early superhumps and ordinary superhumps, as well as persisting superhumps
during six rebrightenings.

\item Shallow eclipses were detected during the entire course of
outbursts in SDSS J0804.  The beat phenomenon was also recorded
with a smaller degree than in HT Cas.

\item Persisting superhumps during the rebrightening phase of SDSS J0804
and WZ Sge (2001) had amplitudes well-correlated with the brightness
of the system.  The relation can be understood assuming the almost
constant luminosity of the superhump light source.  The luminosity
of the light source showed a decrease just before or around
the final rebrightening.

\item The $O-C$ diagrams for three WZ Sge-type dwarf novae SDSS J0804, 
SDSS J1339 and OT J0120 showed a similar pattern to that of GW Lib during
the final fading phase.  In SDSS J0804, a rapid excursion to a shorter
period was recorded.  This phase was probably missed in SDSS J1339,
and the resultant $O-C$ diagrams even looked like a phase reversal
as would be expected for ``traditional'' late superhumps.

\item In SDSS J1339, there was a strong beat phenomenon in the amplitudes
of persistent superhumps even during the post-superoutburst phase.
We interpret this feature as a manifestation of cyclically variable
shape of the eccentric disk with the beat period.

\item The overall behavior in OT J0120 was very similar to that of
WZ Sge (2001), consisting of a post-outburst ``dip'' and multiple
rebrightenings with short intervals.

\item We studied a new superoutburst of the helium dwarf nova V406 Hya
and examined its past outburst activities.

\item Kepler data for V1504 Cyg and V344 Lyr confirmed the presence of
stage A and B superhumps, and the details of stage A evolution, as well as
relation to the precursor phenomenon, were described.

\item In V344 Lyr, secondary maxima of superhumps appeared during the
late stage of superoutbursts, and this signal gradually merged with the
primary maxima to form singly-humped superhumps.
The initial periods of secondary maxima were longer than those of
primary maxima, suggesting that the secondary maxima were formed in
the outer part of the disk.  These superhumps
persisted for at least two cycles of subsequent normal outbursts.
We discussed the role of (1,3)-mode wave in manifestation of doubly-humped
superhumps.

\item In V1504 Cyg, the behavior was different fro V344 Lyr in that
secondary maxima only transiently appeared.

\item In V1504 Cyg, we detected a ``failed superoutburst'' in which
a precursor outburst and a subsequent further rise was recorded.
During this outburst, negative superhumps instead of ordinary (positive)
superhumps were excited.  The appearance of negative superhumps may be
related to premature quenching of the superoutburst.

\item SDSS J0804 (2010) showed a large $P_{\rm dot}$ for a WZ Sge-type
dwarf nova with multiple rebrightenings.  This may have been resulted
from a rapid increase of the superhump period at the end of the plateau
phase.  This phenomenon seems to be commonly seen in eclipsing
WZ Sge-type dwarf novae.

\item We have detected likely negative superhumps during the
post-superoutburst stage of FL Psc in 2004, first times recorded
among WZ Sge-type dwarf novae.  We suggested a possibility that the state
with negative superhumps (possibly a state with a tilted disk) can affect
the development of the superoutburst itself, particularly the presence of
rebrightenings.

\item We have detected orbital modulations in a low-inclination
WZ Sge-type dwarf nova GW Lib during its post-superoutburst in 2007.
We have also identified the orbital period from post-superoutburst
observations of OT J2138 in 2010.

\item The amplitudes of superhumps usually show complex variations.
The amplitudes quickly grow during stage A and reach a maximum
around the stage A--B transition.  Although the amplitude decay
approximately linearly during stage B, many objects show significant
regrowth of amplitudes around or after the transition to stage C.

\item The amplitudes of superhumps are strongly related to the orbital period.
Long-period systems have larger amplitudes.

\item After correcting for the dependence on the orbital period,
there is little indication of dependence on the system inclination
for $i < 80^{\circ}$.

\item In eclipsing systems, the amplitudes of superhumps are larger,
and strongly modulated with the beat period.

\item The beat phenomenon of superhumps in eclipsing systems decrease
with the progress of the superoutburst.  The beat phenomenon is
significantly reduced in stage C superhumps.

\item In WZ Sge-type eclipsing systems, the beat phenomenon is not
as striking as ordinary SU UMa-type dwarf novae.

\item We interpret that the superhumps in non-eclipsing systems
can be approximated by varying surface brightness of the superhump
light source on a relatively geometrically thin disk.
We interpret the behaviors in eclipsing systems as being a result
of vertical extent of the superhump light source.  The dependence
on the system inclinations suggests that the slope of the vertical
extension does not exceed 10$^{\circ}$.

\end{itemize}

\medskip

This work was supported by the Grant-in-Aid for the Global COE Program
``The Next Generation of Physics, Spun from Universality and Emergence"
from the Ministry of Education, Culture, Sports, Science and Technology
(MEXT) of Japan.
The authors are grateful to observers of VSNET Collaboration and
VSOLJ observers who supplied vital data.
We acknowledge with thanks the variable star
observations from the AAVSO International Database contributed by
observers worldwide and used in this research.
We have been benefited by discussions with Makoto Uemura, Akira Imada
and Daisaku Nogami.
This work is deeply indebted to outburst detections and announcement
by a number of variable star observers worldwide, including participants of
CVNET, BAA VSS alert and AVSON networks. 
We are grateful to the Catalina Real-time Transient Survey
team for making their real-time
detection of transient objects available to the public.
We thank the Kepler Mission team and the data calibration engineers for
making Kepler data available to the public.

\section*{Note Added in Proof}

   The following objects have been named in \citet{NameList80b}
during the proofreading period: ASAS J091858$-$2942.6 = DT Pyx,
ASAS J153616$-$0839.1 = QZ Lib, SDSS J080434.20$+$510349.2 = EZ Lyn,
SDSS J133941.11$+$484727.5 = V355 UMa, OT J074727.6$+$065050 = DY CMi.

\end{document}